\documentclass[a4paper,12pt,openright]{book}

\usepackage{amsmath,amsthm,amssymb,bm,braket}
\usepackage[dvips,dvipdfmx]{graphicx} 
\usepackage{cite,caption}
\usepackage{wrapfig}
\usepackage{setspace}
\usepackage{fancybox,fancyhdr}
\usepackage{geometry}
  \renewcommand{\chaptermark}[1]{\markboth{Chapter~\thechapter\; \; #1}{}}
\begin{document}

\renewcommand{\figurename}{Figure}
\renewcommand{\tablename}{Table}
\renewcommand{\thetable}{\arabic{table}}

\newcommand{\bi}[1]{\ensuremath{ \boldsymbol{#1} }}
\newcommand{\oprt}[1]{\ensuremath{ \hat{\mathcal{#1}} }}
\newcommand{\abs}[1]{\ensuremath{ \left| #1 \right| }}
\newcommand{\cgc}[2]{\ensuremath{ \mathcal{C}^{#1}_{#2} }}
\newcommand{\cmd}[1]{\texttt{\symbol{"5C}#1}}
\def \Schr{Schr\"odinger }
\def \CG{Clebsch-Gordan }

\def \beq{\begin{equation}}
\def \eeq{\end{equation}}
\def \beqa{\begin{eqnarray}}
\def \eeqa{\end{eqnarray}}

\def \twop{2{\it p}}
\def \bis{\bi{s}}
\def \bir{\bi{r}}
\def \ubir{\bar{\bi{r}}}
\def \bip{\bi{p}}
\def \ubip{\bar{\bi{r}}}

\renewcommand{\include}[1]{} 
\renewcommand{\documentclass}[2][]{} 

\begingroup \thispagestyle{empty}
\begin{center}
{\Huge Doctoral Thesis} \\
\vspace{20truemm}
\begin{spacing}{3.0} {\Huge \bf Diproton Correlation and Two-Proton Emission from Proton-Rich Nuclei} \end{spacing} 
\vspace{60truemm}
{\Large Tomohiro Oishi} \\
\vspace{20truemm}
{\Large A Dissertation Submitted for the Degree of \\ Doctor of 
Science at Tohoku University} \\
\vspace{20truemm}
{\Large March, 2014}
\end{center}
\endgroup

\newpage
{\bf ABSTRACT}

This is the open-print version of my dissertation thesis in 2014 \cite{2014Oishi}.
In this open-print version, several Figures were inevitably eliminated, because of the copyright and/or file-size problems.
Those figures are indicated with the message ``(Figure is hidden in open-print version.)'',
whereas original files could be available from T. OISHI when requested.
Several minor corrections have been also done, but without changing the scientific discussions, results, and conclusions from the original version \cite{2014Oishi}.

In this thesis, I investigate the two-proton emission in proton-rich nuclei.
The aim of study is to discuss the relation between the observables of this emission and the diproton correlation with quantum entanglement, which is one characteristic phenomenon due to the nuclear pairing interaction.

The pairing correlation between two nucleons plays 
an essential role in nuclear structure. 
Especially, two nucleons near the Fermi surface in nuclei have been 
predicted to make the ``diproton'' and ``dineutron'' correlations. 
These correlations mean 
the spatial localization of two nucleons, associated with 
the dominance of the spin-singlet configuration with entanglement. 
Especially, the dineutron correlation has been actively studied 
recently with a strong connection to the exotic features of neutron-rich nuclei. 
These nuclei far from the beta-stability line have attracted 
novel interests in nuclear physics, and their basic information 
have been extensively investigated for the past few decades. 
On the other hand, the diproton correlation in proton-rich nuclei has been less studied. 

In the former half of this thesis, 
I discuss the universality of the diproton and dineutron correlations, 
with a theoretical study of the $^{17,18}$Ne and $^{18}$O nuclei. 
This study is based on a three-body model, which provides a 
semi-microscopic description for the two nucleons inside nuclei. 
I demonstrate that a diproton correlation exists in the ground state of $^{17,18}$Ne, similarly to a dineutron
correlation in $^{18}$O. 
It is also shown that the Coulomb repulsive force between 
the two protons does not affect significantly this correlation. 
Consequently, the nuclear pairing interaction plays an important 
role to occur the dinucleon correlations inside nuclei. 

The later half of this thesis focuses on the relation 
between the diproton correlation and the two-proton emission. 
Even with theoretical predictions, for the dinucleon correlations, 
there have been no direct experimental evidences, 
because of a difficulty to access the intrinsic structures of nuclei. 
Recently, on the other hand, 
``two-proton emission'' has been focused as a direct 
probe into the diproton correlation with entanglement.
Namely, via this process, a pair of protons is emitted directly from the parent nuclei.
The emitted two protons are expected to provide information about 
the nuclear pairing interaction and the diproton correlation. 
In order to extract such information, one has to consider both the quantum meta-stability and the many-body 
properties in an unified framework. 
For this purpose, I develop a time-dependent three-body model. 
By applying this model to the $^6$Be nucleus, which is the 
lightest two-proton emitter, I investigate several novel 
properties about the role of pairing correlations in the meta-stable state. 
It has been shown that, by considering the Cooper pairing in the initial state of the two protons,
(i) the experimental decay width of $^6$Be is well reproduced, and 
(ii) the two protons are emitted mainly 
as a diproton-like cluster with the spin-singlet configuration, in the early stage of the emission. 
These results strongly suggest the diproton correlation during the emission process.
Namely, the two-proton emission can be an efficient tool to access the diproton correlation with a strong entanglement.

Despite these important theoretical findings, in order to extract the information on the diproton correlation 
from the experimental data, 
there still remain several points to be improved. 
The most important one is to take the final-state interactions into account sufficiently. 
For the two-proton emission from the $^6$Be nucleus, 
my result for the early stage of the emission predicts 
a dominant diproton-like configuration. 
On the other hand, there is only little signal of 
such a configuration in the experimental data of 
the energy-angular correlation pattern, 
which correspond to the late stage of the emission. 
This discrepancy is due to a strong disruption of the diproton 
correlation in the late stage, caused by the final-state 
interactions among all the particles. 
In order to establish the agreement between 
the calculated result and the experimental data, 
I need to expand the model-space so that a longer time-evolution can be carried out.

As a summary of this thesis, the first step 
to investigate the diproton correlation by means of the two-proton 
emission has been established. 
The time-dependent method based on the three-body model 
has provided an intuitive way to 
describe the quantum dynamics of the entangled fermionic pair, with an advantage to understand the role of pairing correlations. 
I expect that the improved time-dependent method will be 
a powerful tool for two-proton emissions, and also for other 
quantum meta-stable processes. 
After further investigations, 
the obtained knowledge of the pairing correlations and the 
meta-stable processes will be an important result 
not only for nuclear physics 
but also for atomic, molecular, condensed matter and 
quantum informative physics.

\tableofcontents

\documentclass[a4paper,12pt]{report}
\include{begin}

\chapter{Introduction} \label{Ch_Intro}
Pairing correlations are characteristic phenomena in many-fermion systems. 
Especially, the nuclear pairing correlation has been a major subject of modern 
nuclear physics \cite{58Bohr,69Bohr,96Doba,05Brink,03Dean_rev,03Bender_rev}. 
After the establishment of the traditional 
mean-field theory for atomic nuclei, 
there have been enormous theoretical, experimental and 
computational developments in this field. 
These developments have led to a deeper insight 
beyond the pure mean-field picture. 

Within the traditional mean-field theory, 
any nucleon inside a nucleus is assumed to be an 
independent particle moving in the mean-field generated by the interactions 
among all the nucleons. 
The traditional mean-field theory was applied to atomic nuclei, 
{\it e.g.} by Mayer {\it et.al.} \cite{49May,49Hax}, 
leading to conclusions about the shell structures and the 
magic numbers, which excellently agree 
with empirical properties of atomic nuclei. 
A more sophisticated definition of the nuclear mean-field is 
given by applying the Hartree-Fock (HF) 
theory \cite{28Hart,30Fock,03Fetter,80Ring} 
\footnote{The HF theory 
itself is a general theory for many-fermion systems. 
As a matter of fact, it was first applied to the electrons in atoms. }. 
Within the HF theory, 
the mean-field for an arbitrary nucleon 
is defined self-consistently by considering 
effective nucleon-nucleon interactions from all the other 
nucleons \footnote{We should notice that this effective interaction 
differs from a nucleon-nucleon interaction in the vacuum, 
which has a strong repulsive core at short distances. 
Except for the Coulomb repulsion, 
the repulsive core is smeared by including the medium effect 
in the effective nucleon-nucleon interaction. }. 
However, this traditional mean-field theory takes into account the 
interaction only on average, and thus misses some parts of the interaction, 
which is called the ``residual interaction''. 

The ``pairing interaction'' is the most important part 
of the residual interaction. 
Taking the pairing interaction into account, 
the traditional mean-field picture is modified to that including 
a collection of two correlated nucleons. 
two nucleons, which are in unnegligible correlations. 
The pairing interaction brings about a significant attraction 
between two nucleons 
when those are coupled to be the spin-singlet state \cite{05Brink,80Ring}. 
Evidences for the pairing correlations can be found, 
for instance, in the fact that 
there is a universal odd-even staggering rule in 
the binding energies. 
That is, even-even nuclei are systematically more bound than the 
neighboring nuclei in the nuclear chart. 
It is also known that 
the even-even nuclei take the spin-parity of $0^+$ in the 
ground state, with no exceptions. 
Similar pairing correlations play important a role not only 
in nuclei, but also 
in several other systems including condensed matters and 
cold fermionic atoms. 

In recent years, the study of nuclear pairing interactions 
and correlations has gained a renewed interest, due to the 
progress of physics of ``unstable nuclei'' 
\footnote{Difference between ``pairing interaction'' and 
``correlation'' is important. 
The pairing interaction means a distinct source of the force 
between nucleons. 
On the other hand, even in the situation where the 
pairing interaction does not exist, 
two nucleons can be kinetically correlated to each other. 
This correlation is mediated by other particles in the system. 
Namely, the paring correlation originates both from the 
pairing interaction and the many-body dynamics. } 
\cite{03Dean_rev,03Bender_rev,03Doba}. 
Unstable nuclei, which have large 
neutron- or proton-excess and locate far from the 
$\beta$-stability line, have been a major topic in 
recent nuclear physics. 
For these nuclei, there are considerably novel features 
which can be connected to the pairing correlation \cite{03Doba}. 
Those include ``dinucleon correlation'', 
which we detail in the next section. 

\section{Dinucleon Correlation}
The diproton and the dineutron correlations are intrinsic 
structural properties of atomic nuclei, caused 
by the pairing interaction. 
As is well known, a diproton or a dineutron is not bound in the vacuum, 
where the only possible bound state of two nucleons is a deuteron. 
However, inside nuclei, 
the situation may be different from that in the vacuum. 
Because of the many-body effect on pairing correlations, 
a possibility of the existence of diproton and dineutron-like 
configurations has been discussed for more than 40 years, 
since the first proposal by Migdal in 1973 \cite{73Mig}. 
These phenomena are called ``dinucleon correlations''. 
The study of dinucleon correlations is expected to provide 
a novel and universal insight into other strongly-correlated 
many-fermion systems. 

Based on the microscopic theory for the pairing correlations, 
{\it e.g.} HF-Bardeen-Cooper-Schrieffer (HF-BCS) or 
HF-Bogoliubov (HFB) theory \cite{80Ring,03Dean_rev,03Bender_rev}, 
it has been known that the pairing correlation 
depends on the surrounding density, $\rho$ 
\cite{05Yama,05Mats,06Mats,07Marg,08Marg,09Yama}. 
For instance, the pairing gap in infinite nuclear matter takes the 
maximum at a density smaller than the normal density. 
This density-dependence is naively due to the many-fermion effects. 
However, the exact origin of this density-dependence has remained unclarified, 
though some candidates have been discussed. 
Those include 
(i) the momentum-dependence of bare nucleon-nucleon forces, 
(ii) the Pauli principle and 
(iii) effects of nuclear three-body forces. 
It should be emphasized that this density-dependence 
causes a variety of pairing correlations. 
In the deeper region of nuclei with normal nuclear matter-density, 
a pair of nucleons is found to be a Cooper pair in the 
HF-BCS theory \cite{05Brink}. 
This pair is in the regime of weak pairing correlations, 
and its spatial distribution is much expanded compared to the typical 
radii of nuclei. 
On the other hand, in the low density region, 
the situation can be altered due to the density-dependence of the 
pairing correlation. 
The effective pairing correlation can be enhanced in that region, 
resulting in the spatial 
localization of two neutrons and protons, 
and also the increase of the probability of the spin-singlet configuration. 
Namely, in this situation, a pair of nucleons plays as a dineutron- or 
diproton-like cluster. 
The existence of this strong correlation can be considered within a wide 
range of the surrounding density, 
$\rho /\rho_0 \sim 0.1 - 0.01$, where $\rho_0 \sim 0.15$ fm$^{-3}$ is the 
nuclear saturation density \cite{06Mats}. 
Finally, if the density becomes infinitesimally small, 
the pairing correlation vanishes and two neutrons or protons 
become unbound. 
This limit is identical to two nucleons in the vacuum. 

From a phenomenological point of view, 
the dinucleon correlation is a kind of phase-crossover 
in many-nucleon systems with dilute 
densities \footnote{Another famous example of similar 
phenomena is the alpha-clustering inside nuclei \cite{13Girod}. 
In this thesis, however, we do not discuss it. }. 
Such a dilute density-situation has been expected to 
occur especially in the valence orbits of weakly-bound neutron-rich nuclei. 
For the past about two decades, 
neutron-rich nuclei with large neutron-excess 
and shorter lifetimes for the beta-decay have been 
extensively studied. 
This is grealy thanks to the experimental achievements which have provided 
the access to these nuclei \cite{85Tani_01, 85Tani_02}. 
In the ground state of these nuclei, 
the valence neutrons should be in the outer orbit far from the core, 
where the surrounding density is not so large that the pairing correlation 
is expected to increase \cite{05Yama,05Mats}. 
This is especially the case for weakly bound nucleons 
\cite{87Hansen,91Bert,05Hagi,07Hagi_01}. 
Various theoretical and experimental studies have been performed to 
investigate the dineutron-correlation in such nuclei, 
in connection to its influence on the nuclear structures and reactions. 
These studies have shown that the correlation may invoke sizable 
effects on 
some phenomena, including the electro-magnetic 
excitations \cite{06Naka,05Hagi,07Bertulani_76}, 
the Coulomb break-up reactions 
\cite{88Koba,04Fuku,01Myo,06Horiuchi,10Kiku,13Kiku} 
and the pair-transfer reactions \cite{73Broglia,91Igarashi,01Oert}. 
On the other hand, for proton-rich nuclei, 
even though a similar diproton correlation can be considered \cite{10Oishi}, 
it has so far been less studied compared to the dineutron correlation. 
Whether the Coulomb repulsion disrupts the diproton-like configuration or not 
in proton-rich nuclei is still an remaining question, though its effect 
has been found to be weak compared to the nuclear 
attraction \cite{10Oishi,11Oishi,02Hila,09Lesi,11Yama}. 
For both dineutron and diproton correlations, 
further quantitative and qualitative investigations are still in 
progress today. 

The prediction of the dinucleon correlation is an important 
conclusion from the recent nuclear theory, 
and its detection will give us a strong constraint 
on the basic properties of our nuclear models. 
However, as we will discuss in Chapter \ref{Ch_2}, 
even with various efforts, 
there have been no direct experimental evidences for the 
dinucleon correlation, mainly because it is an intrinsic structure
which is hard to be detected. 

\section{Two-Proton Decay and Emission}
Given these difficulties mentioned in the previous section, 
``two-proton emission'' and ``two-proton radioactive decay'' 
have been expected to provide a novel way to access the diproton correlation. 
Those are the quantum tunneling phenomena 
that two protons are emitted from 
the proton-rich nuclei beyond the proton-dripline 
\cite{08Blank,12Pfu,09Gri_40}. 
In this process, the decay products can be strongly 
associated with the pairing correlation between two protons. 
The importance of pairing correlations are suggested from, for instance, 
that observed \twop-emitters have the even number of protons 
with no exceptions. 
Thus, emitted two protons are expected to carry information about 
the pairing correlations, probably including the diproton correlations 
in nuclei \cite{07Bertulani_34, 08Bertulani}. 
We focus on these phenomena in the next section. 

The oldest example of the two-proton (\twop-) emitter is the $^6$Be nucleus, 
where its ``alpha+p+p'' resonance has been experimentally 
observed for several decades 
\cite{66Whaling,77Gree,88Ajzen,84Boch,85Boch_377,89Boch,09Gri_677,12Ego}. 
Following $^6$Be, 
similar three-body resonances have been observed in the ground 
state of a few light proton-rich nuclei, such as 
$^{12}$O \cite{78KeKe, 95Kryger} and 
$^{16}$Ne \cite{78KeKe, 83Wood}. 
A typical Q-value and decay-width of these resonances are 
on the order of 100 keV. 
For these nuclei, 
the potential barrier between the core and a proton is mainly 
due to the centrifugal force, 
whereas the Coulomb force is relatively small. 
Because of the low potential barrier, 
the decay width is comparably broad compared to 
the Q-value of these nuclei. 

On the other hand, the \twop-radioactive decay is a novel 
decay-mode of medium-heavy and heavy nuclei outside the 
proton-dripline \footnote{In this thesis, as a criterion of 
``radioactivity'', we adopt a typical lifetime of $10^{-7}$ s \cite{13Olsen}. 
If the considering system or process has a shorter lifetime 
than this criterion, 
we refer to it simply as the \twop-emitter or emission. 
The corresponding decay width to this criterion is about $10^{-14}$ MeV.}. 
A typical lifetime for the \twop-decays of these nuclei is 1-10 ms, 
corresponding to a typical decay width of $10^{-18}$-$10^{-19}$ MeV. 
A typical Q-value is around 1 MeV, similarly to light \twop-emitters. 
The significantly narrow width, compared with light \twop-emitters, 
is due to the higher Coulomb barriers, 
which considerably reduce the tunneling probabilities of two protons. 
In this category, $^{45}$Fe is the most famous example for 
the \twop-radioactivity. 
At the beginning of 2000s, 
the first observation of \twop-radioactivity was made 
for the $^{45}$Fe nucleus \cite{02Pfu,02Gio}. 
After this first discovery, 
the \twop-radioactivity has been confirmed also for $^{54}$Zn and 
possibly for $^{48}$Ni. 

It is also predicted that the \twop-decays and emissions are not limited 
particularly in these nuclides but universally exist along 
the proton-dripline until $Z\leq 82$ \cite{09Gri_40, 13Olsen}. 
Suggested nuclides to have this decay-mode include 
$^{26}_{16}$S \cite{08Blank}, 
$^{30}_{18}$Ar \cite{08Blank, 09Gri_40}, 
$^{34}_{20}$Ca \cite{09Gri_40}, 
$^{38}_{22}$Ti \cite{09Gri_40}, 
$^{41,42}_{24}$Cr \cite{09Gri_40}, 
$^{58}_{32}$Ge \cite{09Gri_40, 13Olsen}, 
$^{62,63}_{34}$Se \cite{09Gri_40, 13Olsen}, 
$^{66}_{36}$Kr \cite{09Gri_40, 13Olsen}, 
$^{102,103}_{52}$Te \cite{13Olsen}, 
$^{109,110}_{56}$Ba \cite{13Olsen}, 
$^{155}_{78}$Pt \cite{13Olsen}, 
$^{159}_{80}$Hg \cite{13Olsen}, and so on. 
Recently, the similar processes but emitting two neutrons, 
namely ``two-neutron emissions or decays'' are 
reported for 
$^{13}_{3}$Li \cite{13Kohley_13Li}, 
$^{16}_{4}$Be \cite{12Spyr} and 
$^{26}_{8}$O \cite{12Lund}. 
Together with the \twop-emitters, 
studies of two-neutron emitters can lead to the universal 
understanding of the two-nucleon radioactivity 
on both proton and neutron-rich sides. 

On the theoretical side, the first prediction of \twop-radioactivity was 
done by Goldansky in 1960 \cite{60Gold, 61Gold}. 
He argued that a ``true \twop-decay'' is allowed only for nuclei 
where the emission of single proton is energetically forbidden. 
The pairing interaction plays an important role to realize this situation, 
by lowering in energy the ground state of even-even parent and daughter 
nuclei with respect of the even-odd intermediate nucleus. 
In this situation, two protons must penetrate the potential 
barriers simultaneously. 
At the earlier stage of study, 
two simple models for the true \twop-radioactivity 
were proposed, namely ``the diproton'' \cite{60Gold, 61Gold, 13Deli} 
and ``the direct decays'' \cite{05Rotu}. 
In these old models, two protons are assumed to decay without passing 
the intermediate core-nucleon resonance. 
The diproton and direct decays correspond to the limits with 
relatively a strong and weak pairing correlations. 
On the other hand, another simple decay-model was also considered 
in different situations. 
It is the ``sequential'', or sometimes called ``cascade \twop-decay'', 
which can exist in nuclei where the one-proton emission is 
energetically available \cite{05Rotu}. 
In this situation, the core-nucleon binary channel becomes dominant, 
whereas the pairing correlations may be not significant. 

However, with various theoretical and experimental developments, 
it has been shown that the actual \twop-decays and emissions are 
more complicated than these simple modes. 
For some \twop-emitters, including $^{45}$Fe and $^6$Be for instance, 
their decaying mechanism cannot be described neither with any of 
these models \cite{07Mie, 09Gri_677, 09Gri_80, 08Muk, 10Muk}. 
It means that the actual \twop-decays and emissions involve several dynamical 
processes in a complicated way. 
From recent studies, 
the structures of material nuclei and the production mechanism 
of the \twop-emitters 
are also shown to be responsible, as well as 
all the final-state interactions among particles \cite{89Boch, 12Ego, 12Gri}. 
The question whether emitted two protons have the diproton-like 
character or not still remains unsolved, 
that critically relates to the diproton correlation. 

As another interest in \twop-emissions, we here briefly mention 
the quantum entanglement \cite{03Bertulani, 08Bertulani}. 
Since \twop-emissions and decays involve a propagation of two fermions, 
analyzing their wave functions may provide another route to approach, 
{\it e.g.} the Bell's inequality \cite{04Bell} or 
the Einstein-Podolski-Rosen paradox \cite{35Einstein}. 
Observation of two protons in spin-entanglements would become 
an examination of the basic quantum mechanics, 
that is complementary to 
other studies performed in quantum optics and atomic physics. 

Obviously, gaining useful information from 
\twop-emissions depends on our ability to describe the 
multi-fermion property and the quantum meta-stability 
simultaneously \cite{61Bardeen,83Cald,10Deli_text}. 
For these quantum resonances and tunneling phenomena, 
there are mainly two theoretical frameworks; 
namely within 
the time-independent framework \cite{28Gamov_01,28Gamov_02,29Gurney,89Bohm} and 
the time-dependent framework \cite{89Bohm,89Kuku,47Kry}. 
The time-independent one is based on non-Hermite quantum mechanics. 
In this framework, 
one solves, {\it e.g.} a Gamow state \cite{28Gamov_01,28Gamov_02}, 
which is assumed to be a purely outgoing wave outside the potential barrier. 
Generally such state must have a complex eigen-energy, in order to 
satisfy the outgoing boundary condition. 
The imaginary part of the complex energy of the Gamow state is related to 
the decay width, 
while the real part corresponds to the resonance energy or the Q-value. 
An advantage of the time-independent approach is that the decay width 
can be calculated with a high accuracy even when it is extremely 
small \cite{09Gri_40, 97Aberg, 00Davis}. 
On the other hand, 
in the time-dependent framework which we will adopt in this thesis, 
resonances or tunnelings are treated as time-developments 
of quantum meta-stable states. 
An advantage of the time-dependent approach, 
compared with the time-independent one, 
is that it provides an intuitive way to understand the 
tunneling mechanism, even though it is difficult to be applied to 
the situation with an extremely small decay width, 
where it needs very long time-evolutions for the meta-stable 
state to decay out. 
Especially, for light \twop-emitters with relatively the broad widths, 
the time-dependent method is expected to provide a complementary 
studies to the time-independent method. 

\section{Aim of This Thesis}
The aim of this thesis is to investigate theoretically the relation between 
the observables in \twop-emissions and the diproton correlation. 
As we wrote, although there have been various predictions, 
direct experimental evidence of diproton and dineutron 
correlations has not been obtained. 
Recently, on the other hand, two-proton decays and emissions have 
attached much attention in order to provide the direct probe 
into the diproton correlation. 
Nevertheless, the relation between the observed data 
and the nuclear intrinsic structures, including the diproton correlation, 
has been little discussed \cite{08Bertulani,12Ego}. 
Thus, our present study is expected to provide a novel insight into 
these important problems. 

In this thesis, we will employ the three-body model 
consisting of the core (daughter) nucleus and two valence nucleons. 
This model can treat the pairing correlations between the valence nucleons 
based on the semi-microscopic picture. 
In order to take the meta-stability into account for the 
\twop-emissions, 
we will adopt the time-dependent framework. 
Though the time-dependent approach has so far been applied only to two-body 
decaying systems, such as $\alpha$-decays or one-proton decays, 
this framework can bring about an useful mean to explore the mechanism of 
many-particle tunnelings, covering the whole stages of the time-evolution. 
We would like to emphasize that this time-dependent model has 
an advantage to distinguish the effect of pairing correlations 
from other results. 
Especially, it is worthwhile to investigate the evolution of 
\twop-wave function inside and outside the potential barriers, 
which can reflect the effect of the diproton correlation on \twop-emissions. 

The thesis is organized as follows. 
In Chapter \ref{Ch_2}, the history of studies about the dinucleon correlation 
is reviewed, with some connections to unstable nuclei. 
We will mention other exotic features of unstable nuclei, closely relating 
to the dinucleon correlation. 
In Chapter \ref{Ch_3body}, in order to describe the dinucleon correlation, 
we formulate the theoretical three-body model. 
In Chapter \ref{Ch_Results1}, 
we will apply this model to $^{17,18}$Ne and $^{18}$O nuclei, 
and discuss the dinucleon correlations in these nuclei. 
Apart from the beta-decays, 
these nuclei are stable against the neutron-, proton-, and alpha-emissions 
and thus provide good testing grounds for the dinucleon correlations 
in bound many-nucleon systems. 
We also discuss the effect of Coulomb repulsions on the 
nuclear pairing correlations, 
and whether the diproton correlation exists similarly to the 
dineutron correlation. 

In Chapter \ref{Ch_5}-\ref{Ch_Results3}, 
we then discuss the diproton correlations in two-proton emissions. 
In Chapter \ref{Ch_5}, the historical overview of 
two-proton emissions and radioactive decays are summarized. 
Chapter \ref{Ch_TDM} is devoted to a formulation of the time-dependent 
method for the quantum meta-stable systems, 
including two-proton emitters. 
In Chapter \ref{Ch_Results2} and \ref{Ch_Results3}, 
the time-dependent three-body model is applied to analyze \twop-emissions of 
$^6$Be and $^{16}$Ne nuclei, for which the three-body treatment is 
expected to be valid. 
These light proton-rich nuclei have relatively large values of the 
\twop-decay width, which are expected to be well described 
within the time-dependent framework. 
We will discuss whether the diproton correlation can be identified in 
the two-proton emissions. 

Finally, the summary of this thesis is present in Chapter \ref{Ch_Summary}. 
Future works towards the further improvements are also proposed. 
\include{end}
\documentclass[a4paper,12pt]{report}
\include{begin}

\chapter{Review of Dinucleon Correlation} \label{Ch_2}
In this Chapter, we briefly summarize the history of studies on the 
dinucleon correlation, and also of some related topics. 
We do not include the two-nucleon emissions and radioactive decays here, 
which will be detailed in Chapter \ref{Ch_5}. 

\section{Dinucleon Correlation in Stable Nuclei} \label{Sec_2_1} 
The first proposal of the dinucleon correlation was made by A.B. Migdal 
for two neutrons inside nuclei \cite{73Mig}. 
He argued that, even a dineutron is not bound in the vacuum, 
there can be a bound state of two neutrons near the surface of 
atomic nuclei, due to the nuclear meanfield confining those. 
After his proposal, several theoretical 
studies have been performed regarding the dineutron correlations. 
The dineutron correlation can be characterized as 
the special localization of two neutrons with, 
a compact distance compared to the total radius of the whole nucleus, 
and a large component of the spin-singlet configuration. 
For the spin-singlet character, 
it has been known from, {\it e.g.} the characteristic 
odd-even staggering of binding energies, that two nucleons 
in the same orbit tend to couple into the spin-singlet state 
due to the pairing correlation. 

Various efforts have been devoted to investigating 
the spatial correlation between two nucleons 
associated with the pairing interaction. 
The paper by Catara {\it et.al.} is worthwhile to be 
mentioned \cite{84Catara}. 
In this paper, the authors discussed the two-neutron spatial correlation 
caused by the pairing interaction in the ground and excited $0^+$ states 
of $^{206}$Pb, based on shell model with a schematic pairing interaction. 
It was shown that the parity-mixing in the partial core-neutron system 
is indispensable to occur the spatial localization of the two neutrons in the 
ground state (see Figure \ref{fig:1984Catara}). 
This parity-mixing is due to the scattering effect due to the pairing 
interaction inside nuclei. 
It was also suggested that the pairing interaction is responsible 
not only for localization of two neutrons, 
but also for an increase of the spin-singlet 
configuration, which cannot be explained within the pure shell 
(mean-field) model. 
At the same time, the authors raised the alarm that contributions 
of the pairing interaction ($\sim 1$ MeV) to the relative distribution of two 
neutrons are not sufficiently 
large to overcome the dominant shell structure. 
They argued that a two-neutron cluster cannot have a 
$\delta$-function-like distribution, 
even if an enormously large model-space is employed. 
\begin{figure*}[h] \begin{center}
(Figure is hidden in open-print version.)
\caption{Figure 2 in Ref.\cite{84Catara}. 
In panels (a) and (b), authors show the density distributions of 
two neutrons without configuration mixing of different parities. 
In panel (c), on the other hand, they show the result 
with configuration mixing, where the localization of two neutrons 
can be seen. } \label{fig:1984Catara}
\end{center} \end{figure*}

Similar calculations but based on different theoretical models 
have also been performed, where their conclusions agree 
with each other \cite{89Lotti, 91Bert, 97Esb, 05Mats, 05Hagi, 11KEnyo}: 
the pairing interaction causes the spatial 
localization with the enhanced spin-singlet configuration, 
which is absent in the pure mean-field model. 
We also touch on the paper \cite{05Mats} by Matsuo and his collaborators. 
In this paper, based on the Hartree-Fock-Bogoliubov theory, 
the authors discussed the pairing and dineutron correlations in 
medium-heavy neutron-rich nuclei. 
It was shown that the mixing of, not only the core-nucleon parities, 
but also higher core-nucleon angular momenta, $l$, are indispensable 
to invoke the spatial localization of two neutrons. 

We also refer to the connection between the dineutron correlations 
and the pair-transfer reactions. 
It has been actively discussed that the dinucleon correlation may 
enhance the cross sections for the simultaneous two-nucleon 
transfer reactions. 
The simplest probe is given with $(t,p)$ and 
$(p,t)$ reactions. 
The pair-transfer strength of nuclides differing by two units 
have been studied extensively in the 
experiments using these reactions \cite{68Bjerr,69Bjerr,73Broglia,01Oert}. 
As a result, the significant increase of transfer cross sections 
for nuclei with even-number nucleons has been found. 
A detailed theoretical studies was also performed in 
\cite{91Igarashi} by Igarashi {\it et.al.} for Pb isotopes. 
They showed that the cross sections of $(p,t)$ reactions are 
increased due to the configuration mixing caused by the pairing interaction, 
that is consistent with the experimental data. 
Following these simple cases, a similar enhancement in 
collisions of two heavy-ions (HIs) has also been predicted 
and observed 
\cite{71Diet,71Kleber,11Shim,13Potel,13Shim_arx,91Speer,99Peter}. 
The enhanced pair-transfer cross sections can be naively understood 
as arising from the transferred dineutron-like cluster, which 
can be associated with collective features, {\it e.g.} the pair-vibrational 
or/and the pair-rotational excitations. 
However, the pair-transfer reaction itself is not only from 
the one-step transfer of spatially localized two nucleons, 
but also from the sequential two-step transfers. 

Thus, in discussing the dinucleon correlations, the second mechanism 
has to be handled with good care. 
Even with many experimental data, 
whether one can extract useful information on the dinucleon correlations 
depends on the theoretical ability to describe its collective effect on 
the pair-transfer reactions in heavy-ion collisions \cite{01Oert}. 
In theoretical calculations, one should treat a change 
of coordinates associated with transferred two nucleons to evaluate 
the reaction cross sections. 

It considerably complicates a theoretical formulation of two-neutron 
transfer reactions, 
if one treats it rigorously. 
At the same time, the results sensitively depend on the wave functions 
of two colliding nuclei, which should be computed by taking 
the pairing correlations into account. 
In order to get a sufficient accuracy, there still remain several 
problems for nuclear structure calculations, 
including the nuclear tensor forces, the core excitations and so on, 
in addition to a theoretical modeling of a complicated pair-transfer 
reaction. 
It is expected that theoretical improvements overcoming these 
difficulties will provide an evidence for the dinucleon correlations. 

\section{Unstable Nuclei}
The dineutron correlation has been attracted a renewed interest 
due to the establishment of the unstable nuclear physics. 
For neutron-rich unstable nuclei, 
the idea of the dinucleon correlations 
has been frequently discussed as one of the exotic features 
associated with the pairing correlation in weakly bound systems. 

The frontier of nuclei in the nuclear chart 
has been expanded enormously for the recent decades. 
This is mainly thanks to the experimental developments enabling one 
to access ``unstable'' nuclei. 
These nuclei have a large proton- or neutron-excess, 
locate far from the $\beta$-stability valley, and 
are significantly short-lived compared with traditional 
radioactive nuclei close to the beta-stability line. 
For any unstable nuclide, 
one should be careful of ``what makes it to be unstable''. 
Most unstable nuclei known today are, in fact, stable against the 
nucleon emission. 
The main source of this instability is thus the weak interactions, 
not the strong interactions. 
On the other hand, by increasing the proton or neutron-excess, 
one can find many nuclides which are unstable against the 
nucleon emission. 
These nuclides define the proton- and neutron- driplines. 
Nuclei near and beyond these driplines can be considered as 
novel and exotic regions in nuclear physics. 
For the past decades, 
studies of these exotic nuclei have brought about deeper insights 
into nuclear physics, 
even though those scarcely exist on earth. 

\subsection{Neutron-Rich Nuclei}
Historically, the earlier interests were focused on 
the neutron-rich side. 
Especially, since the seminal experiments with radioactive isotope (RI) 
beams performed in 1980's \cite{85Tani_01,87Hansen,88Tani}, 
several exotic features in neutron-rich unstable nuclei have been discovered. 
These exotic features mainly due to the weakly binding of valence neutron(s). 
We list them below. 

\begin{enumerate}
\item Dineutron correlation: As mentioned in Chapter \ref{Ch_Intro}, for 
neutron-rich nuclei, the strong pairwise correlation between two neutrons 
has been predicted. 
Its source is the density-dependence of the pairing correlation, and it 
may lead to the dineutron-like clustering inside nuclei. 
We introduce this topic more in detail later. 

\item Halo and skin structures: A Large extension of the density distribution 
has been found for several neutron-rich nuclei, 
which are referred to as ``halo'' or ``skin'' 
nuclei \cite{85Tani_01,88Tani,87Hansen}. 
Famous examples include $^{6}$He and $^{11}$Li. 
For these nuclei, significantly large reaction cross sections 
were observed. 
By analyzing these experimental data with the Glauber model \cite{63Glauber}, 
their neutron radii were shown to be significantly larger than other
isotopes (see Figure \ref{fig:19852006exp}(a)). 
The neutron density was shown to have a long tail from the core nucleus. 
The weakly bound neutron(s) in the valence $(s_{1/2})$- or 
$(p_{3/2,1/2})$-orbit can generate this tail, 
like the halo or the skin around the core. 
With neutron-removal reactions, 
the corresponding narrow momentum distributions have been observed 
in such nuclei \cite{88Koba,90Anne,95Shimoura}. 
Studying these structures can lead to the understanding of 
the loosely bound or the dilute density region of nuclear systems. 

\item Soft multi-pole excitations: A significant increase 
of the probability for the electro-magnetic excitations at the 
lower energies has been observed for several nuclei 
\cite{93Ieki,95Shimoura,99Aumann,04Fuku,06Naka}. 
Especially, as shown in Figure \ref{fig:19852006exp}(b), 
the $E1$-transition strength of $^{11}$Li has a remarkable 
increase at excitation energies around $1$ MeV only. 
This is in marked contrast against normal nuclei, which show 
the $E1$-response at $E=10-20$ MeV due to the 
giant dipole resonance \cite{05Paar,05Terasaki,06Sil}. 
Theoretically, It has been considered that the soft multi-pole excitations 
are due to the relative motion 
between the core and the loosely bound neutron(s) \cite{10Ikeda}. 
Especially, for nuclei with two or more loosely bound neutrons, 
it is expected that the excitation spectra reflect not only the 
core-neutron motion but also the relative motion of two 
neutrons \cite{95Shimoura,06Naka,91Esb,92Esb,95Esb,95Sagawa,97Bona,03Myo,05Mats,05Hagi,06Horiuchi,06Gri,07Hagi_SDE,07Bertulani_76,10Kiku,11Oishi,13Kiku}. 
Geometry of the ingredient particles inside nuclei 
may be also revealed by analyzing these excitations. 
Especially, the opening angle between the valence neutrons is an important 
quantity, which is intimately related to the dineutron 
correlation \cite{06Naka,05Hagi,07Bertulani_76}. 

\item Borromean character: For several nuclides, so called ``Borromean 
character'' has also been discussed \cite{91Bert,93Zhukov,95Esb,05Hagi}. 
A Borromean nucleus is defined as a three-body bound system 
in which any two-body subsystem does not bound alone. 
Famous two-neutron Borromean nuclei are $^6$He $\cong \alpha +n+n$ 
and $^{11}$Li $\cong$ $^{9}$Li $+n+n$, 
where $^5$He, $^{10}$Li and a dineutron have no bound states. 
The pairing interaction between the valence nucleons plays an 
essential role in stabilizing these nuclei \cite{91Bert}. 
A similar character exists in proton-rich nuclei, namely 
a two-proton Borromean nucleus, $^{17}$Ne 
\cite{95Zhukov,96Tim,04Garrido_01,04Garrido_02,05Gri,06Gri}. 
The Borromean character deeply associates with the halo 
structure and the soft multi-pole excitations. 
For $^{6}$He or $^{11}$Li, as mentioned above, 
there have been enormous experiments which 
suggest the extended density-distribution or the enhancement of 
low-lying excitations. 

\item Two-neutron emission: Recently, as we touched on Chapter \ref{Ch_Intro}, 
two-neutron emissions from the ground states have been observed 
in several neutron-rich nuclei \cite{12Spyr,13Kohley_13Li,13Kohley_26O}. 
Because there are no Coulomb barriers for neutrons, 
the main source of these resonances is the centrifugal barriers 
between the core (daughter) nucleus and valence neutrons. 
Similarly to \twop-emissions, two-neutron emissions are promising phenomena 
which can provide the useful means to investigate the 
dineutron correlations. 
In this thesis, however, we do not discuss the two-neutron emissions in detail. 
\end{enumerate}

Of course, these listed properties are entangled to each other. 
Our main interest in this thesis is the dineutron and, as mentioned later, 
the diproton correlation. 
However, except for nuclei with only one weakly bound nucleon, 
we can overlook all of the above properties from a common point 
of view: ``pairing correlation''. 
Therefore, a deep understanding of the dinucleon correlation is 
expected to reveal not only a novel aspect of the pairing correlations, 
but also an universal property covering all the subjects listed above. 
Furthermore, these research achievements may be exported to other 
multi-fermion systems. 
\begin{figure*}[tb] \begin{center}
(Figure is hidden in open-print version.)
\caption{Figure 3 in Ref.\cite{85Tani_01} in the left panel; Figure 3 in Ref.\cite{06Naka} in the right panel.
The left panel: The root-mean-square matter-radii determined from experimental 
data of reaction cross sections. 
Large radii of $^6$He, $^8$He and $^{11}$Li can be seen. 
The right panel: The $E1$-strength distribution 
observed with the Coulomb break-up of $^{11}$Li. 
An enhancement of the strength in lower energy region is present. } 
\label{fig:19852006exp}
\end{center} \end{figure*}

\subsection{Proton-Rich Nuclei}
We also summarize supplementary information unique to the proton-rich side. 
In fact, the exotic features listed in the previous subsection 
can be considered almost equally 
for the proton-rich unstable nuclei. 
For example, the $^{17}$Ne nucleus is a 
\twop-Borromean nucleus 
\cite{95Zhukov,96Tim,04Garrido_01,04Garrido_02,05Gri,06Gri}, and 
also is a famous candidate to have 
the \twop-halo \cite{94Ozawa,95Zhukov,96Tim,05Gri,06Gri} and 
the diproton correlation \cite{04Garrido_02,10Oishi,11Oishi}. 
Nevertheless, compared to the neutron-rich side, 
the proton-rich unstable nuclei have been less studied so far. 
The characteristic problem in proton-rich nuclei is, of course, 
the Coulomb repulsion between the valence protons. 
As a natural consequence of the Coulomb repulsion, 
proton-rich nuclei have less binding energies than those of 
their mirror neutron-rich nuclei. 
Furthermore, even if its mirror partner can be bound, 
a proton-rich nucleus may become unstable against 
proton(s)-emissions. 
Thus, if we restrict our interests in nuclei which are 
stable against nucleon emissions, 
proton-rich side may be, in a sense, ``barren land''. 
This is a symbolic property of the breaking of the mirror-symmetry. 
However, abandoning this restriction, 
breaking of the mirror-symmetry can be interpreted as an useful 
property which produces a variety of phenomena of atomic nuclei, 
some of which can be observed only on the 
proton-rich side \cite{95Doba,96Cole,96Doba}. 

Concerning the pairing properties, it has been 
frequently discussed whether the Coulomb repulsion 
strongly affects the nuclear pairing attraction or not. 
Recent studies suggest that the effect of the Coulomb repulsion on 
binding energies of nuclei is minor, 
and the effect is roughly estimated as an about $10\%$ reduction over 
the nuclear attractions. 
This conclusion can be deduced from several theoretical and experimental 
analysis \cite{02Hila,09Lesi,09Bert,11Yama}. 
Moreover, in our previous studies \cite{10Oishi,11Oishi}, 
it was also suggested that 
the diproton correlation can exist in proton-rich nuclei similarly to 
the dineutron correlation in neutron-rich nuclei, 
due to the minor role of the Coulomb repulsion. 
If the diproton correlation really exists, 
breaking of the mirror-symmetry can provide another route to probe it, 
namely ``two-proton (\twop-) radioactivity''. 
This idea is the basis of this thesis, and we will detail it 
in Chapter \ref{Ch_5}. 

\section{Dinucleon Correlation in Unstable Nuclei} \label{Sec_23}
Because of the recent theoretical and computational developments, 
it has become possible to perform much reliable calculations for 
nuclear pairing correlations. 
This development brought us a point of view to discuss the dinucleon 
correlations in connection to the density-dependence 
of pairing correlations. 
\begin{figure}[tbp] \begin{center}
(Figure is hidden in open-print version.)
\caption{Figure 2 in Ref.\cite{06Mats}. 
The pairing gap in the symmetric and the pure-neutron 
nuclear matters as functions of the density, $\rho$. 
Those are calculated based on the HF-BCS theory with 
some different models of the nuclear interaction. } \label{fig:2006Matsuo_gap}
\end{center} \end{figure}

For this purpose, it is useful to discuss nuclear matter at first. 
There have been considerable studies which reports the 
significant density-dependence of nuclear pairing correlations 
in the nuclear matter \cite{06Mats,07Marg,10XGuang,13Sun}. 
We especially refer to the Ref.\cite{06Mats}, where the author 
applied the HF-BCS approach to the symmetric and pure-neutron 
nuclear matters. 
According to their results, as shown in Figure \ref{fig:2006Matsuo_gap}, 
the pairing gap in both 
symmetric and pure-neutron matters significantly 
depends on the density, $\rho$. 
It takes the maximum value within the 
densities of $\rho/\rho_0 \simeq 0.1 - 0.01$, 
where $\rho_0 \sim 0.15$ fm$^{-3}$ is the nuclear saturation density. 
Furthermore, as shown in Figure \ref{fig:2006Matsuo}, 
it is found that the spatial distribution of the spin-singlet Cooper pair 
of nucleons within a wide range of $\rho/\rho_0 \simeq 0.5 - 10^{-4}$ 
is well localized with a typical distance of $r_{\rm N-N} \leq 5$ fm. 
They also found a compact root-mean-square (rms) radii, $\xi_{\rm rms} \leq 5$ 
fm of two nucleons, 
suggesting the dinucleon correlations in nuclear matters. 
On the other hand, in the saturated or the infinitesimal density-region, 
a Cooper pair loses the dinucleon correlations. 
This result is, of course, the coincidence to the weakening of the 
pairing correlations in the saturated and 
the infinitesimally dilute densities. 
We also note that this variety of the pairing correlations as a function 
of the density
can be connected to the BCS-BEC crossover 
in nuclear matters. 
In the paper \cite{06Mats}, it was suggested that 
the region of $\rho/\rho_0 \simeq 0.1 - 10^{-4}$ 
corresponds to the domain of the BCS-BEC crossover. 
The similar conclusions have been obtained from other studies, 
although there are some quantitative differences in the appropriate 
value of $\rho$ at which the dinucleon correlation and 
the BCS-BEC crossover appear \cite{07Marg,10XGuang,13Sun}. 
\begin{figure*}[tb] \begin{center}
(Figure is hidden in open-print version.)
\caption{Figure 4 in Ref.\cite{06Mats}. 
The wave function of a Cooper neutron-pair in the symmetric and 
pure-neutron matters, calculated 
with the HF-BSC theory. } \label{fig:2006Matsuo}
\end{center} \end{figure*}

The similar studies have been performed 
for finite nuclei, where 
some of those were already introduced in Sec. \ref{Sec_2_1}. 
Furthermore, for unstable nuclei with weakly-bound nucleons, 
the dinucleon correlations have been discussed in connection to 
other exotic features listed in the previous section. 
Especially, $^6$He, $^{11}$Li and $^{17}$Ne nuclei have attracted 
much attentions. 
Theoretical studies in 
Refs.\cite{91Bert,92Esb,93Zhukov,96Tim,96Vinh,97Esb,04Garrido_01,04Garrido_02,05Gri,05Hagi,06Gri,07Bertulani_76,07Hagi_01,07Hagi_SDE,07Hagi_03,08Hagi,10Oishi,10Oishi_err,11Hagi_01,11KEnyo}, 
were dedicated for these problems. 
A popular model, on which almost all of these theoretical studies were based, 
is the nuclear three-body model, where 
one can describe a pair of nucleons in the mean-field generated 
by the core nucleus. 
The density-dependence of pairing correlations is usually taken into account 
in a phenomenological way, such as modifying the pairing interaction 
from that in vacuum. 
According to these mean-field plus pairing model calculations, 
a strong localization of the valence nucleons inside the ground states of 
these nuclei has been predicted 
\cite{91Bert,93Zhukov,96Vinh,04Garrido_01,04Garrido_02,05Hagi,07Bertulani_76,07Hagi_01,07Hagi_03,10Oishi,10Oishi_err}. 
As an example, Figure \ref{fig:2007Hagino} taken from Ref.\cite{07Hagi_01} 
shows this localization. 
This localization often occurs together with an enhancement of 
the spin-singlet configuration, 
identically to the dinucleon correlations. 
We also note that nuclei with weakly bound nucleons are expected 
to be good testing grounds for 
the BCS-BEC crossover in finite nuclei \cite{07Hagi_01} and 
the anti-halo effect of pairing correlations 
\cite{00Benn,11Hagi_AHE,13Sun_arx}. 
These topics are, however, beyond the scope of this thesis. 
\begin{figure*}[tbp] \begin{center}
(Figure is hidden in open-print version.)
\caption{Figure 1 in Ref.\cite{07Hagi_01}. 
The density distribution of the valence two neutrons in $^{11}$Li, 
obtained with $^9$Li$+n+n$ model calculations. 
A localization with $r\cong 2$ fm and $R\cong 3$ fm can be seen. 
} \label{fig:2007Hagino}
\end{center} \end{figure*}

\section{Possible Means to Probe Dinucleon Correlation}
Although there have been various theoretical predictions, 
there have been so far no direct evidences 
for the dinucleon correlation. 
The most serious difficulty is that 
the diproton and dineutron correlations are intrinsic 
phenomena, and are hard to be probed by popular means of 
experiments. 
Especially, for the dinucleon correlations in the bound state, 
it is in principle impossible to probe those without 
disruptions by an external field. 
Thus, we have to change our view to 
``how well do we extract the information on the dinucleon correlations''. 

For the purpose towards this direction, 
a lot of possibilities have been discussed. 
The first one is analyzing 
the pair-transfer reaction in heavy-ion collisions. 
Its basic idea, history and the present difficulties 
have already been introduced in Sec.\ref{Sec_2_1}. 
We should also note that, for unstable nuclei, a theoretical 
analysis for the pair-transfer reactions may become 
even more complicated due to their exotic structures. 
If these problems can be resolved, 
the pair-transfer reaction will be one of the most powerful tools to 
investigate the dinucleon correlations in both stable and 
unstable nuclei. 

The second candidate is using excitations by electro-magnetic interactions. 
The soft multi-pole excitations and the Coulomb break-up reactions belong to 
this category. 
For instance, the momentum distributions observed in 
Coulomb break-up reactions have been discussed frequently 
associated with the dinucleon correlations 
\cite{95Shimoura,06Naka,07Hagi_SDE,07Bertulani_76,10Kiku,13Kiku}. 
However, these experiments are performed by perturbing 
the ground state properties \cite{01Myo,03Myo,10Kiku}. 
Consequently, the experimental results depend not only on the ground 
state, but also on the excited states. 
From recent theoretical studies, 
it is concerned that this inclusion of excited states suppress 
the sensitivity to the dinucleon correlation, bringing a 
serious drawback to the direct access to it \cite{10Kiku}. 
Furthermore, even if there will be a significant signal of 
the dinucleon correlations in the experimental data, 
one must distinguish whether it reflects the dinucleon correlations 
in the ground or in the excited states. 

Another possibility to probe the dinucleon correlation 
is to observe two-nucleon decays and emissions. 
These attempts have been performed intensively since 
the beginning of 2000s, mainly due to the remarkable 
developments in the experimental techniques \cite{08Blank,09Blank,12Pfu}. 
However, the connection between the two-nucleon emissions and 
the dinucleon correlations has not yet been clarified 
\cite{01GCampo,07Bert,08Bertulani,12Maru}. 
In Chapter \ref{Ch_5}, we will summarize 
the history and backgrounds of these topics. 

\section{Summary of this Chapter}
We have introduced the historical background of dinucleon correlations 
and its relevant topics in this Chapter. 
Although it is still a theoretical prediction, 
the dinucleon correlation is one of the important characters of 
multi-nucleon systems, and 
is essentially connected to the nuclear pairing correlations. 
If the dinucleon correlations will be directly detected, it will 
provide strong constraints on the nuclear interactions 
and on the framework for multi-nucleon problems. 
Furthermore, we may extract an universal knowledge in 
other multi-fermion systems from these observations. 

In Chapters \ref{Ch_3body} and \ref{Ch_Results1}, 
we discuss how the dinucleon correlations are theoretically predicted. 
For this purpose, we will employ the three-body model, similarly to Refs. 
\cite{91Bert,92Esb,93Zhukov,96Tim,96Vinh,97Esb,04Garrido_01,04Garrido_02,05Gri,05Hagi,06Gri,07Bertulani_76,07Hagi_01,07Hagi_SDE,07Hagi_03,10Oishi,10Oishi_err}. 
The next Chapter will be dedicated to the formalism of this model. 
Applying this model to several nuclei, we will discuss 
the dinucleon correlations in finite nuclei. 
Those will be summarized in Chapter \ref{Ch_Results1}. 
\include{end}
\documentclass[a4paper,12pt]{report}
\include{begin}

\chapter{Quantum Three-Body Model} \label{Ch_3body}
In this Chapter, we introduce in detail the three-body model of the 
core nucleus + nucleon + nucleon, which we use to describe the 
dinucleon correlations. 
This model is identical to the three-body version of the 
Core-Orbital Shell Model (COSM) \cite{88Suzuki_COSM}. 
With COSM, one starts from considering the core nucleus as a source 
of the mean-field. 
Then one adds one or more valence nucleon(s) around the core. 
In this thesis, we do not care about the core-excitations and thus 
the core plays simply as an inert particle. 
The pairing correlations between the valence nucleons 
can be explicitly included in this model. 
The deviations from the pure mean-field approximation can also 
be discussed, providing the semi-microscopic point of view of 
the pairing correlations. 

In our formalism, 
the coordinates and the spin variables of each nucleon are indicated as 
$\bir_i$ and $\bis_i \phantom{0}(i=1,2)$, respectively. 
We also use $\xi_i \equiv \{ \bir_i, \bis_i \}$ 
for a shortened notation. 
\footnote{Because we only treat systems with the core plus 
two nucleons of the same kind in this thesis, 
the isospin variables are not necessary. } 
The angular variable, that is equivalent to the radial unit vector, 
are indicated by $\ubir$. 
The orbital and the spin-coupled angular momenta are indicated by 
$\bi{l} = \bir \times \bip / \hbar $ and 
$\bi{j} = \bi{l} + \bis$, respectively. 

\section{Three-Body Hamiltonian}
We define the V-coordinates for three-body systems, 
similarly to other papers \cite{91Bert,97Esb,96Vinh}. 
The vector $\bir_i$ indicates the relative coordinates between the core 
nucleus and the $i$-th valence nucleon (see Fig.\ref{fig_3bs}). 
We subtract the center of mass motion of the whole system. 
Thus, apart from the spin variables, 
we need two vectors, $\bir_1$ and $\bir_2$, to fully describe the system. 
The total Hamiltonian reads 
\beqa
 H_{\rm 3b} 
 &=& h_1 + h_2 + \frac{\bip_1 \cdot \bip_2}{A_{\rm c} m} + 
     v_{\rm N-N}(\xi_1, \xi_2 ), \label{eq:H3b} \\
 h_i 
 &=& \frac{\bip_i^2}{2\mu} + V_{\rm c-N}(\xi_i), 
\eeqa
where $h_i$ is the single particle (s.p.) Hamiltonian 
for the relative motion between 
the core and the $i$-th nucleon. 
$\mu \equiv m A_{\rm c} / (A_{\rm c}+1)$ is the 
reduced mass, where $m$ and $A_{\rm c}$ indicate the one-nucleon mass and 
the number of nucleons in the core, respectively. 
The diagonal component of the kinetic energy of the core is 
included in the s.p. Hamiltonians, $h_1+h_2$. 
On the other hand, the off-diagonal component, 
referred to as ``recoil term'' in the following, 
is taken into account as the third term in Eq.(\ref{eq:H3b}) \cite{97Esb,05Hagi}. 
See Appendix \ref{Ap_3body} for a derivation of this Hamiltonian. 

In the Hamiltonian, 
$V_{\rm c-N}$ is the interaction for the core-nucleon subsystem. 
On the other hand, $v_{\rm N-N}$ indicates the pairing interaction for 
the two valence nucleons \footnote{In this thesis, 
we use the subscript N to indicate 
``nucleon'' generally, whereas $p$ and $n$ mean 
``proton'' and ``neutron'', respectively.} \footnote{In this thesis, 
we do not treat a phenomenological three-body force.}. 
It should be mentioned that, 
even if the pairing interaction is zero, 
the pairing correlation does not vanish because of the recoil term, 
$\bip_1 \cdot \bip_2 /A_{\rm c} m$ 
\footnote{Phenomenologically, this correlation can be interpreted as that 
mediated by the core nucleus. }. 
We give explicit forms of these interactions in the next section. 
\begin{figure*}[h] \begin{center}
\fbox{\includegraphics[width = 0.8\hsize]{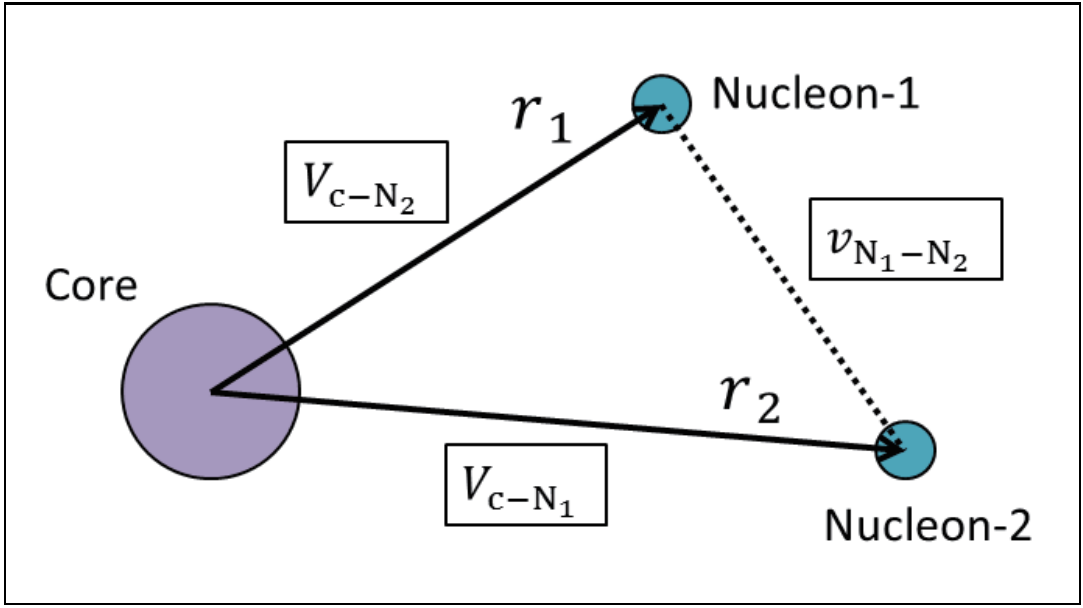}}
\caption{A schematic figure for the three-body model defined 
in the V-coordinates. } \label{fig_3bs}
\end{center} \end{figure*}

\section{Interactions}
In this thesis, we assume that the core-nucleon potential, 
$V_{\rm c-N}$, is spherical and does not depend on the spin variables. 
Apart from the Coulomb interaction for a valence proton, 
we employ the Woods-Saxon potential including the spin-orbit coupling term. 
\beqa
 V_{\rm c-N, Nucl.} (r) 
 &=& \left[ V_0 + V_{ls} r_0^2 (\bi{\ell} \cdot \bi{s}) 
     \frac{1}{r} \frac{d}{dr} \right] f(r), \label{eq:WSP} \\
 &=& \left[ V_0 + V_{ls} r_0^2 
     \left( \frac{j(j+1)-l(l+1)-3/4}{2} \right) 
     \frac{1}{r} \frac{d}{dr} \right] f(r)
\eeqa
with 
\beq
  f(r) = \frac{1}{ 1 + \exp \left( \frac{r-R_{\rm core}}{a_{\rm core}} \right) }, 
\eeq
where $R_{\rm core} = r_0 A_{\rm c}^{1/3}$ is the radius of the core nucleus. 
In the core-proton case, in addition, 
the Coulomb potential of a uniform-charged sphere, whose 
radii and charge are $R_{\rm core}$ and $Z_{\rm c} e$, respectively, is 
also employed. 
\beq
 V_{\rm c-p, Coul.} (r) 
 = \left\{ \begin{array}{cc} 
   \frac{Z_{\rm c} e^2}{4\pi \epsilon_0} \frac{1}{r} & (r > R_{\rm core}) \\
   \frac{Z_{\rm c} e^2}{4\pi \epsilon_0} \frac{1}{2R_{\rm core}} \left( 3 - \frac{r^2}{R_{\rm core}^2} \right) & (r \leq R_{\rm core}) 
   \end{array} \right. \label{eq:cpCoul}
\eeq
Thus, the total core-proton potential is given as 
\beq
 V_{{\rm c-p},lj} = V_{\rm c-p, Nucl.} (r) + V_{\rm c-p, Coul.} (r). 
\eeq
There are four parameters in the core-nucleon potential, 
namely $V_0, V_{ls}, r_0$ and $a_{\rm core}$. 
We determine the values of these parameters for each system, 
as we will explain in 
Chapter \ref{Ch_Results1} and \ref{Ch_Results2}. 

On the other hand, for the nucleon-nucleon pairing interaction, 
we adopt the phenomenological ``density-dependent contact (DDC)'' 
interaction. 
It is formulated as 
\beq
 v_{\rm N-N, Nucl.} (\bir_1,\bir_2) 
 = \delta(\bir_1-\bir_2) 
   \left[ v_0 + 
          \frac{v_{\rho}} 
               {1 + \exp \left( \frac{\abs{(\bir_1+\bir_2)/2}-R_{\rho}}{a_{\rho}} \right)} 
   \right]. \label{eq:DDCP}
\eeq
The first term, $v_0 \delta(\bir_1-\bir_2)$, indicates the nucleon-nucleon 
interaction in vacuum, which is 
approximated to have the zero range. 
The second term is a phenomenological density-dependent part 
which is assumed as the Woods-Saxon form. 
This type of pairing interaction has been employed 
in several nuclear structural 
calculations, with a great advantage that it can dramatically 
reduce the computational cost. 
These calculations have provided reasonable 
results \cite{84Catara,91Bert,97Esb,05Hagi,07Hagi_01}, 
despite the simple form of the pairing interaction. 
Especially, within the three-body model with 
DDC pairing, 
the binding energies and the Borromean properties explained in 
the previous Chapter have been well reproduced 
for $^6$He and $^{11}$Li \cite{91Bert,97Esb,05Hagi}. 
In the case with two protons, 
we also take the Coulomb potential into account. 
\beq
  v_{\rm p-p, Coul.} (\bir_1,\bir_2) \label{eq:ppCoul}
  = \frac{e^2}{4\pi \epsilon_0}  \frac{1}{\abs{\bir_1-\bir_2}}. 
\eeq
For the nuclear part of the pairing interaction, 
there are four parameters in Eq.(\ref{eq:DDCP}), namely 
$v_0, v_{\rho}, R_{\rho}, a_{\rho}$. 
The strength of the bare nucleon-nucleon potential, $v_0$, can be 
defined by solving the nucleon-nucleon scattering problem with 
the bare contact interaction, $v_0 \delta(\bir_1-\bir_2)$. 
As well known, this contact interaction must be treated in a 
truncated space defined with the energy cutoff, 
$\epsilon_{\rm cut}$, or 
it loses physical meanings. 
The strength of the bare interaction, $v_0$, can be 
determined so as to reproduce the empirical scattering length 
$a_{\rm N-N}$ in the nucleon-nucleon scattering \cite{97Esb}. 
For a given cutoff $\epsilon_{\rm cut}$, 
this is formulated as 
\beq
 v_0 = \frac{2\pi^2\hbar^2}{m}\frac{2a_{\rm N-N}}{\pi-2a_{\rm N-N} k_{\rm cut}}, \label{eq:v0}
\eeq
where the relative maximum momentum of two nucleons, $k_{\rm cut}$, is defined as 
\beq
 k_{\rm cut} = \sqrt{{m \epsilon_{\rm cut}}/{\hbar^2}}. 
\eeq
A discussion and derivation of Eq.(\ref{eq:v0}) are summarized as 
Appendix \ref{Ap_Scat_Contact}. 
The empirical scattering length for a neutron-neutron scattering is 
$a_{\rm n-n}=-18.5$ fm \cite{05Baumer}, 
while that for a proton-proton scattering is $a_{\rm p-p}=-7.81$ 
fm \cite{88Berg}. 
The difference between $a_{\rm n-n}$ and $a_{\rm p-p}$ is mainly due to 
the Coulomb repulsion in a two-proton system. 
Since we explicitly include the Coulomb interaction in our calculations, 
we use the neutron-neutron scattering length $a_{\rm n-n}$ to determine the 
strength of the bare interaction, Eq.(\ref{eq:v0}), assuming the 
charge independence of nuclear force. 

Once $v_0$ is determined in this way, the remaining parameters in the 
density-dependent term, $v_\rho, R_\rho$, and $a_\rho$ are adjusted 
to reproduce the three-body binding energy of the considering system. 
The concrete values of these parameters 
used in the actual calculations 
are given in Chapter \ref{Ch_Results1} and \ref{Ch_Results2}.

\section{Single-Particle States}
In order to describe an arbitrary wave function, 
the basis expansion is a popular method. 
We use this method in our three-body problems. 
As the first step, 
we solve the partial core-nucleon wave functions. 
Because we assumed that the core-nucleon potential, 
$V_{\rm c-N}$, is spherical and does not depend on the spin variables, 
the corresponding \Schr equation reads 
\beq
 h_i \phi_{nljm} (\xi_i) = \epsilon_{nlj} \phi_{nljm} (\xi_i). 
 \label{eq:speq} 
\eeq
Here we indicate the radial quantum numbers, 
quantum numbers of 
orbital angular momenta and of 
coupled angular momenta as 
$n,l$ and $j$, respectively. 
We also need $m$ to indicate the magnetic quantum number. 
The wave function of the single nucleon can be separated 
into the radial and the angular parts as 
\beq
 \phi_{nljm} (\xi) = R_{nlj}(r) \mathcal{Y}_{ljm} (\ubir, \bis) 
 = \frac{U_{nlj} (r)}{r} \mathcal{Y}_{ljm} (\ubir, \bis). 
\eeq
where $\ubir \equiv (\theta, \phi)$. 
The function $\mathcal{Y}_{ljm}$ indicates the composite angular part 
of $\bi{j} = \bi{l} + \bis$ coupled to $(j,m)$, that is 
\beq
 \hat{\bi{j}}^2 \mathcal{Y}_{ljm} = j(j+1) \mathcal{Y}_{ljm}, ~~~~~~ 
      \hat{j_z} \mathcal{Y}_{ljm} = m \mathcal{Y}_{ljm}. 
\eeq
Using the \CG coefficients, its explicit form is given as 
\beqa
 \mathcal{Y}_{ljm} (\ubir, \bis) 
 &\equiv & \Braket{ \ubir, \bis \mid (l \oplus 1/2)\phantom{0} j,m} \\
 &=&       \sum_h \sum_v \cgc{j,m}{l,h;1/2,v} Y_{lh} (\ubir) \chi_v (\bis), 
\eeqa
where $Y_{lh}$ and $\chi_{v}$ satisfy 
\beqa
 \hat{\bi{l}}^2 Y_{lh} (\ubir) = l(l+1) Y_{lh}(\ubir), 
 && ~~ \hat{l}_z Y_{lh} (\ubir) = h Y_{lh} (\ubir), \\
 \hat{\bis}^2 \chi_v (\bis) = \frac{3}{4} \chi_v (\bis), 
 && ~~ \hat{s}_z \chi_v (\bis) = v \chi_v (\bis), 
\eeqa
with $h = -l\sim l$ and $v = \pm 1/2$. 
Then Eq.(\ref{eq:speq}) can be reduced to the equation only for the 
radial part $R_{nlj}$. 
That is 
\beq
 \left[ -\frac{\hbar^2}{2\mu} 
        \left( \frac{1}{r} \frac{d^2}{dr^2} r - \frac{l(l+1)}{r^2} \right)
 + V_{{\rm c-N},lj}(r) \right] 
 R_{nlj}(r) = \epsilon_{nlj} R_{nlj}(r), 
\eeq
or equivalently for $U_{nlj}(r)=rR_{nlj}$, 
\beq
 \left[ \frac{d^2}{dr^2} - \frac{l(l+1)}{r^2} 
      - \frac{2\mu}{\hbar^2} 
        \left(  V_{{\rm c-N},lj}(r) - \epsilon_{nlj} \right) 
 \right] U_{nlj}(r) = 0. 
\eeq
In this thesis, 
we solve the radial part $U_{nlj}$ numerically within a discrete 
variable-domain. 
Assuming a radial box with its size $R_{\rm box}$, 
sampling points are distributed in the interval $0 \sim R_{\rm box}$ 
where the distance between two consecutive points is $dr = r_{n-1}-r_n$. 
For the continuum s.p. states with 
$\epsilon_{nlj} > V_{{\rm c-N},lj}(r \to \infty) \equiv 0$, 
we assume the boundary condition with a vanishing wave function 
at $r=R_{\rm box}$. 
That is 
\beq
 U_{nlj} (r=R_{\rm box}) = 0. 
\eeq
Because of this boundary condition, 
the continuum energy spectrum is discretized. 
Either for the bound and the discretized continuum s.p. states, 
their radial wave functions can be calculated numerically. 
The numerical method we employ in this thesis is ``Numerov method'', 
which was developed by B. V. Numerov \cite{93Hairer}. 
A detailed introduction of this method is separately 
summarized as Appendix \ref{Ap_Numerov}. 

\section{Uncorrelated Basis for Two Nucleons}
Using s.p. wave functions $\{ \phi_{nljm} \}$ obtained 
in the previous section, 
the ``uncorrelated basis'' for two-nucleon states can be constructed. 
If two nucleons are coupled to the spin $(J,M)$, 
the uncorrelated states are formulated as 
\beqa
 \Psi^{(J,M)}_{ab} (\xi_1, \xi_2) &=& \Psi^{(J,M)}_{(nlj)_a (nlj)_b} (\xi_1, \xi_2) 
 \equiv \left[ \phi_{(nlj)_a} (\xi_1) \otimes 
                  \phi_{(nlj)_b} (\xi_2) \right]^{(J,M)} \\
 &=& R_{(nlj)_a}(r_1) R_{(nlj)_b}(r_2) \cdot \label{eq:uncbs_03} 
     W^{(J,M)}_{ab} (\ubir_1 \bis_1, \ubir_2 \bis_2), 
\eeqa
where we define the shortened subscripts $(nlj)_a \equiv (n_a,l_a,j_a)$. 
The coupled angular part, $W^{(J,M)}_{ab}$, is defined as 
\beqa
 && W^{(J,M)}_{ab} (\ubir_1 \bis_1, \ubir_2 \bis_2) \label{eq:couplang} 
    \equiv \Braket{ \ubir_1 \bis_1, \ubir_2 \bis_2 \mid (j_a \oplus j_b) ~ J,M } \\
 && ~~~~ = \sum_{m_a,m_b} \cgc{J,M}{j_a,m_a;j_b,m_b} 
           \mathcal{Y}_{(ljm)_a} (\ubir_1,\bis_1) \mathcal{Y}_{(ljm)_b} (\ubir_2,\bis_2). 
\eeqa
This function means that the first and the second valence nucleons 
are in the core-nucleon orbits labeled by 
$(n_a,l_a,j_a)$ and $(n_b,l_b,j_b)$, respectively. 
In actual calculations, we also add another constraint of the 
total parity, by including only those configurations with the 
same value of 
$\pi = (-)^{l_a+l_b}$ in defining basis. 
For two nucleons of the same kind, 
we have to take the anti-symmetrization into account. 
That is 
\beq
 \tilde{\Psi}^{(J,M)}_{ab} (\xi_1, \xi_2) 
 \equiv A_{ab} \label{eq:uncorrebasis}
        \left[ \Psi^{(J,M)}_{ab} (\xi_1,\xi_2) 
             - \Psi^{(J,M)}_{ab} (\xi_2,\xi_1) \right], 
\eeq
where $A_{ab}$ is the normalization factor. 
This is given as 
\beq
 A_{ab} = \label{eq:Aab} \left\{ \begin{array}{cc} 
                1/2 & (n_a=n_b \cap l_a=l_b \cap j_a=j_b) \\
         1/\sqrt{2} & ({\rm otherwise}) \end{array} \right. 
\eeq
If we write it explicitly, Eq.(\ref{eq:uncorrebasis}) is given as 
\beqa
 \tilde{\Psi}^{(J,M)}_{ab} (\xi_1, \xi_2) 
 &=& A_{ab} \left[ R_{(nlj)_a}(r_1) R_{(nlj)_b}(r_2) \cdot W^{(J,M)}_{ab} (\ubir_1 \bis_1, \ubir_2 \bis_2) \right. \nonumber \\
 & & \; \; -\left. R_{(nlj)_a}(r_2) R_{(nlj)_b}(r_1) \cdot W^{(J,M)}_{ab} (\ubir_2 \bis_2, \ubir_1 \bis_1) \right] . \label{eq:53548}
\eeqa
Using the formula of the \CG coefficients; 
\beq
 \cgc{J,M}{j_a,m_a;j_b,m_b} = (-)^{j_a+j_b-J} \cgc{J,M}{j_b,m_b;j_a,m_a}, 
\eeq
the coupled angular part of the second term in Eq.(\ref{eq:53548}) 
can be transformed to 
\beqa
 && W^{(J,M)}_{ab} (\ubir_2 \bis_2, \ubir_1 \bis_1) \nonumber \\
 && ~~~~ = \sum_{m_a,m_b} \cgc{J,M}{j_a,m_a;j_b,m_b} 
    \mathcal{Y}_{(ljm)_a} (\ubir_2,\bis_2) \mathcal{Y}_{(ljm)_b} (\ubir_1,\bis_1) \nonumber \\
 && ~~~~ = (-)^{j_a+j_b-J} \sum_{m_a,m_b} \cgc{J,M}{j_b,m_b;j_a,m_a} 
    \mathcal{Y}_{(ljm)_a} (\ubir_2,\bis_2) \mathcal{Y}_{(ljm)_b} (\ubir_1,\bis_1) \nonumber \\
 && ~~~~ = (-)^{j_a+j_b-J} W^{(J,M)}_{ba} (\ubir_1 \bis_1, \ubir_2 \bis_2). 
\eeqa
Thus we obtain another formula for $\tilde{\Psi}^{(J,M)}_{ab}$. 
\beq
 \tilde{\Psi }^{(J,M)}_{ab} (\xi_1, \xi_2) \label{eq:UNCB}
 = A_{ab} \left[ \Psi^{(J,M)}_{nlj(a),nlj(b)} (\xi_1,\xi_2) 
        - B_{ab} \Psi^{(J,M)}_{nlj(b),nlj(a)} (\xi_1,\xi_2) \right] 
\eeq
with $B_{ab} \equiv (-)^{j_a+j_b-J}$. 
Notice that $\tilde{\Psi}^{(J,M)}_{ab} $ is an eigenstate of the uncorrelated 
Hamiltonian, $h_1 + h_2$. 
Its eigen-equation reads 
\beq
 (h_1 + h_2) \tilde{\Psi}^{(J,M)}_{ab} = (\epsilon_a + \epsilon_b) \tilde{\Psi}^{(J,M)}_{ab}, 
\eeq
where $\epsilon_a$ and $\epsilon_b$ are the eigen-energies of the first and the second 
orbits, respectively. 

We can now expand an arbitrary two-nucleon state 
with $(J,M)$ on the uncorrelated basis. 
That is, 
\beq
 \Phi^{(J,M)} (\xi_1,\xi_2) = \label{eq:exp_ucb} 
 \sum_{a\leq b} \alpha_{ab} \tilde{\Psi}^{(J,M)}_{ab} (\xi_1,\xi_2). 
\eeq
where our model-space is truncated by the cutoff energy for the 
uncorrelated basis, defined as 
$E_{\rm cut}=\epsilon_{\rm cut}(A_{\rm c}+1)/A_{\rm c}$ \cite{97Esb}. 
In practice, we have to introduce also 
the cutoff angular momentum, $l_{\rm cut}$, 
in addition to $E_{\rm cut}$. 
Notice that Matsuo {\it et.al.} have shown 
that the spatial localization cannot 
be reproduced theoretically unless one includes a sufficient number 
of angular momenta. 
Referring to their results, we would have to employ the model-space 
with, at least, up to $l_{\rm max}=5$ in order to take the dinucleon 
correlations into account. 

In the following, for simplicity, 
we omit the subscripts $(J,M)$ unless it is needed. 
For the eigenstates of $H_{\rm 3b}$, namely 
$H_{\rm 3b} \ket{\Phi_N } = E_N \ket{\Phi_N }$, 
the expansion coefficients $\{ \alpha_{ab} \}$ can be obtained 
by diagonalizing the Hamiltonian matrix. 
In the next section, we detail how to calculate these matrix elements.

\section{Matrix Elements with Uncorrelated Basis}
First, for the uncorrelated Hamiltonian, $h_1 + h_2$, 
the matrix elements are trivially given as 
\beq
 \Braket{ \tilde{\Psi}_{cd} \mid (h_1+h_2) \mid \tilde{\Psi}_{ab} } 
 = (\epsilon_a + \epsilon_b) \delta_{cd,ab} , 
\eeq
where $\epsilon_a \equiv \epsilon_{n_a l_a j_a}$. 
For the other parts of the Hamiltonian, 
we need much complicated calculations in general. 
A matrix element (ME) of an arbitrary operator, $\oprt{O}$, 
is decomposed into four terms, 
\beqa
 \Braket{\tilde{\Psi}_{cd} \mid \oprt{O} \mid \tilde{\Psi}_{ab}} 
 &=& A_{cd}A_{ab} \left[ \Braket{ \Psi_{cd} \mid \oprt{O} \mid \Psi_{ab}} 
     + B_{cd}B_{ab} \Braket{ \Psi_{dc} \mid \oprt{O} \mid \Psi_{ba}} \right. \nonumber \\
 & & \left. -B_{ab} \Braket{ \Psi_{cd} \mid \oprt{O} \mid \Psi_{ba}} 
            -B_{cd} \Braket{ \Psi_{dc} \mid \oprt{O} \mid \Psi_{ab}} \right] \label{eq:ME01}
\eeqa
where we have applied Eq.(\ref{eq:UNCB}). 
In the following subsections, we explain how to calculate MEs of several 
important operators. 

\subsection{Single Particle Operators}
This kind of operators is characterized as 
$\oprt{O} = O(\xi_1) + O(\xi_2)$. 
These include the core-nucleon interaction $V_{\rm c-N}(\xi_i)$, 
the s.p. kinetic energy $\bip^2_i / 2\mu = h_i - V_{\rm c-N}(\xi_i)$, 
the radial distance $\abs{\bir_i}^2$, and so on. 
For the operator $\oprt{O}(1) = O(\xi_1)$ which acts only 
on the first particle, 
the first term in Eq.(\ref{eq:ME01}) has the form of 
\beqa
 \Braket{ \Psi_{cd} \mid \oprt{O}(1) \mid \Psi_{ab}} 
 &=& \int d\xi_1 \int d\xi_2 
     \Psi_{cd}^* (\xi_1,\xi_2) O(\xi_1) \Psi_{ab} (\xi_1,\xi_2) \\
 &=& \delta_{d,b} \sum_{m_c,m_a} \nonumber 
     \cgc{J,M*}{j_d,M-m_c;j_c,m_c} \cgc{J,M}{j_b,M-m_a;j_a,m_a} \\
 & & \phantom{000} \times 
     \int d\xi_1 \phi^*_{(nljm)_c}(\xi_1) O(\xi_1) \phi_{(nljm)_a}(\xi_1), 
\eeqa
which vanishes if $n_b \neq n_d \cup l_b \neq l_d \cup j_b \neq j_d$. 
Thus, the only quantity we have to calculate is the integration 
in the last sentence. 
The other terms in Eq.(\ref{eq:ME01}) can be calculated similarly. 
Defining the following symbol; 
\beqa
 O_{ca}^{(1)} &\equiv& \sum_{m_c,m_a} \nonumber 
 \cgc{J,M*}{j_d,M-m_c;j_c,m_c} \cgc{J,M}{j_b,M-m_a;j_a,m_a} \\
 & & \phantom{000} \times \label{eq:spME0}
     \int d\xi_1 \phi^*_{(nljm)_c}(\xi_1) O(\xi_1) \phi_{(nljm)_a}(\xi_1), 
\eeqa
we can represent the matrix element of $\oprt{O}(1) = O(\xi_1)$ 
after the anti-symmetrization as follows. 
\beqa
 \Braket{\tilde{\Psi}_{cd} \mid \oprt{O}(1) \mid \tilde{\Psi}_{ab}} 
 &=& A_{cd}A_{ab} \left[ \delta_{db} O_{ca}^{(1)} 
          + B_{cd}B_{ab} \delta_{ca} O_{bd}^{(1)} \right. \nonumber \\
 & & \left.      -B_{ab} \delta_{cb} O_{da}^{(1)} 
                 -B_{cd} \delta_{da} O_{cb}^{(1)} \right]. \label{eq:spME1}
\eeqa
We also derive the similar formula for the summation of 
$\oprt{O}(1)$ and $\oprt{O}(2)$. 
The result reads 
\beqa
 & & \Braket{\tilde{\Psi}_{cd} \mid \oprt{O}(1)+\oprt{O}(2) \mid \tilde{\Psi}_{ab}} \nonumber \\
 & & \phantom{000} = A_{cd}A_{ab} \nonumber 
     \left[  \delta_{db} (O_{ca}^{(1)} + B_{cd}B_{ab}O_{ca}^{(2)})
            +\delta_{ca} (B_{cd}B_{ab}O_{db}^{(1)} + O_{db}^{(2)}) \right. \\
 & & \phantom{000000000000} 
     \left. -\delta_{cb} (B_{cd}O_{da}^{(1)} + B_{ab}O_{da}^{(2)}) 
            -\delta_{da} (B_{ab}O_{cb}^{(1)} + B_{cd}O_{cb}^{(2)}) \right]. 
\eeqa
We will use this formula to calculate, {\it e.g.} those of $h_1+h_2$ 
or $V_{\rm c-N}(\xi_1)+V_{\rm c-N}(\xi_2)$.

If the operator is spherical; $O(\xi_1)=O(r_1)$, Eq.({\ref{eq:spME0}}) can be 
reduced as the integration only for the radial distance. 
\beqa
 O_{ca}^{(1),spherical} \nonumber 
 &=& \delta_{j_c j_a} \delta_{l_c l_a} \int dr_1 r_1^2 R_{(nlj)_c}(r_1) O(r_1) R_{(nlj)_a}(r_1), \\
 &=& \delta_{j_c j_a} \delta_{l_c l_a} \int dr_1 U_{(nlj)_c}(r_1) O(r_1) U_{(nlj)_a}(r_1), 
\eeqa
where the product of coupled angular parts is given as 
$\delta_{j_c,j_a} \delta_{l_c,l_a} $.

\subsection{Two-Particle Operators}
The Operators in this category are given as 
$\oprt{O} = O(\xi_1,\xi_2)$. 
These include, for instance, the pairing interaction, $v_{\rm N-N}$, 
the recoil term, $\bip_1 \cdot \bip_2 / A_{\rm c} m$, and the 
opening angle between 
two nucleons, $\cos \theta_{12}$. 
In order to calculate Eq.(\ref{eq:ME01}), it is often necessary to know 
the following quantity. 
\beq
 \Braket{\mathcal{Y}_{(ljm)_c} \mid Y_{lh} \mid \mathcal{Y}_{(ljm)_a}} 
 = \int d\ubir \int d\bis \mathcal{Y}^*_{(ljm)_c}(\ubir,\bis) Y_{lh}(\ubir) \mathcal{Y}_{(ljm)_a}(\ubir,\bis)
\eeq
For this purpose, one can use the Wigner-Eckart 
theorem found in, {\it e.g.} the textbook 
by Edmonds \cite{60Edm}, 
\beqa
 && \Braket{\mathcal{Y}_{(ljm)_c} \mid Y_{lh} \mid \mathcal{Y}_{(ljm)_a}} \nonumber \\
 && ~~~ = (-)^{j_a-m_a-2h} \frac{\cgc{l,h}{j_c,m_c;j_a,m_a} }{\sqrt{2l+1}} 
    \Braket{j_c(l_c,1/2) \Vert Y_l \Vert j_a(l_a,1/2)}, 
\eeqa
where the reduced matrix element is written with the $6j$-symbols as 
\beqa
 \Braket{j_c(l_c,1/2) \Vert Y_l \Vert j_a(l_a,1/2)} &=& (-)^{l_c+l_a+j_a+l} \sqrt{(2j_c+1)(2j_a+1)} \nonumber \\
 & & \times \left\{ \begin{array}{ccc} l_c & j_c & 1/2 \\ j_a & 1/2 & l \end{array} \right\} 
     \Braket{l_c \Vert Y_l \Vert l_a}, \nonumber \\
 \Braket{l_c \Vert Y_l \Vert l_a} 
 &=& (-)^{l_c+l} \frac{\sqrt{(2l_c+1)(2l_a+1)}}{4\pi} \nonumber 
     \cgc{l,0}{l_c,0;l_a;0}. 
\eeqa
When the operator is scalar and does not include spin variables, 
it can be generally represented by the multi-pole expansion. 
Namely, 
\beq
 O(\bir_1,\bir_2) = \sum_{l=0}^{\infty} O_l(r_1,r_2) \sum_{h=-l}^{l} 
 Y_{l,h} (\ubir_1) \cdot (-)^h Y_{l,-h} (\ubir_2). \label{eq:39634}
\eeq
Then, we can formulate each component in Eq.(\ref{eq:ME01}). 
For the $l$-th term in Eq.(\ref{eq:39634}), 
the radial part becomes 
\beqa
 && rad.part_{(l)} \left[ \Braket{\Psi_{cd} \mid \oprt{O}(1,2) \mid \Psi_{ab}} \right] \nonumber \\
 && ~= \iint dr_1 dr_2 
       R^*_{(nlj)_c}(r_1) R^*_{(nlj)_d}(r_2) O_l(r_1,r_2) R_{(nlj)_a}(r_1) R_{(nlj)_b}(r_2), 
\eeqa
whereas the angular part is given as 
\beqa
 && ang.part_{(l)} \left[ \Braket{\Psi_{cd} \mid \oprt{O}(1,2) \mid \Psi_{ab}} \right] \nonumber \\
 && ~~~= \left< (j_c\oplus j_d)J,M \mid \sum_h Y_{lh}(1) (-)^h Y_{l,-h}(2) \mid (j_a\oplus j_b)J,M \right> \nonumber \\
 && ~~~= \sum_{all~m} \cgc{*J,M}{j_c,m_c;j_d,m_d} \cgc{J,M}{j_a,m_a;j_b,m_b} \nonumber \\ 
 && ~~~~~~~ \sum_h \Braket{\mathcal{Y}_{(ljm)_c} \mid Y_{l, h} \mid \mathcal{Y}_{(ljm)_a}} 
            (-)^h  \Braket{\mathcal{Y}_{(ljm)_d} \mid Y_{l,-h} \mid \mathcal{Y}_{(ljm)_b}}. \label{eq:ME12a} 
\eeqa
By performing a few calculations for the angular momenta, 
Eq.(\ref{eq:ME12a}) can be simplified as 
\beqa
 && ang.part_{(l)} \left[ \Braket{\Psi_{cd} \mid \oprt{O}(1,2) \mid \Psi_{ab}} \right] \nonumber \\
 && ~~ = (-)^{j_a+j_d-J} 
    \left\{ \begin{array}{ccc} j_a & l & j_c \\ j_d & J & j_b \end{array} \right\} \nonumber \\
 && \phantom{00000} \times \Braket{j_c(l_c,1/2) \Vert Y_l \Vert j_a(l_a,1/2)} \cdot \Braket{j_d(l_d,1/2) \Vert Y_l \Vert j_b(l_b,1/2)}, 
\eeqa
where the summations over the magnetic quantum numbers do not appear \cite{60Edm}. 
Consequently, we can write down the general formula for the ME of 
a two-particle operator as 
\beqa
 && \Braket{\Psi_{cd} \mid \oprt{O}(1,2) \mid \Psi_{ab}} \nonumber \\
 && ~~ = \sum_l \iint dr_1 dr_2 \nonumber 
       R^*_{(nlj)_c}(r_1) R^*_{(nlj)_d}(r_2) O_l(r_1,r_2) R_{(nlj)_a}(r_1) R_{(nlj)_b}(r_2) \\
 && \phantom{00000} \times (-)^{j_a+j_d-J} \left\{ \begin{array}{ccc} j_a & l & j_c \\ j_d & J & j_b \end{array} \right\} \nonumber \\
 && \phantom{0000000} \Braket{j_c(l_c,1/2) \Vert Y_l \Vert j_a(l_a,1/2)} \cdot \Braket{j_d(l_d,1/2) \Vert Y_l \Vert j_b(l_b,1/2)}. \label{eq:ME12g}
\eeqa
We mention that the orbital angular momenta, $l$, 
must be truncated in actual calculations. 
Thus, the summation over $l$ is also truncated as 
$\sum_{l=0}^{\infty} \rightarrow \sum_{l=0}^{l_{\rm max}}$. 

We also mention how to derive the $O_{l}(r_1,r_2)$. 
For the pairing interaction, the two-particle operator 
depends only on the relative distance, 
\beq
 r_{12}=\abs{\bir_1-\bir_2}=\sqrt{r_1^2+r_2^2-2r_1r_2\cos \theta_{12}}. 
\eeq
The multi-pole expansion for an arbitrary function of $r_{12}$ satisfies 
\beqa
 f(r_{12}) &=& \sum_{l=0}^{l_{\rm max}} (2l+1) g_l (r_1,r_2) P_{l} (\cos \theta_{12}) \\
 &=& \sum_{l=0}^{l_{\rm max}} g_l (r_1,r_2) 4\pi \sum_{h=-l}^{l} 
     Y_{l,h}(\ubir_1) (-)^h Y_{l,-h}(\ubir_2), 
\eeqa
with 
\beq
  g_l (r_1,r_2) = \frac{1}{2} \int_0^{\pi} f(r_{12}) P_{l} (\cos \theta_{12}) \sin \theta_{12} d\theta_{12}, 
\eeq
where $P_l$ is the Legendre polynomial. 
We list below concrete forms of the functions used for 
the pairing interaction. 
\begin{enumerate}
\item a delta function; 
      \beq
        f(r_{12}) = \delta(\abs{\bir_1-\bir_2}) = \frac{\delta(r_1-r_2)}{r_1 r_2} \sum_l \sum_{h=-l}^{l} Y_{l,h}(\ubir_1) (-)^h Y_{l,-h}(\ubir_2). 
      \eeq
\item an inverse function; 
      \beq
        f(r_{12}) = \frac{1}{\abs{\bir_1-\bir_2}} = \sum_l \frac{r_<^l}{r_>^{l+1}} \frac{4\pi}{2l+1} \sum_{h=-l}^{l} Y_{l,h}(\ubir_1) (-)^h Y_{l,-h}(\ubir_2), 
      \eeq
      where $r_>$ ($r_<$) indicates the larger (smaller) one between $r_1$ and $r_2$. 
\end{enumerate}
On the other hand, for the recoil term; $\bip_1 \cdot \bip_2 / A_{\rm c} m$, 
we first use the formula of the spatial differentiation, that is 
\beq
 \frac{1}{2} \left[ \nabla^2, \bir \right] = \nabla \Longleftrightarrow 
 \nabla = \ubir \left( \frac{d}{dr} + \frac{1}{r} \right) - \frac{1}{2r} \left[ \hat{\bi{l}}^2,\ubir \right]. 
\eeq
Thus, for the product $\nabla_1 \cdot \nabla_2$, 
its ME before the anti-symmetrization takes the form of 
\beqa
 && \Braket{\Psi_{cd} | \nabla_1 \cdot \nabla_2 | \Psi_{ab}} \nonumber \\
 && ~~~ = \left< R_c R_d \mid \left\{ (\frac{d}{dr_1}+\frac{1}{r_1}) - \frac{1}{2r_1}(l_c(l_c+1)-l_a(l_a+1)) \right\} \right. \nonumber \\
 && ~~~~~~~~ \left. \left\{ (\frac{d}{dr_2}+\frac{1}{r_2}) - \frac{1}{2r_2}(l_d(l_d+1)-l_b(l_b+1)) \right\} \mid R_a R_b \right> \nonumber \\
 && ~~~~~ \times \Braket{W_{cd} \mid \ubir_1 \cdot \ubir_2 \mid W_{ab}}, 
\eeqa
where the radial part can be calculated with the first derivatives. 
In the angular part, we expand the function, $\ubir_1 \cdot \ubir_2$, 
as follows. 
\beq
 \ubir_1 \cdot \ubir_2 = \cos \theta_{12} = P_{l=1}(\cos \theta_{12}) 
 = \frac{4\pi}{3} \sum_{h=-1}^{1} Y_{1,h} (\ubir_1) (-)^h Y_{1,-h} (\ubir_2). 
\eeq
Consequently, the MEs of this operator can be calculated 
by means of the dipole expansion. 
\beqa
 && \Braket{W_{cd} \mid \ubir_1 \cdot \ubir_2 \mid W_{ab}} \nonumber \\
 && ~~ = \frac{4\pi}{3} (-)^{j_a+j_d-J} 
    \left\{ \begin{array}{ccc} j_a & 1 & j_c \\ j_d & J & j_b \end{array} \right\} \nonumber \\
 && \phantom{00000} \times \Braket{j_c(l_c,1/2) \Vert Y_1 \Vert j_a(l_a,1/2)} \cdot 
    \Braket{j_d(l_d,1/2) \Vert Y_1 \Vert j_b(l_b,1/2)}. \label{eq:3050}
\eeqa
Obviously, the recoil term mixes the uncorrelated basis which satisfy $\abs{l_c-l_a}=1$. 
If we limit the model space with $(-)^{l_a}=odd$ or $even$ only, 
the recoil term does not contribute.

\section{Density Distribution}
We also derive the formulas for the density distributions of two nucleons. 
With our uncorrelated basis, the two-nucleon state can be expanded as 
Eq.(\ref{eq:exp_ucb}). 
Its density distribution is obviously given by 
\beqa
\rho (\xi_1,\xi_2) 
&=& \abs{\Phi(\xi_1,\xi_2)}^2 
    = \sum_{34} \sum_{12} \alpha_{34}^* \alpha_{12} 
      \tilde{\Psi}_{34}^* \cdot \tilde{\Psi}_{12} (\xi_1,\xi_2) \\
&=& \sum_{34} \sum_{12} \alpha_{34}^* \alpha_{12} A_{34} A_{12} \nonumber \\
& & \times \left[ \pi_{34,12} (\xi_1,\xi_2) + B_{34} B_{12} \pi_{43,21} 
  - B_{34} \pi_{43,12} -                      B_{12} \pi_{34,21} \right], \label{eq:rho_expand} 
\eeqa
with $B_{ab} \equiv (-)^{j_a+j_b-J}$. 
Each component $\pi_{cd,ab}(\xi_1,\xi_2)$ can be written as 
\beqa
  \pi_{34,12} (\xi_1,\xi_2) &\equiv& \Psi_{34}^* \cdot \Psi_{12} (\xi_1,\xi_2) \label{eq:denspi} \\
  &=& \left[ \sum_{m_3} \cgc{J,M}{j_3,m_3;j_4,M-m_3} 
            \phi_{(nljm)_3} (\xi_1) \phi_{(nljm)_4} (\xi_2) \right]^* \times \nonumber \\
  & & \left[ \sum_{m_1} \cgc{J,M}{j_1,m_1;j_2,M-m_1} 
            \phi_{(nljm)_1} (\xi_1) \phi_{(nljm)_2} (\xi_2) \right] \nonumber \\
  &=& R^*_{(nlj)_3}(r_1) R^*_{(nlj)_4}(r_2) \cdot R_{(nlj)_1}(r_1) R_{(nlj)_2}(r_2) \nonumber \\
  & & \times ~ W_{34}^{*(J,M)}(\ubir_1 \bis_1, \ubir_2 \bis_2) \cdot W_{12}^{(J,M)}(\ubir_1 \bis_1, \ubir_2 \bis_2). 
\eeqa
where we used the Eqs(\ref{eq:uncbs_03}) and (\ref{eq:couplang}).

\section{Spin-Orbit Decomposition}
It will be also helpful to formulate the decomposition of two-nucleon states 
into those of the spin-singlet and triplet configurations. 
For this purpose, at first, we have to discuss some mathematics of angular momenta. 
In Eq.(\ref{eq:couplang}), to fix the final angular momentum, $(J,M)$, 
we first couple $\bi{l}_a$ and $\bis_1$ to $\bi{j}_a$, 
and then couple $\bi{j}_a$ and $\bi{j}_b$ to $\bi{J}$. 
That is, 
\beq
  \left. \begin{array}{c} 
    \bi{l}_a \oplus \bis_1 = \bi{j}_a \\ 
    \bi{l}_b \oplus \bis_2 = \bi{j}_b 
  \end{array} \right\} \longrightarrow \bi{j}_a \oplus \bi{j}_b = \bi{J}, 
\eeq
where $s_i=\abs{\bis_i}=1/2$. 
Within this coupling scheme, we got the coupled angular part, 
$W_{ab}^{(J,M)}=W_{l_a l_b j_a j_b}^{(J,M)}(\ubir_1 \bis_1, \ubir_2 \bis_2)$. 
On the other hand, another coupling scheme 
can be considered as 
\beq
  \left. \begin{array}{c} 
    \bi{l}_a \oplus \bi{l}_b = \bi{L} \\ 
      \bis_1 \oplus \bis_2   = \bi{S} 
  \end{array} \right\} \longrightarrow \bi{L} \oplus \bi{S} = \bi{J}.
\eeq
Those two coupling schemes can be related to each other by 
the unitary transformation. 
Namely, we can write down 
\beqa
 & & W_{l_a l_b j_a j_b}^{(J,M)}(\ubir_1 \bis_1, \ubir_2 \bis_2) \nonumber \\
 & & ~~~~~~~ = 
     \sum_{L=|l_a-l_b|}^{l_a+l_b} \sum_{S=0,1} 
     D_J (j_a j_b;l_a l_b s_1 s_2;LS) 
     \cdot \Xi_{l_a l_b LS}^{(J,M)}(\ubir_1 \bis_1, \ubir_2 \bis_2), 
\eeqa
with the $LS$-coupled angular part; 
\beqa
 & & \Xi_{l_a l_b LS}^{(J,M)}(\ubir_1 \bis_1, \ubir_2 \bis_2) 
     = \sum_{M_S = \pm 1} \cgc{J,M}{L,M-V ; S,M_S} \nonumber \\
 & & ~~~~~~~~~~~~~ \times 
     \left[ Y_{l_a}(\ubir_1) \otimes Y_{l_b} (\ubir_2) \right]^{(L,M-V)} 
     \left[ \chi(\bis_1) \otimes \chi(\bis_2) \right]^{(S,V)}, 
\eeqa
and the expansion coefficients including the $9j$-symbol; 
\beqa
 && D_J (j_a j_b;l_a l_b s_1 s_2;LS) \nonumber \\
 && ~~~~~~~ \equiv \sqrt{(2L+1)(2S+1)(2j_a+1)(2j_b+1)} 
    \left\{ \begin{array}{ccc} 
         l_a & l_b & L \\ 
         s_1 & s_2 & S \\
         j_a & j_b & J \end{array} \right\}. \label{eq:3iarg}
\eeqa
Using these formulas, the anti-symmetrized uncorrelated basis can be 
decomposed into the spin-singlet and triplet configurations as follows. 
\beqa
 \tilde{\Psi}_{ab} (\xi_1,\xi_2) &=& 
 \tilde{\Psi}_{ab,S=0} (\xi_1,\xi_2) + \tilde{\Psi}_{ab,S=1} (\xi_1,\xi_2), \\
 \tilde{\Psi}_{ab,S} (\xi_1,\xi_2) &=& 
     A_{ab} \left[ \Psi_{ab,S}(\xi_1,\xi_2) - B_{ab} \Psi_{ba,S}(\xi_1,\xi_2) \right], 
\eeqa
with 
\beqa
 && \Psi_{ab,S}(\xi_1,\xi_2) = R_{(nlj)_a}(r_1) R_{(nlj)_b}(r_2) \nonumber \\
 && ~~~~~~ \times \sum_{L=|l_a-l_n|}^{l_a+l_b} D_J (j_a j_b;l_a l_b s_1 s_2;LS) \cdot \Xi_{l_a l_b LS}^{(J,M)}(\ubir_1 \bis_1, \ubir_2 \bis_2). 
\eeqa
Notice that the normalization of each basis function reads 
\beqa
 1 &=& \int d\xi_1  \int d\xi_2  \abs{\tilde{\Psi}_{ab} (\xi_1,\xi_2)}^2 \\
   &=& \int d\bir_1 \int d\bir_2 \left\{ \abs{\tilde{\Psi}_{ab,S=0} (\bir_1,\bir_2)}^2 + \abs{\tilde{\Psi}_{ab,S=1} (\bir_1,\bir_2)}^2 \right\} \nonumber \\
   &=& \abs{A_{ab}}^2 \sum_S \sum_L \left\{ \abs{D_J (j_a j_b;l_a l_b s_1 s_2;LS)}^2 + \abs{D_J (j_a j_b;l_a l_b s_1 s_2;LS)}^2 \right. \nonumber \\
   & & -B_{ab} D^*_J(j_b j_a;l_b l_a s_2 s_1;LS) D_J(j_a j_b;l_a l_b s_1 s_2;LS) \delta_{n_b,n_a} \nonumber \\
   & & \left. -B_{ab} D^*_J(j_a j_b;l_a l_b s_1 s_2;LS) D_J(j_b j_a;l_b l_a s_2 s_1;LS) \delta_{n_b,n_a} \right\}, 
\eeqa
where the radial integrations in the cross terms become $\delta_{n_b,n_a}$. 
We also introduce a similar decomposition for the density distribution. 
Namely, Eq.(\ref{eq:denspi}) can be decomposed as 
\beqa
  \pi_{34,12} (\xi_1,\xi_2) 
  &\equiv & \Psi_{34}^* \cdot \Psi_{12} (\xi_1,\xi_2) \nonumber \\
  &=& R^*_{(nlj)_3}(r_1) R^*_{(nlj)_4}(r_2) \cdot R_{(nlj)_1}(r_1) R_{(nlj)_2}(r_2) \nonumber \\
  & & \times \sum_{L',S'} \left[ D_J(j_3 j_4;l_3 l_4 s_3 s_4 ;L'S') \cdot \Xi_{l_3 l_4 L'S'}^{(J,M)}(\ubir_1 \bis_1, \ubir_2 \bis_2) \right]^* \nonumber \\
  & & \times \sum_{L, S } \left[ D_J(j_1 j_2;l_1 l_2 s_1 s_2 ;L S ) \cdot \Xi_{l_1 l_2 L S }^{(J,M)}(\ubir_1 \bis_1, \ubir_2 \bis_2)\right], 
\eeqa
where $s_1 \sim s_4 = 1/2$. 
Substituting this equation into Eq.(\ref{eq:rho_expand}), 
we can also decompose 
the total density into the spin-singlet and triplet terms. 
The cross terms of the spin-singlet and triplet components are, indeed, 
irrelevant because those can be vanished by integrating over 
the spin variables. 
We use this technique in order to derive the spin-integrated density 
as we show in the next subsection. 

\subsection{Spin-Integrated Density}
In practice, we often need to integrate the density 
over the spin variables. 
From the orthogonality between the spin-singlet and 
triplet configurations, 
\beqa
 \Braket{S',M'_S|S,M_S} &=& \int d\bis_1 \int d\bis_2 
 \left[ \chi(\bis_1) \otimes \chi(\bis_2) \right]^{(S',M'_S) \dagger} 
 \left[ \chi(\bis_1) \otimes \chi(\bis_2) \right]^{(S ,M_S )} \nonumber \\
 &=& \delta_{S'S} \delta_{M'_S M_S}, 
\eeqa
a component of the spin-integrated density, 
$d_{34,12}(\bir_1,\bir_2)$, can be represented as 
\beqa
 && d_{34,12}(\bir_1,\bir_2) \equiv \int d\bis_1 \int d\bis_2 \pi_{34,12}(\xi_1,\xi_2), \label{eq:3117} \\
 && = R^*_{(nlj)_3}(r_1) R^*_{(nlj)_4}(r_2) \cdot R_{(nlj)_1}(r_1) R_{(nlj)_2}(r_2) \sum_{S=0,1} \sum_{M_S = -S}^{S} \nonumber \\
 && ~ \times \sum_{L'} \left[ D_J(j_3 j_4;l_3 l_4 s_3 s_4 ;L'S) \cgc{J,M}{L',M-M_S;S,M_S} \left[ Y_{l_3}(\ubir_1) \otimes Y_{l_4} (\ubir_2) \right]^{(L',M-M_S)} \right]^* \nonumber \\
 && ~ \times \sum_{L } \left[ D_J(j_1 j_2;l_1 l_2 s_1 s_2 ;L S) \cgc{J,M}{L ,M-M_S;S,M_S} \left[ Y_{l_1}(\ubir_1) \otimes Y_{l_2} (\ubir_2) \right]^{(L, M-M_S)} \right]. \nonumber \\ && \\
 && = d_{34,12,S=0}(\bir_1,\bir_2) + d_{34,12,S=1}(\bir_1,\bir_2). 
\eeqa
Therefore, we can finally formulate the spin-integrated density, 
$\rho(\bir_1,\bir_2)$, as below. 
\beqa
 \rho(\bir_1,\bir_2) &\equiv& \int d\bis_1 \int d\bis_2 \rho(\xi_1,\xi_2) \label{eq:rhod} \\
 &=& \int d\bis_1 \int d\bis_2 \sum_{cd} \sum_{ab} \alpha_{cd}^* \alpha_{ab} \tilde{\Psi}^*_{cd} \cdot \tilde{\Psi}_{ab} (\xi_1,\xi_2) \nonumber \\
 &=& \sum_{cd} \sum_{ab} \alpha_{cd}^* \alpha_{ab} A_{cd} A_{ab} \nonumber \\
 & & ~ \sum_{S=0,1} \left[ d_{cd,ab,S}(\bir_1,\bir_2) + B_{cd} B_{ab} d_{dc,ba,S}(\bir_1,\bir_2) \right. \nonumber \\
 & & ~~~ \left.   - B_{cd} d_{dc,ab,S}(\bir_1,\bir_2) -        B_{ab} d_{cd,ba,S}(\bir_1,\bir_2) \right], \\
 &=& \rho_{S=0}(\bir_1,\bir_2) + \rho_{S=1}(\bir_1,\bir_2) \label{eq:rhos01}. 
\eeqa
Note that the normalization is given as 
\beq
 \int d\bir_1 \int \bir_2 \rho(\bir_1,\bir_2) 
 = \sum_{ab} \abs{\alpha_{ab}}^2 = 1, 
\eeq
since $\int d\bir_1 \int \bir_2 d_{cd,ab}(\bir_1,\bir_2)=\delta_{ca} \delta_{db}$. 

\section{Matrix Diagonalization}
In this Chapter, 
we have derived the basic formulas for the three-body model. 
With the uncorrelated basis, one can represent the 
eigen-states of the Hamiltonian with a spin $(J,M)$, namely 
$H_{\rm 3b} \ket{E_N ^{(J,M)}} = E_N \ket{E_N ^{(J,M)}}$, as follows. 
\beq
 \ket{E_N ^{(J,M)}} = \sum_K U_{NK} \ket{\tilde{\Psi}_K^{(J,M)}}, 
\eeq
where $K \equiv \left\{ (nlj)_a (nlj)_b \right\}$. 
In this expansion, there are also continuum basis with 
$\epsilon_a + \epsilon_b > 0$. 
One should notice that, even for a bound three-body state with $E_N \leq 0$, 
the wave function includes continuum s.p. states. 
The expansion coefficients $\{ U_{NK} \}$ can be obtained 
by diagonalizing the Hamiltonian matrix, 
$\Braket{\tilde{\Psi}_{K'} | H_{\rm 3b} | \tilde{\Psi}_K}$. 
Since we consider the pure Hermite space, all the MEs are real numbers. 
Thus, in order to diagonalize the Hamiltonian matrix, 
we employ ``Jacobi method'' 
for real, symmetric matrices \cite{95Kelley}. 
A typical dimension of our Hamiltonian is about 
from $100 \times 100$ to $1000 \times 1000$. 
The dimension actually depends on the cutoff 
parameters which we will explain later. 


In the next Chapter, we will apply the formalism presented in this Chapter 
to the pairing and dinucleon correlations in particle-bound nuclei, 
whereas an application to \twop-emitters will be discussed in 
Chapter \ref{Ch_Results2} and \ref{Ch_Results3}. 
\include{end}
\documentclass[a4paper, 12pt]{report}
\include{begin}

\chapter{Diproton Correlation in Light Nuclei} \label{Ch_Results1}
Before we discuss the two-proton (\twop-) emission, 
we first discuss in this Chapter the pairing and dinucleon correlation 
in particle-bound systems. 
To this end, 
we apply the three-body model to several light nuclei. 
Similar theoretical studies have been carried out 
especially for $^6$He and $^{11}$Li, 
which are well known as $2n$-halo as well as $2n$-Borromean nuclei. 
In these light and weakly bound neutron-rich nuclei, 
it has been shown that the pairing correlation plays an important role 
in generating the dineutron correlation, including 
a spatial concentration of two neutrons and the enhancement of the spin-singlet 
configuration \cite{87Hansen, 91Bert, 93Zhukov, 96Vinh, 97Esb, 05Hagi}. 

It is important to notice that the dineutron correlation itself 
can be considered 
even in deeper bound valence neutrons. 
Based on this idea, in this Chapter, we consider 
the $^{18}$O nucleus, 
in which the three-body picture should be reasonable. 
Additionally, in connection to the two-proton radioactivity, 
we will also discuss light proton-rich nuclei, 
$^{18}$Ne and $^{17}$Ne \cite{04Garrido_01,04Garrido_02,05Gri,10Lay}. 
We particularly discuss the following two points; 
(i) whether the diproton correlation exists similarly to the 
dineutron correlation, and 
(ii) whether the dinucleon correlations are limited only for 
weakly bound nucleons or not. 
For the point (i), the main attention will be paid to 
the effect of the Coulomb 
repulsion between two protons, which may break the diproton-like 
configurations to some extent. 
For the point (ii), 
the universality between strongly and weakly bound nucleons will be 
a key issue.

\section{Dinucleon Correlation in $^{16}$O+N+N Systems: 
$^{18}$Ne and $^{18}$O} \label{Sec_4q1y3}
We start our discussions with applying our three-body model to 
the ground states of $^{16}$O+N+N nuclei, which are expected to give a 
good testing ground for the dinucleon correlations. 
In the following, we only treat pairs of identical 
nucleons in valence orbits. 
Thus, the corresponding systems are $^{18}$O and $^{18}$Ne, 
with N=n and N=p, respectively. 
In their ground states, these nuclei have the spin-parity of $0^+$. 
The core nucleus, $^{16}$O, consists of eight protons and eight neutrons, 
building the doubly-closed nuclear shell-structure (a double-magic nucleus). 
Because of the double-magic nature, the assumption of a rigid core 
is expected to be reasonable for $^{16}$O, and thus the behaviors of 
the two valence nucleons should be 
well described within the three-body model. 
Indeed, the first excited state of $^{16}$O locates at $6.05$ MeV, 
which is higher enough than the single-nucleon energies of valence orbits in 
$^{17}$O$=^{16}$O+n and $^{17}$F$=^{16}$O+p, namely 
$0.87$ MeV $(2s_{1/2})$ and $5.08$ MeV $(1d_{3/2})$ in $^{17}$O, and 
$0.49$ MeV $(2s_{1/2})$ and $5.00$ MeV $(1d_{3/2})$ in $^{17}$F, measured 
from their ground states with a $(1d_{5/2})$-valence neutron and proton, 
respectively \cite{NNDCHP}. 
We assume that the core is always in its ground state 
and has the spin-parity of $0^+$. 
\begin{table}[tb] \begin{center}
  \catcode`? = \active \def?{\phantom{0}} 
  \begingroup \renewcommand{\arraystretch}{1.2}
  \begin{tabular*}{\columnwidth}{ @{\extracolsep{\fill}} ccccc ccc} \hline \hline
                            && \multicolumn{2}{c}{$^{17}$F}         && \multicolumn{2}{c}{$^{17}$O}        & \\ \cline{3-4} \cline{6-7}
                            && calc.     & Exp.\cite{NNDCHP} && calc.     & Exp.\cite{NNDCHP} & \\ \hline
  $\epsilon(1d_{5/2})$ (MeV)  && $-0.601$  & $-0.600$         && $-4.199$  & $-4.143$           & \\
  $\epsilon(2s_{1/2})$ (MeV)  && $-0.106$  & $-0.105$         && $-3.235$  & $-3.273$           & \\
  \hline \hline \end{tabular*}
  \endgroup
  \catcode`? = 12 
  \caption{ The energies of the $(2s_{1/2})$ and $(1d_{5/2})$ orbits 
in $^{17}$F and $^{17}$O, 
calculated with the core-nucleon two body model. 
For the comparison, the experimental values are also shown \cite{NNDCHP}. 
All the values are measured from the one-proton or one-neutron 
separation thresholds. 
Note that the experimental errors are only the order of 
$1$ keV or smaller. } \label{tb:A17_18}
\end{center} \end{table}

\subsection{Core-Nucleon Subsystems}
We first solve the core-nucleon two-body states. 
For the core-nucleon interaction, 
we use $r_0=1.22$ fm and $a_{\rm core}=0.65$ fm for the 
Woods-Saxon potential (Eq.(\ref{eq:WSP})). 
The parameters of the potential depth are defined as $V_0 = -55.06$ MeV and 
$V_{ls} = 16.71$ ${\rm MeV \cdot fm^2}$, both for $^{17}$F and $^{17}$O. 
These parameters well reproduce the measured energies of 
the $(2s_{1/2})$ and $(1d_{5/2})$ orbits, as shown in Table \ref{tb:A17_18}. 
In Figure \ref{fig:VCNs_18}, the core-nucleon potentials in $(s_{1/2}),(p_{3/2})$ 
and $(d_{5/2})$ channels are plotted. 
\begin{figure*}[tb]
(a) $V_{\rm c-p}$ for $^{17}$F \\
\fbox{\includegraphics[width=0.4\hsize, scale=1.0, trim= 60 50 0 0]{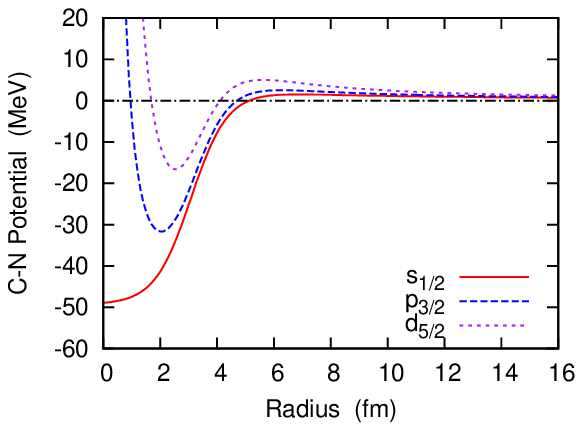}}\\
(b) $V_{\rm c-n}$ for $^{17}$O \\
\fbox{\includegraphics[width=0.4\hsize, scale=1.0, trim= 60 50 0 0]{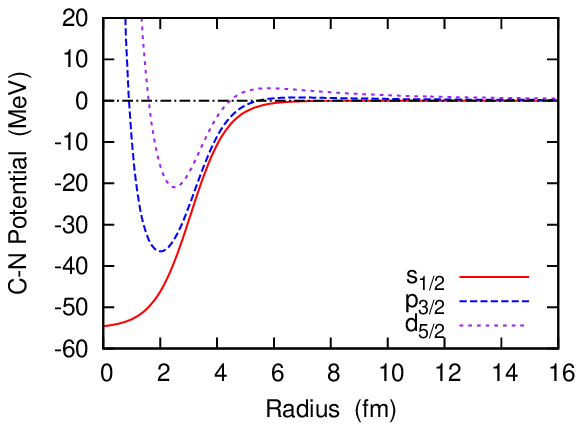}}\\
\caption{The core-nucleon potentials in the $(s_{1/2}),(p_{3/2})$ and $(d_{5/2})$ channels in $^{17}$F$\equiv ^{16}$O+p and $^{17}$O$\equiv ^{16}$O+n. } \label{fig:VCNs_18}
\end{figure*}

\subsection{Uncorrelated Basis} \label{Sec_4183}
The spin-parity of the ground states of $^{18}$Ne and $^{18}$O is $0^+$. 
On the other hand, as we noted, the core $^{16}$O is assumed to have $0^+$. 
Thus, for the uncorrelated two-nucleon basis, we only need the 
$(J,M)^{\pi}=(0,0)^+$ subspace, 
\beq
 \sum_K \ket{\tilde{\Psi}_K^{(J,M)^{\pi}}} \longrightarrow 
 \sum_K \ket{\tilde{\Psi}_K^{(0,0)^{+}}}, 
\eeq
where $K\equiv \left\{ (nlj)_a,(nlj)_b \right\}$ and $\pi=(-)^{l_a+l_b}$. 
From the basic properties of the angular momenta, 
the condition of $J=0$ and $\pi=+$ leads to $j_a=j_b$ and $l_a=l_b$. 
In other words, apart from the radial quantum numbers, two nucleons must 
have the same angular momenta. 
We represent these bases as $\ket{\tilde{\Psi}_{n_a n_b lj}}$ in the following, 
omitting the superscripts $(0,0)^{+}$ for simplicity. 
Using Eqs.(\ref{eq:Aab}) and (\ref{eq:UNCB}), 
the explicit form of uncorrelated wave functions can be written as 
\beqa
 \tilde{\Psi}_{n_a n_b lj}(\bir_1,\bir_2) 
 &=& \frac{1}{\sqrt{2(1+\delta_{n_a,n_b})}} \left[ \sum_m \cgc{0,0}{j,m:j,-m}\phi_{n_a lj,m}(\bir_1)\phi_{n_b lj,-m}(\bir_2) \right. \nonumber \\
 & & \left. -\sum_{m'} \cgc{0,0}{j,m':j,-m'}\phi_{n_a lj,m'}(\bir_2) \phi_{n_b lj,-m'}(\bir_1) \right] \\
 &=& \tilde{\Pi}_{n_a n_b lj}(r_1,r_2) \sum_m \cgc{0,0}{j,m:j,-m} \nonumber \\
 & & \phantom{00000000000000000} \mathcal{Y}_{lj,m}(\ubir_1 \bis_1) \mathcal{Y}_{lj,-m}(\ubir_2 \bis_2), 
\eeqa
where we defined 
\beq
 \tilde{\Pi}_{n_a n_b lj}(r_1,r_2) \equiv \frac{1}{\sqrt{2(1+\delta_{n_a,n_b})}} 
 \left[ R_{n_a lj}(r_1)R_{n_b lj}(r_2) + R_{n_b lj}(r_1)R_{n_a lj}(r_2) \right]. 
\eeq

In the calculations shown in this Chapter, 
the single particle (s.p.) states are solved within the radial box of 
$R_{\rm box} = 30$ fm, with the radial mesh of $dr = 0.1$ fm. 
We take all the s.p. states up to $l_{\rm max}=5$ into account. 
Namely, we include the uncorrelated partial waves from 
$(l_a j_a)\otimes(l_b j_b)=(s_{1/2})^2$ to $(h_{11/2})^2$. 
In order to truncate the model space, the energy cutoff is also introduced. 
We use $\epsilon_a + \epsilon_b \leq E_{\rm cut} = 30$ MeV, 
where $\epsilon_a$ means the energy of the $a$-th s.p. state. 
According to these constraints, 
we adopt about 360 uncorrelated states in our model space. 
This means that the dimension of the total Hamiltonian matrix is about 
$360\times 360$ for $^{18}$Ne and $^{18}$O. 

\subsection{Parameters for Pairing Interaction}
As introduced in the previous Chapter, we employ the density-dependent 
contact (DDC) interaction for the nuclear part of the pairing interaction, 
\beq
 v_{\rm N-N, Nucl.} (\bir_1,\bir_2) 
 = \delta(\bir_1-\bir_2) 
   \left[ v_0 + \frac{v_{\rho}}{1 + \exp \left( \frac{\abs{(\bir_1+\bir_2)/2}-R_{\rho}}{a_{\rho}} \right)} 
   \right]. \label{eq:DDCP_2}
\eeq
Since $E_{\rm cut}=30$ MeV and $a_{\rm nn}=-18.5$ fm, the parameter $v_0$ is fixed as 
$-875.34$ MeV from Eq.(\ref{eq:v0}). 
For the remaining parameters, we use $a_{\rho}=0.65$ fm 
and $R_{\rho}=1.22\cdot 16^{1/3}\cong 3.07$ fm, 
which are equal to those in the Woods-Saxon function 
of $V_{\rm c-N}$ (see Sec.\ref{Sec_4183}). 
The strength of the phenomenological density-dependent part, $v_{\rho}$, 
is adjusted so that the calculated two-nucleon 
binding energies are consistent to the experimental values, 
$S_{\rm 2p}=4.52$ and $S_{\rm 2n}=12.19$ MeV, for $^{18}$Ne and $^{18}$O, 
respectively. 
This condition yields $v_{\rho}=-1.104v_0$ and $-1.159v_0$ for 
$^{18}$Ne and $^{18}$O, respectively. 

Notice that the density-dependent term decreases the pairing attraction 
inside the core 
($\abs{\bir_1+\bir_2} /2 \lesssim R_{\rho}$), 
compared with the bare pairing attraction 
($\abs{\bir_1+\bir_2} /2 \longrightarrow \infty$). 
It corresponds to taking into account the medium effect on the pairing interaction. 

\subsection{Energy Expectational Values}
We now calculate and diagonalize the matrix elements of the total Hamiltonian 
(Eq.(\ref{eq:H3b})), in the way which was explained in the previous Chapter. 
The obtained wave function for the ground state is given as a 
superposition of the $0^+$ uncorrelated basis, 
\beq
 \Phi_{g.s.}(\bir_1,\bir_2) = 
 \sum_{ab} \alpha_{ab} 
 \tilde{\Psi}_{n_a n_b lj}(\bir_1,\bir_2). \label{eq:gsexp}
\eeq
The two nucleon binding, or equivalently, 
separation energies of $^{18}$Ne and $^{18}$O are 
given as the expectation value of the total Hamiltonian, 
\beq
 -S_{\rm 2N} = \Braket{H_{\rm 3b}} \equiv \Braket{\Phi_{g.s.} |H_{\rm 3b} |\Phi_{g.s.}}. 
\eeq
These values calculated by our parameters are shown in the 
first row of Table \ref{tb:A18_1}. 
\begin{table}[t] \begin{center}
  \catcode`? = \active \def?{\phantom{0}} 
  \begingroup \renewcommand{\arraystretch}{1.2}
  \begin{tabular*}{\columnwidth}{ @{\extracolsep{\fill}} c c c c c c} \hline \hline
                                        && $^{18}$Ne && $^{18}$O & \\ \hline
    $\Braket{H_{\rm 3b}}=-S_{\rm 2N}$ (MeV) && $-4.52$      && $-12.19$  & \\
                                        &&          &&        & \\
    $\Braket{v_{\rm N-N}}$ (MeV)          && $-4.25$      && $-4.89$   & \\
    $\Braket{v_{\rm N-N, Nucl.}}$ (MeV)     && $-4.80$      && $-4.89$   & \\
    $\Braket{v_{\rm N-N, Coul.}}$ (MeV)     && $??0.55$      && $??0.??$   & \\
    $\Braket{\rm recoil}$ (MeV)         && $-0.45$      && $-0.56$   & \\
    $\Braket{h_1 + h_2}$ (MeV)           && $??0.18$      && $-6.74$   & \\
    $\Braket{V_{\rm c-N_1}+V_{\rm c-N_2}}$ (MeV)  && $-14.25$   && $-22.03$   & \\
                                        &&          &&        & \\
    $\Braket{h_{\rm N-N}}$ (MeV)          && $??6.17$      && $??6.78$   & \\
    $\Braket{h_{\rm c-NN}}$ (MeV)         && $-10.69$      && $-18.97$   & \\
  \hline \hline \end{tabular*}
  \endgroup
  \catcode`? = 12 
  \caption{ The energy-expectation values for $^{18}$Ne $\equiv ^{16}$O+p+p and 
$^{18}$O $\equiv ^{16}$O+n+n, calculated with the three-body model. 
The label ``recoil'' means $\bip_1 \cdot \bip_2 / mA_{\rm c}$. 
The experimental two-nucleon separation energies are 
$S_{\rm 2p}=4.52$ and $S_{\rm 2n}=12.19$ MeV for $^{18}$Ne and $^{18}$O, 
respectively \cite{NNDCHP}. 
Notice that $H_{\rm 3b} = h_1 + h_2 + v_{\rm N-N} + ({\rm recoil})$ 
and $= h_{\rm N-N} + h_{\rm c-NN}$. } \label{tb:A18_1}
\end{center} \end{table}

In Table \ref{tb:A18_1}, we summarize several energy-expectation 
values for the these three-body systems. 
According to Eq.(\ref{eq:H3b}), 
the total energies can be decomposed into 
the expectation values of the uncorrelated Hamiltonian, 
the pairing interaction and the recoil term. 
That is, 
\beq
 \Braket{H_{\rm 3b}} = 
 \Braket{h_1+h_2} + \Braket{v_{\rm N-N}} + \Braket{\frac{\bip_1 \cdot \bip_2}{A_{\rm c} m} }. 
\eeq
Note that the uncorrelated Hamiltonian can further be decomposed as 
\beq
 \Braket{h_1+h_2} = \Braket{V_{\rm c-N_1}+V_{\rm c-N_2}} 
                  + \Braket{\frac{\bip_1^2}{2\mu}+\frac{\bip_2^2}{2\mu}}, 
\eeq
where we show only the potential term in Table \ref{tb:A18_1}. 
It is also useful to decompose the total Hamiltonian into two 
relative components. 
One is the Hamiltonian between the core and a pair of 
nucleons, $h_{\rm c-NN}$, 
whereas the other is that between the two nucleons, $h_{\rm N-N}$. 
That is, 
\beqa
 H_{\rm 3b} &=& h_{\rm c-NN} + h_{\rm N-N} \\
 &=& \left[ \frac{p_{\rm c-NN}^2}{2\mu_{\rm c-NN}} 
     + V_{\rm c-N_1}(\bir_1) + V_{\rm c-N_2}(\bir_2) \right] 
     + \left[ \frac{p_{\rm N-N}^2}{2\mu_{\rm N-N}} 
     + v_{\rm N-N}(\abs{\bir_1-\bir_2}) \right], 
\eeqa
with $\mu_{\rm c-NN} = m A_{\rm c}/(A_{\rm c}+2)$ and $\mu_{\rm N-N}=m/2$. 
Notice that, indeed, there is still a coupling between the core-2N and N-N 
subsystems in $h_{\rm c-NN}$, due to the core-nucleon potentials. 
The relative momenta, $\left\{ \bip_{\rm c-NN}, \bip_{\rm N-N} \right\}$, 
can be related to the original momenta in the V-coordinates, 
$\left\{ \bip_1,\bip_2 \right\}$, by the transformation below. 
\beqa
 \bip_{\rm c-NN} &=& \bip_1+\bip_2, \\
 \bip_{\rm N-N}  &=& (\bip_1-\bip_2)/2. 
\eeqa
The expectational values of $h_{\rm N-N}$ and $h_{\rm c-NN}$ are 
also shown in Table \ref{tb:A18_1}. 

As one see in Table \ref{tb:A18_1}, 
the total binding energies are quite different 
between $^{18}$Ne and $^{18}$O. 
This difference is mainly due to the Coulomb interactions 
in $v_{\rm N-N}$ and $V_{\rm c-N}$. 
These Coulomb repulsions are also affected the $\Braket{h_{\rm N-N}}$ and 
$\Braket{h_{\rm c-NN}}$ values in $^{18}$Ne. 
However, apart from the Coulomb repulsions, $\Braket{v_{\rm N-N,Nucl.}}$ and 
$\Braket{\rm recoil}$ have similar values both in $^{18}$Ne and $^{18}$O. 
It means that the pairing correlations 
caused by the nuclear force and the recoil effect 
are not sensitive to the total binding energy, as long as we consider the same 
valence orbits (In these two nuclei, 
the major valence orbit is $(d_{5/2})^2$, as we will 
discuss in the next subsection). 
It is also notable that the ratio of $\Braket{v_{\rm p-p,Coul.}}$ and 
$\Braket{v_{\rm p-p,Nucl.}}$ is about $-0.11$. 
It shows that the Coulomb repulsion reduces the pairing energy by about 10\%. 
This result is consistent to what has been found with, 
a non-empirical pairing energy-density functional for proton 
pairing gaps \cite{09Lesi}, HFB calculations \cite{09Bert, 11Yama} and the 
three-body model calculations \cite{10Oishi, 10Oishi_err}. 
Accordingly, we can conclude that 
the degrees of pairing correlations, 
indicated by $\Braket{v_{\rm N-N}}+\Braket{\rm recoil}$, 
significantly depend neither on the total binding energy, 
nor the existence of the Coulomb repulsions. 

We can estimate the relative momentum between the two nucleons by using 
\beq
 \Braket{h_{\rm N-N}} = \Braket{\frac{p_{\rm N-N}^2}{2\mu_{\rm N-N}}}
                        + \Braket{v_{\rm N-N}}. 
\eeq
From Table \ref{tb:A18_1}, this equation yields $\sqrt{\Braket{p_{\rm N-N}^2}}=98.9$ 
and $\sqrt{\Braket{p_{\rm N-N}^2}}=104.6$ MeV/c for $^{18}$Ne and $^{18}$O, respectively. 
Because of these similar values of $\sqrt{\Braket{p_{\rm N-N}^2}}$, it is expected 
that the spatial distance between the two nucleons are also similar in $^{18}$Ne and $^{18}$O. 
We will confirm this point in the next subsection. 
One should notice, however, even if the diproton or dineutron correlation is confirmed 
in these nuclei, it does not mean the existence of the 
bound subsystem of the two nucleons, 
since the expectational value of $h_{\rm N-N}$ is positive in both systems.

\subsection{Density Distributions}
We next study the structural properties of the ground states of 
these three-body systems. 
We summarize the results in Table \ref{tb:A18_2}. 
In this Table, $\Braket{r_i} \equiv \sqrt{\Braket{r^2_i}}$ is the 
expectation value of the averaged distance between 
the core and the $i$-th nucleon. 
Likewise, 
\beqa
 && \sqrt{\Braket{r_{\rm N-N}^2}}  = \sqrt{ \Braket{r_1^2+r_2^2 -2r_1 r_2 \cos \theta_{12}} }, \label{eq:rNtoN} \\
 && \sqrt{\Braket{r_{\rm c-NN}^2}} = \sqrt{ \Braket{r_1^2+r_2^2 +2r_1 r_2 \cos \theta_{12}} }/2, \label{eq:rcto2N} 
\eeqa
mean the mean relative distances between the two nucleons and between the core and 
the center of two nucleons, respectively. 
We also show 
$\theta_{12} \equiv \cos^{-1} \left(\Braket{\cos \theta_{12}} \right)$ 
in the 4th row. 
The probability of each angular channel, 
\beq
 \sum_{n_a,n_b} \abs{\alpha_{n_a n_b lj}}^2, 
\eeq
where $\alpha_{n_a n_b lj}$ are the expansion coefficients given by Eq.(\ref{eq:gsexp}), 
is also computed. 
Those of $(s_{1/2})^2$, $(d_{5/2})^2$, $(p_{3/2})^2$ and $(p_{1/2})^2$ 
channels are listed in the 5-8th rows of Table \ref{tb:A18_2}, 
whereas those of the other channels are summarized as ``others''. 
In the last row, we show the ratio of the spin-singlet configuration of 
the two valence nucleons, which can be calculated as 
\beq
 P(S_{12}=0) = \iint d\bir_1 d\bir_2 \rho_{S_{12}=0}(\bir_1,\bir_2), 
\eeq
where $\rho_{S_{12}=0}$ is the spin-singlet density 
given by Eq.(\ref{eq:rhos01}). 
Of course, the ratio of the spin-triplet configuration is given 
by $P(S_{12}=1)=1-P(S_{12}=0)$. 
\begin{table}[tb] \begin{center}
  \catcode`? = \active \def?{\phantom{0}} 
  \begingroup \renewcommand{\arraystretch}{1.2}
  \begin{tabular*}{\columnwidth}{ @{\extracolsep{\fill}} c c c c c c} \hline \hline
                       && $^{18}$Ne && $^{18}$O & \\ \hline
    $\Braket{r_1}=\Braket{r_2}$ (fm) && 3.62 && 3.48 & \\
    $\sqrt{\Braket{r^2_{\rm N-N}}}$ (fm)   && 4.62     && 4.37    & \\
    $\sqrt{\Braket{r^2_{\rm c-NN}}}$ (fm)  && 2.79     && 2.70    & \\
    $\Braket{\theta_{12}}$ (deg) && 81.1     && 79.7    & \\
                       &&          &&         & \\
    $(s_{1/2})^2$ (\%)  && ?6.88    && ?5.75    & \\
    $(d_{5/2})^2$ (\%)  && 86.30    && 86.90    & \\
    $(p_{3/2})^2$ (\%)  && ?0.55    && ?0.54    & \\
    $(p_{1/2})^2$ (\%)  && ?0.14    && ?0.14    & \\
    others,$(l=even)^2$ (\%)  && ?3.36   && ?3.41    & \\
    others,$(l=odd)^2?$ (\%)  && ?2.77   && ?3.26    & \\
                       &&          &&         & \\
    $P(S_{12}=0)$ (\%)  && 79.08    && 78.84   & \\
  \hline \hline \end{tabular*}
  \endgroup
  \catcode`? = 12 
  \caption{ The structural properties of $^{18}{\rm Ne}$ and $^{18}{\rm O}$, calculated 
with the two-nucleon wave functions. 
The radius of the core nucleus is assumed to be 
$R_0=1.22\cdot 16^{1/3}\cong 3.074$ in the Woods-Saxon potential. } \label{tb:A18_2}
\end{center} \end{table}

In Figures \ref{fig:4181} and \ref{fig:4182}, 
we exhibit the density distributions, 
$\rho(\bir_1,\bir_2)=\abs{\Phi_{g.s.}(\bir_1,\bir_2)}$, 
obtained from the wave functions for the three-body systems. 
We integrate the density for the spin variables, as explained 
with Eqs.(\ref{eq:3117}) and (\ref{eq:rhod}). 
Because of the symmetry in these systems, 
the angular part of the density depends only on 
the opening angle between the valence nucleons, 
$\theta_{12}=\abs{\ubir_2-\ubir_1}$. 
Therefore, for the plotting purpose, we can fix $\ubir_1=\bar{\bi{z}}$ 
without lacking the general information. 
The integrations for the angular variables are replaced as 
\beq
 \iint d\ubir_{1} d\ubir_{2} \longrightarrow 8\pi^2 \int_{0}^{\pi} \sin \theta_{12} d\theta_{12}. 
\eeq
Thus, the density distribution is normalized as 
\beqa
 1 &=& \iint d\bir_1 d\bir_2 \rho(\bir_1,\bir_2) \\
   &=& \int_{0}^{R_{\rm box}} dr_1 \int_{0}^{R_{\rm box}} dr_2 \int_{0}^{\pi} d\theta_{12} 
       \bar{\rho}(r_1,r_2,\theta_{12}), 
\eeqa
with 
\beqa
 && \bar{\rho}(r_1,r_2,\theta_{12}) = 8\pi^2 r_1^2 r_2^2 \sin \theta_{12} \rho(r_1,r_2,\theta_{12}), \\
 && \rho(r_1,r_2,\theta_{12}) = \abs{\Phi_{g.s.}(r_1,r_2,\theta_{12})}^2. \label{eq:456}
\eeqa
The density distribution, $\abs{\Phi_{g.s.}(r_1,r_2,\theta_{12})}^2$, 
can be decomposed into the spin-singlet and the spin-triplet components. 
After some calculations, we get 
\beqa
 \abs{\Phi_{g.s.}(r_1,r_2,\theta_{12})}^2_{S=0} 
 &=& \frac{1}{4\pi} \sum_{l',j'} \sum_{l,j} Q^*_{l'j'}\cdot Q_{lj}(r_1,r_2) 
     \frac{(-)^{l'+l}}{4} \sqrt{\frac{2j'+1}{2l'+1}} \sqrt{\frac{2j+1}{2l+1}} \nonumber \\
 & & \times 2Y^*_{l',0}(\ubir_2) Y_{l,0}(\ubir_2) 
\eeqa
for the spin-singlet, and 
\beqa
 \abs{\Phi_{g.s.}(r_1,r_2,\theta_{12})}^2_{S=1} 
 &=& \frac{1}{4\pi} \sum_{l',j'} \sum_{l,j} Q^*_{l'j'}\cdot Q_{lj}(r_1,r_2) 
     \frac{(-)^{j'+j}}{4} \sqrt{2-\frac{2j'+1}{2l'+1}} \sqrt{2-\frac{2j+1}{2l+1}} \nonumber \\
 & & \times \left[ Y^*_{l',1}(\ubir_2) Y_{l,1}(\ubir_2) + Y^*_{l',-1}(\ubir_2) Y_{l,-1}(\ubir_2) \right] 
\eeqa
for the spin-triplet\footnote{Notice that these formulas are valid 
only for a state with $J^{\pi}=0^+$.} \cite{91Bert,97Esb}. 
Here, we have defined the radial density for each angular channel, 
$Q_{lj}(r_1,r_2)$, as 
\beq
 Q_{lj}(r_1,r_2) \equiv \sum_{n_a>n_b} \alpha_{n_a n_b lj} \tilde{\Pi}(r_1,r_2). 
\eeq
For the angular part, we can use the following formula. 
\beqa
 Y^*_{l',m}(\ubir_2) Y_{l,m}(\ubir_2) &=& (-)^m \sum_{L=|l'-l|}^{l'+l} 
 \sqrt{\frac{(2l'+1)(2l+1)(2L+1)}{4\pi}} \nonumber \\
 && \phantom{000} \times 
 \left( \begin{array}{ccc} l'&l&L \\ 0&0&0 \end{array} \right)
 \left( \begin{array}{ccc} l'&l&L \\-m&m&0 \end{array} \right) Y_{L,0}(\theta_{12}), 
\eeqa
where it depends only on the opening angle $\theta_{12}$. 
\begin{figure*}[t] \begin{center}
 \begin{tabular}{c} 
  ``$^{18}$Ne (g.s.)'' \\
  \begin{minipage}{0.48\hsize}
     (a) \\ \fbox{ \includegraphics[height=42truemm,scale=1.0, trim = 60 50 0 0]{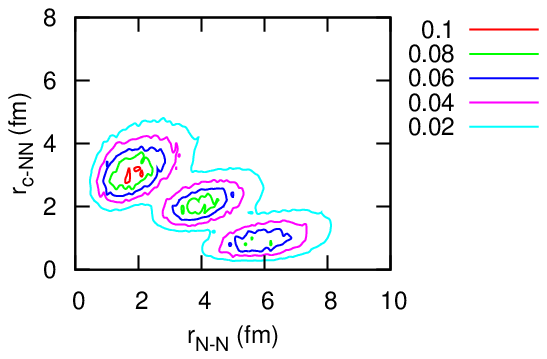}}\\ 
     (c) \\ \fbox{ \includegraphics[height=42truemm,scale=1.0, trim = 50 50 0 0]{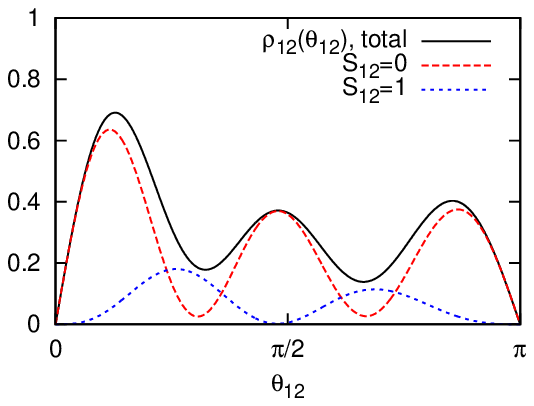}} 
  \end{minipage}
  \begin{minipage}{0.48\hsize}
     (b) \\ \fbox{ \includegraphics[height=42truemm,scale=1.0, trim = 50 50 0 0]{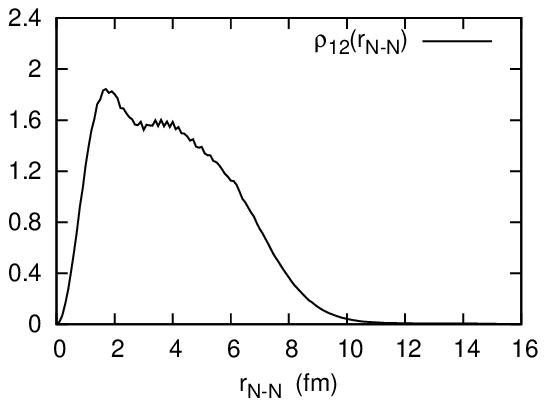}}\\ 
     (d) \\ \fbox{ \includegraphics[height=42truemm,scale=1.0, trim = 50 50 0 0]{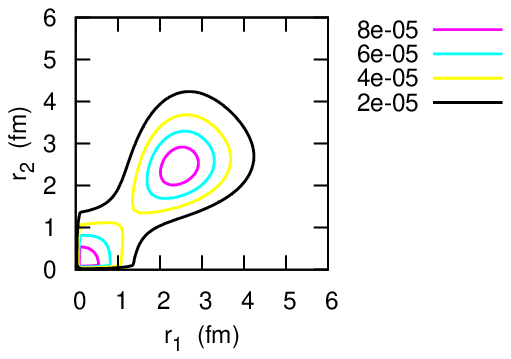}} 
  \end{minipage}
 \end{tabular}
 \caption{The two-nucleon density distribution of $^{18}$Ne, 
$\rho$, calculated for the ground state within the three-body model. 
Those are plotted for several sets of coordinates as follows. 
(a) with $r_{\rm N-N}$ and $r_{\rm c-NN}$. 
(b) with $r_{\rm N-N}$, integrated for $r_{\rm c-NN}$. 
(c) with the opening angle $\theta_{12}$ between the valence 
nucleons, integrated for $r_1$ and $r_2$. 
(d) with $r_1$ and $r_2$, integrated for $\theta_{12}$. 
In panel (d), the radial weight $r_1^2r_2^2$ is omitted 
to emphasize the peak(s). } \label{fig:4181}
\end{center} \end{figure*}
\begin{figure*}[t] \begin{center}
 \begin{tabular}{c} 
  ``$^{18}$O (g.s.)'' \\
  \begin{minipage}{0.48\hsize}
     (a) \\ \fbox{ \includegraphics[height=42truemm,scale=1.0, trim = 60 50 0 0]{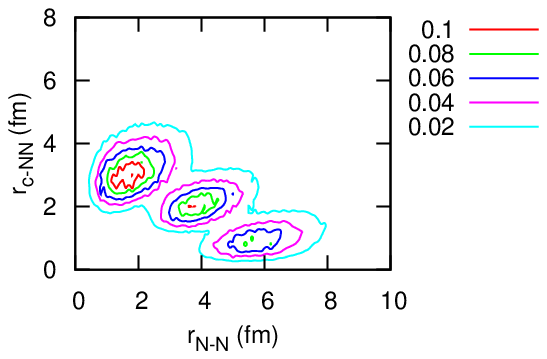}} \\
     (c) \\ \fbox{ \includegraphics[height=42truemm,scale=1.0, trim = 50 50 0 0]{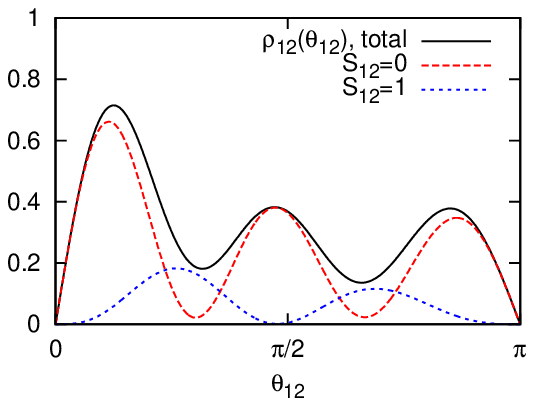}} 
  \end{minipage}
  \begin{minipage}{0.48\hsize}
     (b) \\ \fbox{ \includegraphics[height=42truemm,scale=1.0, trim = 50 50 0 0]{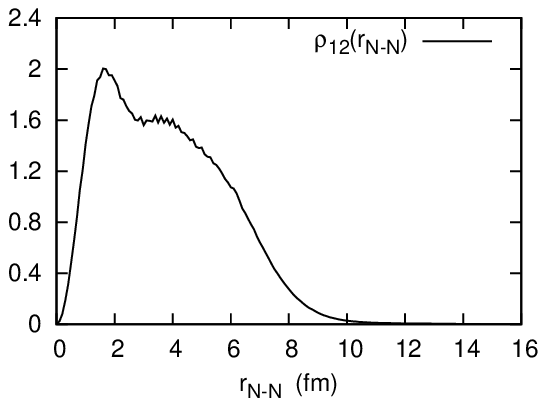}} \\
     (d) \\ \fbox{ \includegraphics[height=42truemm,scale=1.0, trim = 50 50 0 0]{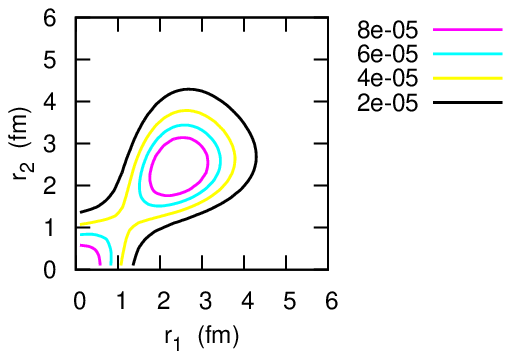}} 
  \end{minipage}
 \end{tabular}
 \caption{The same to Figure \ref{fig:4181} but of $^{18}$O. } \label{fig:4182}
\end{center} \end{figure*}

In Figure \ref{fig:4181}, we show the density distribution of $^{18}$Ne, 
plotted within several sets of coordinates. 
In panel (a), $\rho(r_1,r_2,\theta_{12})$ is plotted as a function of the 
relative distances, $r_{\rm N-N}$ and $r_{\rm c-NN}$ 
given by Eqs.(\ref{eq:rNtoN}) and (\ref{eq:rcto2N}). 
In panel (b), this function is integrated for $r_{\rm c-NN}$, and plotted only with $r_{\rm N-N}$. 
Conversely, in panel (c), we integrate $\rho$ for $r_1$ and $r_2$, 
and plot it as a function of the opening angle, $\theta_{12}$. 
We also plot the spin-singlet and triplet components separately in this panel. 
Finally, in panel (d), we integrate $\rho$ for the opening angle, and plot it 
as a function of $r_1$ and $r_2$. 
In this plotting, in order to clarify the peak(s), we omit 
the radial weight, $r_1^2 r_2^2$ in Eq.(\ref{eq:456}). 
We show similar plots for $^{18}$O in Figure \ref{fig:4182}. 

As general aspects, 
from Figures \ref{fig:4181}, \ref{fig:4182} and Table \ref{tb:A18_2}, 
we can see the similarity of the two-nucleon 
configurations in $^{18}$Ne and $^{18}$O, consistently to 
the similarity shown in Table \ref{tb:A18_1}. 
It means that the reduction of pairing energies caused by the 
Coulomb repulsion, which is evaluated as about $-10$\% reduction, 
does not affect significantly the two-nucleon densities. 
Because of the weakly binding due to Coulomb repulsions, 
the density of $^{18}$Ne is slightly extended compared with $^{18}$O. 
This tendency is intuitively understood by comparing Figs. 
\ref{fig:4181}(b),(d) and \ref{fig:4182}(b),(d). 
Correspondingly, the expectation values of distances 
in the 1st-4th rows of Table \ref{tb:A18_2} 
show larger values in the case of $^{18}$Ne. 
It is also shown from the probabilities of the angular channels 
that the $(d_{5/2})^2$ wave is 
dominant in both two cases, 
whereas the $(s_{1/2})^2$ wave has also considerable contributions. 
The distinct three peaks in panels (a) and (c) are mainly due to 
the $(d_{5/2})^2$ component, 
although the mixing of the other waves occurs with the pairing correlations, 
where the Coulomb repulsion plays a minor role. 

We note that the mean distance between the two nucleons, $r_{\rm N-N}$, 
shows a considerably smaller value, compared with the total diameter of 
the whole nucleus, estimated as $\simeq 2r_{\rm c-NN} \simeq 5.5$ fm. 

\subsection{Diproton and Dineutron Correlations}
In the ground states of both $^{18}$Ne and $^{18}$O, 
the dinucleon correlation, 
or at least, its tendency can be seen. 
The spatial localization is apparent at 
($r_{\rm N-N},r_{\rm c-NN}$) $\simeq $ (2fm, 3fm) in 
Figs. \ref{fig:4181}(a) and \ref{fig:4182}(a). 
It corresponds to the two nucleons confined in $r_{\rm N-N} \leq 2$ fm, 
which corresponds to the first peak of $\rho_{12}(r_{\rm N-N})$ shown in 
Figs. \ref{fig:4181}(b) and \ref{fig:4182}(b). 
We also find that $\rho_{12}(r_{\rm N-N})$ has almost 
all the components inside $r_{\rm N-N} \leq 8$ fm. 
According to the Ref.\cite{06Mats}, this result 
may be connected to the pairing densities in 
the nuclear matter at $\rho/\rho_0 = 0.1-0.01$ 
(see Fig.\ref{fig:2006Matsuo}), 
even though the assumption of nuclear matters cannot 
be translated directly to the conditions in finite nuclei. 

In Figs. \ref{fig:4181}(c) and \ref{fig:4182}(c), 
the corresponding angular distributions take the asymmetric forms, 
and have the highest peak at the small opening angle, 
$\theta_{12} \simeq \pi/6$. 
Indeed, this asymmetry is an important character of the 
dinucleon correlations: 
it is caused by the mixing of different parities of the core-nucleon 
partial system \cite{84Catara}. 
If we exclude this parity-mixing, 
the angular distributions have the perfect 
symmetric forms. 
We will check this point in the next section. 

We also note that in the asymmetric angular distributions, 
the most localized peak is mainly from the spin-singlet 
configuration, consistently to the definition of the 
dinucleon correlations. 
In both two nuclei, the spin-singlet has a 
probability of about 80\%. 
On the other hand, if we take the naive mean-field approximation, 
the two nucleons have the pure $(d_{5/2})^2$ wave, 
where the contribution from the spin-singlet 
is determined exactly as 60\% from the properties of the 
CG and the $9j$-coefficients. 
Thus, the pairing correlation works to enhance the spin-singlet 
configuration compared to the pure $(d_{5/2})^2$ wave. 

As interim conclusions, we have confirmed that the diproton correlation, 
which is characterized as the spatial localization of two-proton 
density mainly carried out by the spin-singlet configuration, 
is able to occur similarly to the dineutron correlation in 
the mirror nucleus. 
This result is consistent to the minor effect of the Coulomb repulsion, 
estimated by about -10\% reduction of the pairing energy. 

\section{Diproton Correlation in $^{17}$Ne}
In the previous section, it has been suggested that the structural 
properties of the two nucleons are insensitive 
to the total binding energy. 
In order to investigate the dependence of the dinucleon correlations 
on the binding energy, 
we next study the 
$^{17}$Ne nucleus, which has been famous as a \twop-Borromean nucleus. 
This nucleus is also a candidate to have the \twop-halo structure, 
due to the loosely bound two protons \cite{95Zhukov,05Gri}. 
In this system, two valence protons are bound with significantly 
small binding energies, $S_{\rm 2p} = 0.93$ MeV \cite{NNDCHP}. 
Therefore, it provides another testing ground to investigate 
the diproton correlation in a weakly bound system, 
in comparison with the diproton correlation in a deeply 
bound nucleus, $^{18}$Ne. 
In this section, 
we will also perform case-studies with different theoretical conditions, 
in order to gain a deeper understanding of the diproton correlation. 
Although these theoretical conditions may not 
correspond to realistic situations, 
those will be helpful to know what is the essential point in the diproton and 
dineutron correlations. 

\subsection{Set up for Calculations}
It is known that there is no bound state in $^{16}$F $\cong ^{15}$O+p, but 
four resonances in the low-lying region. 
These low-lying levels are shown in Figure \ref{fig:17Nelv}. 
The ground state of $^{15}$O has $1/2^-$, whereas its first excited state 
is located at $5.18$ MeV above the ground state. 
Because this excited energy is sufficiently high, 
the first and the second low-lying resonances in $^{16}$F 
at 0.536 and 0.729 MeV can be interpreted 
as the coupled states of $^{15}$O$_{g.s.}(1/2^-)$+$p(s_{1/2})$. 
Likewise, the 3rd and the 4th resonances 
at 0.960 and 1.257 MeV can be interpreted 
as those of $^{15}$O$_{g.s.}(1/2^-)$+$p(d_{5/2})$. 
\begin{figure*}[htbp] \begin{center}
\fbox{\includegraphics[width = 0.4\hsize]{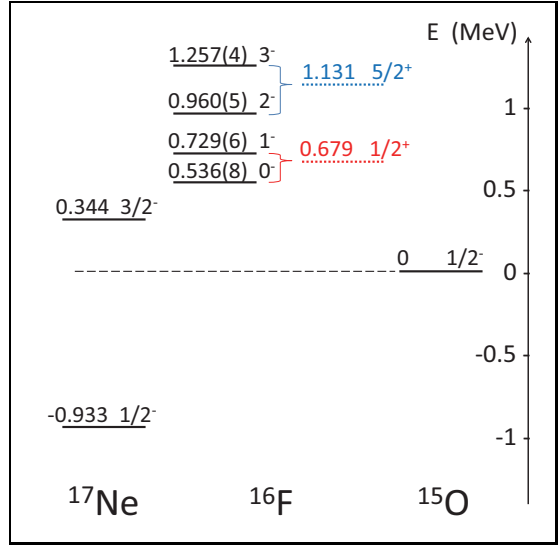}}
\caption{The level scheme of the $^{17}$Ne nucleus and its isotones. 
All the experimental values, printed by black letters, are quoted 
from the database \cite{NNDCHP}, 
except for the first excited state of $^{17}$Ne \cite{06Gri}. 
For the decay widths of $^{16}$F, the experimental data are 
$\Gamma(0^-)=40 \pm 20$ keV and 
$\Gamma(1^-)<40$ keV for the lower two levels, whereas 
$\Gamma(2^-)=40 \pm 30$ keV and 
$\Gamma(3^-)<15$ keV for the upper two levels \cite{NNDCHP}. 
Note that all these levels decay via one-proton emission, where 
the branching ratios to other decay-modes are negligible. 
The values printed by the red and blue letters for $^{16}$F 
indicate the spin-averaged s.p. energies. 
The decay widths of these levels are theoretically computed as 
$\Gamma_0(s_{1/2})=63$ keV and 
$\Gamma_0(d_{5/2})=8.2$ keV, respectively. } \label{fig:17Nelv}
\end{center} \end{figure*}

In this case, we neglect the internal spin of the core, and 
fit the parameters in the core-proton potential 
to the spin-averaged s.p. energies of 
$(s_{1/2})$ and $(d_{5/2})$ states. 
These averaged levels are shown in Figure \ref{fig:17Nelv} 
by the red and blue letters. 
To this end, we calculate the phase shift, $\delta_{lj}(E)$, 
and its derivative for the energy $E$. 
The calculated result is fitted with a function, which consists 
of a pure Breit-Wigner distribution and a smooth background. 
That is, 
\beq
 \frac{d\delta_{lj}(E)}{dE} = \frac{\Gamma_0/2}{\Gamma_0^2/4 + (E_0-E)^2} 
                            + \frac{dC_{lj}(E)}{dE}, \label{eq:sigde}
\eeq
where the right-hand side is the empirical formula. 
How to derive Eq.(\ref{eq:sigde}) and calculate $\delta_{lj}(E)$ 
in the left-hand side is summarized as Appendix \ref{Ap_Scat_2body}. 

The calculated results and fitted functions are shown in 
Figure \ref{fig:409378}. 
At this moment, the smooth background is neglected. 
By fitting the right-hand side in Eq.(\ref{eq:sigde}) 
to the calculated left-hand side, 
we can extract the resonant energy, $E_0$ and 
the decay width, $\Gamma_0$ of considering resonances. 
With $r_0=1.22$ fm, $a_{\rm core}=0.65$ fm, $V_0 = -53.68$ MeV and 
$V_{ls} = 15.06$ ${\rm MeV \cdot fm^2}$ for the $^{15}$O-proton 
potential, we obtained 
$E_0=0.679$ MeV with $\Gamma_0=63$ keV for the $(s_{1/2})$-resonance, and 
$E_0=1.131$ MeV with $\Gamma_0=8.2$ keV for the $(d_{5/2})$-resonance. 
Obtained values of $E_0$ are consistent 
with the empirical resonant energies shown in Fig. \ref{fig:17Nelv}. 
Notice that the values of $r_0$ and $a_0$ are similar to those 
used for $^{18}$Ne and $^{18}$O. 
The other s.p. states are also solved within this potential. 
\begin{figure*}[t] \begin{center}
 \begin{tabular}{c}
  ``$d\delta_{lj} /dE$ of $^{15}$O-proton'' \\
  \begin{minipage}{0.48\hsize} \begin{center}
     (a) for $(s_{1/2})$ \\ \fbox{ \includegraphics[height=48truemm,scale=1, trim = 50 50 10 0]{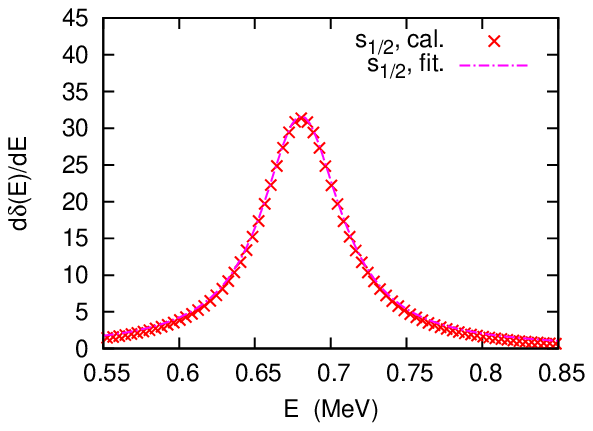}} \\
  \end{center} \end{minipage}
  \begin{minipage}{0.48\hsize} \begin{center}
     (b) for $(d_{5/2})$ \\ \fbox{ \includegraphics[height=48truemm,scale=1, trim = 50 50 10 0]{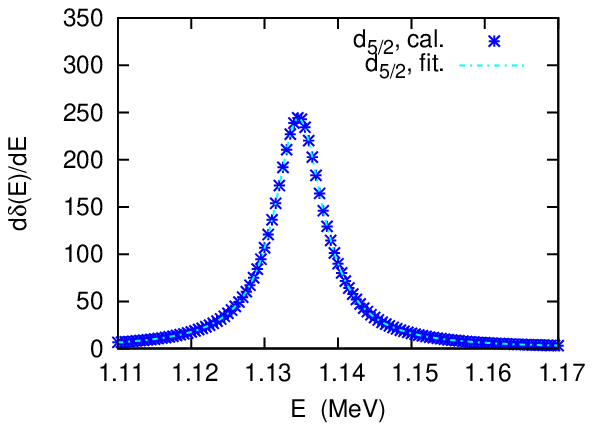}} \\
  \end{center} \end{minipage}
 \end{tabular}
 \caption{The derivative of the phase shift, $\delta_{lj}(E)$ 
for the energy $E$ in the scattering of $^{15}$O-proton. 
The calculated results are shown with symbols, whereas the fitted 
functions given by Eq.(\ref{eq:sigde}) are 
plotted with lines. } \label{fig:409378}
\end{center} \end{figure*}

In order to solve the ground state of $^{17}$Ne, 
we employed the similar setting to that for $^{18}$Ne and $^{18}$O. 
Namely, we take all the s.p. states up to $l_{\rm max}=5$ into account. 
Since the ground state of $^{17}$Ne has the same spin-parity to $^{15}$O, 
which is $(1/2^-)$, 
the two-proton uncorrelated basis can be reduced only to the $0^+$ subspace. 
Consequently, we adopt from $(s_{1/2})^2$ to $(h_{11/2})^2$ partial waves. 
We use $\epsilon_a + \epsilon_b \leq E_{\rm cut} = 30$ MeV as the energy cutoff 
for the uncorrelated basis, providing about 360 bases. 
For the pairing interaction, except for $v_{\rho}$ 
in the density-dependent part, 
we adopt the same parameters as those for $^{18}$Ne and $^{18}$O. 
In order to reproduce the empirical binding energy of the two protons, 
$S_{\rm 2p}=0.93$ MeV \cite{NNDCHP}, we use $v_{\rho}=-1.131v_0$. 

In the setting of the calculations introduced above, 
we treat the pairing correlations as fully as possible, by mixing 
all the uncorrelated bases up to $(h_{11/2})^2$. 
We often refer to this condition as ``full-mixing'' or just ``full'' 
in the following. 
In addition to this ``full-mixing'' case, 
we perform two other sets of calculations with different conditions 
explained below, 
which reveal the essential aspect of the dinucleon correlations. 
\begin{itemize}
 \item {\it Limitation of Core-proton Parity}: \\
As shown in Table \ref{tb:A18_2}, the contributions from 
the partial waves with $(l=odd)^2$ are quite small in $^{18}$Ne and $^{18}$O. 
This situation is expected to occur also in $^{17}$Ne, where the two protons 
may mainly have the $(d_{5/2})^2$ and $(s_{1/2})^2$ configurations. 
Thus, we examine the case by excluding the $(l=odd)^2$ partial waves 
from the calculation, for comparison with the full-mixing case. 
It corresponds to a situation where the parity-mixing in the partial 
core-proton system is prohibited. 
In order to reproduce the \twop-binding energy, we tune the parameter 
$v_{\rho}$ as $v_{\rho}=-0.9981v_0$. 
It meas that, without the parity mixing, we need 
a stronger pairing attraction than that in the full case. 
Notice also that the recoil term, which couples two bases satisfying 
$\abs{l'-l}=1$, do not contribute in this case 
(see Eq.(\ref{eq:3050}) also). 
In the following, we call this setting as ``$(l=even)^2$'' case. 

 \item {\it No Pairing}: \\
In this case, we completely omit the pairing correlations. 
It means that we ignore all the nuclear attraction, 
the Coulomb repulsion and 
the recoil term between the two protons in Eq.(\ref{eq:H3b}). 
Thus, the total Hamiltonian is only 
the uncorrelated Hamiltonian, $h_1+h_2$, 
and the two-proton wave functions becomes identical to 
one of the uncorrelated bases with 
single angular channel. 
We take the $(d_{5/2})^2$ state, which is expected to be the 
major channel in the full-mixing case. 
Because of the lack of the pairing correlations, 
we cannot reproduce the empirical 
binding energy, $S_{\rm 2p}=0.93$ MeV, 
with the original core-proton interaction used in 
the ``full'' and ``$(l=even)^2$ only'' cases. 
Therefore, we inevitably modify $V_{\rm c-N}$. 
We use $V_0=-57.663$ MeV, which yields the {\it bound} s.p. state 
in the $(d_{5/2})$ channel with $\epsilon(d_{5/2})\simeq -0.93/2$ MeV. 
Notice that it is no longer possible to reproduce the Borromean character 
in this case. 
In the following, we call this setting as ``no pairing''. 
\end{itemize}

By comparing among these three cases, 
we will make an attempt to extract the essential character of 
the dinucleon correlation. 
The results and discussions are summarized below. 

\subsection{Energy Expectation Values}
We first discuss the energetic properties tabulated in Table \ref{tb:A17_1}. 
In the full-mixing case, these expectation values show similar 
results to those for $^{18}$Ne shown in Table \ref{tb:A18_1}, 
except for $\Braket{h_1+h_2}$ and $\Braket{h_{\rm c-NN}}$. 
This difference can be interpreted as an effect of the 
weak attractive potential in the core-proton subsystem. 
On the other hand, it is implied that the effects of the pairing 
correlations, 
as well as the relative energy between the two protons, 
are not significantly dependent on the total binding energy. 
The effect of the Coulomb repulsion in the pairing correlation is 
estimated again as a $10\%$ reduction over the nuclear attraction. 
These conclusions are similar to those obtained in Sec. \ref{Sec_4q1y3}. 

In the $(l=even)^2$ case, the situation is significantly different. 
Even though we employ a stronger pairing attraction than in 
the full case, the pairing energy, $\Braket{v_{\rm N-N}}$ has a 
higher value. 
On the other hand, the expectation value of the energy of 
the proton-proton subsystem, $\Braket{h_{\rm N-N}}$ becomes lower. 
Consequently, 
the relative proton-proton kinetic energy, 
$\Braket{p^2_{\rm N-N}/2\mu_{\rm N-N}}=\Braket{h_{\rm N-N}}-\Braket{v_{\rm N-N}}>0$, 
has the lower value than that in the full-mixing case. 
The lower value of $\Braket{p^2_{\rm N-N}/2\mu_{\rm N-N}}$ suggests that 
the spatial distribution between the two protons is 
further expanded, and possibly deviated from a 
diproton-like configuration. 

Finally, in the no pairing case, it is worthwhile to point out that 
$\Braket{h_{\rm N-N}}$ and $\Braket{h_{\rm c-NN}}$ have similar 
values to those in the full case. 
From this result, one may infer that the pairing correlations are 
well mocked up in the mean-field, $V_{\rm c-N}$. 
However, compared with the full-mixing case, 
a discussion on the \twop-configuration is not straight forward, 
because 
the modification of $V_{\rm c-N}$ makes the proton-proton 
subsystem considerably different in the two cases. 
Thus, we will check directly the difference in the spatial 
\twop-distributions in the next subsection. 
\begin{table}[tb] \begin{center}
  \catcode`? = \active \def?{\phantom{0}} 
  \begingroup \renewcommand{\arraystretch}{1.2}
  \begin{tabular*}{\columnwidth}{ @{\extracolsep{\fill}} ccccc c} \hline \hline
                                        && \multicolumn{3}{c}{$^{17}$Ne} & \\ \cline{3-5}
                                        && full   & $(l=even)^2$ only & no pairing, deeper $V_{\rm c-p}$ & \\ \hline
    $\Braket{H_{\rm 3b}}=-S_{\rm 2N}$ (MeV) && $-0.93$  & $-0.93$        & $-0.93$  & \\
    &&&&& \\
    $\Braket{v_{\rm N-N}}$ (MeV)          && $-4.00$  & $-3.86$        & $??0.??$  & \\
    $\Braket{v_{\rm N-N, Nucl.}}$ (MeV)     && $-4.53$  & $-4.34$        & $??0.??$  & \\
    $\Braket{v_{\rm N-N, Coul.}}$ (MeV)     && $??0.53$ & $??0.48$       & $??0.??$  & \\
    $\Braket{\rm recoil}$ (MeV)         && $-0.40$  & $??0.??$       & $??0.??$  & \\
    $\Braket{h_1 + h_2}$ (MeV)           && $??3.47$ & $??2.93$       & $-0.93$  & \\
    $\Braket{V_{\rm c-N_1} + V_{\rm c-N_2}}$ (MeV) && $-10.64$ & $-11.75$       & $-12.82$  & \\
    &&&&& \\
    $\Braket{h_{\rm N-N}}$ (MeV)          && $??5.61$ & $??3.02$        & $??5.57$  & \\
    $\Braket{h_{\rm c-NN}}$ (MeV)         && $-6.54$  & $-3.95$         & $-6.50$  & \\
  \hline \hline \end{tabular*}
  \endgroup
  \catcode`? = 12 
  \caption{ The energy expectation values for the ground state of 
$^{17}$Ne, calculated with the three-body model of $^{15}$O+p+p. 
See the text for the details of each calculational setting. 
The experimental two-proton separation energy is 
$S_{\rm 2p}=0.93$ MeV \cite{NNDCHP}. 
All quantities are evaluated in the same manner as 
in Table \ref{tb:A18_1}. } \label{tb:A17_1}
\end{center} \end{table}

\subsection{Structural Properties}
The results for the structural properties of $^{17}$Ne are shown 
in Table \ref{tb:A17_2} and 
Figs. \ref{fig:441}, \ref{fig:442} and \ref{fig:443}. 
All the quantities are evaluated and plotted in the same manner as 
those in Table \ref{tb:A18_2} and Figure \ref{fig:4181}. 
\begin{table}[tb] \begin{center}
  \catcode`? = \active \def?{\phantom{0}} 
  \begingroup \renewcommand{\arraystretch}{1.2}
  \begin{tabular*}{\columnwidth}{ @{\extracolsep{\fill}} ccccc c} \hline \hline
                       && \multicolumn{3}{c}{$^{17}$Ne}    & \\ \cline{3-5}
                       && full  & $(l=even)^2$ only & no pairing, deeper $V_{\rm c-p}$ & \\ \hline
    $\sqrt{\Braket{r^2_1}}=\sqrt{\Braket{r^2_2}}$ (fm) && 3.92  & 3.70  & 3.67 & \\
    $\sqrt{\Braket{r^2_{\rm N-N}}}$ (fm)        && 4.98  & 5.24  & 5.20 & \\
    $\sqrt{\Braket{r^2_{\rm c-NN}}}$ (fm)       && 3.03  & 2.62   & 2.60 & \\
    $\Braket{\theta_{12}}$ (deg)      && 81.2  & 90.0  & 90.0 & \\
    &&&&& \\
    $(s_{1/2})^2$ (\%)        && 14.63 & 13.24 & ??0.?  & \\
    $(d_{5/2})^2$ (\%)        && 77.97 & 82.64 & 100.?  & \\
    $(p_{3/2})^2$ (\%)        && ?0.79 & ?0.?? & ??0.?  & \\
    $(p_{1/2})^2$ (\%)        && ?0.23 & ?0.?? & ??0.?  & \\
    others,$(l=evev)^2$ (\%) && ?3.87 & ?4.12 & ??0.? & \\
    others,$(l=odd)^2?$ (\%) && ?2.51 & ?0.?? & ??0.? & \\
    &&&&& \\
    $P(S_{12}=0)$ (\%)        && 82.62 & 82.39 & 60.00 & \\
  \hline \hline \end{tabular*}
  \endgroup
  \catcode`? = 12 
  \caption{The structural properties in the ground state of 
$^{17}$Ne, calculated with the three-body model of $^{15}$O+p+p. 
See the text for the details of each calculational setting. 
The radius of the core nucleus is estimated 
as $R_0=r_0A_c^{1/3}=1.22\cdot 15^{1/3}\cong 3.009$ fm. } \label{tb:A17_2}
\end{center} \end{table}

In the full-mixing case, we first find that the general features 
of $^{17}$Ne are similar to those of $^{18}$Ne, although the binding energy 
is remarkably smaller. 
This smaller energy yields the sizable extension of 
the \twop-density distribution, 
shown in Figs. \ref{fig:441}(a), (b) and (d). 
Consequently, the expectation values of the radial parameters 
become larger as one can see in Table \ref{tb:A17_2}. 
This extension of the \twop-wave function is consistent to an increment 
of the $(s_{1/2})^2$ wave, which has a long tail outside 
the core-nucleon potential. 
We also stress that the major components in the \twop-wave function are 
$(s_{1/2})^2$ and $(d_{5/2})^2$, 
reflecting the existence of the $(s_{1/2})$- 
and $(d_{5/2})$-resonant states in the core-proton subsystem. 
The dominance of the $(d_{5/2})^2$ wave in the ground state can be 
understood from Eq.(\ref{eq:ME12g}), 
which indicates that the matrix element 
of a coupled operator, $O(\xi_1,\xi_2)$, has a larger value for the 
uncorrelated basis with larger $j$. 
Therefore, to gain a deeper binding energy, the two protons 
tend to occupy the $(d_{5/2})^2$. 
\begin{figure*}[t] \begin{center}
  \begin{tabular}{c} 
    ``$^{17}$Ne (g.s.), $v_{\rm N-N}=$DDC+Coul., full-mixing'' \\
    \begin{minipage}{0.48\hsize}
     (a) \\ \fbox{ \includegraphics[height=42truemm,scale=1, trim = 60 50 0 0]{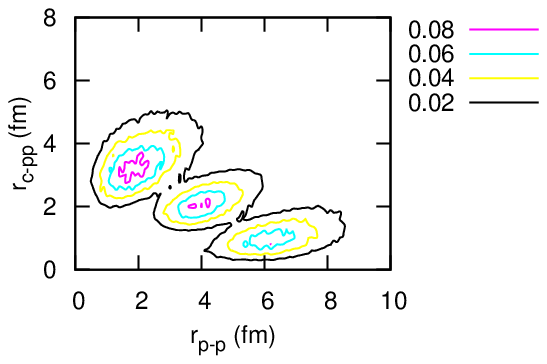}}\\ 
     (c) \\ \fbox{ \includegraphics[height=42truemm,scale=1, trim = 50 50 0 0]{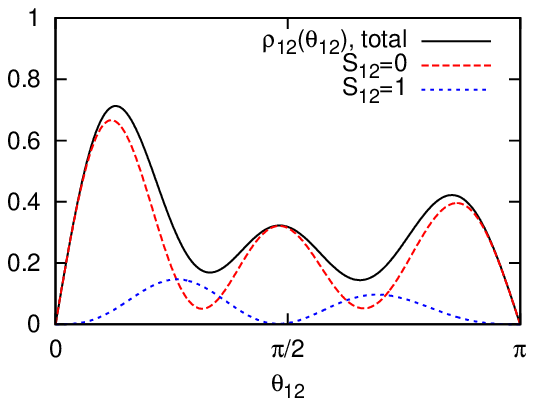}}
    \end{minipage}
    \begin{minipage}{0.48\hsize}
     (b) \\ \fbox{ \includegraphics[height=42truemm,scale=1, trim = 50 50 0 0]{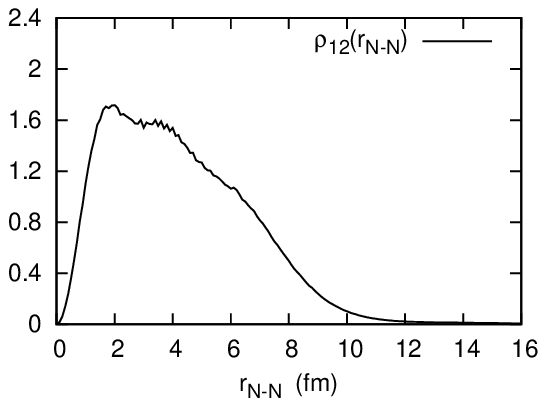}}\\ 
     (d) \\ \fbox{ \includegraphics[height=42truemm,scale=1, trim = 50 50 0 0]{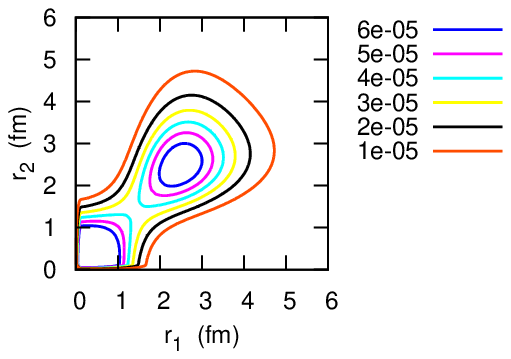}}
    \end{minipage}
  \end{tabular}
\caption{The density distribution of the 
valence two protons in $^{17}$Ne, computed with 
the three-body model of $^{15}$O+p+p. 
In this case, all the uncorrelated bases up to $(h_{11/2})^2$ are 
fully taken into 
account (the full-mixing case). 
The coordinates for all the panels are defined in a 
similar way as Figure \ref{fig:4181}. } 
\label{fig:441} 
\end{center} \end{figure*}

In order to discuss the diproton correlation, it is useful to 
compare the results obtained with the three settings for 
calculations. 
First, in panels (a), (b) and (c) of Fig. \ref{fig:441} for the 
full-mixing case, 
the localization of the density 
at small values of $r_{\rm p-p}$ and of $\theta_{12}$ 
can be seen, 
whereas the localization 
cannot be observed in both 
Fig. \ref{fig:442} for the $(l=even)^2$ only case and 
Fig. \ref{fig:443} for the no-pairing case. 
Notice that this localization in the full-mixing case 
is mainly due to the spin-singlet configuration, 
as shown in Fig. \ref{fig:441}(c). 
This implies the existence of the diproton correlation, 
or at least, its tendency in the ground 
state of weakly bound $^{17}$Ne. 

We also point out that the spatial localization of the two protons 
in $^{17}$Ne is less significant than that in $^{18}$Ne. 
For example, by comparing panels (b) and (d) 
in Figs. \ref{fig:4181} and \ref{fig:441}, 
one can find that the \twop-density distribution shows a larger 
extent in $^{17}$Ne. 
The expectational values, 
$\Braket{r_{N-N}}$ and $\Braket{\theta_{12}}$, 
also have larger values in $^{17}$Ne. 
This result can be interpreted as the effect of the density-dependence 
of the pairing correlation. 
If the density is too low, 
the pairing correlation decreases, 
and eventually vanishes in the zero-density limit \cite{06Mats}. 
In $^{17}$Ne, the valence two protons possibly feel 
the surrounding density, 
which is rather low due to the weakly bound system, and 
causes the diproton correlation to be weaker. 
Whether this tendency leads to the reduction of the diproton correlation 
from unbound systems or not will be a critical point when we analyze 
\twop-emissions. 
We will discuss this point in Chapter \ref{Ch_Results2}. 
\begin{figure*}[t] \begin{center}
  \begin{tabular}{c} 
    ``$^{17}$Ne (g.s.), $v_{\rm N-N}=$DDC+Coul., $(l=even)^2$ only'' \\
(Figure is hidden in open-print version.)
  \end{tabular}
\caption{The same as Figure \ref{fig:441} but for the $(l=even)^2$ case. } 
\label{fig:442} 
\end{center} \end{figure*}
\begin{figure*}[htb] \begin{center}
  \begin{tabular}{c} 
    ``$^{17}$Ne (g.s.), no pairing'' \\
(Figure is hidden in open-print version.)
  \end{tabular}
  \caption{The same as Figure \ref{fig:441} but without the 
pairing correlations, 
for which a modified core-proton potential is employed. 
Notice that the two protons have a pure $(d_{5/2})^2$ configuration. } 
\label{fig:443} 
\end{center} \end{figure*}

In the $(l=even)^2$ case, 
the density $\rho_{12}(r_{\rm N-N})$ shown in Fig. \ref{fig:442}(b) 
shows a larger extent than that in the full case. 
This is consistent with the larger value of $\sqrt{\Braket{r^2_{\rm N-N}}}$ 
in Table \ref{tb:A17_2}. 
The angular distribution in Fig. \ref{fig:442}(c) has a completely 
symmetric form, yielding $\Braket{\theta_{12}}=90$ (deg). 
From these results, we can conclude that the parity-mixing in the 
core-nucleon subsystem is indispensable to induce the spatial 
localization of two nucleons. 
In other words, if this parity-mixing is forbidden or excessively suppressed, 
two nucleons cannot be localized even with a strong pairing interaction. 

One should remember that, even though the two protons are not 
localized, the $(l=even)^2$ case does not mean the complete 
lack of the pairing correlations. 
In Table \ref{tb:A17_2}, the enhancement of the spin-singlet configuration 
can be seen, as well as in the full-mixing case. 
Comparing Fig. \ref{fig:442}(c) with Fig. \ref{fig:443}(c), 
one can find that the peaks at $\theta_{12} \simeq \pi/6$ and $5\pi/6$ 
become more significant in the $(l=even)^2$ case 
than those in the no pairing case. 
These show that a part of the pairing correlations is taken into account, 
even though the diproton correlation is missing.

\section{Interaction-Dependence of Diproton Correlation}
It is also useful to check a model-dependence of the results in 
the previous section, which showed the possibility of the dinucleon 
correlations in weakly bound systems. 
In order to clarify this point, 
in this section, we repeat the same calculations 
for $^{17}$Ne, but employing a different pairing interaction. 

\subsection{Minnesota Potential}
To this end, we adopt the ``Minnesota potential'' for the 
pairing interaction instead of 
the DDC potential. 
This potential was originally proposed by Thompson {\it et.al.}, 
in order to solve nucleon-nucleus scattering problems within the microscopic 
``resonating group method'' \cite{77Thom}. 
For the proton-proton and the neutron-neutron systems, 
the potential is given as 
\beq
 v_{\rm N-N}(r_{12}) = v_0 e^{-b_0 r^2_{12}} - v_1 e^{-b_1 r^2_{12}} \label{eq:Minne}
 + \alpha \hbar c \frac{e^2}{r_{12}} \frac{(1+\hat{t}^{(3)}_{1}) (1+\hat{t}^{(3)}_{2})}{4}, 
\eeq
with $r_{12} \equiv \abs{\bir_1 - \bir_2}$ and $\alpha$ is the 
fine structure constant. 
The Coulomb part is necessary only in the proton-proton case. 
For this interaction, the nuclear part has finite rages, 
in contrast to the zero-range DDC pairing. 
The first term is a phenomenological repulsive part, 
whereas the second term describes the pairing attractions. 
The original parameters in Eq.(\ref{eq:Minne}) were given as 
\beq
 v_0 = 200.~{\rm MeV},~~~~~b_0 = 1.487~{\rm fm}^2, 
\eeq
for the repulsive part, whereas 
\beq
 -v_1 = \left\{ \begin{array}{cc} -178.~{\rm MeV},&(S_{12}=0) \\
                                 -91.85~{\rm MeV},&(S_{12}=1) \end{array} \right. ~~~~~
  b_1 = \left\{ \begin{array}{cc} 0.639~{\rm fm}^2,&(S_{12}=0) \\
                                  0.465~{\rm fm}^2,&(S_{12}=1) \end{array} \right. \label{eq:Minne3}
\eeq
for the attraction, which depends on the total spin of the two nucleons. 
In Eq.(\ref{eq:Minne3}), the parameters for the spin-singlet 
configuration were determined so as to reproduce the proton-proton, 
spin-singlet, $s$-wave scattering properties. 
On the other hand, for the spin-triplet configuration, 
those parameters were determined consistently to the neutron-proton, 
spin-triplet, $s$-wave scattering 
properties \footnote{In the neutron-proton, spin-triplet case, 
Thompson {\it et al.} modified the potential from Eq.(\ref{eq:Minne}) by 
using an additional parameter \cite{77Thom}. 
We do not use this n-p potential in this thesis. }. 
In several theoretical studies based on few-body models for 
finite nuclei, the Minnesota potential has been employed 
with reasonable successes \cite{77Thom,01Myo,04Suzu,07Hagi_03,10Myo}. 
In Figure \ref{fig:vmin}, the potentials for the proton-proton and 
the neutron-neutron channels are shown. 
\begin{figure*}[tb] \begin{center}
\begin{tabular}{c} 
  \begin{minipage}{0.48\hsize} \begin{center}
     (a) $S_{12}=0$ \\ \fbox{ \includegraphics[height=50truemm,scale=1, trim = 60 50 10 0]{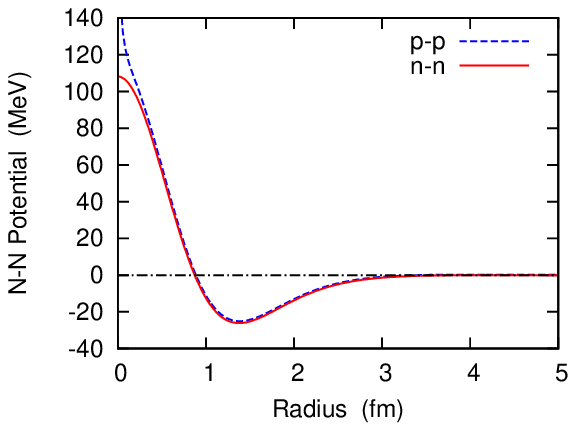}}\\ 
  \end{center} \end{minipage}
  \begin{minipage}{0.48\hsize} \begin{center}
     (b) $S_{12}=1$ \\ \fbox{ \includegraphics[height=50truemm,scale=1, trim = 60 50 10 0]{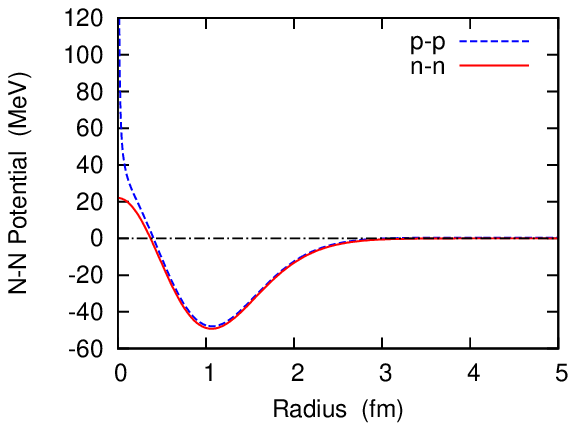}}\\ 
  \end{center} \end{minipage}
\end{tabular}
\caption{The original Minnesota potentials in the $S_{12}=0$ 
(the left panel) and the $S_{12}=1$ (the right panel) channels. 
In the proton-proton case, the Coulomb term is also included. 
The dashed curves are for the proton-proton, while the solid 
lines are for the neutron-neutron. } \label{fig:vmin}
\end{center} \end{figure*}

In our calculations for $^{17}$Ne, however, the original set of 
parameters underestimates the empirical \twop-separation energy. 
We thus weaken $v_0$ to $178.1$ MeV, which effectively 
enhances the pairing attraction. 
One should be conscious of that, in this case, a stronger 
pairing attraction is adopted inside nuclei, compared with 
the bare pairing attraction in the vacuum. 
It is quite contrary to the previous case with the DDC pairing interaction, 
where we needed to reduce the pairing attraction inside nuclei to reproduce 
the \twop-binding energy. 
We do not know exactly the origin for the difference, 
but one possibility is due to the range of the pairing attraction. 

To calculate the matrix elements 
of the Minnesota potential based on Eq.(\ref{eq:ME12g}), 
we have to know the multi-pole expansion formula of the Gaussian function. 
This was given by Swiatecki \cite{51Swi}, and thus we do not 
show it here. 
We stress that, except using the Minnesota pairing, our calculations 
were performed within the same assumption to that 
in the full mixing case with DDC pairing. 

\subsection{Results and Comparison}
The results obtained with the Minnesota pairing are summarized in Table 
\ref{tb:17Ne_min} and Figure \ref{fig:461}, 
in the same manner as in the previous section. 
Qualitatively, energy expectational values are independent of the 
choice of the pairing interaction. 
Indeed, $\Braket{v_{\rm N-N,Nucl.}}$ is similar to one another, 
even it is evaluated slightly lower than 
that in the DDC+Coul. case, which 
maybe due to a character of the finite-range potential. 
\begin{table}[tb] \begin{center}
  \catcode`? = \active \def?{\phantom{0}} 
  \begingroup \renewcommand{\arraystretch}{1.2}
     \begin{tabular*}{\columnwidth}{c|c} \hline \hline
     \multicolumn{2}{c}{$^{17}$Ne (g.s.), $v_{\rm N-N}=$Minne.+Coul., full-mixing} \\ \hline
     \begin{minipage}{0.45\hsize} \begin{center}
         \begin{tabular*}{\columnwidth}{ @{\extracolsep{\fill}} cc}
         $\Braket{H_{\rm 3b}}=-S_{\rm 2N}$ (MeV) & $-0.93$ \\
         & \\
         $\Braket{v_{\rm N-N}}$ (MeV) & $-3.52$ \\
         $\Braket{v_{\rm N-N, Nucl.}}$ (MeV) & $-4.03$ \\
         $\Braket{v_{\rm N-N, Coul.}}$ (MeV) & $??0.51$ \\
         $\Braket{\rm recoil}$ (MeV) & $-0.31$ \\
         $\Braket{h_1 + h_2}$ (MeV) & $??2.90$ \\
         $\Braket{V_{\rm c-N_1} + V_{\rm c-N_2}}$ (MeV) & $-11.1$ \\
         & \\
         $\Braket{h_{\rm N-N}}$ (MeV) & $??5.41$ \\
         $\Braket{h_{\rm c-NN}}$ (MeV) & $-6.34$ \\ & \\ & \\ \end{tabular*}
     \end{center} \end{minipage} 
     &
     \begin{minipage}{0.45\hsize} \begin{center}
         \begin{tabular*}{\columnwidth}{ @{\extracolsep{\fill}} cc}
         $\sqrt{\Braket{r^2_1}}=\sqrt{\Braket{r^2_2}}$ (fm) & 3.81 \\
         $\sqrt{\Braket{r^2_{\rm N-N}}}$ (fm) & 4.95 \\
         $\sqrt{\Braket{r^2_{\rm c-NN}}}$ (fm) & 2.89 \\
         $\Braket{\theta_{12}}$ (deg) & 83.37 \\
         & \\
         $(s_{1/2})^2$ (\%) & 12.12 \\
         $(d_{5/2})^2$ (\%) & 84.99 \\
         $(p_{3/2})^2$ (\%) & ?0.42 \\
         $(p_{1/2})^2$ (\%) & ?0.10 \\
         others,$(l=evev)^2$ (\%) & ?0.82 \\
         others,$(l=odd)^2?$ (\%) & ?1.55 \\
         & \\
         $P(S_{12}=0)$ (\%) & 72.73 \\ \end{tabular*}
     \end{center} \end{minipage} \\ \hline \hline
     \end{tabular*}
  \endgroup \catcode`? = 12 
  \caption{The properties of the ground state of $^{17}$Ne, 
obtained with the Minnesota interaction. 
All the uncorrelated bases up to $(h_{11/2})^2$ are taken into account. 
All the quantities are evaluated in a similar way as in 
Table \ref{tb:A17_1} and \ref{tb:A17_2}. } \label{tb:17Ne_min}
\end{center} \end{table}

Even though the relative proton-proton energy is less evaluated 
with the Minnesota potential, 
the tendency of the diproton correlation can be apparent again 
in this case. 
The structural properties shown in Table \ref{tb:17Ne_min} and 
the \twop-density distribution shown in Figure \ref{fig:461} 
exhibit the similar behaviors to those in the DDC+Coul. case. 
The spatial localizations at 
$r_{\rm p-p}\simeq 2$ fm in Fig. \ref{fig:461}(a), 
and at $\theta_{12}\simeq \pi/6$ in Fig. \ref{fig:461}(c) are 
clearly seen 
by taking the core-proton parity-mixing into account. 
The spin-singlet configuration carries the main part of 
this localization as one can see in Fig. \ref{fig:461}(c). 
One may concern the spin-singlet ratio which is slightly smaller 
than that in the DDC+Coul. case. 
However, it is significantly larger than $P(S_{12}=0)=60$ \% 
in the no pairing case, and the enhancement 
of the spin-singlet configuration remains qualitatively 
also with the Minnesota potential. 
\begin{figure*}[t] \begin{center}
  \begin{tabular}{c} 
    ``$^{17}$Ne (g.s.), $v_{\rm N-N}=$Minne.+Coul., full-mixing'' \\
    \begin{minipage}{0.48\hsize}
     (a) \\ \fbox{ \includegraphics[height=42truemm,scale=1, trim = 60 50 0 0]{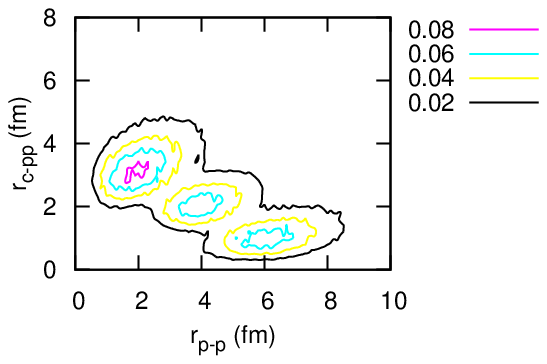}}\\ 
     (c) \\ \fbox{ \includegraphics[height=42truemm,scale=1, trim = 50 50 0 0]{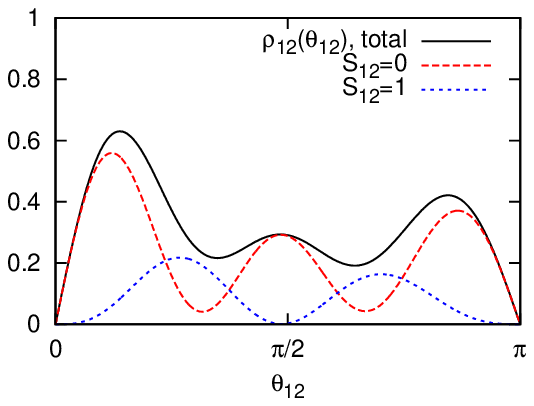}} 
    \end{minipage}
    \begin{minipage}{0.48\hsize}
     (b) \\ \fbox{ \includegraphics[height=42truemm,scale=1, trim = 50 50 0 0]{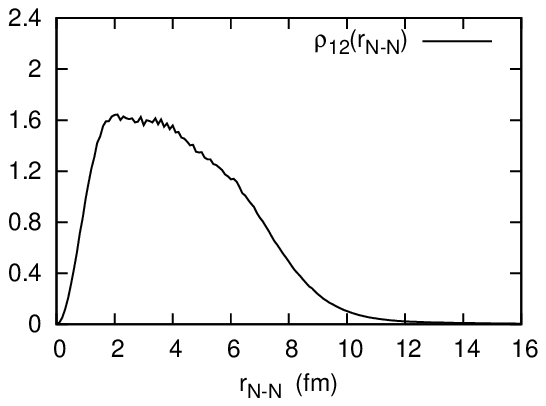}}\\ 
     (d) \\ \fbox{ \includegraphics[height=42truemm,scale=1, trim = 50 50 0 0]{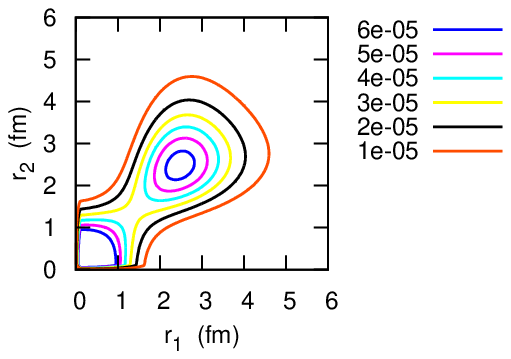}} 
    \end{minipage}
  \end{tabular}
  \caption{The same as Figure \ref{fig:441}, 
but obtained with the Minnesota interaction. } \label{fig:461}
\end{center} \end{figure*}

\section{Summary of this Chapter}
We demonstrated the appearance of the dinucleon correlations in 
the ground states of several light nuclei based on the 
core plus two-nucleon model. 
It is found that the Coulomb repulsive force plays a minor 
role in the pairing correlation, and the diproton correlation 
can be realized in proton-rich nuclei, 
in a similar way as the dineutron correlation in neutron-rich nuclei. 
That is, 
our evaluation of the Coulomb effect, which is about a 10\% 
reduction against the nuclear pairing attraction, is 
not sufficient to affect the spatial localization of the two protons. 
We also confirmed that these correlations are not significantly 
dependent on the total binding energy of nuclei. 
In other words, the dinucleon correlations are present 
not only in weakly bound but also in stable nuclei, as long as 
the pairing correlation is sufficiently large. 

As mentioned in Chapter \ref{Ch_2}, 
the dinucleon correlations in the bound, ground states are not 
easy to be directly probed. 
The sensitivity of observables to these intrinsic structures is 
still not evident, although various possibilities 
have been explored. 
Facing on this situation, we propose a possibility to verify 
the diproton correlation with the two-proton emissions and 
radioactive decays. 
Because the valence two protons are spontaneously 
emitted without no disturbance from the external fields, 
observing their wave functions may be a 
direct probe into the diproton correlation in a resonant state. 
Based on this idea, we will extend our analysis to 
a meta-stable three-body system after this Chapter. 
In the next Chapter, we will summarize the historical back-ground of 
the two-proton emissions and radioactive decays. 
We will introduce the time-dependence into our three-body model 
in order to describe the two-proton emissions in Chapter \ref{Ch_TDM}. 
Employing this model, 
we will the diproton correlation associated with 
the two-proton emissions, whose results will be present 
in Chapters \ref{Ch_Results2} and \ref{Ch_Results3}.

\include{end}
\documentclass[a4paper,12pt]{book}
\include{begin}

\chapter{Review of Two-Proton Decay and Emission} \label{Ch_5}
The two-proton decay and emission are characteristic decay-modes of 
nuclei beyond the proton-dripline. 
We review in this Chapter the theoretical and experimental studies 
of these phenomena, with their relevant topics. 

In nuclear physics, 
there have been five major radioactive processes in which 
one or several nucleons are emitted from the parent nuclei 
\footnote{There are also known radioactive processes by 
the weak or the electro-magnetic interactions. 
In this thesis, however, we push aside these topics. }. 
Those are (i) alpha decay, 
(ii) one-proton and one-neutron decays, 
(iii) two-proton and two-neutron decays, 
(iv) heavier cluster decay, and 
(v) nuclear fission. 
All these processes belong to the quantum meta-stable phenomena by 
the nuclear interaction. 

Needless to say, the alpha-decay is one of the most famous 
nuclear radioactive processes. 
In many standard textbooks of nuclear physics, 
this problems is discussed as an tunneling problem of a point-like 
alpha-particle. 
However, it is also known that the emitted alpha-particle is 
a composite system of four nucleons. 
Therefore, to describe the alpha-decay properly, 
one would need a microscopic framework including 
many-body effects. 
There have been several theoretical studies based this 
consideration \cite{70Sasa,79Tono,94Varga,12Betan}. 
However, mainly because of a difficulty to handle with many-body 
correlations, there have been no quantitatively successful works yet. 
We also note that the physics of meta-stable states with intrinsic 
degrees of freedom, or of many particles 
are one of the major subjects in modern physics. 
They occupy the essential positions not only in nuclear physics, 
but also in molecular, condensed matter and astro-nuclear physics. 
Famous examples include, e.g. nuclear fissions and fusions, 
resonances of cold atoms and Jossefson effects. 
A unified study of multi-fermion meta-stable systems in different 
scales might be useful in gaining a deeper 
understanding of our world. 

In the following, we mainly focus on the \twop-decay and emission, 
whereas other processes will be briefly or never mentioned. 
In these processes, two protons are emitted simultaneously or 
sequentially from the parent nucleus with an even-number of protons. 
Because of the remarkable developments in the experimental 
techniques \cite{08Blank,09Blank,09Gri_40,12Pfu}, for recent about 10 years, 
two-proton (\twop-) emitters have been one of the main topics in radioactive 
nuclear physics, and knowledge about \twop-emissions and 
radioactive decays have been accumulated. 
Recently, furthermore, 
the two-proton (\twop-) emission has attracted 
much attention as an useful tool to probe the diproton correlation. 
We detail this history in the following. 

\section{History of Two-Proton Decay and Emission}
Comparing with alpha-decays and fissions, 
the two-proton emissions and decays are much simpler. 
Moreover, those are the dynamical phenomena, including 
the pairing correlations which are unique in multi-fermion systems. 
By studying \twop-decays, it has been expected to provide 
the benchmark for the quantum meta-stability of many fermions. 
\begin{figure*}[t] \begin{center}
  \begin{tabular}{c} 
     \begin{minipage}{0.48\hsize} \begin{center}
        (a) true \twop-emitter \\
        \fbox{ \includegraphics[width=0.9\hsize, clip, trim = 0 0 0 0]{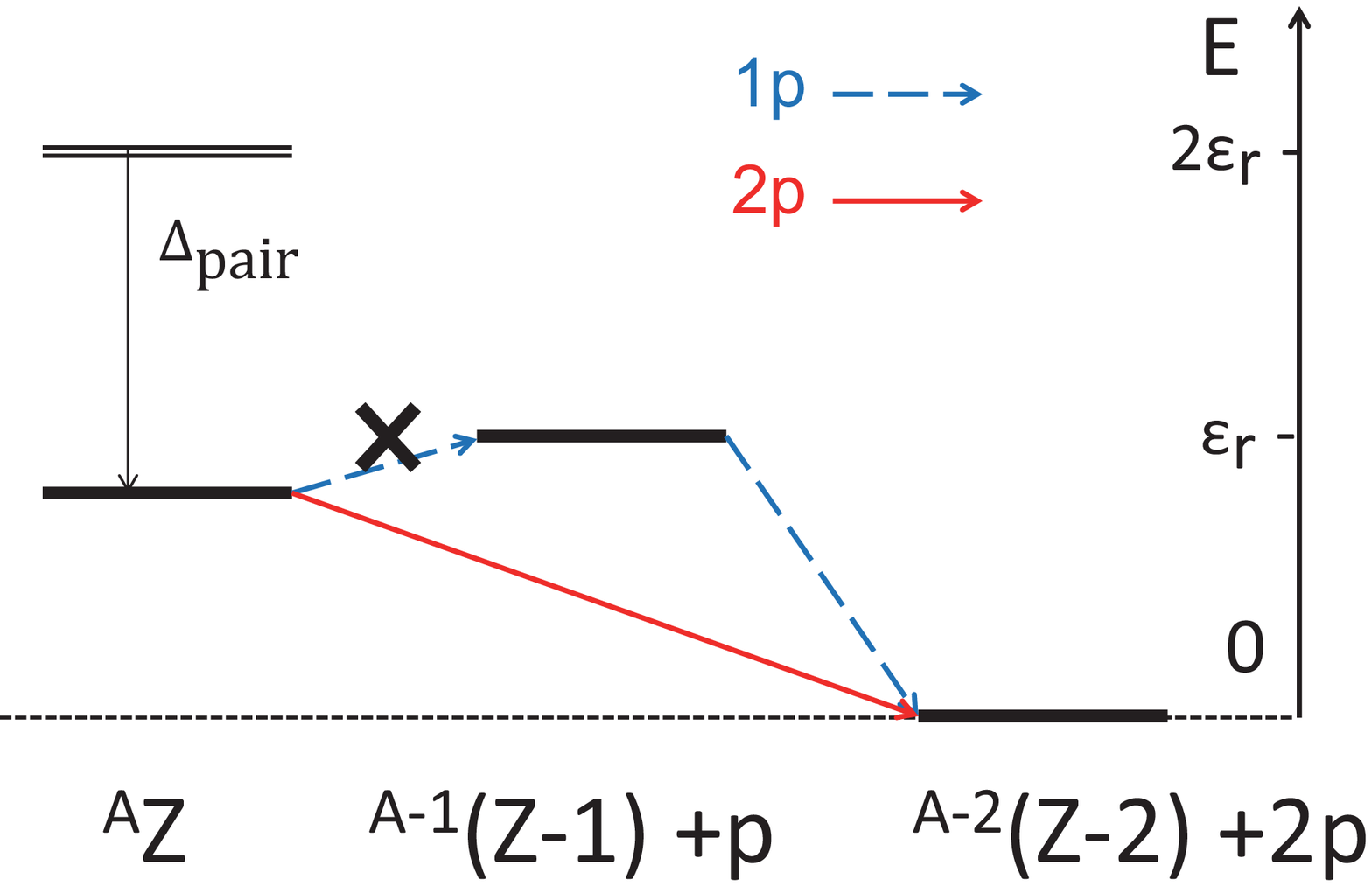}}
     \end{center} \end{minipage}
     \begin{minipage}{0.48\hsize} \begin{center}
        (b) sequential \twop-emitter \\
        \fbox{ \includegraphics[width=0.9\hsize, clip, trim = 0 0 0 0]{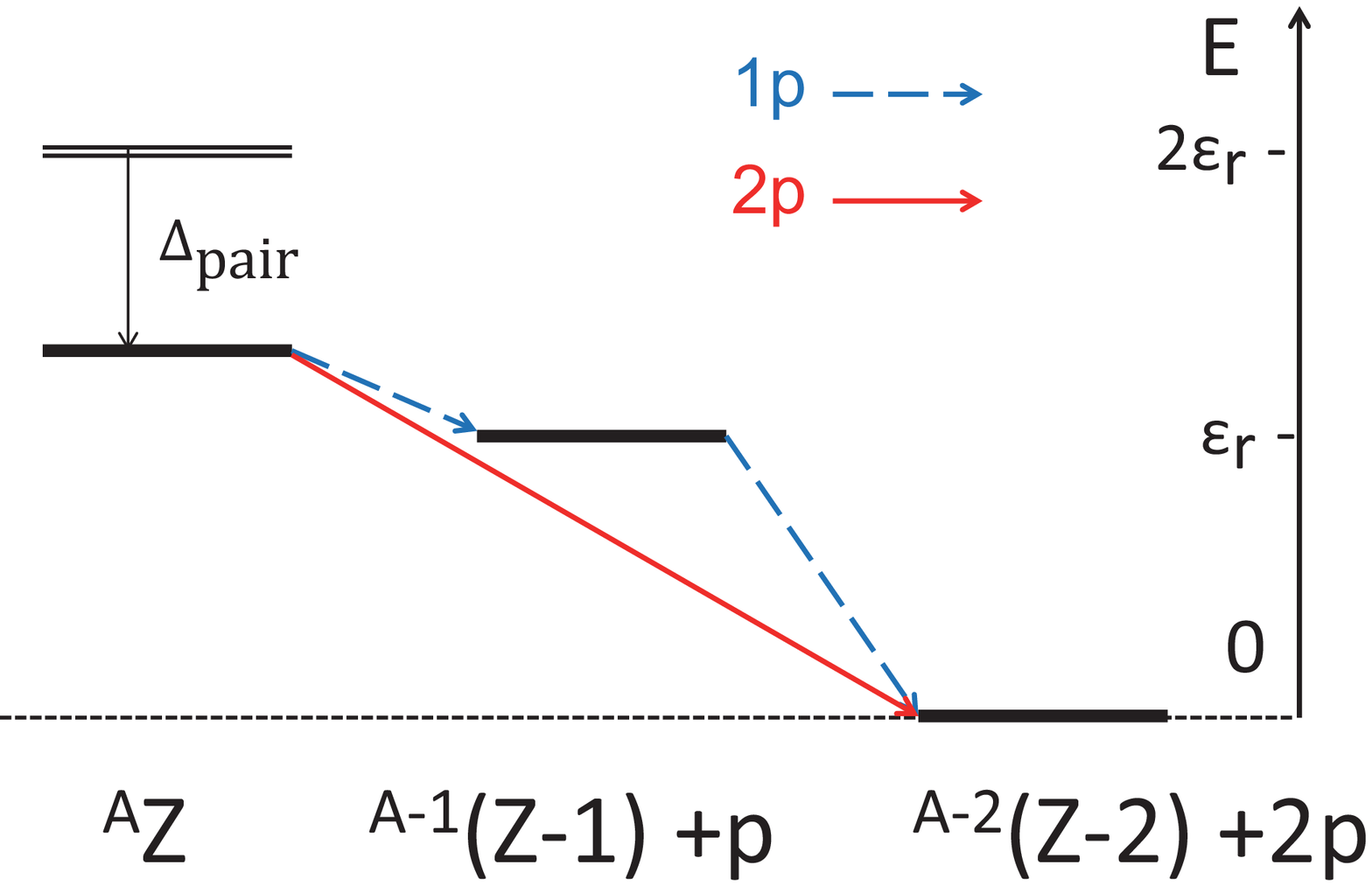}}
     \end{center} \end{minipage}
  \end{tabular}
\caption{Schematic figures for the energy conditions of the true \twop-emitters (the left panel) and 
the sequential \twop-emitters (the right panel). 
The parent nucleus is indicated as $^AZ$. 
Each level is measured from the ground state of 
the daughter nucleus, $^{A-2}(Z-2)$. 
It is assumed that there are no bound $1p$- and $2p$-states in the 
$^{A-1}(Z-1)$ and $^AZ$ nuclei, respectively. 
The symbol $\Delta_{\rm pair}$ means the pairing energy gap due to the 
pairing attraction. 
The decay widths are not considered in these schematic figures. } \label{fig:51}
\end{center} \end{figure*}

\subsection{Early Studies: before the 21th Century}
The first prediction of \twop-decays was done by 
V. I. Goldansky \cite{60Gold,61Gold}. 
In his theoretical study, 
he considered two different, 
simple situations for \twop-emissions: ``true'' and ``sequential'' 
\twop-emitters. 
In Fig. \ref{fig:51}, we schematically describe these two situations, 
as the energy conditions for a parent nucleus 
with respect to the nuclei after $1p$- and \twop-emissions. 

In the true \twop-emitter, 
the energy pf the \twop-resonance of a parent nucleus 
is lower than the $1p$-resonance of its isotone after the $1p$-emission. 
In this situation indicated in Fig. \ref{fig:51}(a), 
only the simultaneous \twop-emission is allowed, whereas the emission 
of single proton is forbidden. 
In the pure mean-field approximation, 
this situation is never realized as long as 
the intermediate nucleus, $^{A-1}(Z-1)$, 
has no bound single-proton states: 
the \twop-resonance has the energy of $2\epsilon_r$, 
where the $\epsilon_r>0$ is the single-proton resonance energy in the 
intermediate nucleus. 
Thus, in order to realize the true \twop-emission, 
the pairing energy gap must be large enough to pull down 
the \twop-resonance under the single-proton resonance. 
This energy gap is, of course, caused by the pairing interaction. 
Based on the quasi-classical formulas, Goldansky showed that, 
in the true \twop-emission, the decay width is sharply reduced 
from that of common binary decays \cite{60Gold,61Gold}. 

In Goldansky's pioneer works, he proposed two types of 
the true \twop-emissions. 
The first one is ``diproton'' emission: 
if the pairing attraction is much stronger than the 
Coulomb repulsion, 
two protons are emitted almost as a diproton. 
He showed that, in the case of two protons restricted to the relative 
$(s_{1/2})$-orbit from the core, the penetrability is identical to that 
of a diproton with the total spin $S=0$. 
This observation has been a basic idea of the diproton emission. 
On the other hand, he also proposed ``direct'' emission, 
where the pairing correlation is rather weak. 
In the direct \twop-emission, the observables are mainly governed by the 
core-proton interactions, where the disruption by the pairing 
interaction can be negligible 
(Notice that even if the pairing energy gap is large enough, 
the effect of pairing correlations may 
be minor for the emitted particles). 
Consequently, the diproton and direct decays or emissions 
correspond to the extreme conditions with relatively 
strong and weak pairing correlations, respectively. 
The binary channel of the proton-proton and the core-proton plays 
a dominant role in the diproton and the direct emissions, respectively. 

However, it is important to notice that 
the actual true \twop-emissions are not so simple that 
they usually show an intermediate character between the diproton and 
direct emissions. 
It is often necessary to deal with three particles without 
any discrimination. 
This process is referred to as ``democratic'' emission 
\cite{89Boch,09Gri_40,12Ego}, 
where there are no autocrat binary channels. 
Namely, the actual true \twop-emissions are essentially the three-body 
problems, and they cannot be approximated as the product 
of partial two-body problems. 

In the sequential \twop-emitter indicated in Fig. \ref{fig:51}(b), 
on the other hand, 
the $1p$-emission becomes dominant in a parent nucleus. 
This situation can be understood by the mean-field plus a 
small pairing gap, which is insufficient to realize the true 
\twop-emission. 
Thus, the decay process to the final daughter nucleus 
occurs as a sequence of two $1p$-emissions. 
According to this picture, sometimes the process is 
also called ``cascade'' decay. 
In the sequential \twop-emissions, of course, the pairing correlations 
should be minor. 
Observables are almost explained only with the core-proton interactions. 
The total penetrability should be well approximated as 
the bi-product of the penetrabilities for the first and second protons. 
However, one should also notice that the direct \twop-emission 
can takes place also in this situation, if the first and second 
$1p$-emissions occur at the same time. 

After Goldanski's works, these three types of $\twop$-emissions 
have been the major assumptions in almost all the theoretical 
works \cite{64Gali,91Brown,96Nazare,97Woods}. 
We also note that there were less theoretical 
works before 2000. 
The reason for this poor crop may be a shortage of observed 
examples of \twop-emitters, which can be analyzed within the 
theoretical models in those days. 

In the experiments, 
two categories of $\twop$-emissions, namely 
the \twop-emissions from the ground state nuclei and the beta-delayed 
\twop-emissions, have been known. 
The first one includes the $^{6}$Be nucleus, which is the 
simplest \twop-emitter interpreted as the $\alpha$-particle 
with two valence protons 
\cite{66Whaling,77Gree,88Ajzen,84Boch,85Boch_377,87Boch,88Boch,89Boch,92Boch}. 
Its decay width, $\Gamma_{\rm 2p} \cong 92$ keV, was measured 
more than 30 years ago. 
Especially, the works performed by Bochkarev {\it et.al.} have 
presented the benchmark results 
\cite{84Boch,85Boch_377,87Boch,88Boch,89Boch,92Boch}. 
In their results, it was already suggested that the assumption of the 
diproton emission is not valid: 
it leads to an unrealistic property that 
the relative proton-proton kinetic energy is extremely small. 
Thus, the necessity of considering the democratic emission has been 
extensively discussed. 
It is worthwhile to note that the $\alpha$+N+N three-body decay 
from the $2^+$ excited states of $^{6}$He and $^{6}$Li has been 
also observed as well as the $^{6}$Be \cite{85Boch_374,87Boch,94Boch}. 
For these 2N-emissions, the isobaric symmetry in the meta-stable states 
has been discussed. 
This is still an open problem even at present. 

We also mention that similar three-body resonances have been known 
in $^{12}$O \cite{78KeKe,95Kryger} and 
$^{16}$Ne nuclei \cite{83Wood}. 
Compared with $^{6}$Be, these nuclei have comparable or larger 
decay widths of the order of $100$ keV. 
Phenomenologically, investigations of these nuclei is important 
for the universal understanding of \twop-emissions along the 
proton-dripline. 
Nevertheless, they have been less studied in the past. 
Recently, several improvements have been done both 
in the theoretical and experimental studies of these 
nuclei \cite{02Gri,09Gri_40,08Muk,10Muk,12Jager}. 
However, at the same time, 
a new and serious problem has been realized 
that the decay widths of these nuclides are 
too broad to be reproduced 
within the simple three-body model \cite{02Gri}. 
This is still an open problem at present. 
We will mention again this discrepancy in Chapter \ref{Ch_Results3}, 
where the $^{16}$Ne nucleus will be treated within our model. 
\begin{figure}[tb] \begin{center}
(Figure is hidden in open-print version.)
\caption{Figure 14 in Ref.\cite{84Cable}. 
The decay scheme of $^{22}$Al, which is one of the beta-delayed 
\twop-emitters. } \label{fig:1984Cable}
\end{center} \end{figure}

The second category of the \twop-emissions 
is that after the $\beta$-decay 
of a parent nuclei (the beta-delayed \twop-emission) 
\cite{84Cable,91Detraz,92Moltz,98Mukha,00Fynbo,03Fynbo}. 
The most famous example may be the $^{22}$Al \cite{84Cable}. 
In Figure \ref{fig:1984Cable} taken from the Ref.\cite{84Cable}, 
the decay scheme of $^{22}$Al is shown. 
The ground state of this nuclide undergoes the $\beta^+$-decay 
to the $4^+$ state of the 
$^{22}$Mg, which has 12 protons. 
Note that the branching ratio for this decay is very small, being 
about a few percent \cite{84Cable}. 
Apart from the de-excitations and $\alpha$-decays, 
the generated $4^+$ state is unstable against the $1p$- and 
the associated sequential $2p$-emissions. 
We must pay attention to that, in this decay scheme, there are several 
intermediate resonances in the $^{20}$Ne-proton binary channel 
for the $1p$-emission. 
Furthermore, the $^{21}$Na nucleus has the {\it bound} 
single-proton states. 
Thus, there are two destinations of the proton(s)-emissions from the 
$^{22}$Mg: 
one is the bound $1p$-states of the $^{21}$Na 
reached by the $1p$-emission, whereas 
the other is the bound \twop-states of $^{20}$Ne through 
the sequential $2p$-emission. 
Similar complexities in the decay scheme are also in other 
$\beta$-delayed \twop-emitters, such as 
$^{26}$P \cite{84Cable,91Detraz}, $^{31}$Ar \cite{98Mukha,00Fynbo} 
and $^{39}$Ti \cite{92Moltz}. 
Therefore, the theoretical treatments are not simple for these nuclides. 
The effect of pairing correlations may not be significant, 
due to the dominant core-proton binary channels. 
Because of these complexities, 
we do not treat the beta-delayed \twop-emissions in this thesis. 
\begin{table}[t] \begin{center}
  \begingroup \renewcommand{\arraystretch}{1.3}
  \begin{tabular*}{\columnwidth}{ @{\extracolsep{\fill}} ccccc c} \hline \hline
  nuclide           & decay & $E^*$ (keV) & $Q_{\rm 2N}$ (keV) & $\Gamma_{\rm 2N}$ (keV) & other refs. \\ \hline
  $^{6}_{4}$Be$(0^+)$ & $\alpha$+2p & g.s. & 1371(5) & 92(6) \cite{88Ajzen,02Till} & \cite{89Boch},\cite{09Gri_80,09Gri_677}$^c$ \\
  $^{12}_{8}$O$(0^+)$ & $^{10}$C+2p & g.s. & 1790(40) & 578(205) \cite{95Kryger} & \cite{78KeKe,12Jager} \\
  $^{16}_{10}$Ne$(0^+)$ & $^{14}$O+2p & g.s. & 1400(20) & 110(40) \cite{83Wood} & \cite{78KeKe},\cite{08Muk,10Muk}$^c$ \\
  $^{17}_{10}$Ne$(3/2^-)$ & $^{15}$O+2p & 1288(8) & 344(8) & $7.6^{+4.9}_{-3.7}\times 10^{-6}$ \cite{97Chro} & \cite{93Till} \\
  $^{19}_{12}$Mg$(1/2^-)$ & $^{17}$Ne+2p & g.s. & 750(50) & $1.1^{+1.4}_{-0.25}\times 10^{-7}$ \cite{07Muk} & \cite{08Muk}$^c$ \\
  $^{45}_{26}$Fe$(3/2^+)$ & $^{43}$Cr+2p & g.s. & 1154(16) & [$2.8^{+1.0}_{-0.7}$ ms] \cite{05Doss} & \cite{02Pfu,02Gio},\cite{07Mie}$^c$ \\
  $^{48}_{28}$Ni$(0^+)$ & $^{46}$Fe+2p & g.s. & 1350(20) & [$8.4^{+12.8}_{-7.0}$ ms] \cite{05Doss} & \cite{11Pomo} \\
  $^{54}_{30}$Zn$(0^+)$ & $^{52}$Ni+2p & g.s. & 1480(20) & [$3.7^{+2.2}_{-1.0}$ ms] \cite{05Blank} & \cite{11Asch} \\
  &&&&& \\
  $^{6}_{2}$He$(2^+)$ & $\alpha$+2n & 1797(25) & 825 & 113(20) \cite{94Boch} & \cite{87Boch} \\
  $^{16}_{4}$Be$(0^+)$ & $^{14}$Be+2n & g.s. & 1350(100) & $800^{+100}_{-200}$ \cite{12Spyr} & \cite{03Audi} \\
  $^{26}_{8}$O$(0^+)$ & $^{24}$C+2n & g.s. & 150$^{+50}_{-150}$ & ? \cite{12Lund} & \cite{13Kohley_26O,13Caesar} \\ \hline \hline \end{tabular*}
  \endgroup
  \caption{ Table of nuclides in which two-nucleon emissions or 
radioactive decays have been experimentally observed. 
Similar tables can be found in the Refs.\cite{09Gri_40,12Pfu}. 
The 1st column is for the parent nucleus and its spin-parity in the 
reference state. 
The 2nd column indicates decay-modes. 
The 3th column is for the excited energy of the corresponding state, 
measured from the ground state. 
The 4th and 5th columns are for the Q-value and the decay width, 
respectively. 
Q-values are respect to the ground states of the daughter nuclides. 
For some long-lived nuclides, their lifetimes are shown 
instead of the decay widths. 
The 6th column lists the references other than that listed in the 5th column. 
Those which report the \twop-correlation measurements are indicated 
by the superscript $c$. } \label{tb_ch5_1}
\end{center} \end{table}

Even with several examples introduced above, however, there had not 
any observed nuclides, which have a long lifetimes enough to 
characterize the \twop-radioactivity. 
The breakthrough was made at the beginning of the 21th century in the 
experimental side, as we will review in the next subsection. 

\subsection{Modern Studies: after the 21th Century}
At the beginning of 21th century, a great development was made in the 
study of \twop-radioactivity. 
In 2002, the first observation of the true \twop-radioactivity in the 
$^{45}$Fe nucleus was made independently by two experimental groups, 
headed by M. Pf\"{u}tzner \cite{02Pfu} and by B. Blank \cite{02Gio}. 
In these experiments, the $^{45}$Fe nucleus was created by the 
projectile fragmentation with the primary beam of $^{58}$Ne. 
The decay products, namely $^{43}$Cr and two protons, 
were implanted into silicon detectors, where the total energy 
release in the decay can be determined experimentally. 
On the other hand, the identification of $^{43}$Cr was done by 
means of the energy-loss and the time-of-flight measurements. 
From the measured distributions of the energy release, 
the half-life of $^{45}$Fe was determined 
as $T_{1/2} = 3.2^{+2.6}_{-1.0}$ ms \cite{02Pfu} and 
$4.7^{+3.4}_{-1.4}$ ms \cite{02Gio}, which is long enough 
to be characterized as the \twop-radioactivity. 
We also note that, for these experiments, 
theoretical works \cite{91Brown,96Ormand,96Cole,97Ormand} 
played an helpful roles to infer the candidates of the true 
\twop-radioactive nuclides. 
\begin{table}[t] \begin{center}
  \begingroup \renewcommand{\arraystretch}{1.3}
  \begin{tabular*}{\columnwidth}{ @{\extracolsep{\fill}} ccccc c} \hline \hline
   nuclide           & decay & $J^{\pi}_{\rm core}$ & $Q_{\rm 1N}$ (keV) & $\Gamma_{\rm 1N}$ (keV) & other refs. \\ \hline
  $^{5}_{3}$Li$(3/2^-)$ & $\alpha$+p & $0^+$ & 1960(50) & $\simeq 1500$ \cite{88Ajzen} & \cite{00Hoef,02Till} \\
  $^{11}_{7}$N$(1/2^-)$ & $^{10}$C+p & $0^+$ & 2200(100) & 740(100) \cite{95Kryger} & \cite{78KeKe} \\
  $^{15}_{9}$F$(1/2^+)$ & $^{14}$O+p & $0^+$ & 1370(180) & 530(300) \cite{78KeKe} & \cite{03Peters,10Mukhamed} \\
  $^{16}_{9}$F$(0^-)$   & $^{15}$O+p & $1/2^-$ & 535(8) & 40(20) \cite{93Till} & \\
  $^{18}_{11}$Na$(1^-)$ & $^{17}$Ne+p & $1/2^-$ & 1250(110) & $\simeq 700$ \cite{NNDCHP} & \\
  $^{44}_{25}$Mn$(2^-)$ & $^{43}$Cr+p & $3/2^-$ & 1700(600) & [$<151$ ns] \cite{NNDCHP} & \\
  $^{47}_{27}$Co$(?)$    & $^{46}$Fe+p & $0^+$ & 2000(9000) & ? \cite{NNDCHP} & \\
  $^{53}_{29}$Cu$(3/2^-)$ & $^{52}$Ni+p & $0^+$ & $>350$ & [$<188$ ns] \cite{13Blank_49} & \\
  &&&&& \\
  $^{5}_{2}$He$(3/2^-)$ & $\alpha$+n & $0^+$ & 735(20) & 600(20) \cite{88Ajzen} & \cite{02Till} \\
  $^{15}_{4}$Be$(3/2^+)$ & $^{14}$Be+n & $2^+$(?) & $>1540$ & ? \cite{11Spyr} & \cite{03Audi} \\
  $^{25}_{8}$O$(3/2^+)$ & $^{24}$O+n & $0^+$ & $770^{+10}_{-10}$ & 172(30) \cite{08Hoffman} & \cite{13Caesar} \\ \hline \hline \end{tabular*}
  \endgroup
  \caption{ The core-nucleon subsystems of the two-nucleon emitters listed 
in the Table \ref{tb_ch5_1} are summarized. 
All the listed states are the ground states as 1N-resonances. 
The 1st column is for the core-nucleon system and its spin-parity 
in the reference state. 
The 2nd column indicates decay-modes. 
The 3th column indicates the spin-parity of the core nucleus. 
The 4th and 5th columns are for the Q-value and the decay width. 
Q-values are respect to the ground states of the daughter nuclides. 
For some long-lived nuclides, 
their lifetimes are shown instead of the decay widths. 
The 6th column is for the references other than that listed 
in the 5th column. } \label{tb_ch5_2}
\end{center} \end{table}

Since the memorable works for $^{45}$Fe, experimental efforts 
have been continued, in order to detect other \twop-radioactive 
nuclides and also to increase the accuracy of data. 
The novel \twop-emitters observed in this period include 
$^{19}$Mg, $^{48}$Ni, $^{54}$Zn and so on. 
In Table. \ref{tb_ch5_1}, we summarize the up-to-date properties 
of the observed \twop-emitters, and also of $2n$-emitters. 
As the additional information of the parent nuclei 
shown in Table \ref{tb_ch5_1}, 
we tabulate the properties of their core-nucleon subsystems 
in Table \ref{tb_ch5_2}. 
By comparing the corresponding 2N- and 1N-resonance energies, 
one can infer whether the interested nucleus is a true 
2N-emitter or not (the resonance energies are indicated as $Q_{\rm 2N}$ 
and $Q_{\rm 1N}$ in these Tables). 
For instance, in the case of $^{6}$Be, the data show 
that the \twop-resonance energy of $^6$Be is lower than the $1p$-resonance 
energy of its core-proton subsystem, $^5$Li. 
Thus, $^6$Be is expected to be a true \twop-emitter. 

For the \twop-emitters listed in Table. \ref{tb_ch5_1}, one can find that 
there is a broad gap between the lifetimes of the lighter and 
the heavier nuclides. 
In the lighter \twop-emitters, such as $^6$Be, $^{12}$O and $^{16}$Ne, 
the Coulomb barrier plays a minor role and the resonance is 
mainly stabilized by the centrifugal barriers between the core and 
the valence protons. 
Consequently, their typical decay widths are on the same order 
among those nuclei, namely about $100$ keV. 
On the other hand, in the heavier \twop-emitters, the 
Coulomb barrier is higher, which reduces the penetrability of two protons, 
and the lifetimes become considerably longer. 
In recent studies, searching intermediate long-lived 
\twop-emitters, which may locates between $14\leq Z \leq 24$, 
has been a challenging task. 
Also notice that there have been no heavier $2n$-radioactive nuclides 
observed than those listed in Table \ref{tb_ch5_2}. 
Whether the $2n$-radioactive nuclide with 20 or more neutrons 
exists or not is still an open question. 
\begin{figure}[t] \begin{center}
(Figure is hidden in open-print version.)
\caption{Figure 1 in Ref.\cite{07Mie}. 
A photograph of the \twop-radioactive decay of $^{45}$Fe 
obtained with the optical time-projection chamber. 
A track of a $^{45}$Fe ion entering the chamber from left is seen. 
The two bright, short tracks are protons emitted after 
the implantation of $^{45}$Fe on the detector. } \label{fig:2007Mie}
\end{center} \end{figure}

It is worthwhile to mention that the kinematics of the emitted 
two protons has been measured in the recent experiments, 
especially owing to the time-projection chamber. 
This device yields the 
photographs of the \twop-decays in a real-time 
regime, and the complete kinematics in most cases can 
be reconstructed \cite{07Mie_TPC,07Mie,07Gio,08Blank}. 
The photograph in Figure \ref{fig:2007Mie} displays 
this kinematics. 
\begin{figure}[t] \begin{center}
(Figure is hidden in open-print version.)
\caption{Figure 1 in Ref.\cite{09Gri_677}. 
The energy-angular correlation pattern of the 
\twop-emission from the $^{6}$Be nucleus. 
The upper two panels show the theoretical results in two different 
coordinates, whereas the lower two panels show the experimental results. 
See the original paper \cite{09Gri_677} for the definition of 
the variables. } \label{fig:2009Gri_06Be}
\end{center} \end{figure}

It should be noticed that, 
for the two-particle decay processes including the alpha-decays and 
the $1p$-emissions, the kinetic properties is completely determined with the 
total energy release. 
On the other hand, for three or more particle decays, 
even if the total energy is identified, 
one needs additional degrees of freedom to fully understand the process. 
In the studies of the \twop-emissions, as a tradition, 
(i) the total energy release and 
(ii) the opening angle between two relative momenta, 
corresponding to the two relative coordinates in the three-body system, 
have been often employed for this 
purpose \cite{01Gri_I,10Gri_V,09Gri_80,09Gri_677,12Ego}. 
Owing to the recent developments of experimental techniques, 
for several nuclides, their energy-angular distributions 
have been measured \cite{09Gri_80,09Gri_677,08Muk,10Muk,07Mie}. 
The observed distributions usually 
show the characteristic correlation pattern of the two protons, 
which can be interpreted as the dynamical character of each nuclide. 
In Figures \ref{fig:2009Gri_06Be} and \ref{fig:2009Gri_45Fe} taken from 
the Ref.\cite{09Gri_677}, for instance, the correlations in the 
decay of $^{6}$Be 
and $^{45}$Fe in the energy-angle plane are shown. 
In the 6th column of Table \ref{tb_ch5_1}, we list the references 
which report these measurements. 
It has also been expected that the qualitative information 
during the decay process can be extracted from 
these correlation patters. 
To investigate them could clarify, 
for instance, the density-dependence of 
the nuclear force, the pairing correlations in loosely or quasi-bound systems, 
and possibly the diproton correlation \cite{08Bertulani}. 
\begin{figure}[t] \begin{center}
(Figure is hidden in open-print version.)
\caption{Figure 2 in Ref.\cite{09Gri_677}. 
The same as Figure 1 in Ref.\cite{09Gri_677}, but for the \twop-radioactive decay from the $^{45}$Fe nucleus. } \label{fig:2009Gri_45Fe}
\end{center} \end{figure}

As shown in Fig. \ref{fig:2009Gri_45Fe}, 
in the \twop-radioactivity of $^{45}$Fe, the measured correlation 
pattern suggests that there are considerable probabilities 
for the diproton-decay, characterized by the strong correlation 
between the emitted two protons. 
On the other hand, in $^6$Be shown in Fig. \ref{fig:2009Gri_06Be}, 
the diproton decay is less significant, 
and the correlation pattern shows a more extended 
and complicated distributions. 
It means that all the interactions in the final state take comparable 
contributions in this system to each other. 
Consequently, in light \twop-emitters, the observed 
quantities may be strongly affected by the final state interactions. 
It remains an open question how to extract the information on the 
diproton correlation from the experimental observables of the 
\twop-decays, which should be addressed by theoretical approaches. 

On the theoretical side, making synergy with the experiments, 
there have been remarkable developments established. 
The theoretical works by L. V. Grigorenko {\it et. al.} should be especially 
introduced \cite{93Zhukov,01Gri_I,03Gri_II,07Gri_III,07Gri_IV,10Gri_V,02Gri_15,02Gri,03Gri,09Gri_80,09Gri_677,09Gri_40,12Ego,12Gri}. 
Their works until 2009 are well summarized in the Ref.\cite{09Gri_40}. 
They have investigated \twop-decays based on 
the Three-body scattering equation, 
which is the basic formalism of scattering problems in 
quantum three-body systems. 
To solve the three-body Three-body scattering equation, they have developed 
the Hyper-spherical Harmonics (HH-) method within the 
non-Hermite framework. 
The HH-functions were originally proposed as an efficient basis to 
solve the quantum few-body problems \cite{93Zhukov}. 
Additionally, they have carefully treated the asymptotic properties 
of the Coulomb three-body problems. 
Notice that even the asymptotic solutions cannot be analytically 
obtained for this problem. 
Thus, within an approximate asymptotic conditions, they have 
employed an enormously large model space, 
which guarantees the saturation of results. 
Up to date, their calculations have been remarkably successful in 
reproducing the experimental results of both the decay widths (the lifetime) 
and the \twop-correlation patterns for several nuclides 
\cite{09Gri_40,09Gri_80,09Gri_677} (see Figs. \ref{fig:2009Gri_06Be} and 
\ref{fig:2009Gri_45Fe}). 
They have also shown that the final-state interactions lead to a crucial 
effect on the democratic \twop-emission, especially for $^{6}$Be 
\cite{09Gri_80,09Gri_677,12Ego}. 
Recently, they have also discussed the effect of 
the initial configurations of two protons before 
the barrier penetration \cite{12Ego,12Gri}. 
However, the relations between the \twop-emission and the 
diproton correlation is still not investigated. 

Other theoretical efforts based on the microscopic 
picture of \twop-decays 
have also been devoted \cite{05Rotu,06Rotu,13Deli,13Olsen}. 
As a notable progress, it was predicted that 
the \twop-radioactive nuclides can exist widely along 
the proton-dripline, up to the proton number of $Z\leq82$ \cite{13Olsen} 
(the upper limit of $Z$ is owing to the dominance of 
the $\alpha$-decay). 
This prediction is an inspiring work towards the further 
exploration of the \twop-emitters. 
Today, predicting and discovering the novel information of 
\twop-emissions are hot interests in both theoretical 
and experimental sides. 
We also mention that, if a full microscopic theory of 
\twop-radioactivity is established, it can be naturally extended to 
other processes, such as $2n$- and $\alpha$-decays \cite{12Betan}. 
However, in these microscopic models, the equal treatment of both 
the true and sequential processes is a challenging task. 
Furthermore, there remains a serious problem, 
associated with the computational resources. 

\section{Theoretical Frameworks for Quantum Meta-stability}
Theoretically, there are two main frameworks for quantum 
meta-stability. 
One is the non-Hermite, time-independent framework, 
whereas the other is the time-dependent framework. 
In this section, we detail the advantages and shortcomings of 
these methods, 
regarding their applications to the \twop-emissions. 
From a theoretical point of view, the \twop-emissions are 
quantum-mechanical phenomena, dominated critically by 
the tunneling effect coupled to the continuum region. 
Additionally, in contrast to the two-body decay processes 
including the alpha-decays and $1p$-emissions from the spherical 
parent nuclei, 
the many-body properties with the nuclear and Coulomb interactions 
must be treated on equal footing in the \twop-decays. 

\subsection{Time-Independent Framework}
Up to present, almost all the theoretical works of 
\twop-emissions have been based on the time-independent, or 
equivalently on the non-Hermite framework. 
The original idea of this method was proposed by Gamow 
to understand the $\alpha$-particle 
decays \cite{28Gamov_01,28Gamov_02,29Gurney,89Bohm}. 
With the time-independent method, 
a meta-stable state is solved as a time-independent 
eigen-state of the Hamiltonian with a {\it complex} eigen-energy, 
corresponding to the boundary condition that the wave function 
should be asymptotically connected to the out-going wave. 
In actual calculations, one solves this non-Hermite 
eigen-state by, {\it e.g.} complex-scaling the coordinates 
in the wave function so as to yield the complex eigen-energy 
in the continuum region \cite{06Aoyama}. 
The imaginary part of the eigen-energy corresponds to the decay width, 
while the real part corresponds to the total energy release of the 
decay (the Q-value). 
An advantage of this method is that one can solve the meta-stable 
states in almost the same way as the stable states. 
The decay width can be calculated with a high accuracy even if 
it is extremely small \cite{97Aberg,00Davis}. 

As already introduced in the previous section, for \twop-emissions, 
the results by Grigorenko {\it et.al.} within the 
time-independent method show the excellent agreement with the 
experiments for the observed momentum and angular correlations. 
On the other hand, this method is somewhat difficult to extract the 
essential cause of phenomena. 
The correspondence between the wave functions with complex 
energies and the 
real phenomena is not completely recognized. 
Although the obtained results have well reproduced the experimental 
data for the \twop-decay, the mechanism to yield this agreement 
has not been sufficiently clarified. 
The connection between the diproton correlation and the decay 
observables has yet to be revealed. 

\subsection{Time-Dependent Framework}
In contrast to the time-independent method, 
the time-dependent method treats the quantum 
resonances or tunnelings as the temporal developments 
of meta-stable states, maintaining the Hermiticy in the 
framework \cite{89Bohm,89Kuku,47Kry}. 
These approaches have been applied to several quantum tunneling phenomena 
\cite{94Serot, 98Talou, 99Talou_60, 00Talou, 11Garc, 11Campo, 12Pons}, 
with an advantage that it can provide an intuitive way to understand 
the tunneling mechanism. 
However, there have been no applications of this method 
to the \twop-decays, except for that for the dynamics 
in the classically allowed region after 
the tunneling stage \cite{08Bertulani}. 

In applications of the time-dependent method to \twop-emissions, 
the initial \twop-state should be defined as a quasi-bound 
state inside the Coulomb and centrifugal potential barriers. 
For instance, one modifies the potential barrier at $t=0$ so that 
the initial state can be prepared as a quasi-bound state of 
the original Hamiltonian. 
The modified potential is then suddenly changed to the original one, 
and the initial state evolves in time 
to the final state where all the particles 
are separated along the time-evolution with the original Hamiltonian. 
The decay width can be determined from 
the survival probability of the initial state. 
Furthermore, the tunneling process can be intuitively 
understood by monitoring the time-development of the wave function 
and thus of the density distribution. 
The sensitivity of the \twop-emissions to, {\it e.g.} the diproton 
correlation, can be translated to the dependence on 
the initial configuration of two protons inside a parent nucleus. 
A drawback of this method is that it does not practically work 
when the decay width is extremely small. 
It often needs a great amount of computational resources to 
obtain the final results. 

In this thesis, from a complementary point of view to the works 
in past based on the time-independent method, 
we employ the time-dependent method. 
We only focus on the light \twop-emitters with comparably short 
lifetimes, to which the time-dependent method practically works. 
This method can be a powerful tool to reveal 
the relation between the diproton correlation and the \twop-emission 
by making full use of its intuitive nature. 
We stress that these problems have seldom been studied in literature 
in the past, and our present study is expected to provide 
a novel insight into the multi-nucleon meta-stable 
systems and their decays. 
\include{end}
\documentclass[a4paper,12pt]{book}
\include{begin}

\chapter{Time-Dependent Method} \label{Ch_TDM}
We extend our three-body model to the time-dependent (TD) one, 
in order to treat a meta-stable state 
of the three-body system and its decay 
via the emission of the two valence particles. 
General formalism of the time-dependent method for 
quantum meta-stable phenomena is summarized in Appendix \ref{Ap_TDM}. 
Therefore, in the following, we mainly describe how to apply 
the time-dependent method to the 
two-proton (\twop-) decays and emissions.

Within our TD three-body model, the two-proton decays and emissions 
can be described as 
dynamical processes driven by the static Hamiltonian, $H_{\rm 3b}$. 
Furthermore, in many cases, a proton in the three-body system does not 
have a sufficient energy to get over the potential barriers from 
other particle(s). 
Thus, the quantum tunneling effect plays an 
essential role in these processes. 
We emphasize that this tunneling effect can be naturally taken 
into account by solving 
the time-dependent Schr\"{o}dinger equation. 
Our formalism will not assume whether the 
two protons are either emitted sequentially or simultaneously. 
In other words, our method includes all the possible configurations in the 
emission process on equal footing. 

\section{Discretized Continuum Space}
Assuming the \twop-emission as a time-dependent process, 
we carry out the time-evolving calculations for 
the three-body system. 
First we have to prepare the initial state, 
$\ket{\Phi (0)}$, defined consistently to 
the realistic emissions in order for our calculations to be valid. 
Phenomenologically, the initial state, $\ket{\Phi (0)}$, should reflect 
the configuration of two protons confined inside 
the potential barrier. 
The \twop-density for such initial state should 
have almost no amplitudes 
outside the potential barrier. 
For this purpose, we employ ``confining potential method'' 
in this thesis. 
A concrete form of the confining potential will be given 
in the next Chapter, because the definition of it 
is critical to our final results. 

Before we do this definition, however, we would like to introduce 
some formulas which will be used in the actual calculations 
of the \twop-emissions. 
In performing the time-dependent calculations, we discretize 
the continuum energy space. 
Let us expand the initial state as a confined wave-packet 
on the discretized continuum eigen-space of 
the Hamiltonian, namely 
\beq
 \ket{\Phi (0)} = \sum_N F_N (0) \ket{E_N}, \label{eq:expand_E} 
\eeq
where $H_{\rm 3b} \ket{E_N} = E_N \ket{E_N}$. 
We note that the more general formalism of 
the time-dependent calculation without 
the continuum-descretization is summarized in Appendix \ref{Ap_TDM}. 
That formalism is, however, not useful for 
the numerical calculations. 

In Eq.(\ref{eq:expand_E}), one can obtain the discretized 
eigen-states $\ket{E_N}$ by, {\it e.g.} solving 
the three-body Hamiltonian within a large box. 
We stress that all the eigen-energies are real numbers: 
$E_N \in \mathbb{R}$. 
Namely, we consider within the pure Hermit framework, in contrast to other 
non-Hermite frameworks frequently used for quantum meta-stable 
processes \cite{10Deli_text,28Gamov_01,28Gamov_02,71Aguilar,71Balslev,06Aoyama,09Gri_40,09Gri_80}. 
Each eigenstate, $\ket{E_N}$, is expanded on 
the anti-symmetrized uncorrelated basis 
given by Eq.(\ref{eq:uncorrebasis}). 
Namely, 
\beq
 \ket{E_N} = \sum_K U_{NK} \ket{\tilde{\Psi}_K}, 
\eeq
where we use simplified labels as 
$K \equiv \left\{ n_a,l_a,j_a,n_b,l_b,j_b \right\}$. 
Note that the expansion coefficients, $\left\{ U_{NK} \right\}$, 
are obtained by diagonalizing the Hamiltonian matrix, 
$\left\{ \Braket{\tilde{\Psi}_{K'}| H_{\rm 3b} |\tilde{\Psi}_K} \right\}$, 
and are independent of time, $t$. 

The state at an arbitrary time, $t$, can be expanded on 
the uncorrelated basis, or equivalently, 
on the eigen-states of the Hamiltonian, $H_{\rm 3b}$. 
Those are represented as 
\beqa
 \ket{\Phi (t)} 
 &=& \exp \left[ -it \frac{H_{\rm 3b}}{\hbar} \right] \ket{\Phi (0)} \nonumber \\
 &=& \sum_{N} F_N (t) \ket{E_N}, \label{eq:ex_E} \\
 &=& \sum_{K} C_K (t) \ket{\tilde{\Psi}_K}, 
 \label{eq:tdse}
\eeqa
with the expansion coefficients given by 
\beqa
 F_N (t) &=& e^{-itE_N/\hbar} F_N(0) \label{eq:excf_E} , \\
 C_K (t) &=& \sum_N F_N (t) U_{NK} \label{eq:excf_UNCB} . 
\eeqa
The Q-value of the \twop-emission is given as the expectation 
value of the total Hamiltonian. 
From Eq.(\ref{eq:expand_E}) and (\ref{eq:excf_E}), 
it is shown that the Q-value is conserved during the time-evolution, 
that is, 
\beq
 Q_{2p} 
 \equiv \Braket{\Phi (t) | H_{\rm 3b} | \Phi (t)} 
 = \sum_N E_N \abs{F_N(t)}^2 
 = \sum_N E_N \abs{F_N(0)}^2. \label{eq:Q_conserve} 
\eeq
We also note that the norm of the \twop-state is normalized 
at any time: 
\beq
 \Braket{\Phi(0) | \Phi(0)} = \Braket{\Phi(t) | \Phi(t)} = 
 \sum_N \abs{F_N(0)}^2 = 1. 
\eeq

\section{Decay State and Width}
In order to extract the information on the emission, 
it is useful to  define the ``{\it decay state}'', $\ket{\Phi_d (t)}$, 
by projecting out to the initial state \cite{08Bertulani}. 
That is, 
\beq
 \ket{\Phi_d (t)} \equiv 
 \ket{\Phi (t)} - \beta(t) \ket{\Phi (0)}, \label{eq:decaystate}
\eeq
where $\beta (t) = \Braket{\Phi (0) | \Phi (t)}$ is the 
overlap coefficient. 
Because we prepare the initial state which has almost no amplitude 
outside the potential barrier, 
the decay state is mostly an outgoing wave, and its density has 
non-zero values almost only outside the potential barrier. 
The decay probability is given by its norm, 
\beq
 N_d (t) \label{eq:603DComp}
 = \Braket{\Phi_d (t) | \Phi_d (t)} = 1 - \abs{\beta (t)}^2 . 
\eeq
Notice that $N_d(0) = 0$ since $\beta (0) = 1$. 
Noticing that the quantity $\abs{\beta (t)}^2$ is identical to 
the survival probability for the initial state, 
the decay width can be defined from 
$N_d (t)$ \cite{94Serot,98Talou,99Talou_60,00Talou}. 
That is 
\beqa
 \Gamma (t) 
 &\equiv & -\hbar \frac{d}{dt} \ln \left[ 1-N_d(t) \right] \\
 &=& \frac{\hbar }{1-N_d(t)} \frac{d}{dt} N_d(t). \label{eq:width} 
\eeqa
If the time-evolution converges to the well-known exponential 
decay process, such that 
\beq
 \left[ 1-N_d(t) \right] = e^{-t/\tau}, 
\eeq
then $\Gamma(t)$ obviously corresponds to the lifetime, 
$\Gamma = \hbar / \tau$. 
This is the situation in which the energy spectrum, 
defined with $\{ \abs{F_N(0)}^2 \}$ in the discrete continuum space, 
is well approximated by the Breit-Wigner 
distribution \cite{89Kuku, 89Bohm, 09Konishi}. 
For the relation between the exponential decay-rule and 
the Breit-Wigner distribution, 
see also Appendix \ref{Ap_TDM}. 

It is also helpful to define the ``{\it partial decay width}'' 
to discuss the tunneling properties. 
The purpose is to decompose the total decay width into the 
widths for partial components labeled by $s$, such as 
\beq
 \Gamma (t) = \sum_s \Gamma_s (t). 
\eeq
The corresponding expansion for the decay state on the 
partial components, $\{ \ket{s} \}$, can be defined as 
\beq
 \ket{\Phi_d (t)} = \sum_s a_s(t) \ket{s}, 
 \label{eq:part_expand} 
\eeq
where all the partial components are orthogonal 
to each other: $\Braket{s' | s} = \delta_{s's}$. 
Using Eq.(\ref{eq:width}), we can write 
\beq
 \Gamma_s (t) \equiv \frac{\hbar }{1-N_d(t)} \frac{d}{dt} N_{d,s}(t). 
 \label{eq:pwidth} 
\eeq
where $N_{d,s} = \abs{a_s(t)}^2$. 

We note that Eqs.(\ref{eq:part_expand}) and (\ref{eq:pwidth}) 
can be defined generally for any choice of the partial components 
as long as they are orthogonal. 
For example, one can employ the components which have different 
energies, angular momenta or spin-parities. 
In the next Chapter, we will apply these formulas in order to 
calculate the spin-singlet and triplet widths in the \twop-emission of $^6$Be. 
We also emphasize that our formulas themselves in this subsection 
are not limited to the three-body framework, but can be extended to 
further complex systems. 

\section{Time-Dependent Density Distribution}
By integrating over the spin-variables, 
similarly to Chapters \ref{Ch_3body} and \ref{Ch_Results1}, 
we can obtain the spatial density distribution, 
parametrized by the radial 
distances $\{ r_1,r_2 \}$ and the opening angle $\theta_{12}$ 
from the symmetry. 
It is formulated as 
\beqa
 && \bar{\rho}_{2p} (t;r_1,r_2,\theta_{12}) = 8\pi^2 r_1^2 r_2^2 \sin \theta_{12} \rho_{2p} (t;r_1,r_2,\theta_{12}), \\
 && \rho_{2p} (t;r_1,r_2,\theta_{12}) = \abs{\Phi (t;r_1,r_2,\theta_{12})}^2, 
\eeqa
where $\bar{\rho}_{2p}$ is normalized at any time as 
\beq
 \int_0^{R_{\rm box}} dr_1 \int_0^{R_{\rm box}} dr_2 \int_{0}^{\pi} d\theta_{12} 
 \bar{\rho}_{2p}(t) = 1. 
\eeq
However, for the emission process, 
it is often more useful to discuss the density of the decay state 
defined by Eq.(\ref{eq:decaystate}). 
This is given by 
\beq
 \bar{\rho}_d (t;r_1,r_2,\theta_{12}) 
 = 8\pi^2 r_1^2 r_2^2 \sin \theta_{12} 
   \abs{\Phi_d (t;r_1,r_2,\theta_{12})}^2. 
\eeq
Because of the definition of the decay state, 
this quantity represents the components which 
have penetrated the potential barrier. 
For a presentation purpose, 
we often renormalize the $\rho_d(t)$ so that 
its integration become unity at each time: 
\beq
 \bar{\rho}_d (t) \longrightarrow \frac{\bar{\rho}_d (t)}{N_d (t)}, 
\eeq
where $N_d (t)$ is the decay probability given by Eq.(\ref{eq:603DComp}). 

In our discussions after this Chapter, 
we make full use of this decay density, in order to investigate, 
{\it e.g.} the effect of diproton correlations, 
the competition between the true and the sequential emissions, 
the spatial distributions of two protons and so on. 
It will provide a great advantage to intuitively understand 
the \twop-emissions and its relation to the diproton correlation. 

\section{Test of Time-Dependent Method: One-Proton Emission}
To check that the time-dependent method can correctly describe the 
decay of a quantum meta-stable state, 
we apply it to a problem of the one-proton ($1p$-) emission. 
This is a two-body problem of a core nucleus and a valence proton, 
with a spherical potential, $V_{lj}(r)$. 
Thus, the Hamiltonian is given by 
\beq
 h = \frac{\bip^2}{2\mu} + V_{lj}(r). \label{eq:sph61}
\eeq
Taking the relative wave function as 
$\Psi_{ljm}(\bir,\bis) = U_{lj}(r)/r \cdot \mathcal{Y}_{ljm}(\ubir,\bis)$, 
the Schr\"{o}dinger equation is given as 
\beq
  \left[ -\frac{\hbar^2}{2\mu}\left\{ \frac{d^2}{dr^2} 
         -\frac{l(l+1)}{r^2} \right\} + V_{lj}(r) - E \right] U_{lj} (r) 
  = 0, 
\eeq
with the relative energy larger than the threshold of 
the $1p$-emission in this system: 
\beq
 E > \lim_{r \rightarrow \infty} V_{lj}(r) \equiv 0. 
\eeq
We adopt the Woods-Saxon plus the Coulomb 
potential of a uniform-charged sphere for $V_{lj}(r)$. 
That is, 
\beqa
  V_{lj}(r) &=& V_{lj,{\rm Nucl.}}(r) + V_{{\rm Coul.}}(r) \\
  &=& \left[ V_0 + V_{lj} r_0^2 (\bi{\ell} \cdot \bi{s}) 
      \frac{1}{r} \frac{d}{dr} \right] f(r) + V_{{\rm Coul.}}(r), \label{eq:vorig64}
\eeqa
with
\beq
    f(r) = \frac{1}{ 1 + \exp \left( \frac{r-R_c}{a_c} \right)}, 
\eeq
and 
\beq
 V_{\rm Coul.} (r) 
 = \left\{ \begin{array}{cc} 
   \frac{Z_{\rm c} e^2}{4\pi \epsilon_0} \frac{1}{r} & (r > R_c), \\
   \frac{Z_{\rm c} e^2}{4\pi \epsilon_0} \frac{1}{2R_c} \left( 3 - \frac{r^2}{R_c^2} \right) & (r \leq R_c). 
   \end{array} \right. 
\eeq
The parameters are taken to be $A_c=4$, $Z_c=2$, 
$r_0=1.12$ fm, $R_c=r_0\cdot A_c^{1/3}$ fm, 
$a_c=0.755$ fm, $V_0 =-58.7$ MeV, and 
$V_{lj} r_0^2 = 51.68$ MeV$\cdot {\rm fm}^2$. 
We will use the similar parameters 
to study the $^{6}$Be nucleus in the next Chapter. 
For the angular channel, we only discuss the $(p_{3/2})$-channel. 
As we will show, this channel has a resonant state within 
the present core-proton potential. 
\begin{figure}[tb] \begin{center}
\fbox{\includegraphics[width=0.5\hsize, scale=1, trim = 50 50 0 0]{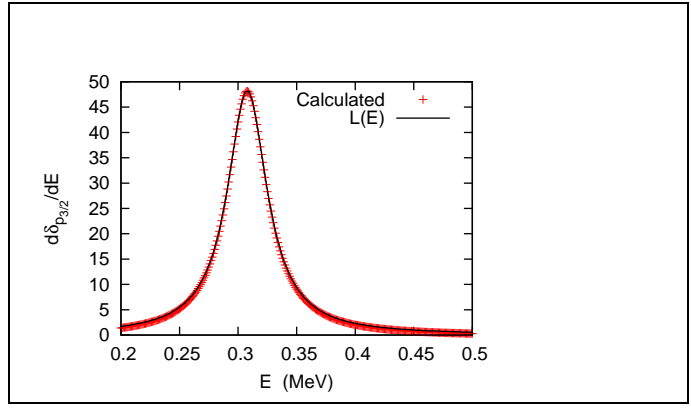}}
\caption{The calculated energy-derivative of the phase-shift, $\delta_{p_{3/2}}$(E). 
For the fitting purpose, the pure Breit-Wigner distribution, 
$L(E)=\frac{\Gamma_0/2}{\Gamma_0^2/4 + (E_0-E)^2}$ is assumed. } \label{fig:05LI_3t_1}
\end{center} \end{figure}

By calculating and fitting the phase-shift according to 
the formalism in Appendix \ref{Ap_Scat_2body}, 
we obtain the resonant energy and width as 
$E_0=308$ keV and $\Gamma_0=41$ keV, respectively. 
These values are obtained using 
the fitting function as the pure Breit-Wigner distribution: 
\beq
 \frac{d\delta_{lj}(E)}{dE} \equiv 
 \frac{\Gamma_0/2}{\Gamma_0^2/4 + (E_0-E)^2}, \label{eq:psch64}
\eeq
for $l=1$ and $j=3/2$, based on the two-body scattering formalism. 
The calculated result and its fit are presented in Fig. \ref{fig:05LI_3t_1}. 

On the other hand, we can calculate the resonant energy and width 
by another method, namely by the time-dependent method. 
With this method, we first have to prepare 
the initial state for the $1p$-emission. 
For this purpose, we adopt the ``confining potential'' method. 
That is, we assume the modified Hamiltonian, 
\beq
 h_{(p_{3/2})}^{conf} = \frac{\bip^2}{2\mu} + V_{(p_{3/2})}^{conf}(r), 
\eeq
with the confining potential for the $(p_{3/2})$-channel as 
\beq
  V_{(p_{3/2})}^{conf}(r) \label{eq:vconf64} 
 = \left\{ \begin{array}{cc} 
            V_{(p_{3/2})}(r) & (r \leq R_b), \\
            V_{(p_{3/2})}(R_b) & (r > R_b), \end{array} \right. 
\eeq
with $R_b = 8.2$ fm. 
The original and confining potentials 
are shown in Fig. \ref{fig:05LI_3t_2}(b). 
The initial state, $\Phi(t=0;\bir,\bis)$, is defined as an 
eigen-state of this modified Hamiltonian, namely, 
\beq
 h_{(p_{3/2})}^{conf} \ket{\Phi(0)} = E^{conf} \ket{\Phi(0)}. 
\eeq
On the other hand, we also consider the eigen-states of the original 
Hamiltonian as 
\beq
 h_{(p_{3/2})} \ket{E_N} = E_N \ket{E_N}, 
\eeq
with the discretized continuum energies, $\{ E_N \}$. 
In order to discretize the continuum, we assume the radial box of 
$R_{\rm box}=120$ fm in this case, and 
impose a boundary condition that 
the wave function satisfies $U_{(p_{3/2})}(r=R_{\rm box}) = 0$. 
The energy cutoff is employed as $E_{\rm cut} = 40$ MeV. 
As a result, we have $N_{\rm max}=47$ bases. 
\begin{figure*}[tb] \begin{center}
  \begin{tabular}{c} 
     \begin{minipage}{0.48\hsize} \begin{center}
        (a) \\ \fbox{ \includegraphics[height=44truemm, scale=1, trim = 50 50 0 0]{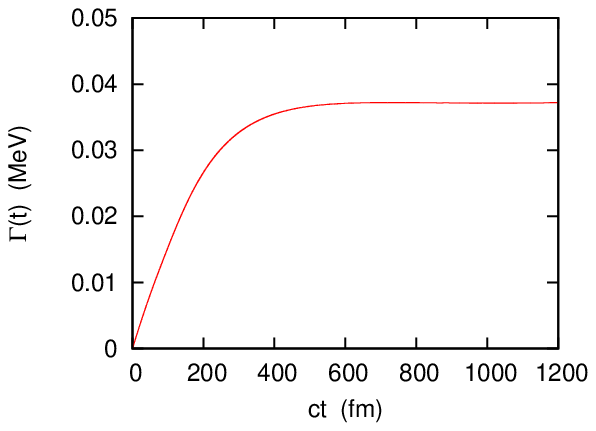}}
     \end{center} \end{minipage}
     \begin{minipage}{0.48\hsize} \begin{center}
        (b) \\ \fbox{ \includegraphics[height=44truemm, scale=1, trim = 50 50 0 0]{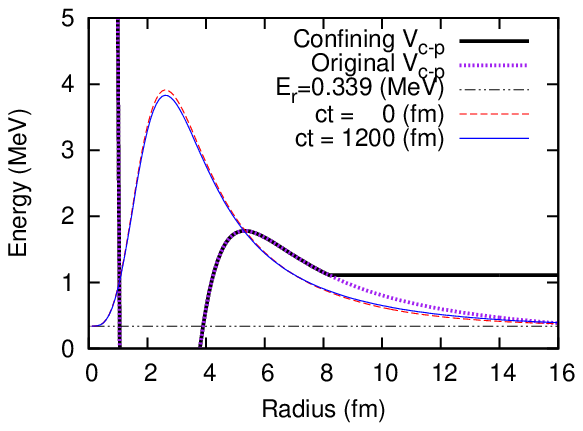}}
     \end{center} \end{minipage}
  \end{tabular}
\caption{(Left panel) The calculated decay width within the confining potential, Eq.(\ref{eq:vconf64}). 
(Right panel) The original and confining core-proton potentials given by 
Eqs.(\ref{eq:vorig64}) and (\ref{eq:vconf64}). 
The Q-value, $E_0=0.339$ MeV, and the density distributions at 
$ct=0$ and $1200$ fm are also shown. } \label{fig:05LI_3t_2}
\end{center} \end{figure*}

By diagonalizing the $47\times 47$ matrix, 
$\{ \Braket{E_M | h_{(p_{3/2})}^{conf} | E_N} \}$, 
the initial state can be represented as the expansion on the 
original eigen-states. 
That is, 
\beq
 \ket{\Phi(0)} = \sum_N F_N(0) \ket{E_N}. 
\eeq
This equation has an identical form to Eq.(\ref{eq:expand_E}). 
Thus, by performing the calculations according to Eqs.(\ref{eq:ex_E}), 
(\ref{eq:Q_conserve}) and (\ref{eq:width}), 
we can determine 
the Q-value, $E_0 = \sum_N E_N \abs{F_N(0)}^2$ and 
the decay width, $\Gamma(t)$. 
We obtain $E_0=339$ keV. 
The calculated decay width is shown in Figure \ref{fig:05LI_3t_2}(a). 
After a sufficient time-evolution, the decay width well converges to a 
constant value, corresponding to the exponential decay-rule. 
We have obtained $\Gamma_0=37$ keV at $ct=1200$ fm, when the 
decay width is sufficiently converged. 
These results are consistent to those obtained by calculating the 
two-body scattering phase-shift, 
justifying the time-dependent approach. 
Furthermore, 
in Figure \ref{fig:05LI_3t_2}, we also show the density distribution, 
$\abs{U_{(p_{3/2})}(r)}^2$ at $ct = 0$ and $1200$ fm. 
Although these two functions have almost the same form, 
the later one shows a larger amplitude outside the potential barrier. 
This indicates the penetration of the valence proton. 
Consequently, we can observe the time-development of the emitted 
particle(s) based on this method. 
This will be a great advantage providing an intuitive view 
of the decay process, 
when we will apply this method to 
the two-proton emissions, in order to discuss the effect of 
the diproton correlation. 
\include{end}
\documentclass[a4paper,12pt]{book}
\include{begin}

\chapter{Two-Proton Emission of $^{6}$Be} \label{Ch_Results2}
We now apply the time-dependent method to the ground state of 
$^{6}$Be nucleus, assuming the three-body structure of $\alpha$+p+p. 
Because the $\alpha$-particle can be well assumed as a rigid core, 
this system provides a good testing ground for our method. 
We also note that, as shown in Figure \ref{fig:054}, 
this is one of the closest systems to a true \twop-emitter: 
the experimental Q-value of the \twop-emission is 
$1.37$ MeV \cite{89Boch,88Ajzen,02Till}, 
which is lower than the one-proton resonance energy in $^5$Li, 
being about $2$ MeV with a broad width \cite{88Ajzen, 02Till}. 
Thus, the sequential process via $\alpha + p$ subsystem is considered 
to be suppressed, and the two protons should penetrate the potential 
barrier simultaneously \footnote{Actually, due to the broad 
width of the $\alpha$-p subsystem, 
there can be a non-negligible possibility of 
the sequential \twop-emission. }. 
In this simultaneous \twop-emission, 
the association between the dinucleon correlations and 
the \twop-emissions, might be clarified. 
\begin{figure*}[htbp] \begin{center}
\fbox{\includegraphics[width=0.6\hsize]{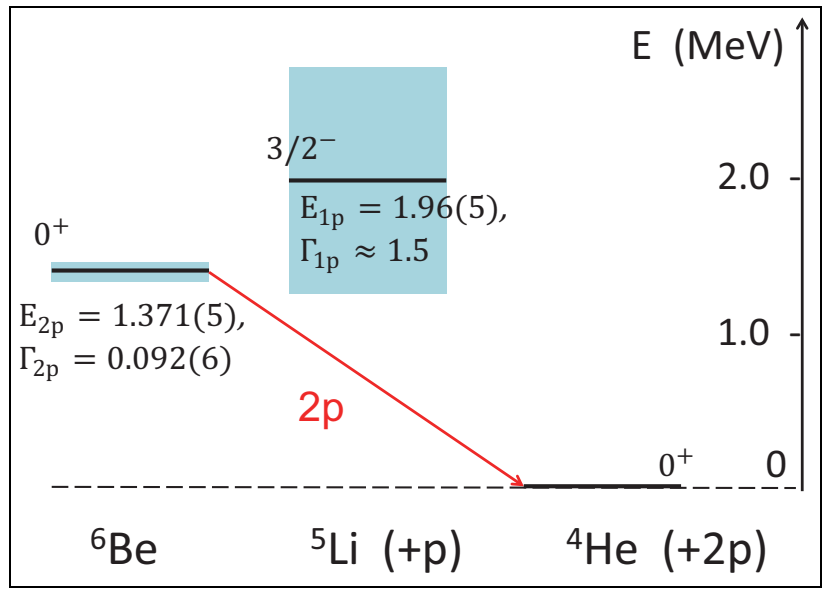}}
\caption{The experimental level scheme of $^{6}$Be and its isotones. 
The values for $^{5}$Li are quoted from 
refs.\cite{88Ajzen,09Shirokov}, whereas 
those for $^{6}$Be are quoted from 
refs.\cite{88Ajzen,02Till}. 
The color-box of each level indicates its decay width. } \label{fig:054}
\end{center} \end{figure*}

\section{Set up for Calculations}
Our three-body model consists of $\alpha$-particle 
as a structureless core (daughter) nucleus and two valence protons. 
We employ the V-coordinates similarly to Chapters \ref{Ch_3body} 
and \ref{Ch_Results1}. 
That is, 
\beqa
 H_{\rm 3b} 
 &=& h_1 + h_2 + \frac{\bip_1 \cdot \bip_2}{A_{\rm c} m} + 
     v_{\rm p-p}(\bir_1, \bir_2), \\
 h_i 
 &=& \frac{\bip_i^2}{2\mu} + V_{\rm c-p}(r_i) \; \; \; \; (i=1,2), 
\eeqa
where $h_i$ is the single particle (s.p.) Hamiltonian for the 
relative motion between 
the core and the $i$-th proton. 
We assume that the $\alpha$-p potential is spherical, and independent of 
the spin variables. 
We also assume that the $\alpha$-particle 
always remains in the ground state with the spin-parity of $0^+$. 
Thus, similarly in Chapter \ref{Ch_Results1}, 
we need uncorrelated bases only for the $0^+$ configuration 
since the ground state of $^6$Be also has the spin-parity of $0^+$. 
That is, 
\beqa
 \tilde{\Psi}_{ab} (\xi_1, \xi_2) &\longrightarrow& 
 \tilde{\Psi}^{(0^+)}_{n_a n_b lj} (\xi_1, \xi_2) \nonumber \\
 & & = \frac{1}{\sqrt{2(1+\delta_{n_a,n_b})} } 
       \sum_m \cgc{0,0}{j,m;j,-m} \nonumber \\
 & & \phantom{000} \left[ \phi_{n_a ljm} (\xi_1) \phi_{n_b lj-m} (\xi_2) 
                        + \phi_{n_a ljm} (\xi_2) \phi_{n_b lj-m} (\xi_1) 
                  \right]. \label{eq:601basis_0p} 
\eeqa
In the following, for simplicity, 
we use simplified labels for the uncorrelated bases: 
$\ket{\tilde{\Psi}_K}$ where $K=\left\{ n_a, n_b, l,j \right\}$. 
Then, the time-dependent three-body state, except for 
the center of mass motion of the whole system, 
can be expanded as 
\beq
 \ket{\Phi(t)} = \sum_{K} C_K (t) \ket{\tilde{\Psi}_K}, 
\eeq
where the coefficients $C_K (t)$ are given by 
Eqs. \ref{eq:excf_E} and \ref{eq:excf_UNCB}. 
All our calculations presented below 
are performed in the truncated space 
defined by the energy-cutoff: 
$\epsilon_a + \epsilon_b \leq E_{\rm cut} =40$ MeV. 
We have confirmed that our conclusions do not change 
even if we employ a larger value of $E_{\rm cut}$. 

For the angular momentum channels, 
we include from $(s_{1/2})^2$ to $(h_{11/2})^2$ partial waves, similarly to 
Chapter \ref{Ch_Results1}. 
In order to take into account the effect of the Pauli principle, 
we exclude the bound $(s_{1/2}$ state from $\phi_{nljm}$ in 
Eq.(\ref{eq:601basis_0p}), 
that is occupied by the protons in the core nucleus. 
The continuum states are discretized within the radial box of 
$R_{\rm box}=80$ fm. 
Even though this model space might be not sufficient to fully describe 
the \twop-emission of $^{6}$Be, 
increasing $l_{\rm max}$ or $R_{\rm box}$ causes a serious 
rise of computational costs, which makes our calculations 
practically impossible. 
Also note that $R_{\rm box}$ limits the interval for time-evolution, 
because the wave functions are inevitably reflected 
once it reaches at $r=R_{\rm box}$. 
This reflection as an calculational artifact causes 
the deviation from reality at the 
late-time region. 
A typical maximum time in our time-dependent calculations is 
$ct_{\rm max} \sim 2000$ fm, corresponding to 
the earlier stage of \twop-emissions.

\subsection{Interactions}
We describe the interaction between $\alpha$ and a valence proton 
using the nuclear Woods-Saxon potential and 
the Coulomb potential, similarly in the previous Chapters. 
That is, 
\beq
 V_{\rm c-p}(r_i) = V_{\rm c-p, Nucl.} (r_i) + V_{\rm c-p, Coul.} (r_i), 
 \label{eq:cp_pot6}
\eeq
where the nuclear and Coulomb terms are 
\beqa
 V_{\rm c-N, Nucl.} (r) &=& 
   \left[ V_0 + V_{ls} r_0^2 (\bi{\ell} \cdot \bi{s}) 
   \frac{1}{r} \frac{d}{dr} \right] f(r), \label{eq:cp_WS} \\
 f(r) &=& 
   \frac{1}{ 1 + \exp \left( \frac{r-R_{\rm core}}{a_{\rm core}} \right) }, 
\eeqa
and 
\beq
 V_{\rm c-p, Coul.} (r) 
 = \left\{ \begin{array}{cc} 
   \frac{Z_{\rm c} e^2}{4\pi \epsilon_0} \frac{1}{r} 
    & (r > R_{\rm core}), \\
   \frac{Z_{\rm c} e^2}{4\pi \epsilon_0} \frac{1}{2R_{\rm core}} \left( 3 - \frac{r^2}{R_{\rm core}^2} \right) 
    & (r \leq R_{\rm core}). 
   \end{array} \right. \label{eq:cp_C}
\eeq
For the Coulomb term, we adopt the one of a uniform-charged sphere 
with the charge radius of the $\alpha$-particle, 
$R_{\rm core} = r_c = 1.68$ fm. 
For the Woods-Saxon part, we use 
$R_{\rm core} = r_c$ and $a_0$ = 0.615 fm, whereas 
strength parameters are fixed as $V_0 = -58.7$ MeV and 
$V_{ls} = 46.3$ ${\rm MeV(fm)^2}$. 
These parameters reproduce the measured resonance energy and width 
for the $(p_{3/2})$-channel of $\alpha$-p scattering \cite{02Till}: 
it yields $E_r (p_{3/2}) = 1.96$ MeV and $\Gamma_r (p_{3/2}) = 1.56$ MeV. 
We calculate and fit the derivative of the phase-shift, according to 
Eq.(\ref{eq:apcps}), to get this result. 
We note that this resonance is quite broad and there are large 
ambiguities in the observed 
decay width \cite{88Ajzen,00Hoef,02Till,09Shirokov}, 
as summarized in Table \ref{tb:709425}.
\begin{figure}[t] \begin{center}
$V_{\rm c-p}(r)$ and $V_{\rm c-p}^{conf}(r)$ \\
\fbox{\includegraphics[width=0.5\hsize,scale=1,trim = 50 50 0 0]{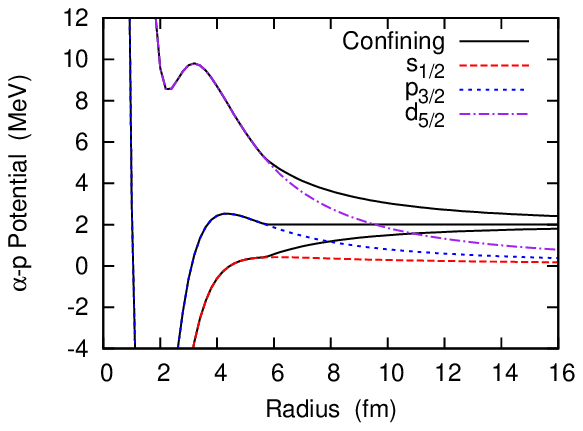}}
\caption{The original and confining potentials for 
$(s_{1/2})$, $(p_{3/2})$ and $(d_{5/2})$ channels in the $\alpha$-p subsystem. 
The border radius for modifying the potential is 5.7 fm for all the channels. } \label{fig:4001}
\end{center} \end{figure}
\begin{table}[b] \begin{center}
  \catcode`? = \active \def?{\phantom{0}} 
  \begingroup \renewcommand{\arraystretch}{1.2}
  \begin{tabular}{ccccc c} \hline \hline
  &                               && $E_r$ (MeV)  & $\Gamma_r$ (MeV) & \\ \hline
  & This work                     && 1.96????    & 1.56??           & \\
  & Ref.\cite{88Ajzen,09Shirokov} && 1.96(5)?    & $\simeq$ 1.5     & \\
  & Ref.\cite{00Hoef}             && 2.90(20)    & 1.0(2)           & \\
  & Ref.\cite{02Till}             && 1.69????    & 1.06??           & \\ \hline \hline
  \end{tabular}
  \endgroup
  \catcode`? = 12 
  \caption{The resonant energy and width of the $^{5}$Li 
nucleus in the $(p_{3/2})$-channel. } \label{tb:709425}
\end{center} \end{table}

For the proton-proton interaction, we use the Minnesota potential, 
in which the Coulomb term is explicitly included. 
\beq
 v_{\rm p-p}(r_{12}) = v_0 e^{-b_0 r^2_{12}} - v_1 e^{-b_1 r^2_{12}} 
 + \alpha \hbar c \frac{e^2}{r_{12}}. \label{eq:Minnepp}
\eeq
For $b_0, b_1$ and $v_1$ in Eq.(\ref{eq:Minnepp}), 
we use the same parameters introduced in the original paper \cite{77Thom}. 
On the other hand, the strength of the repulsive core, $v_0$, 
is adjusted so as to reproduce the empirical Q-value 
for two protons, $Q_{2p}=1.37$ MeV \cite{88Ajzen,02Till}. 

\section{Initial State}
As mentioned in Chapter \ref{Ch_TDM}, 
the initial configuration of the two protons is characterized such that 
the density distribution is localized around the core nucleus and 
has almost no amplitude outside the core-proton potential barrier. 
In order to generate such state, we employ 
the confining potential method \cite{87Gurv, 88Gurv, 04Gurv}. 
The confining potential for the initial \twop-state is defined as follows. 
Because the $\alpha$-p subsystem has a resonance at 
$E_0=1.96$ MeV in the $(p_{3/2})$-channel, 
the two protons in $^6$Be are expected to have a large component for 
the $(p_{3/2})^2$ configuration. 
Thus, we first modify the core-proton potential for the 
$(p_{3/2})$-channel at $t=0$ in order to 
generate a quasi-bound state as follows: 
\beq
 V_{{\rm c-p},~(p_{3/2})}^{conf}(r) \nonumber 
 = \left\{ \begin{array}{cc} 
            V_{{\rm c-p},~(p_{3/2})}(r) & (r \leq R_b), \\
            V_{{\rm c-p},~(p_{3/2})}(R_b) & (r > R_b), \end{array} \right.
\eeq
with $R_b=5.7$ fm. 
For the other s.p. channels, we define the confining potential as 
\beq
 V_{\rm c-p}^{conf}(r) \nonumber 
 =\left\{ \begin{array}{cc} 
          V_{\rm c-p}(r) \phantom{0000} & (r \leq R_b), \\
          V_{\rm c-p}(r) + V_b(r) & (r > R_b), \end{array} \right.
\eeq
where $V_b(r) = V_{{\rm c-p},~(p_{3/2})}(R_b) - V_{{\rm c-p},~(p_{3/2})}(r)$. 
The original and confining potentials for the 
$(s_{1/2})$, $(p_{3/2})$ and $(d_{5/2})$ channels are shown in 
Fig. \ref{fig:4001}. 
We note that, for this system, the core-proton barrier 
is mainly due to the centrifugal potential in the 
$(p_{3/2})$ channel, 
rather than the Coulomb potential. 
This situation is quite different from heavy \twop-emitters 
with a large proton-number, such as $^{45}$Fe. 
\begin{figure*}[tb] \begin{center}
  $^{6}$Be (g.s.), $t=0$, ``full'' \\
  \begin{tabular}{c} 
     \begin{minipage}{0.48\hsize} \begin{center}
        \fbox{ \includegraphics[height=45truemm, scale=1, trim = 50 50 0 0]{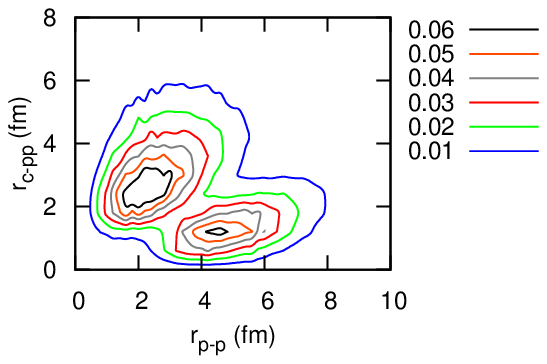}}
     \end{center} \end{minipage}
     \begin{minipage}{0.48\hsize} \begin{center}
        \fbox{ \includegraphics[height=45truemm, scale=1, trim = 50 50 0 0]{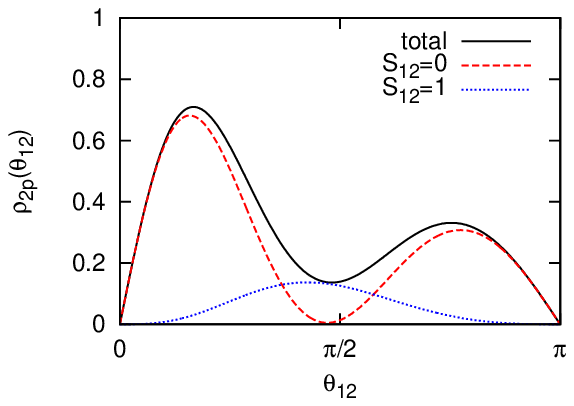}}
     \end{center} \end{minipage}
  \end{tabular}
  \caption{The left panel: 
The \twop-density distribution at $t=0$ for the ground state of $^6$Be. 
It is obtained by including all the partial waves up to $(h_{11/2})^2$, 
and is plotted as a function of 
$r_{\rm c-pp} = (r_1^2 + r_2^2 + 2r_1r_2\cos \theta_{12})^{1/2}/2$ and 
$r_{\rm p-p} = (r_1^2+r_2^2-2r_1r_2\cos \theta_{12})^{1/2}$. 
The right panel: The corresponding angular distribution 
obtained by integrating $\bar{\rho}_{2p}$ over $r_1$ and $r_2$. } \label{fig:2}
\end{center} \end{figure*}
\begin{figure*}[tb] \begin{center}
  $^{6}$Be (g.s.), $t=0$, ``$(l=odd)^2$ only'' \\
  \begin{tabular}{c} 
   \begin{minipage}{0.48\hsize} \begin{center}
     \fbox{ \includegraphics[height=45truemm, scale=1, trim = 50 50 0 0]{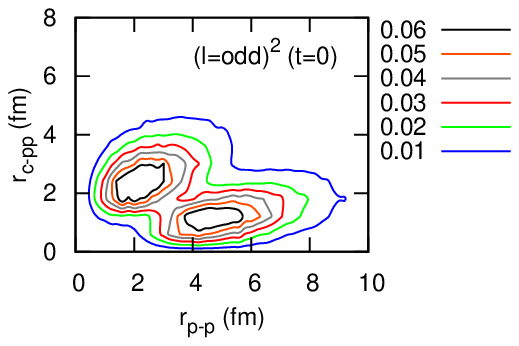}}
   \end{center} \end{minipage}
   \begin{minipage}{0.48\hsize} \begin{center}
     \fbox{ \includegraphics[height=45truemm, scale=1, trim = 50 50 0 0]{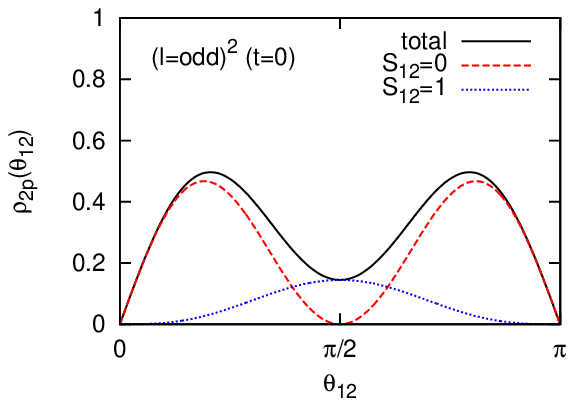}}
   \end{center} \end{minipage}
  \end{tabular}
  \caption{The same as Fig. \ref{fig:2} but for the case 
with only the partial waves of $(l=odd)^2$. } \label{fig:3}
\end{center} \end{figure*}

The initial state for the \twop-emission 
is solved by diagonalizing the modified 
Hamiltonian including $V^{conf}_{\rm c-p}(r)$. 
In Fig. \ref{fig:2}, we display the density distribution of the 
initial state obtained in this way. 
\beq
 \bar{\rho}_{2p}(t=0;r_1,r_2,\theta_{12}) 
 = 8\pi^2 r_1^2 r_2^2 \sin \theta_{12} 
   \abs{\Phi(t=0;r_1,r_2,\theta_{12})}. 
\eeq
In the left panel of Fig. \ref{fig:2}, $\bar{\rho}_{2p}$ is plotted as a 
function of the distance between the core and the center of mass 
of two protons: 
$r_{\rm c-pp} = \sqrt{r_1^2 + r_2^2 + 2r_1r_2\cos \theta_{12}}/2$, 
and the relative distance between two protons: 
$r_{\rm pp} = \sqrt{r_1^2 + r_2^2 - 2r_1r_2\cos \theta_{12}}$. 
In the right panel of Fig. \ref{fig:2}, we also show 
the angular distributions obtained by integrating $\bar{\rho}_{2p}$ 
for $r_1$ and $r_2$. 

It is clearly seen that the wave function 
is confined inside the potential barrier at $r \cong 4$ fm 
(see Fig. \ref{fig:4001} again). 
Furthermore, the \twop-density is concentrated 
near $r_{\rm p-p} = 2$ fm, 
corresponding to the diproton correlation in the bound nuclei. 
The corresponding angular distribution becomes asymmetric 
and has the higher peak at the opening angle $\theta_{12} \cong \pi /6$. 
This peak is almost due to the spin-singlet configuration, 
being analogous to the dinucleon correlation. 

As we discussed in Chapter \ref{Ch_Results1}, 
the parity-mixing in the subsystem of the core and a nucleon 
plays an essential role in generating the dinucleon correlation. 
In order to confirm the similar effect in the \twop-emission, 
we have performed the same calculations but only with 
$(p,f,h)^2=(l=odd)^2$ partial waves. 
In this case, pairing correlations are partially taken into 
account only among the s.p. states with the same parity, 
although the parity-mixing in the core-nucleon subsystem 
is perfectly ignored. 
In Fig. \ref{fig:3}, 
we show the initial configuration obtained only 
with $(l=odd)^2$ partial waves. 
In the left panel of Fig. \ref{fig:3}, 
there are two comparable peaks at $r_{\rm p-p} = 2$ and $5$ fm 
whereas, in the right panel, 
the corresponding angular distribution has a symmetric form. 
This result is in contrast with that in the full mixing 
case (see Fig. \ref{fig:2}), 
where the parity-mixing is fully taken into account (``full mixing''). 

The empirical Q-value for the \twop-emission is 1.37 MeV 
for $^6$Be \cite{88Ajzen,02Till}. 
However, the original parameters of the Minnesota potential 
overestimates this value, for instance, by about 50\% in 
the full mixing case. 
We therefore modify the parameter $v_0$ in Eq.(\ref{eq:Minnepp}) 
so as to yield Q = 1.37 MeV. 
Note that we use the modified $v_0$ not only at $t=0$ but also 
during the time-evolutions. 
\begin{table}[tb] \begin{center}
  \catcode`? = \active \def?{\phantom{0}} 
  \begingroup \renewcommand{\arraystretch}{1.0}
  \begin{tabular*} {\hsize} { @{\extracolsep{\fill}} ccccc ccc} \hline \hline
  && \multicolumn{3}{c}{$^{6}$Be, $t=0$}         && $^{6}$He & \\ \cline{3-5} \cline{7-7}
  && full & $(l=odd)^2$ only & $(p_{3/2})^2$ only && full    & \\ \hline
   $\Braket{H_{\rm 3b}}$ (MeV) && 1.37 & 1.37 & 1.37 && $-0.975$ & \\
  &&&&& && \\
   $\Braket{r_{\rm N-N}}$ (fm)   && 4.92 & 5.29 & 5.16 && 4.67 & \\
   $\Braket{r_{\rm c-NN}}$ (fm)  && 3.43 & 2.64 & 2.58 && 3.18 & \\
   $\Braket{\theta_{12}}$ (deg) && 75.9 & 90.0 & 90.0 && 78.0 & \\
  &&&&& && \\
   $(p_{3/2})^2$ (\%)        && 88.9  & 97.1 & 100. && 92.7 & \\
   $(p_{1/2})^2$ (\%)        && ?3.1  & ?2.8 & ??0. && ?1.6 & \\
   $(s_{1/2})^2$ (\%)        && ?2.2  & ?0.? & ??0. && ?1.3 & \\
   others,$(l=even)^2$ (\%) && ?5.2  & ?0.? & ??0. && ?4.2 & \\
   others,$(l=odd)^2$ (\%)  && ?0.6  & ?0.1 & ??0. && ?0.2 & \\
  &&&&& && \\
   $P(S_{12}=0)$ (\%) && 82.2 & 80.6 & 66.6 && 78.1 & \\
  &&&&& && \\
   $v_0$ (MeV) && 156.0 & 88.98 & 66.69 && 212.2 & \\ \hline \hline
  \end{tabular*}
  \endgroup
  \catcode`? = 12 
  \caption{Calculated properties for the initial state of $^6$Be 
and $^6$He. 
The results with all the uncorrelated basis 
from $(s_{1/2})^2$ to $(h_{11/2})^2$ are 
labeled by ``full''. 
Those obtained only with the $(l=odd)^2$ and $(p_{3/2})^2$ 
bases are also shown. 
The values of $v_0$ (the strength of the repulsive part) for the 
nucleon-nucleon interaction (Eq.(\ref{eq:Minnepp})) 
are tabulated in the last row. 
The original value is $v_0 = 200.$ MeV \cite{77Thom}. } \label{tb:71111}
\end{center} \end{table}

In Table \ref{tb:71111}, properties of the initial state are summarized. 
In this Table, for comparison, we also perform the same 
calculation but in the 3rd case, namely 
only with the $(p_{3/2})^2$ partial wave. 
The values of $v_0$ in the Minnesota potential 
are tabulated in the last row in Table \ref{tb:71111}. 
It is clearly seen that, 
in the full mixing case, the main component is $(p_{3/2})^2$, 
reflecting that the $(p_{3/2})$ channel has a 
resonance in the $\alpha$-p subsystem. 
The mixing of different partial waves are due to the 
off-diagonal matrix elements of $H_{\rm 3b}$, 
corresponding to the pairing correlations. 
The spin-singlet configuration is remarkably enhanced 
for the full mixing case compared to 
that in the $(p_{3/2})^2$ case. 
On the other hand, in the case only with $(l=odd)^2$, 
a comparable enhancement of the $S_{12}=0$ configuration exists, 
even though there is no localization of the two protons 
as shown in Fig. \ref{fig:3}. 
Notice also that in the $(l=odd)^2$ case, we have to assume a stronger 
pairing attraction in order to reproduce the empirical Q-value, 
as compared to the full-mixing case. 

\subsection{Comparison with $^6$He}
From the point of view of the isobaric symmetry in nuclei, 
it is interesting to compare the initial state of $^6$Be with the 
ground state of its mirror nucleus, $^6$He. 
Assuming the $\alpha$+n+n structure, 
we perform the similar calculation but for the ground state of $^6$He. 
For the $\alpha$-n system, there is an observed 
resonance of $(p_{3/2})$ at 
$E_r=0.735(20)$ MeV with its width, 
$\Gamma_r=0.600(20)$ \cite{NNDCHP,88Ajzen}. 
In order to reproduce this resonance, we exclude the Coulomb term from 
Eq.(\ref{eq:cp_pot6}) and modify the depth parameter 
to $V_0 = -61.25$ MeV in Woods-Saxon potential. 
The pairing interaction is adjusted to reproduce 
$\Braket{H_{\rm 3b}}=-S_{\rm 2n}=-0.975$ MeV \cite{NNDCHP}, 
yielding $v_0=212.2$ MeV in Eq.(\ref{eq:Minne}). 
One may concern the difference of $v_0$ between $^6$Be and $^6$He. 
This might be due to an ambiguity in $V_{\rm c-p}$ for $^6$Be 
originated from a broad resonance in the core-proton subsystem. 
Improving $V_{\rm c-p}$ in $^6$Be can lead to the more consistent 
set of parameters among $V_{\rm c-N}$ and $v_{\rm N-N}$. 
We note that this ambiguity does not affect our qualitative discussions. 
\begin{figure*}[htb] \begin{center}
  $^{6}$He (g.s.), ``full'' \\
  \begin{tabular}{c} 
     \begin{minipage}{0.48\hsize} \begin{center}
        \fbox{ \includegraphics[height=45truemm, scale=1, trim = 50 50 0 0]{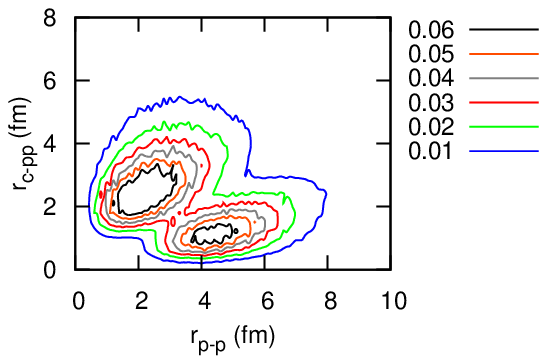}}
     \end{center} \end{minipage}
     \begin{minipage}{0.48\hsize} \begin{center}
        \fbox{ \includegraphics[height=45truemm, scale=1, trim = 50 50 0 0]{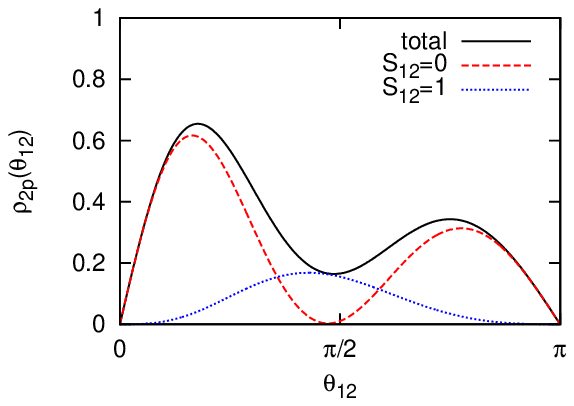}}
     \end{center} \end{minipage}
  \end{tabular}
\caption{The density distribution of the valence two neutrons, $\bar{\rho}_{2n}$, in the ground state of $^6$He. 
Those are plotted in the same manner as in the left and right panels 
of Figure \ref{fig:2}. 
The partial waves up to $(h_{11/2})^2$ are included. } \label{fig:22}
\end{center} \end{figure*}

In Fig. \ref{fig:22}, the two-neutron density distribution is 
shown in the same manner as in Fig. \ref{fig:2}. 
The energetic and structural properties are tabulated in the last column 
of Table \ref{tb:71111}. 
Obviously, the two-neutron wave function in $^6$He 
has a similar distribution to 
the \twop-wave function in $^6$Be. 
Because the two neutrons are bound in this system, the spatial 
distribution is less expanded in $^6$He. 
This is why both $\Braket{r_{\rm N-N}}$ and $\Braket{r_{\rm c-NN}}$ have 
smaller values than those of $^{6}$Be. 
The dinucleon correlation is present 
also in $^6$He, characterized as the spatial localization with the 
enhanced spin-singlet component \cite{05Hagi}. 
Consequently, the confining potential which we employ 
provides the initial state 
of $^6$Be, which can be interpreted as the isobaric analogue state of $^6$He. 
\begin{figure}[tb] \begin{center}
 \fbox{ \includegraphics[width=0.5\hsize, scale=1, trim = 50 50 0 0]{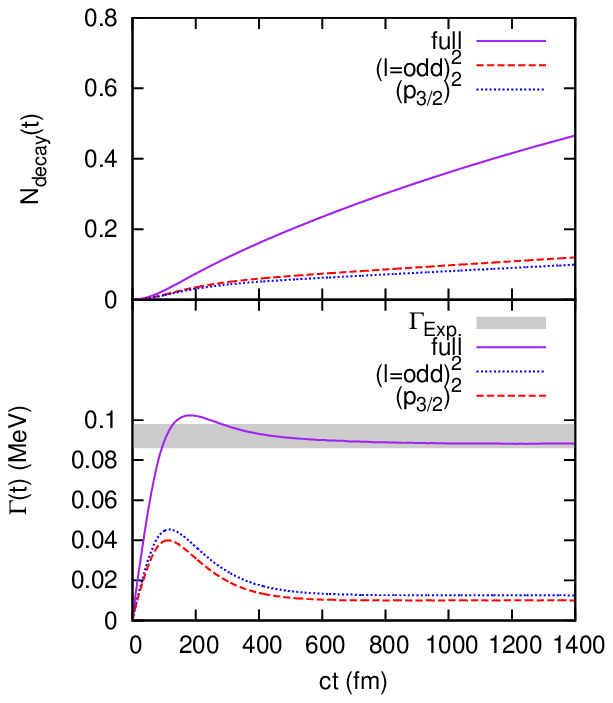}}
 \caption{The decay probabilities and the decay widths 
for the \twop-emissions from $^6$Be, obtained with the time-dependent method. 
The result with all the partial waves in the model-space 
(full mixing) is plotted by the solid line. 
For c$t \geq 1000$ (fm), the decay width well converges to a constant 
value of $88.2$ keV in the full-mixing case. 
The experimental value, $\Gamma_{\rm Exp} = 92 \pm 6$ keV \cite{02Till}, 
is marked by the bold line. 
The results obtained only with $(p_{3/2})^2$ (dotted line) 
and $(l=odd)^2$ (broken line) partial waves are also shown, 
where the calculated decay widths are clearly 
underestimated. } \label{fig:694}
\end{center} \end{figure}

\section{Decay Width}
Starting from the initial state obtained in the previous section, 
we perform the time-evolving calculations for the first $0^+$ 
resonance of $^{6}$Be. 
We show the results of the decay-component $N_d(t)$ 
and width $\Gamma (t)$ 
(see the Eqs.(\ref{eq:603DComp}) and (\ref{eq:width}) in 
the previous Chapter) 
obtained with the time-evolution in Fig. \ref{fig:694}. 

In Fig. \ref{fig:694}, the calculation is carried out up to 
$ct = 0-1400$ fm. 
We have confirmed that the artifact due to the reflection 
at $r=R_{\rm box}$ is negligible in this time-interval. 
One can clearly see that, 
after a sufficient time-evolution, the decay width converges 
to a constant value for all the cases, 
and the exponential decay-rule is realized. 
Furthermore, the result for the full case yields the saturated value 
of $\Gamma (t) \cong 88.2$keV, which 
reproduces the experimental decay width, 
$\Gamma = 92 \pm 6$ keV \cite{88Ajzen,02Till}. 
On the other hand, 
the limitation of the partial waves only to $(l=odd)^2$ or $(p_{3/2})^2$ 
significantly underestimates the decay width. 
This is caused by an increase of the pairing attraction: 
With the $(l=odd)^2$ or $(p_{3/2})^2$ waves only, 
to reproduce the empirical Q-value, 
we needed a stronger pairing attraction (see Table \ref{tb:71111}). 
The two protons are then strongly bound to each other 
and are difficult to go outside, even they have a similar energy release 
of that in the full mixing case. 

From these studies, we can conclude that the parity-mixing 
in the core-proton subsystem is indispensable in order to 
reproduce simultaneously the Q-value and the decay 
width of the \twop-emission. 
This result supports the assumption of the diproton 
correlation at $t=0$. 
\begin{figure*}[htb] \begin{center}
  \begin{tabular}{c} 
     \begin{minipage}{0.48\hsize} \begin{center}
     (a) full \\ 
     \fbox{ \includegraphics[height=42truemm, scale=1, trim = 50 50 0 0]{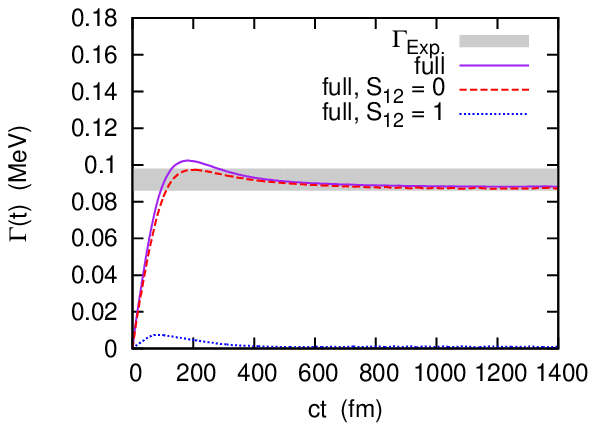}}
     \end{center} \end{minipage}
     \begin{minipage}{0.48\hsize} \begin{center}
     (b) $(l=odd)^2$ only \\ 
     \fbox{ \includegraphics[height=42truemm, scale=1, trim = 50 50 0 0]{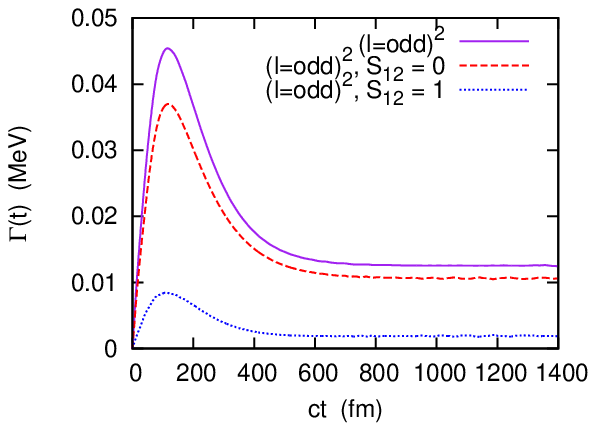}}
     \end{center} \end{minipage}
  \end{tabular}
  \caption{(a) The total and the partial decay widths for the spin-singlet 
and triplet configurations of $^6$Be. 
The partial waves from $(s_{1/2})^2$ to $(h_{11/2})^2$ are fully included. 
(b) The same as panel (a) but for the case with 
only $(l=odd)^2$ partial waves. } \label{fig:5-12}
\end{center} \end{figure*}

For the cases of full mixing and only $(l=odd)^2$ partial waves, 
we also calculate the partial decay widths for the 
spin-singlet and triplet configurations. 
The corresponding formula to Eq.(\ref{eq:pwidth}) is given as 
\beq
 \Gamma_{S_{12}}(t) \equiv \frac{\hbar }{1-N_d(t)} \frac{d}{dt} N_{d,S_{12}}(t), 
\eeq
where 
\beqa
 N_{d,S_{12}}(t) &\equiv& \Braket{\Phi_{d,S_{12}}(t) | \Phi_{d,S_{12}}(t)} \nonumber \\
 &=& \int_0^{R_{\rm box}} dr_1 \int_0^{R_{\rm box}} dr_2 \int_{0}^{\pi} d\theta_{12} 
     8\pi^2 r_1^2 r_2^2 \sin \theta_{12} \abs{\Phi_{d,S_{12}} (t;r_1,r_2,\theta_{12})}^2, \\
 \ket{\Phi_d (t)} &\equiv& 
 \ket{\Phi (t)} - \beta(t) \ket{\Phi (0)}, \label{eq:dst}
\eeqa
with $\beta (t) = \Braket{\Phi (0) | \Phi (t)}$. 
The results are shown in Fig. \ref{fig:5-12}. 
Clearly, the spin-singlet configuration almost exhausts 
the decay width in the full mixing case shown in Fig. \ref{fig:5-12}(a). 
This suggests that the emitted two protons from the ground state of $^6$Be 
have mostly the configuration of $S_{12}=0$, like a diproton. 
On the other hand, from Fig. \ref{fig:5-12}(b), 
one can see that 
the spin-triplet configuration occupies a considerable amount 
of the total decay width when we exclude $(s,d,g)^2$ partial waves. 
\begin{table}[htb] \begin{center}
  \catcode`? = \active \def?{\phantom{0}} 
  \begingroup \renewcommand{\arraystretch}{1.2}
  \begin{tabular*}{\hsize} { @{\extracolsep{\fill}} ccccc cc} \hline \hline
    & & full & $(l=odd)^2$ only & $(l=odd)^2 \oplus (s_{1/2})^2$ & no pairing & exp.data \\
    & & & & & ($ct=3000$ fm) & \cite{88Ajzen,02Till} \\ \hline
    & $\Gamma_{\rm total}$ (keV) & 88.2 & 12.5 & 35.8 & 348. & 92(6) \\
    & $\Gamma_{S_{12}=0}$ (keV) & 87.1 & 10.7 & 34.3 & 232. & - \\
    & $\Gamma_{S_{12}=1}$ (keV) & ?1.1 & ?1.8 & ?1.5 & 116. & - \\ \hline \hline
  \end{tabular*}
  \endgroup
  \catcode`? = 12 
  \caption{The contributions from the spin-singlet and triplet 
configurations to the decay width of $^6$Be. 
The values are evaluated at $ct=1200$ fm, except for the ``no pairing'' case, 
whose values are evaluated at $ct=3000$ fm. 
Note that in the all cases, the Q-value of the \twop-emission 
is reproduced consistently to the experimental value, 
$1.37$ MeV \cite{88Ajzen,02Till}. } \label{tb:6473}
\end{center} \end{table}

In the 2nd and the 3rd columns of Table \ref{tb:6473}, 
we tabulate the total and partial widths in the full and 
the $(l=odd)^2$ cases, respectively. 
The values are estimated at $ct=1200$ fm, where the total widths 
sufficiently converge. 
Clearly, there is a significant increase of the spin-singlet width 
in the full mixing case, 
by about one-order magnitude larger than that in the case of $(l=odd)^2$. 
On the other hand, we get similar values of the spin-triplet width 
in the full and $(l=odd)^2$ cases. 
From this result, 
we can conclude that the core-nucleon parity-mixing is responsible 
for the enhancement of the spin-singlet emission, 
although the dominance of the spin-singlet configuration 
in the initial state is apparent in both the 
two cases (see Table \ref{tb:71111}). 

The dominance of the spin-singlet 
configuration is due to the $(s_{1/2})^2$ channel. 
Considering the coupled orbit, $L_{12} \equiv l_1 \oplus l_2$, 
from the coupling rule to the spin-parity of $0^+$, 
the $(s_{1/2})^2$ channel leads to $S_{12} = L_{12} = 0$ 
with $l_1 = l_2 = 0$. 
Because there is no centrifugal barrier in this channel, 
the spin-singlet emission can be dominant. 
On the other hand, for the spin-triplet configuration, 
the only $L_{12} = 1$ are permitted in order to have the 
total angular momentum $0^+$. 
Thus, there is a centrifugal barrier for all the channels in 
the spin-triplet configuration. 
Consequently, apart from the reduction due to the stronger 
pairing attraction, 
the spin-triplet width has similar values to each other 
in the full-mixing and $(l=odd)^2$ cases. 
\begin{figure*}[htb] \begin{center}
  \begin{tabular}{c} 
     $(l=odd)^2 \oplus (s_{1/2})^2$ \\ 
     \fbox{ \includegraphics[height=42truemm, scale=1, trim = 50 50 0 0]{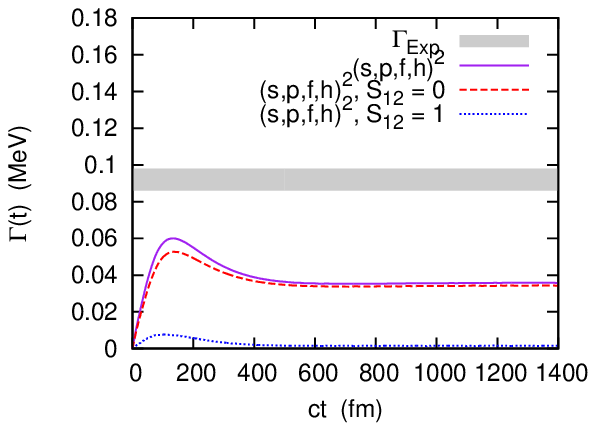}}
  \end{tabular}
  \caption{The same as Fig. \ref{fig:5-12} but for the 
$(l=odd)^2$ plus $(s_{1/2})^2$ case. } \label{fig:7ljfg}
\end{center} \end{figure*}

In order to check this effect of the $(s_{1/2})^2$ waves directly, 
we perform the same calculation but including the 
uncorrelated bases with $(l=odd)^2$ and 
$(s_{1/2})^2$ configurations. 
Namely, we add only the $(s_{1/2})^2$ waves to the $(l=odd)^2$ case. 
In this case, we use the same parameters for the calculation as those 
for the full-mixing and the $(l=odd)^2$ cases, 
except for the $v_0$ in the Minnesota potential: 
we use $v_0=99.14$ MeV in order to reproduce the Q-value, 
$Q_{2p}=1.37$ MeV for $^6$Be. 
The result is shown in Figure \ref{fig:7ljfg} and in the 4th column 
in Table \ref{tb:6473} in the same manner as the former two cases. 
One can see that the spin-singlet width is significantly increased 
due to the existence of the $(s_{1/2})^2$ channel, whereas the 
spin-triplet width has a similar value to those in the 
full-mixing and $(l=odd)^2$ cases. 
This result supports our former speculation about the role 
of the $(s_{1/2})^2$ channel in the \twop-emission. 
Notice also that, because of the stronger pairing attraction, 
the total width in Figure \ref{fig:7ljfg} is still underestimated 
than the experimental data. 

\section{Time-Evolution of Decay State}
In order to discuss the emission process, 
we show the density distribution of the decay state, 
\beqa
 && \bar{\rho}_{d}(t) = 8\pi^2 r_1^2 r_2^2 \sin \theta_{12} \rho_d(t), \\
 && \rho_d(t) = \abs{\Phi_{d} (t;r_1,r_2,\theta_{12})}^2. 
\eeqa
The most of the amplitude of the decay state exists 
outside the potential barrier, 
because we prepare the initial state, which is orthogonal to the decay state, 
so as to have no amplitude in that region. 
For the presentation, 
we renormalize the $\bar{\rho}_d(t)$ so that 
its integration become unity at each time: 
\beq
  \bar{\rho}_d (t) \longrightarrow \frac{\bar{\rho}_d (t)}{N_d (t)}, 
\eeq
where $N_d (t)$ is the decay probability given by Eq.(\ref{eq:603DComp}). 
We adopt three sets of radial coordinates in the following. 
(i) The first set includes 
$r_{\rm c-pp} = (r_1^2 + r_2^2 + 2r_1r_2\cos \theta_{12})^{1/2}/2$ and 
$r_{\rm p-p} = (r_1^2+r_2^2-2r_1r_2\cos \theta_{12})^{1/2}$, 
similarly to the left panel of Figure \ref{fig:2}. 
(ii) In the second set, we integrate $\bar{\rho}_d$ with respect to 
the opening angle, $\theta_{12}$, 
and plot it as a function of $r_1$ and $r_2$. 
In order to see the peak-structure clearly, 
we omit the radial weight $r_1^2 r_2^2$ in 
$\bar{\rho}_d$ in the second setting. 
(iii) Within the third set, on the other hand, 
we integrate $\bar{\rho}_d(t)$ over radial distances, 
and plot it as a function of $\theta_{12}$. 
\begin{figure*}[tp] \begin{center}
  \begin{tabular}{c} 
     \begin{minipage}{0.48\hsize} \begin{center}
        (a) diproton \\
     \fbox{ \includegraphics[height=48truemm, clip, trim = 0 20 0 10]
 {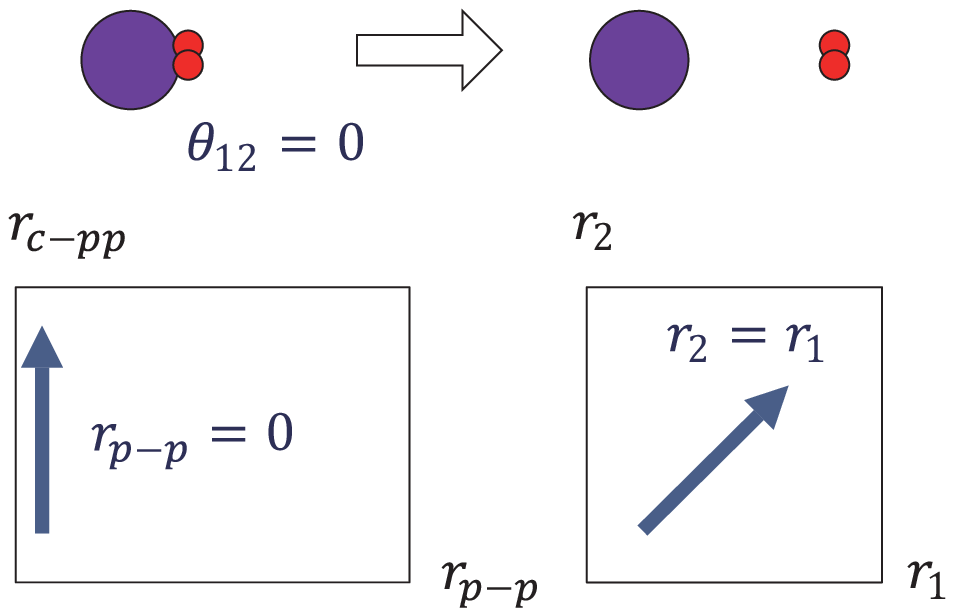}} \vspace{10pt} \\
        (c) simultaneous, $\theta_{12}=0$ \\
     \fbox{ \includegraphics[height=48truemm, clip, trim = 0 20 0 10]
 {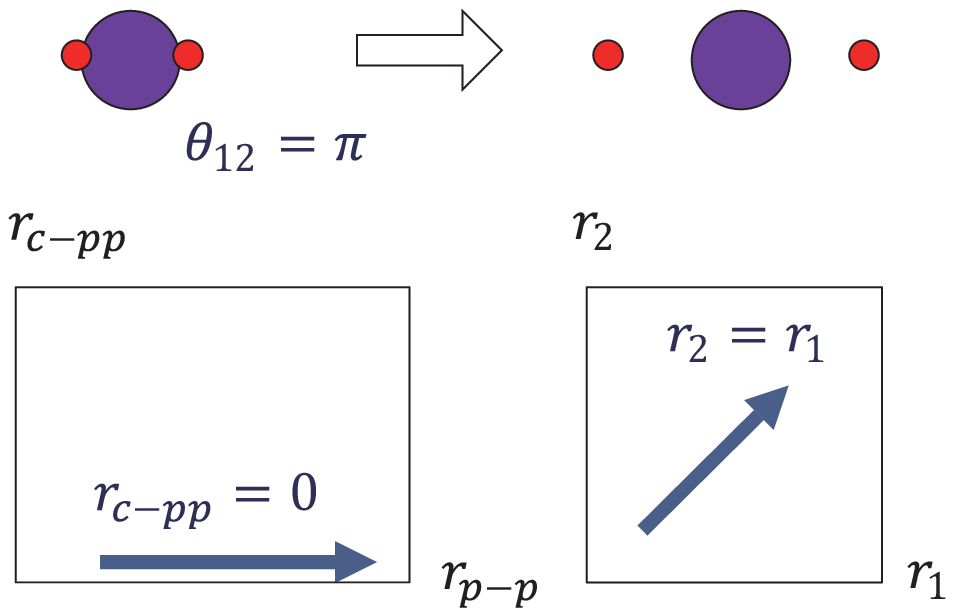}} \vspace{10pt} \\
        (e) correlated \\
     \fbox{ \includegraphics[height=48truemm, clip, trim = 0 20 0 10]
 {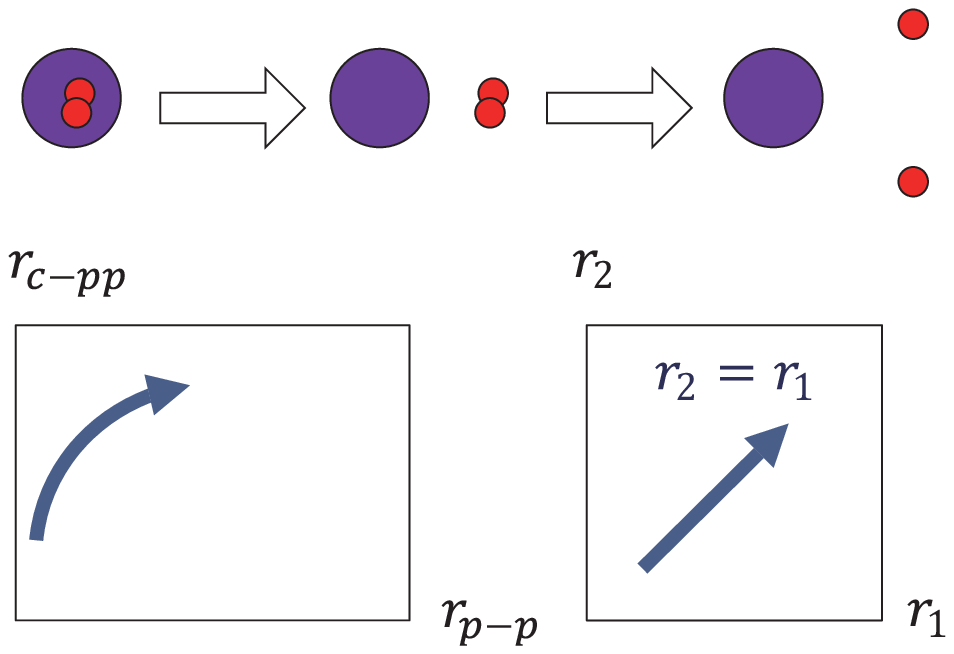}} \vspace{10pt} \\
     \end{center} \end{minipage}

     \begin{minipage}{0.48\hsize} \begin{center}
        (b) simultaneous, $\theta_{12}=\pi/2$ \\
     \fbox{ \includegraphics[height=48truemm, clip, trim = 0 20 0 10]
 {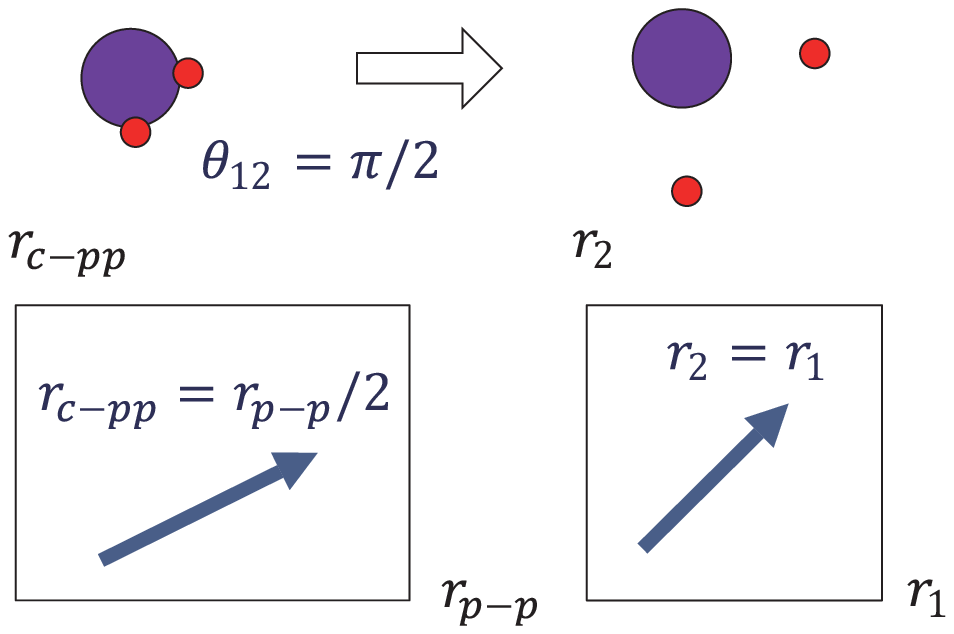}} \vspace{10pt} \\
        (d) one-proton \\
     \fbox{ \includegraphics[height=48truemm, clip, trim = 0 20 0 10]
 {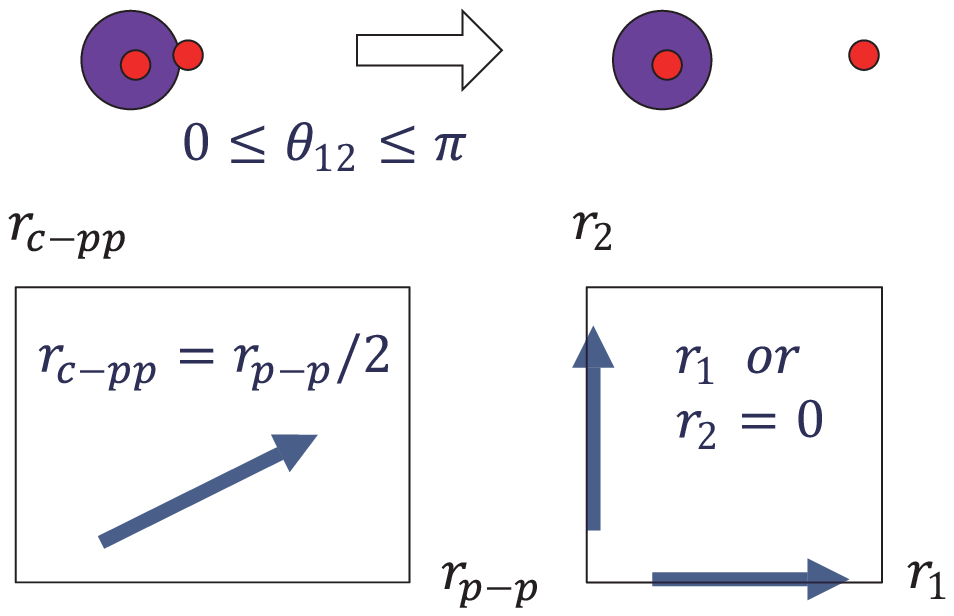}} \vspace{10pt} \\
        (f) sequential \\
     \fbox{ \includegraphics[height=48truemm, clip, trim = 0 20 0 10]
 {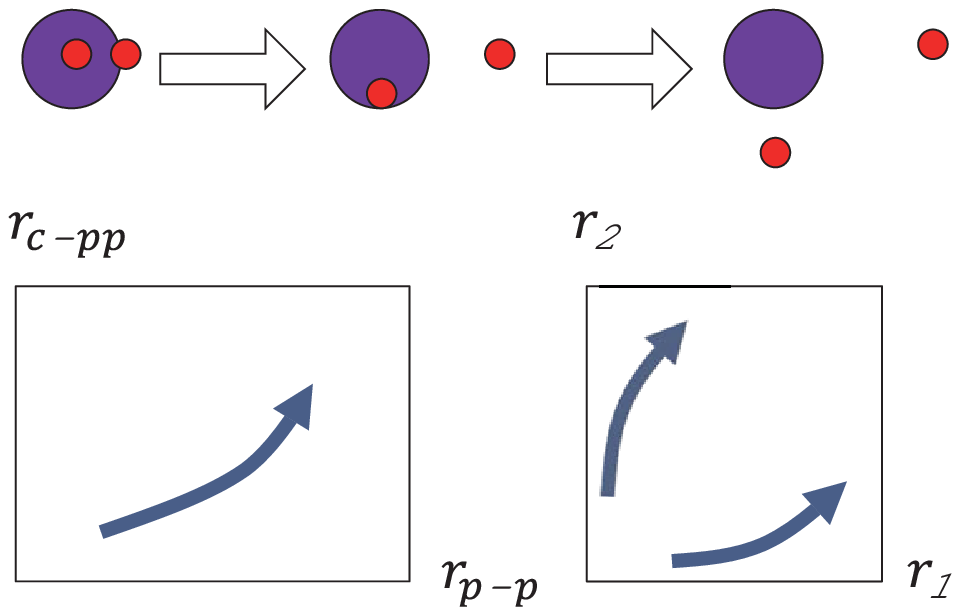}} \vspace{10pt} \\
     \end{center} \end{minipage}
  \end{tabular}
  \caption{Schematic illustrations for the trajectories of 
different \twop-emission modes. } \label{fig:80}
\end{center} \end{figure*}

Before we show the results of the actual calculations, 
we schematically illustrate the dynamic of the 
\twop-emissions in Fig. \ref{fig:80}. 
From the geometry, one can distinguish two modes: 
``simultaneous two-proton'' and ``one-proton ($1p-$)'' emissions. 
The diproton emission is a special case in the first category. 
The second category corresponds to the case 
where only one proton penetrate the barrier. 
The trajectories of three simultaneous \twop- and a $1p$-emissions 
are schematically shown in 
Fig. \ref{fig:80}(a), (b), (c) and (d). 

In the simultaneous emissions, 
two protons are emitted simultaneously with their opening 
angle remaining from 
$\theta_{12}=0$ to $\pi$, where $\theta_{12}=0$ corresponds to 
the diproton emission. 
Fig. \ref{fig:80}(a), (b) and (c) correspond to $\theta_{12}=0,\pi/2$ 
and $\pi$. respectively. 
In these cases, the density in the $(r_1,r_2)$-plane shows the same 
patterns in these figures, and is concentrated along $r_1\cong r_2$. 
The simultaneous emissions with different opening angles 
 can be distinguished only in the $(r_{\rm p-p}, r_{\rm c-pp})$-plane: 
for instance, in the diproton emission, 
the probability shows mainly along the line with 
$r_{\rm c-pp} \gg r_{\rm p-p}$, 
while it is along the line with $r_{\rm c-pp}=0$ for $\theta_{12}=\pi$. 
In the one-proton emission shown in Fig.\ref{fig:80}(d), 
only one of the two protons goes through while 
the other proton remains inside the core nucleus. 
This is seen as the increment along $r_{\rm c-pp} \cong r_{\rm p-p}/2$ 
and $r_1$ or $r_2 \cong 0$ lines. 
\begin{figure*}[tp] \begin{center}
  \begin{tabular}{c} 
   \begin{minipage}{0.32\hsize} \begin{center}
     \fbox{ \includegraphics[height=31truemm, scale=0.9, trim = 60 55 0 0]
{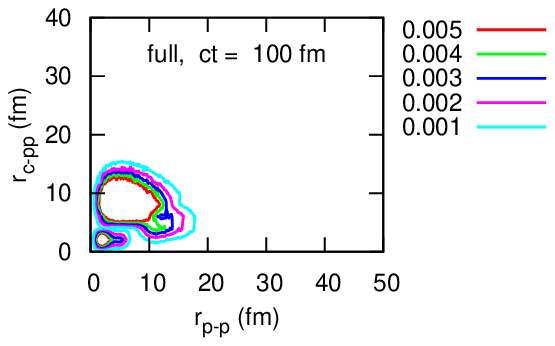}} \\
     \fbox{ \includegraphics[height=31truemm, scale=0.9, trim = 60 55 0 0]
{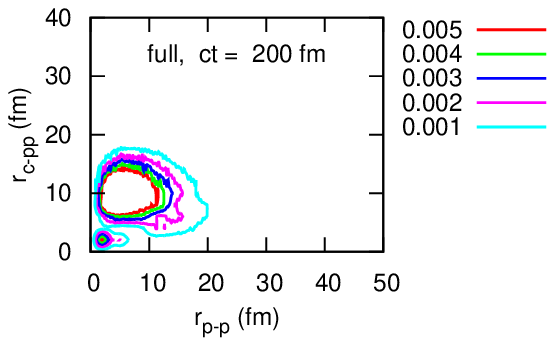}} \\
     \fbox{ \includegraphics[height=31truemm, scale=0.9, trim = 60 55 0 0]
{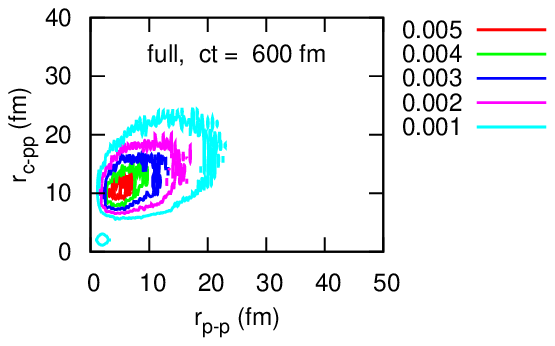}} \\
     \fbox{ \includegraphics[height=31truemm, scale=0.9, trim = 60 55 0 0]
{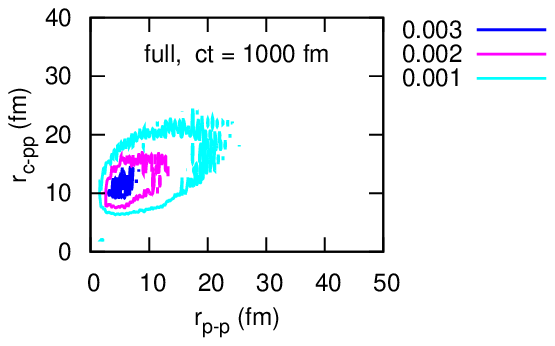}} \\
   \end{center} \end{minipage}

   \begin{minipage}{0.32\hsize} \begin{center}
     \fbox{ \includegraphics[height=31truemm, scale=0.9, trim = 60 55 0 0]
{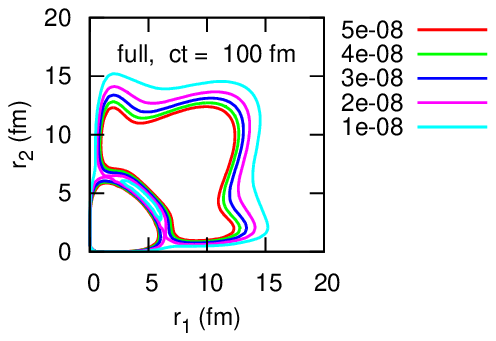}} \\
     \fbox{ \includegraphics[height=31truemm, scale=0.9, trim = 60 55 0 0]
{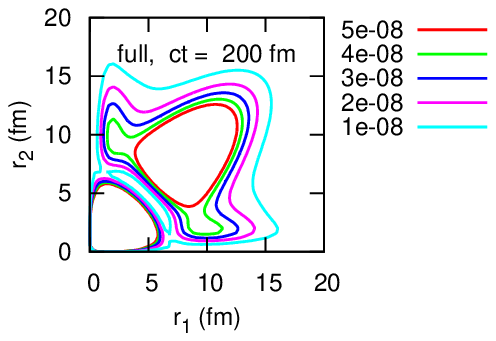}} \\
     \fbox{ \includegraphics[height=31truemm, scale=0.9, trim = 60 55 0 0]
{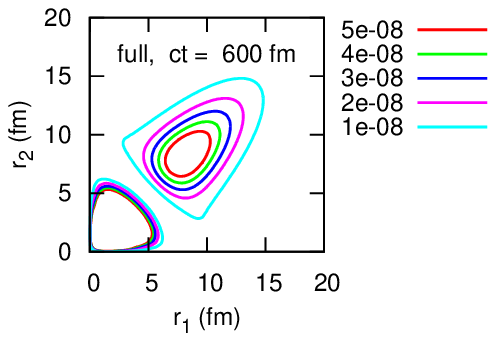}} \\
     \fbox{ \includegraphics[height=31truemm, scale=0.9, trim = 60 55 0 0]
{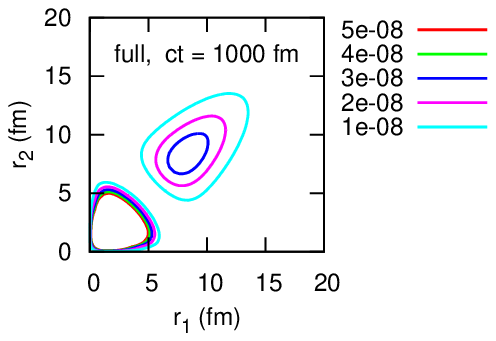}} \\
   \end{center} \end{minipage}

   \begin{minipage}{0.32\hsize} \begin{center}
     \fbox{ \includegraphics[height=31truemm, scale=0.7, trim = 60 55 0 0]
{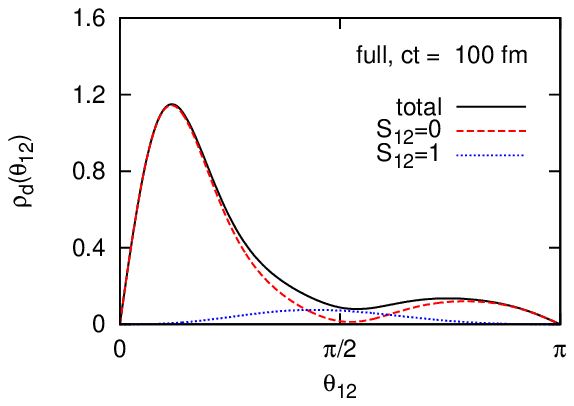}} \\
     \fbox{ \includegraphics[height=31truemm, scale=0.7, trim = 60 55 0 0]
{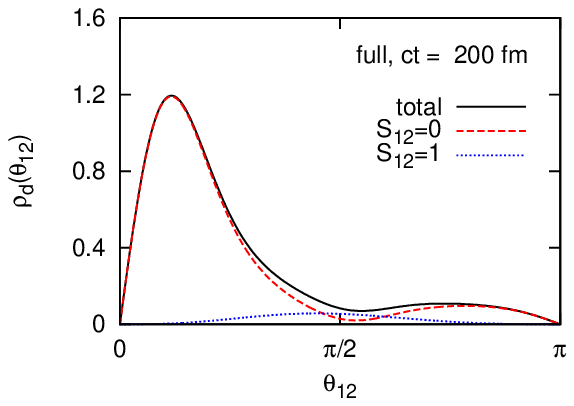}} \\
     \fbox{ \includegraphics[height=31truemm, scale=0.7, trim = 60 55 0 0]
{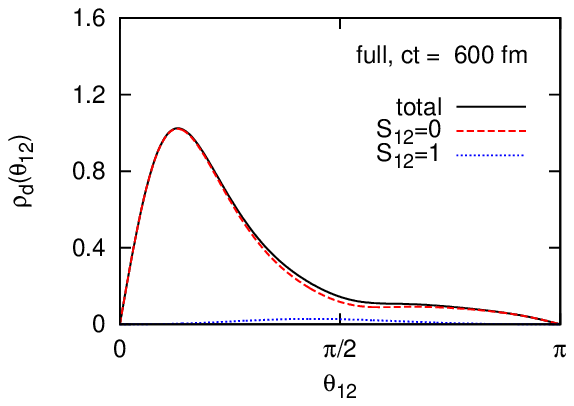}} \\
     \fbox{ \includegraphics[height=31truemm, scale=0.7, trim = 60 55 0 0]
{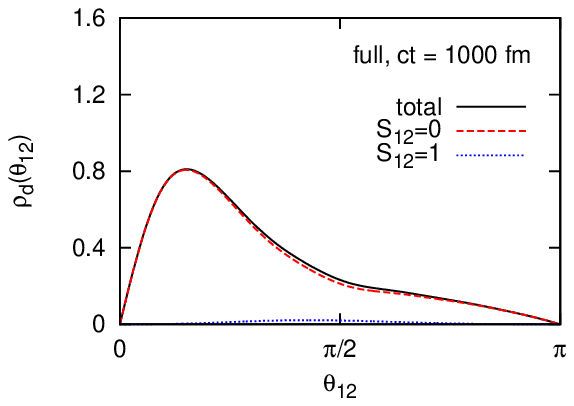}} \\
   \end{center} \end{minipage}
  \end{tabular}
\caption{The \twop-density distribution for the decay states, 
$\bar{\rho}_d(t)$, obtained with the time-evolving calculations. 
All the uncorrelated partial waves up to $(h_{11/2})^2$ are included. 
(The left column): These distributions are plotted as a function of 
$r_{\rm c-pp} = (r_1^2 + r_2^2 + 2r_1r_2\cos \theta_{12})^{1/2}/2$ and 
$r_{\rm p-p} = (r_1^2+r_2^2-2r_1r_2\cos \theta_{12})^{1/2}$. 
(The middle column): The same as the left column but 
as a function of $r_1$ and $r_2$, 
obtained by integrating $\bar{\rho}_d$ for $\theta_{12}$. 
In order to clarify the peak(s), 
the radial weight $r_1^2 r_2^2$ is omitted. 
(The right column): The angular distributions of the decay state 
plotted as a function of the opening angle $\theta_{12}$ 
between the two protons. 
It is obtained by integrating $\bar{\rho}_d(t)$ 
for the radial coordinates. 
Beside the total distribution, the spin-singlet and triplet 
components are also plotted. } \label{fig:81}
\end{center} \end{figure*}

In Fig. \ref{fig:80}(e) and (f), 
we additionally illustrate the two hybrid processes. 
The first one is a ``correlated emission'', 
shown in Fig. \ref{fig:80}(e). 
In the correlated emission, 
the two protons are emitted simultaneously 
to almost the same direction, 
holding the diproton-like configuration. 
In this mode, at the earlier stage of tunneling, 
the density distribution 
has a larger amplitude in the region with 
$r_1 \cong r_2$ and small $\theta_{12}$. 
In the $(r_{\rm p-p}, r_{\rm c-pp})$-plane, 
It corresponds to the increment of the probability 
in the region of $r_{\rm p-p} \ll r_{\rm c-pp}$. 
After the barrier penetration, the 
two protons separate from each other 
mainly due to the Coulomb repulsion, 
increasing $r_{\rm p-p}$. 

The second hybrid process is a ``sequential emission'', 
which is shown in Fig. \ref{fig:80}(f). 
In this mode, 
there is a large possibility of that one proton is emitted 
whereas the other proton remains around the core. 
The density distribution shows high peaks along 
$r_1 \gg r_2$ and $r_1 \ll r_2$. 
In the $(r_{\rm p-p}, r_{\rm c-pp})$-plane, it corresponds to 
the increment along the line of 
$r_{\rm c-pp} \cong r_{\rm p-p}/2$. 
Being different from the pure one-proton emission, 
the remaining proton eventually goes 
through the barrier also 
when the core-proton subsystem is unbound. 
\begin{figure*}[tp] \begin{center}
  \begin{tabular}{c} 
(Figure is hidden in open-print version.)    
  \end{tabular}
\caption{The same as Fig.\ref{fig:81} but for the case with 
only $(l=odd)^2$ waves. 
Notice a different scale in the left column 
from that in Fig.\ref{fig:81}. } \label{fig:82}
\end{center} \end{figure*}

\subsection{Full-Mixing Case}
We now show the results of the time-dependent calculations 
for the \twop-emission of $^6$Be. 
We first discuss the full-mixing case which is the closest 
assumption to reality. 
The density distribution for the decay state 
along the time-evolution is shown in Fig. \ref{fig:81}. 
The left, middle and right columns correspond to the coordinate sets 
(i), (ii) and (iii), respectively. 
The 1st to 4th panels in each column show the decay-density at 
$ct=100,200,600$ and $1000$ fm, respectively. 
For a presentation purpose, we normalize $\bar{\rho}_d$ 
at any step of time. 

In the left and middle columns of Fig. \ref{fig:81}, 
it can be seen that the process in this case 
is likely the correlated emission shown in Fig. \ref{fig:80}(e). 
Contributions from the other modes shown 
in Fig. \ref{fig:80} are small. 
In the middle column of  Fig. \ref{fig:81}, 
during the time-evolution, 
there is a significant increment of 
$\bar{\rho}_d$ along the line with $r_1 \cong r_2$. 
The corresponding peak in the left column is at 
$r_{\rm p-p} \ll r_{\rm c-pp} \cong 10$ fm, which means 
a small value of $\theta_{12}$. 
It should also be noted that, after the barrier penetration, 
the two protons lose their diproton-like configuration 
due to the Coulomb repulsion 
increasing $r_{p-p}$. 
Thus, for $r_{\rm c-pp} \geq 10$ fm which is 
a typical position 
of the potential barrier from the core, 
the density distribution extends around 
the $r_{\rm c-pp} \cong r_{\rm p-p}$ region. 
In this process, the pairing correlation plays an important role 
to generate the significant diproton-like configuration before 
the end of the barrier penetration, 
similarly to the dinucleon correlations. 

In the right column of Fig. \ref{fig:81}, the distributions 
are also displayed as a function of 
the opening angle, $\theta_{12}$. 
We can clearly see that the decay state has a high peak at 
$\theta_{12} \cong \pi/6$, and thus 
the emitted two protons should show the opening angles 
close to this value. 
However, this result may appear 
somewhat inconsistent to the experiments, 
in which the correlation is much weaker in the observed 
angular distribution of $^{6}$Be \cite{09Gri_80, 09Gri_677} 
(see Fig. \ref{fig:2009Gri_06Be}). 
A reason for this discrepancy is due to the final-state 
interactions (FSIs) at the late stage of propagation of 
the two protons. 
In the experiments, 
the observed spectra and the correlation patterns 
correspond to those at the 
late-time region, 
where the two protons have been much 
influenced by FSIs. 
On the other hand, in this thesis, we mainly discuss the 
earlier stage of the \twop-emission with a 
small value of $R_{\rm box}$. 
By taking the FSIs into account at the late stage, 
we expect that we achieve a better agreement 
between the theoretical and experimental results. 
For this purpose, however, we would have to 
expand the model space defined 
with $R_{\rm box}$ and $l_{\rm max}$, 
which would severely increase the 
computational costs. 

\subsection{Case of $(l=odd)^2$ Waves}
We next discuss the case only with $(l=odd)^2$ waves 
(Fig. \ref{fig:82}). 
Even though the experimental $Q_{\rm 2p}$ and $\Gamma_{\rm 2p}$ 
are not simultaneously represented in this case 
(see Fig. \ref{fig:5-12}), 
it is still useful to discuss the density distribution in order 
to know what happens when the pairing correlation 
between the parity-plus and minus states in the core-proton 
subsystem is absent. 
In Fig. \ref{fig:82}, 
the decay density shows strong patterns as the sequential emission 
introduced in Fig. \ref{fig:80}(f): 
significant increments occur 
along the lines with $r_{\rm c-pp} \cong r_{\rm p-p}/2$ and 
$r_1 \gg r_2$ or $r_1 \ll r_2$. 
Notice that the contribution from the simultaneous 
emissions also exists, especially in the earlier time region. 
As a result, the decay state has widely spread amplitudes 
as a mixture of these emission modes. 
However, the simultaneous mode is quite minor compared with 
the full mixing case. 
Notice that the character of a true \twop-emitter exists also 
in this case: 
the core-proton resonance is located at $1.96$ MeV 
which is above $Q_{\rm 2p}1.37$ MeV. 
Even with the strong pairing attraction and the energy 
condition of the true \twop-emitter, 
the process hardly becomes the correlated emission when the 
parity-mixing is forbidden or extensively suppressed. 
On the other hand, 
the angular distribution shows exactly the symmetry form, 
and is almost invariant during the time-evolution. 
In this calculation, 
we exclude the pairing correlation 
between the parity-plus and minus states in the core-proton, 
not only at $t=0$ but also during the time-evolution. 
In other words, there are almost no FSIs to alter the shape of 
the angular distribution. 

\subsection{Without Pairing Correlation}
For a comparison with the above two cases, 
we also perform similar calculations but by 
completely neglecting the pairing correlation. 
In this case, we only consider the uncorrelated 
Hamiltonian, $h_1 + h_2$. 
Because of the absence of the non-diagonal components 
in the Hamiltonian matrix, 
it can be proved that, if the s.p. resonance is at 
an energy $\epsilon_0$ with its width $\gamma_0$, 
the \twop-resonance should be at $2 \epsilon_0$ with its 
width $2 \gamma_0$ since there are no 
couplings between the two protons. 
The \twop-wave function is expanded on the uncorrelated 
basis with a single set of angular quantum numbers. 
Namely, 
\beq
 \ket{\Phi_{(lj)}(t)} = \sum_{n_a,n_b} C_{n_a,n_b,l,j} 
                        \ket{\tilde{\Psi}_{n_a,n_b,l,j}}, 
\eeq
where $(lj)=(p_{3/2})$ for $^6$Be. 
In order to reproduce the empirical Q-value of $^6$Be, 
we inevitably modify the core-proton potential. 
We employ $V_0=-68.65$ MeV instead of that in the full mixing case 
to yield the s.p. resonance at $\epsilon_0(p_{3/2}) = 1.37/2 = 0.685$ MeV, 
with which the core-proton scattering data are not reproduced and 
the character of a true \twop-emitter disappears. 
With this potential, we get the s.p. resonance with 
a broad width: $\gamma_0(p_{3/2}) \cong 170$ keV. 
\begin{figure}[t] \begin{center}
    \fbox{ \includegraphics[height=55truemm, scale=1, trim = 50 50 0 0]{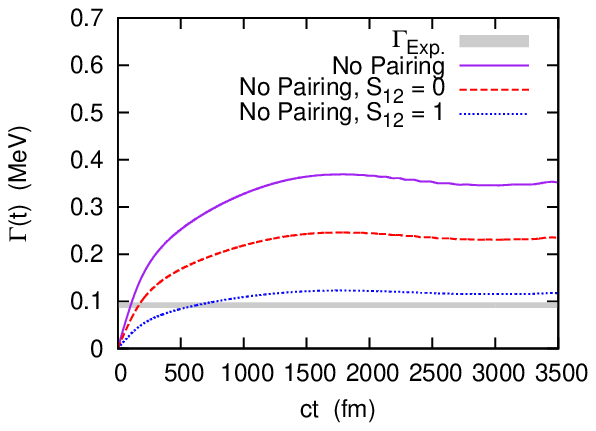}}
\caption{The same as Fig. \ref{fig:5-12} but for the case 
without the pairing correlations. } \label{fig:5-3}
\end{center} \end{figure}

The result for the \twop-decay width is shown in Fig. \ref{fig:5-3} 
and in the last column of Table \ref{tb:6473}. 
To get the saturated result, 
we somewhat need a relatively longer time-evolution than that 
in the full mixing and the $(l=odd)^2$ cases. 
Thus, in Table \ref{tb:6473}, we estimate the decay 
at $ct3000$ fm by width the result well converges 
in this case. 
We also expand the radial box to $R_{\rm box}=200$ fm in order 
to neglect the artifact due to the reflection 
in the longer time-evolution. 
After a sufficient time-evolution, 
the total decay width, $\Gamma(t)$, converges to about $340$ keV 
which is consistent to that expected from 
the s.p. resonance, $\gamma_0(p_{3/2})$. 
During the time-interval shown in Fig. \ref{fig:5-3}, 
there still remain some oscillations in $\Gamma(t)$. 
This is a characteristic behavior of the broad resonance, 
namely the oscillatory deviation from the exponential decay-rule. 
For the spin-singlet and triplet configurations, 
their contributions have exactly the ratio of $2:1$. 
This result is simply due to including 
only $(p_{3/2})^2$ partial waves, 
and is proved by calculating the coefficient $D_J$ in Eq.(\ref{eq:3iarg}). 
\begin{figure*}[tbp] \begin{center}
\begin{tabular}{c} 
(Figure is hidden in open-print version.)
\end{tabular}
\caption{The same as Fig. \ref{fig:81} but for the case 
without the pairing correlations 
and a deeper $V_{\rm c-p}$. } \label{fig:83}
\end{center} \end{figure*}

By comparing the results with those in 
the full mixing case, where the pairing correlations are fully 
taken into account, 
we can clearly see a decisive role of the pairing correlations 
in \twop-emissions. 
Assuming the empirical Q-value, 
if we explicitly consider the pairing correlations, 
the decay width becomes narrow and agrees with 
the experimental data. 
On the other hand, in the no pairing case, 
we need a modified core-proton interaction to reproduce the 
empirical Q-value, and 
the core-proton resonance properties become inconsistent 
with the experimental data. 
Even though the Q-value is adjusted in this way, 
the calculated \twop-decay width 
is significantly overestimated in this case. 
Namely, we cannot simultaneously reproduce 
the experimental Q-value and the decay width with 
the no pairing assumption. 
If one is focused to reproduce them simultaneously, 
one may need unphysical assumptions 
for the core-proton interactions. 
In the next Section, we will present further investigations about 
this problem. 

In Fig. \ref{fig:83}, we show the density distribution 
of the decay state during the time-evolution. 
Obviously, the process is the sequential or, moreover, like 
the one-proton emission in this case. 
There is a significant increase of the density 
along the lines with $r_{\rm c-pp} \cong r_{\rm p-p}/2$ and, consistently, 
with $r_1 \gg r_2$ and $r_1 \ll r_2$ (see Fig. \ref{fig:80} again). 
On the other hand, the probability for the simultaneous and 
correlated emissions are 
negligibly small. 
We emphasize that this is quite different from that in the 
full mixing case, where the correlated emission is apparent. 
Notice that, 
with a disagreement with the experimental decay width, 
this result should not correspond to 
the \twop-emission of $^6$Be in reality. 
This situation can be interpreted as the limit where the 
core-proton resonance plays an excessively dominant role. 

\section{Role of Pairing Correlation}
In order to discuss the role of the pairing correlations
in the \twop-emission, 
we calculate the \twop-decay width for different Q-values, 
for the full-mixing and the no pairing cases. 

To this end, the Q-value is 
varied by modifying the parameter $V_0$ in 
the core-proton potential (Eq.(\ref{eq:cp_WS})). 
In the previous calculations, 
we used $V_0=-58.7$ and $V_0=-68.65$ MeV 
in the full mixing and the no pairing cases, respectively. 
These original values yield the empirical Q-value, $Q_{\rm 2p}=1.37$ MeV. 
In addition to these original values, 
we change the value of $V_0$ as 
$V_0 \pm 0.5$ and $V_0 \pm 1.0$ MeV. 
The calculated decay widths are well converged after 
a sufficient time-evolution in all the cases. 
We note that, in the full mixing case, we adopt the same pairing 
interaction as in the previous calculation. 
\begin{figure*}[htbp] \begin{center}
  \begin{tabular}{c} 
\fbox{\includegraphics[width=0.5\hsize, scale=1, trim = 50 50 0 0]{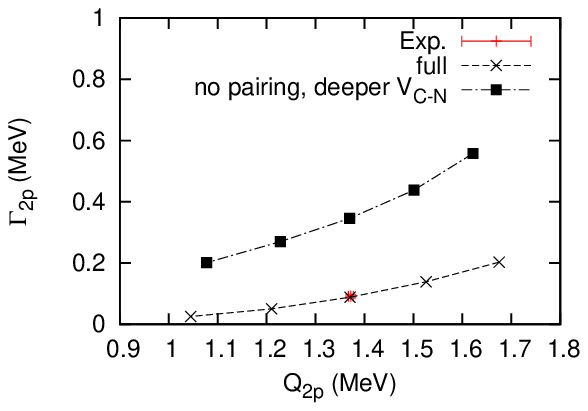}}
  \end{tabular}
\caption{The calculated decay widths for the \twop-emission of $^{6}$Be, as a function of the Q-value. 
The Q-value is varied by modifying the core-proton potential. 
The experimental values are indicated as the point at 
$Q_{\rm 2p}=1.37$ and $\Gamma_{\rm 2p}=0.092(6)$ MeV. } \label{fig:5qp}
\end{center} \end{figure*}

In Fig. \ref{fig:5qp}, 
the decay width is plotted as a function of the Q-value. 
The decay width in each case is evaluated at 
$ct=1200$ and $3000$ fm in the full mixing and 
the no pairing cases, respectively. 
Clearly, the no pairing calculation overestimates the decay width, 
in all the region of $Q_{\rm 2p}$. 
Namely, the three-body system becomes easier to decay without 
the pairing correlations compared to the full-mixing case, 
even if we consider the same value of the total energy (Q-value). 
In other words, the pairing correlation plays an essential role in the 
meta-stable state, stabilizing it against particle emissions. 
Moreover, as we have confirmed in the previous section, 
the emission modes with and without the 
pairing correlations are essentially different to each other: 
the correlated emission is suggested if the pairing correlation 
is fully considered, 
whereas omitting it yields the sequential emission. 
Of course, this result can be associated with the character of $^6$Be 
as a true \twop-emitter. 
Consequently, we conclude that the pairing correlation must be 
treated explicitly in the meta-stable states, 
or one would miss the essential effect on the 
dynamical phenomena. 

\section{Summary of this Chapter}
We have applied the time-dependent three-body model 
to $^6$Be, which has a close 
character to the true \twop-emitter. 
The initial state of $^6$Be has the diproton correlation, 
similarly to the 
dineutron correlation in the ground state of $^6$He. 
The empirical relation between 
$Q_{\rm 2p}$ and $\Gamma_{\rm 2p}$ is well 
reproduced by fully including the pairing correlation 
(in the full mixing case). 
We have also showed that the decay process at 
its earlier stage is mainly 
the correlated emission, 
in which the two protons are emitted 
to the same direction with $S_{12}=0$, like a diproton. 
The dominance of the spin-singlet decay width is 
explained as the effect of the $(s{1/2})^2$ wave. 

We have performed the calculations by switching off a part of 
the pairing correlation in order to study 
its role in the \twop-emissions. 
First, we excluded the parity-mixing in the core-proton subsystem, 
equivalently to forbidding the diproton correlation 
in particle-bound nuclei. 
Notice that the character of a true \twop-emitter exists 
also in this assumption. 
In this case, the decay width and its spin-singlet ratio are 
remarkably underestimated compared to the full mixing case. 
The decay process has a large component of the sequential emission, 
which is quite different from the correlated emission. 
From this result, we can infer that the diproton correlation 
is essential in describing the \twop-emission. 
Second, we completely omitted the pairing correlation, and 
adjusted the mean-field between the core and a proton to 
reproduce the Q-value of the emitted two protons. 
The character of a true \twop-emitter no longer exists in this case. 
It was shown that the pairing correlation plays an essential role 
in the meta-stable states: 
omitting the pairing correlation leads to a largely overestimated 
decay width, and almost the perfect sequential emission which scarcely 
exists in the full mixing case. 

At this moment, the dependence of \twop-emissions on the initial 
diproton correlation is strongly suggested, 
but this has not yet been proved. 
Indeed, the FSIs must be taken into account at the late stage 
of the time-evolution, 
in order to probe the diproton correlation with the 
experimental observables. 
Towards this goal, we plan to expand our model space defined with 
$R_{\rm box}$ and $l_{\rm max}$, enabling us to perform the longer 
time-evolution where the FSIs play a dominant role. 
The sensitivity to the diproton correlation is translated to the 
initial-configuration dependence of observables. 
If the $Q_{\rm 2p}$ and $\Gamma_{\rm 2p}$ are by no means 
reproduced simultaneously by excluding 
the diproton correlation at $t=0$, 
we will be able to conclude the presence of the 
diproton correlation. 
Possibly, for instance, we will also infer that the observed 
signals associated with the diproton-emission \cite{09Gri_80, 09Gri_677} 
reflect the survived components originally emerged at $t=0$. 
The time-dependent method, which can distinguish the cause and 
the effect in the observables, 
will be a powerful tool in these discussions. 
We also mention that the other approaches within 
complex-energy framework 
is hard to separately discuss the early and late time regions, or 
equivalently, the cause and the effect. 
Therefore, our studies will produce a complementary point of view to 
the \twop-emission and possibly the diproton correlation. 

The expansion of the model space, however, will lead to a serious increment 
of computational costs. 
To overcome this difficulty, we will have to adopt an improved boundary 
condition which does not emerge the reflection of the wave function at 
the edge of the radial box, 
or/and more efficient bases which can reduce the dimension 
of the Hamiltonian matrix. 
Additionally, we should also concern the pairing interaction. 
The pairing interaction employed in this thesis 
should be regarded as an effective interaction, 
since it is inconsistent to the scattering 
problem of \twop in vacuum due to our modification of $v_0$. 
Within further expanded model space, it may cause the unphysical result. 
One may also introduce a three-body force, 
which works only if three particles 
are close to each other \cite{95Aoyama,01Myo,10Kiku}. 
The effect of this three-body force on decay processes is an 
important topic, in regard to whether such an interaction is 
really just a phenomenological one 
or has a physical meaning beyond the two-body force. 
\include{end}
\documentclass[a4paper,12pt]{book}
\include{begin}

\chapter{Two-Proton Emission of $^{16}$Ne} \label{Ch_Results3}
For further investigation of \twop-emissions in this Chapter, 
we take up another \twop-emitter, $^{16}$Ne. 
To this end, we assume the 
three-body system of $^{14}$O and two valence protons. 
The quantum meta-stability is treated within 
the time-dependent framework. 
In the case of $^6$Be discussed in the previous Chapter, 
our time-dependent three-body model 
well reproduced the experimental $Q_{\rm 2p}$ and $\Gamma_{2p}$ 
simultaneously. 
It suggests that the three-body assumption is valid for this nucleus. 
On the other hand, it has been a serious problem that a similar 
theoretical three-body model within the complex-energy framework 
do not simultaneously reproduce the $Q_{\rm 2p}$ and $\Gamma_{2p}$ of 
other light \twop-emitters \cite{02Gri}. 
Thus, it is worthwhile to 
check whether our model works or not for these nuclei. 
The application to the $^{16}$Ne nucleus to be discussed in this Chapter 
is one example of studies towards this direction. 
\begin{figure}[htb] \begin{center} 
\fbox{\includegraphics[width=0.6\hsize]{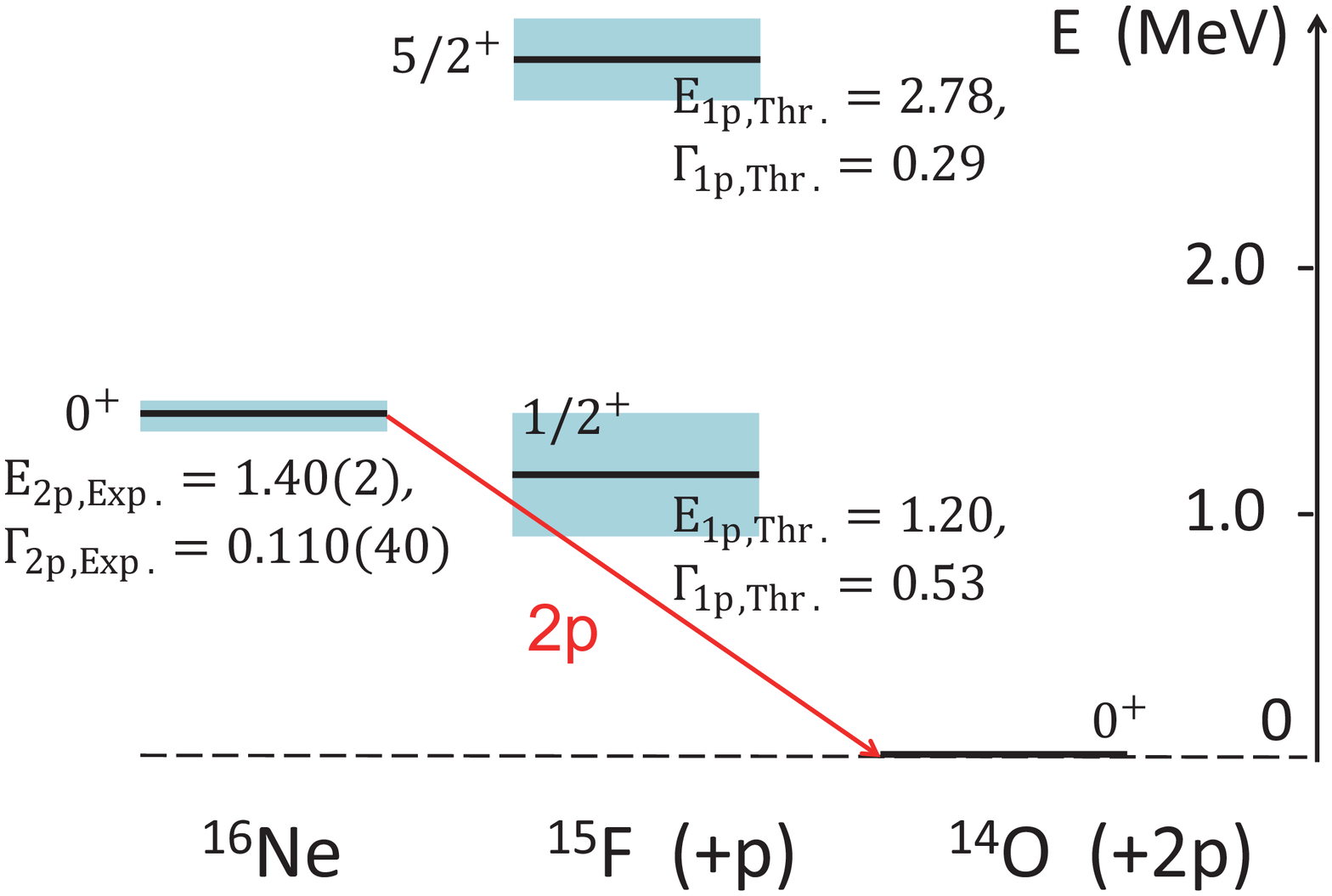}}
\caption{The level scheme of $^{16}$Ne and its isotones. 
The printed values for $^{16}$Ne are taken from 
the Ref. \cite{83Wood}. 
Those for $^{15}$F are calculated with the core-proton 
interaction which is originally introduced in 
the Ref. \cite{10Mukhamed}. 
The color-box of each level indicates its decay width. } \label{fig:801}
\end{center} \end{figure}

In Figure \ref{fig:801}, we show the level scheme of $^{16}$Ne 
and its daughter nuclei after the $1p$- and \twop-emissions, 
measured from the ground state of $^{14}$O. 
Indeed, the $1p$-resonance of the $^{14}$O+p in the $(s_{1/2})$-channel 
is located near the \twop-resonance of $^{16}$Ne. 
Its width, $\Gamma_{\rm 1p} \simeq 500$ keV 
is so large that one may wonder whether the resonance character truly 
exists in this system or not. 
The sequential emission through the core-proton channel is 
expected to be minor, due to its broad width. 
We also stress that there are still ambiguities in the 
experimental data of the first resonance 
of $^{15}$F \cite{78KeKe,03Peters,10Mukhamed,91Ajzen,03Szily,04Goldberg,05Guo}. 
Consequently, it is still unclear whether 
the $^{16}$Ne nucleus is a true \twop-emitter or not. 
In this work, we will assume a relatively low energy and 
narrow width in the $(s_{1/2})$-channel, as detailed in the 
next section. 

\section{Set up for Calculations}
General assumptions for the numerical calculations are similar to those 
for the $^6$Be in the previous Chapter. 
We assume that the core nucleus, $^{14}$O is a structureless 
particle with the spin-parity of $0^+$. 
Because the first resonance state of $^{16}$Ne also has the spin-parity 
of $0^+$, we only need the $0^+$ uncorrelated basis 
for the valence two protons. 
The calculations are performed in the truncated space 
defined by the energy-cutoff: 
$\epsilon_a + \epsilon_b \leq E_{\rm cut} =40$ MeV. 
The continuum states are discretized within a radial box of $R_{\rm box}=80$ fm. 
For the angular momentum channels, we take $l_{\rm max}=5$, that is, 
we include all the partial waves from $(s_{1/2})^2$ to $(h_{11/2})^2$. 
\begin{figure}[t] \begin{center}
$V_{\rm c-p}(r)$ and $V_{\rm c-p}^{conf}(r)$ \\
\fbox{\includegraphics[width=0.5\hsize,scale=1, trim = 50 50 0 0]{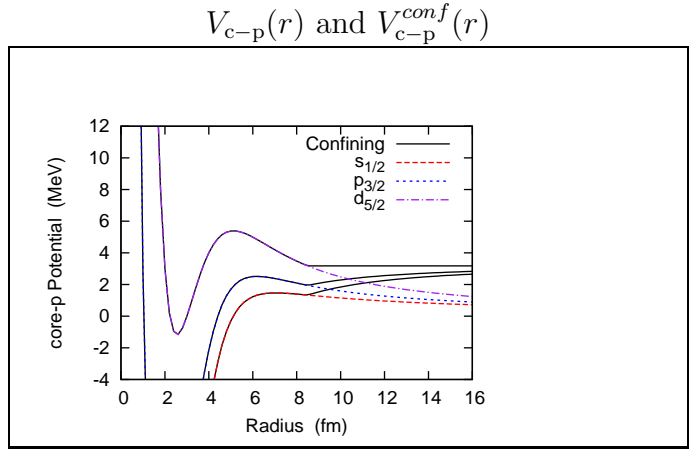}}
\caption{The original and confining potentials for the 
$(s_{1/2})$, $(p_{3/2})$ and $(d_{5/2})$ channels in the $^{14}$O-p subsystem. 
The border radius for modifying the potential is 8.5 fm for all the channels. 
} \label{fig:812}
\end{center} \end{figure}
\begin{figure*}[htb] \begin{center}
  $^{16}$Ne (g.s.), $t=0$, ``full'' \\
  \begin{tabular}{c} 
     \begin{minipage}{0.48\hsize} \begin{center}
        \fbox{ \includegraphics[height=45truemm,scale=1,trim = 50 50 0 0]{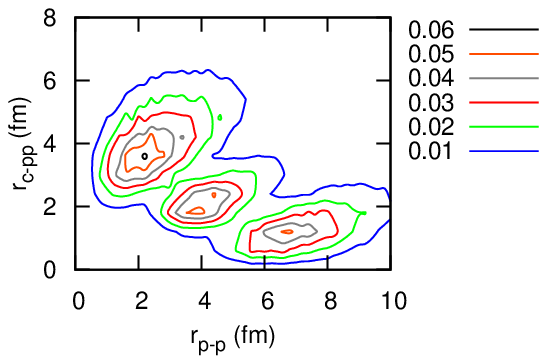}}
     \end{center} \end{minipage}
     \begin{minipage}{0.48\hsize} \begin{center}
        \fbox{ \includegraphics[height=45truemm,scale=1,trim = 50 50 0 0]{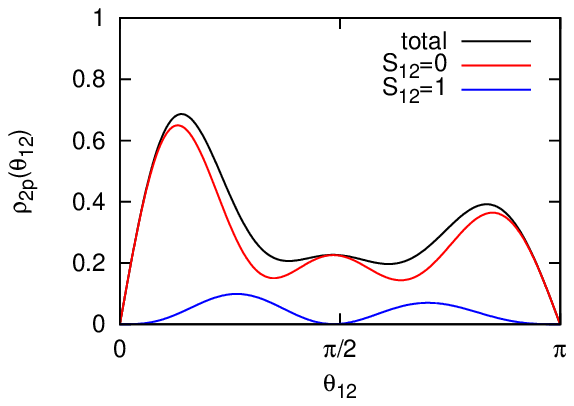}}
     \end{center} \end{minipage}
  \end{tabular}
  \caption{(The left panel) 
The \twop-density distribution at $t=0$ for the ground state of $^{16}$Ne. 
It is obtained by including all the partial waves up to $(h_{11/2})^2$, 
and plotted as a function of 
$r_{\rm c-pp} = (r_1^2 + r_2^2 + 2r_1r_2\cos \theta_{12})^{1/2}/2$ and 
$r_{\rm p-p} = (r_1^2+r_2^2-2r_1r_2\cos \theta_{12})^{1/2}$. 
(The right panel) The angular distribution at $t=0$ obtained 
by integrating $\bar{\rho}_{2p}(t=0)$ with $r_1$ and $r_2$. } \label{fig:813}
\end{center} \end{figure*}

We assume the Woods-Saxon and Coulomb potentials between 
the core and a proton, similarly to Eq.(\ref{eq:cp_pot6}). 
We employ the same parameters as those in the 
Ref. \cite{10Mukhamed}, in which the authors discussed 
the scattering problem of $^{14}$O+p theoretically. 
The first and second resonances obtained with these parameters are 
shown in Fig. \ref{fig:801}. 
We calculate and fit the derivative of the phase-shift, according to 
Eq.(\ref{eq:apcps}), to get $E_{\rm 1p}$ and $\Gamma_{\rm 1p}$. 
These values are consistent to several experimental 
results \cite{78KeKe,04Goldberg,05Guo}. 

For the proton-proton pairing interaction, 
we use the Minnesota potential given by Eq.(\ref{eq:Minne}) 
in this case. 
In order to reproduce the Q-value of \twop-emission, 
$Q_{\rm 2p} \equiv \Braket{H_{\rm 3b}}=1.40$ MeV, 
we adjust the strength of the repulsive part as $v_0=126.2$ MeV. 
The other parameters in the Minnesota potential are fixed to 
the original values in ref.\cite{77Thom}. 

The initial \twop-state for the time-evolution is defined as 
a quasi-bound state obtained with the confining potentials, 
similarly 
to the previous calculations for $^6$Be. 
The confining potentials for $^{16}$Ne are defined as follows. 
In Chapter \ref{Ch_Results1}, we have 
confirmed that in $^{17,18}$Ne nuclei, 
the valence two protons are mainly in the $(d_{5/2})^2$-orbit. 
Because $^{16}$Ne is an isotope of these nuclei, 
the first resonance of $^{16}$Ne is also expected to have a large 
component of $(d_{5/2})^2$-configuration. 
According to this consideration, 
for the single particle (s.p.) $(d_{5/2})$-channel, 
we modify the core-proton potential at 
$t=0$ in order to fix the quasi-bound state as follows. 
\beq
 V_{{\rm c-p},~(d_{5/2})}^{conf}(r) \label{eq:8dcv}
 = \left\{ \begin{array}{cc} 
            V_{{\rm c-p},~(d_{5/2})}(r) & (r \leq R_b), \\
            V_{{\rm c-p},~(d_{5/2})}(R_b) & (r > R_b), \end{array} \right.
\eeq
with $R_b=8.5$ fm in this case. 
For other s.p. channels, we define it as 
\beq
 V_{\rm c-p}^{conf}(r) \label{eq:8scv}
 =\left\{ \begin{array}{cc} 
          V_{\rm c-p}(r) \phantom{0000} & (r \leq R_b), \\
          V_{\rm c-p}(r) + V_b(r) & (r > R_b), \end{array} \right.
\eeq
where $V_b(r) = V_{{\rm c-p},~(d_{5/2})}(R_b) - V_{{\rm c-p},~(d_{5/2})}(r)$. 
The original and confining potentials for the 
$(s_{1/2})$, $(p_{3/2})$ and $(d_{5/2})$ channels are 
shown in Fig. \ref{fig:812}. 

By diagonalizing the modified Hamiltonian including $V_{\rm c-p}^{conf}(r)$, 
we obtain the initial \twop-state. 
In Figure \ref{fig:813}, we show the initial density, 
$\rho_{2p}(t=0)$, and its angular distribution. 
One can clearly see a significant diproton correlation, 
characterized by 
the localization of the two protons in the $S_{12}=0$ configuration. 
The spin-singlet ratio is obtained as $P(S_{12}=0)=87.9$\%. 
The prominent three peaks are due to the dominant $(d_{5/2})^2$ wave, 
with its probability of $49.56$\% in this state. 
In addition, the $(s_{1/2})^2$ wave 
has a comparable probability of $44.70$\% 
in this state. 
The other waves with $(l=odd)^2$ and $(l=even)^2$ carry the 
probabilities of $3.61$\% and $2.13$\%, respectively. 
It is worthwhile to compare this result with that of $^{17}$Ne 
obtained in Chapter \ref{Ch_Results1}. 
By comparing the left panel of Fig. \ref{fig:813} with 
Fig. \ref{fig:461}(a), 
it can be seen that the spatial distribution 
is a little more extended in the $^{16}$Ne nucleus. 
This is consistent with the increment of the $(s_{1/2})^2$ wave, 
which has a long tail outside the core-proton potential. 
While this result can be interpreted as a characteristic difference 
between the bound and meta-stable \twop-states, 
anyway, the diproton correlation is still suggested in the 
initial state of $^{16}$Ne. 
If our time-dependent calculation yields the decay width that 
is consistent with the experiments, 
we can associate the behavior of the emitted two protons 
(at the earlier stage) with the diproton correlation. 
\begin{figure}[tb] \begin{center}
\fbox{\includegraphics[height=80truemm,scale=1,trim = 50 50 0 0]{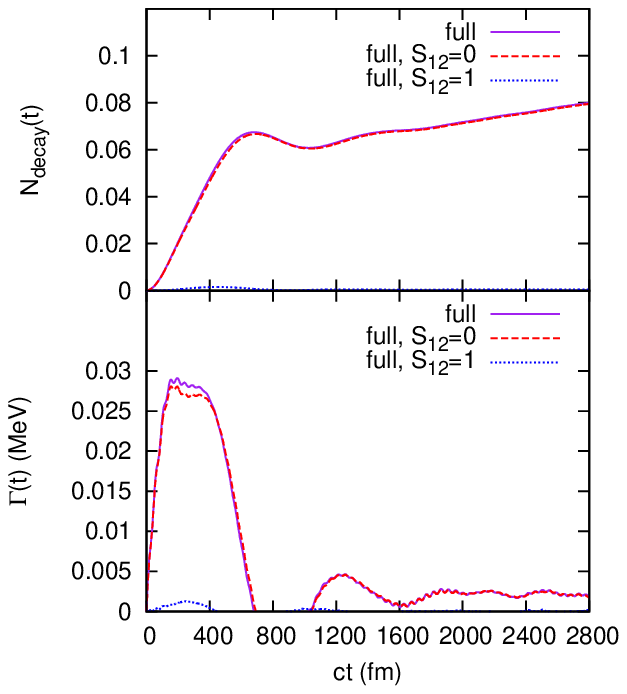}}
\caption{The decay probabilities and the decay width 
of \twop-emissions from $^{16}$Ne, obtained with the time-dependent method. 
Those for the spin-singlet and triplet configurations are also plotted. 
In calculations, all the partial waves up to $(h_{11/2})^2$ are included. 
The parameters of the pairing interaction are adjusted to reproduce 
the experimental Q-value, $Q_{\rm 2p, Exp.}=1.40(2)$ MeV \cite{83Wood}. 
Note that the experimental decay width, 
$\Gamma_{\rm 2p,Exp.}=110\pm 40$ keV \cite{83Wood} 
is too higher to be indicated 
in the lower panel. } \label{fig:914}
\end{center} \end{figure}

\section{Decay Width}
The results for the decay probability and the decay width are shown 
in Fig. \ref{fig:914}. 
The calculation is carried out up to $ct=2800$ fm 
where the reflection at $R_{\rm box}$ can be neglected. 
Unfortunately, there is a large discrepancy 
between the calculated and the experimental decay widths. 
In Fig. \ref{fig:914}, after a sufficient time-evolution, 
the calculated decay width approximately converges to 
$\Gamma_{\rm 2p,Thr.}\simeq 2-3$ keV, which is underestimated 
against the experimental value, 
$\Gamma_{\rm 2p,Exp.}=110\pm 40$ keV \cite{83Wood}. 
This discrepancy would not be attributed to our small 
model space, or the uncertainties in the time-dependent calculation 
since a similar three-body model calculation also yielded a 
similar discrepancy \cite{02Gri,09Gri_40}. 
Additionally, in the earlier time region, 
the decay probability shows a big bump, causing a large 
oscillation in the decay width, $\Gamma (t)$. 
We do not know exactly whether this bump is just an 
artifact, or originates from the initial configuration, 
including the diproton correlation. 

In order to investigate a possible cause of the underestimated 
decay width, 
we carry out similar calculations but with different values of $v_0$ 
in the Minnesota pairing interaction, 
intuitively discarding the fine set up for the Q-value. 
In Fig. \ref{fig:933}, we show the results obtained with 
$v_0=200.0,168.0$ and $126.2$ MeV. 
The first value is identical to the original parameter \cite{77Thom}, 
whereas the third value is that used in Fig. \ref{fig:914}. 
The decay width is reproduced if we take 
$v_0=168$ MeV, which yields $Q_{\rm 2p,Thr.}=2.04$ MeV. 
In this case, compared with the previous case with $Q_{\rm 2p,Thr.}=1.40$, 
the two protons have a larger energy 
to overcome the potential barriers, and the decay width 
becomes also larger, 
leading to the agreement with the experimental value. 
It means that our naive three-body model leads to 
an over stabilization against the \twop-emission of $^{16}$Ne. 

Notice also that there remains sizable oscillations in $\Gamma(t)$ 
even after a sufficient time-evolution. 
We conjecture that the mixing of the two resonances, namely those in 
the $(s_{1/2})$ and $(d_{5/2})$-channels of the core-proton subsystem, 
is responsible for this result. 
However, we do not explore deeply into this problem here. 
Indeed, we will rather discuss what causes the 
over stabilization of the two protons. 
\begin{figure}[tb] \begin{center}
\fbox{\includegraphics[height=50truemm,scale=1,trim = 50 50 0 0]{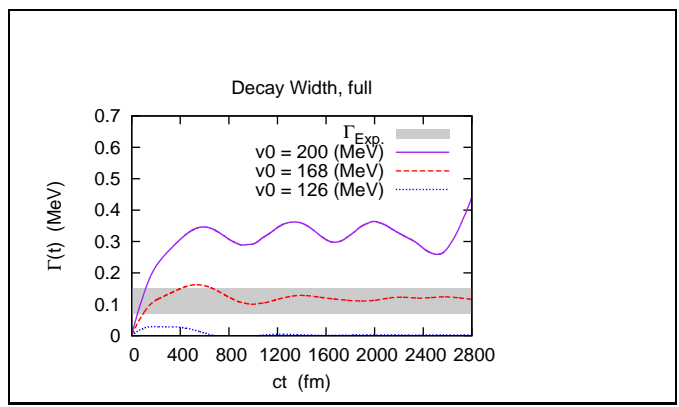}}
\caption{The decay widths obtained for different values of $v_0$ in the 
Minnesota pairing attraction. 
The corresponding Q-values are $Q_{\rm 2p}=2.48$, $2.04$ and $1.40$ MeV for 
$v_0 = 200$, $168$ and $126.2$ MeV, respectively. 
The experimental decay width, 
$\Gamma_{\rm 2p,Exp.} = 110 \pm 40$ keV \cite{83Wood}, 
is also indicated by the shaded area. } \label{fig:933}
\end{center} \end{figure}

\section{Possibilities of Improvements}
The first possible cause of the over stabilization of $^{16}$Ne is 
a lack of core excitations in the present theoretical model. 
In other words, the intrinsic degrees of freedom of the core nucleus 
may not be neglected for $^{16}$Ne. 
One may wonder why the present model works well for $^6$Be, but 
does not for $^{16}$Ne. 
This is due to the stability of the core nuclei, 
namely $alpha$-particle and $^{14}$O for $^6$Be and 
$^{16}$Ne, respectively. 
The first excited state of $\alpha$ is located at $E=20.2$ MeV, 
which is much higher than the first excited state of $^{14}$O 
at $E=5.1$ MeV \cite{NNDCHP}. 
Thus, the core excitation may be relatively important 
in $^{16}$Ne compared to $^{6}$Be. 

Indeed, from recent studies on weakly-bound nuclei, 
it is expected that excitations of the core nucleus play an important 
role in the halo structure and the electro-magnetic excitations of these 
nuclei \cite{93Sagawa,95Esb_2,99Tost,01Shyam,08Myo,12Moro,13Moriguchi}. 
For the meta-stable processes, 
it has been recognized that 
a coupling of the valence particle 
with these degrees of freedom might 
enhance the tunneling probability and thus 
increase the decay width \cite{83Cald,00Esb,00Ferre,02Hagino_1p}. 
The similar effect of the core excitation is expected to exist 
in the \twop-emission, which might restore an discrepancy 
between the calculated and the experimental decay widths. 
The most possible source of these excitations is 
the deformation of the core nucleus. 
However, even if the core is not deformed, 
the considerable component of these excitations exist, 
namely the ``two-particle and two-hole (2p2h-)'' type of excitations 
from the naive shell structure, due to the pairing correlations. 
It includes the 2p2h-excitations by two protons, two neutrons 
and a proton-neutron pair. 
The former two are caused by the ordinary pairing correlations, 
whereas the last one is associated with the tensor force \cite{08Myo}. 
Taking these core excitations into account means 
an extension of the inert-core model, 
and may lead to a relaxation of over 
stabilization of \twop-emissions. 
However, for this purpose, we must expand the model space, 
which would increase the cost of calculations. 
To treat it correctly will be a challenging task in our future works. 

The second possibility is due to a model-dependence of the 
pairing interaction, as partially discussed in the previous Chapter. 
In this thesis, we have adopted the simple Minnesota interaction between 
two protons. 
With other interactions, which include the spin-orbit or the 
momentum-dependence, might reproduce the experimental 
$Q_{\rm 2p}$ and $\Gamma_{\rm 2p}$ simultaneously. 
Furthermore, in order to reproduce the total Q-value, 
we have intuitively modified the parameter of the Minnesota 
potential in this work. 
However, by modifying the parameter, 
two-nucleon scattering property at infinitely far from the core 
is no longer reproduced. 
This deviation may affect the calculated results, 
especially for the meta-stable processes, in which the final-state 
interactions play an important role even far from the core. 
To improve this point, we will have to install the density-dependence 
into the pairing interaction, or employ the phenomenological 
three-body force, which works only when all the particles are 
localized in a small region. 
A work towards this direction is in the progress now. 

\include{end}
\documentclass[a4paper, 12pt]{report}
\include{begin}

\chapter{Summary of Thesis} \label{Ch_Summary}
We have theoretically investigated the diproton correlation and 
its effect on the two-proton emission. 
These are both the exotic features of proton-rich nuclei which are located  
far from the beta-stability line, and thus are the novel interests 
in current nuclear physics. 
Furthermore, these two phenomena can be strongly connected to each other, 
because they have a common basis of physics, namely the nuclear 
pairing correlation. 
The diproton and dineutron correlations are the intrinsic structures 
characterized by a spatial localization of two nucleons of the same kind with 
a large component of the spin-singlet configuration. 
These are exotic features which cannot be reproduced within the 
pure mean-field theory of nuclei, 
and are strongly related to 
the density-dependence of pairing correlations, 
which is an important prediction from the 
modern theory for nuclear structures. 
Recently, the two-proton decays and emissions have attracted 
much interests as an efficient tool to probe the diproton correlation. 
In the observables of the emitted two protons, information on the diproton 
correlation may be reflected. 
In order to establish this idea, further investigations were 
still necessary with a realistic assumptions for calculations. 
Revealing their fundamental relation is expected to provide another 
way to probe the nuclear pairing phenomena 
and the dinucleon correlations, simultaneously 
a development of the advanced nuclear 
theory which covers both the low and high density-regions, and 
both bound and meta-stable systems. 
Nevertheless, few people have discussed this possibility within a realistic 
assumptions of calculations \cite{08Bertulani,12Maru}. 

For this purpose, we have carried out the quantum three-body model 
calculations in this thesis. 
Our calculations provide semi-microscopic 
descriptions for the nuclear pairing correlations. 
By performing model calculations and analyzing their results, 
we have obtained several important conclusions. 
\vspace{24pt}

In the former half of this thesis, we have discussed the diproton correlation 
in bound proton-rich nuclei. 
By calculating $^{18}$Ne, $^{18}$O and $^{17}$Ne nuclei, 
we have confirmed that the diproton correlation exists 
in the ground state of these nuclei, similarly to the dineutron correlation 
in neutron-rich nuclei. 
In these systems, a prominent localization of the two protons and 
neutrons are predicted. 
The spin-singlet configuration takes the major contribution 
to this localization. 
It has also been shown that the Coulomb repulsive force between the two
protons does not affect significantly this correlation. 
Even though this repulsion extends 
the density distribution of two nucleons and 
weakens the binding energy, 
its effect is not sufficiently strong 
to destroy the diproton correlation. 
Our calculations have indicated that the effect 
of the Coulomb force reduces 
the pairing energy gap only by about 10 \%, 
being consistent to 
other theoretical studies. 
We have also found that whether the dinucleon 
correlations exist or not 
is insensitive to the total binding energy 
of a three-body systems. 
Namely, the dinucleon correlations can be 
considered not only in 
loosely bound nuclei, 
but also in deeper bound systems. 

From these results, we can conclude that the dinucleon correlations exist 
almost independently of (i) whether the pair consists of protons or neutrons, 
and of (ii) whether that the pair is loosely or 
deeply bound to the core nucleus. 
Eventually, the dinucleon correlations should be discussed as a 
common property of both stable and unstable nuclei. 
\vspace{24pt}

In the latter half of this thesis, 
we have focused on the relation between 
the diproton correlation and the two-proton emissions. 
In two-proton emissions, a pair of protons are emitted directly from 
the parent nucleus. 
This process is a typical meta-stable phenomenon governed by the 
quantum tunneling effect, and can be an promising tool to 
probe the diproton correlation. 
It can be expected that if the valence two protons have 
the diproton correlation 
inside the potential barrier, its effect can be reflected in the 
decay observables. 
However, in order to extract information on the diproton correlation 
from the decay observables, 
one has to treat both the quantum meta-stability and 
the many-body property on an equal footing. 

For this purpose, we have developed 
the time-dependent three-body model. 
In this model, the initial \twop-state is defined as a quasi-bound 
state within a phenomenological confining potential. 
The quantum tunneling process can be treated by solving 
the time-dependent Schr\"{o}dinger equation. 
The sensitivity of \twop-emissions to the diproton correlation 
has been discussed by studying its dependence on the initial 
configuration of the two protons, 
with or without a diproton-like clustering. 
We would like to emphasize that our time-dependent approach has an advantage 
to treat the quantum meta-stable processes, enabling us to distinguish 
the essential cause of phenomena. 
Especially, for two-proton emissions, 
several theoretical works have already 
been done, most of which have been based on 
the time-independent formalism. 
However, the relation between the observables 
in the \twop-emissions and the nuclear 
intrinsic structures, 
including the diproton correlation, 
has not been discussed. 
Thus, our present study provides a novel insight 
into these important problems. 

By applying this model to the $^6$Be nucleus, 
which is the simplest two-proton emitter, 
We have obtained several results suggesting 
that the diproton correlation is 
reflected in the decay observables. 
To be more specific, 
first, we have confirmed that the experimental two-proton 
decay-width of 6Be is well reproduced only 
by assuming the diproton correlation in the initial state. 
Furthermore, the decay width is mostly from the spin-singlet configuration. 
Second, the emitted two protons are expected to have a diproton-like cluster 
at the early stage of emissions, due to the pairing correlations 
(that is, the correlated emission). 
We have also performed the same calculations but 
based on the pure mean-field model, 
completely ignoring the pairing correlations. 
In such calculations, the decay width is severely overestimated, 
and the emission process shows mostly the pure sequential emission, 
differing from the case with the pairing correlations. 
These results suggest the importance of the 
diproton correlation in the \twop-emissions. 
\vspace{24pt}

At this moment, the strong dependence of the \twop-emission on 
the diproton correlation is suggested. 
It means that the 2p-emission can be an effective tool to evince 
the diproton correlation. 
In order to prove it completely, however, there still remain several 
open problems listed below. 

\begin{enumerate}
\item Final-state interactions: 
Our present results have predicted 
a significant correlated emission, 
including the diproton-like clustering 
in the early stage of the two-proton 
emissions of $^{6}$Be. 
However, on the other hand, there is no significant signal 
of the correlated emission in the experimental angular and 
energy correlation patterns. 
A reason for this discrepancy may be the 
final-state interactions (FSIs). 
In the experimental data, 
there is a strong modification of the correlation pattern 
by the FSIs among all the particles. 
Especially, the long-ranged Coulomb forces can extensively affect the 
two protons during their propagation. 
Consequently, the observed data correspond to the late stage of 
the emission process. 
It has yet to be clarified 
how the diproton correlation at the initial and the earlier stages 
are reflected in the experiments. 
In order to address this question, 
by taking the FSIs sufficiently into account, 
we would need to expand our model-space so that 
a longer time-evolution can be carried out. 
However, at the same time, 
it inevitably leads to a serious 
increment of computing costs, and 
one would need to develop more economic 
procedures for the three-body model calculations 
to resolve this problem. 
Such procedures include, {\it e.g.} 
assuming a sophisticated boundary condition to avoid 
the reflection of wave functions at the edge 
of the box, and employing an 
efficient bases to reduce the dimension 
of the Hamiltonian matrix. 

\item Pairing interaction in the asymptotic region: 
In the present study, we have used the nuclear potential 
between two nucleons, by modifying its parameter 
to reproduce the total Q-value of the \twop-emissions. 
However, such modified potential is not consistent to the 
two-nucleon scattering property in vacuum, 
and may lead to a serious 
error in the calculated results. 
Especially, in order to reproduce the 
angular distributions of the two protons, 
we might have to weaken the pairing attraction in the 
asymptotic region. 
For this purpose, one should install the density-dependence 
into the pairing potential, 
or employ a three-body force which works 
as a short-ranged attraction between three particles. 
The effect of the three-body force on the two-nucleon emissions and 
decays is an important problem, as well as whether such an three-body 
force has an physical meaning or not. 

\item Core excitations: 
For the $^6$Be nucleus, our three-body model well reproduces 
the experimental data of the Q-value and decay-width. 
However, for another light two-proton emitter, $^{16}$Ne, 
our calculations have not been successful in reproducing them consistently: 
the \twop-decay width is considerably underestimated even if we employ the 
appropriate parameters for the total Q-value. 
We note that this problem is not only in our calculations but also in other 
studies based on a similar three-body model to ours \cite{02Gri,09Gri_40}. 
Given this discrepancy, we anticipate a limitation 
of the simple three-body model assuming 
an inert, structure-less core, and the importance of the 
core excitations. 
Similar problems have been reported in other studies of, 
{\it e.g.} the nuclear meta-stable processes 
\cite{83Cald,94Varga,00Esb,00Ferre,02Hagino_1p} and 
the structures of nuclei far from the beta-stability line 
\cite{93Sagawa,95Esb_2,99Tost,01Shyam,08Myo,12Moro,13Moriguchi}, 
suggesting that the core excitation plays an important role 
in these phenomena. 
To discard the assumption of an inert core 
may resolve the discrepancy between the 
calculated and the observed \twop-widths. 
Treating it correctly will be an important task in our future works. 
We also note that the explicit treatment of the core excitations 
would be connected to the role of the tensor force, 
because the tensor force causes 
the 2p2h-type of excitations of the core \cite{08Myo}. 
\end{enumerate}

After these improvements, our time-dependent method will be more 
sophisticated and will be able to 
reveal the essential relation between the diproton correlation and 
the two-proton emissions. 
Moreover, similar time-dependent approaches can be 
applied to describe other quantum meta-stable processes 
in few-body systems. 
Especially, considering the dinucleon correlations, 
the most important one may be the two-neutron ($2n$-) emissions. 
In analogy to the relation between the diproton correlation and 
the \twop-emissions, 
the $2n$-emissions can be a powerful tool 
to examine the dineutron correlation 
in neutron-rich nuclei. 
Because of the absence the long-ranged Coulomb FSIs, 
a theoretical treatment may be easier 
than that for the \twop-emissions, 
although the problems of an asymptotic pairing 
interaction and of the 
core excitations still remain. 
Work towards this direction is also an challenging task in the future. 
\vspace{24pt}

The quantum meta-stability plays an important role in various 
situations in our world. 
Especially, those of three or more fermions, or with many degrees of 
freedom, have been one of the most important 
subjects in modern physics. 
However, in spite of its importance, 
there remain a lot of unknown aspects 
of the quantum meta-stability. 
The theoretical treatment still needs a further development, 
where the two complementary (or competing) frameworks coexist at present. 
Atomic nuclei, which show various radioactive processes, 
are one of the most suitable fields to discuss these physics. 
The knowledge gained in this field can be extended to other 
quantum meta-stable phenomena with many fermions or 
with strong correlations. 
Understanding the quantum meta-stability 
among different scales will be a great 
benchmark in future physics. 
\include{end}
\documentclass[a4paper,12pt]{book}
\include{begin}

\chapter*{Acknowledgements}
\addcontentsline{toc}{chapter}{Acknowledgements}

I would like to give my sincere thanks to my supervisor, 
Prof. Kouichi Hagino. 
His guidance and suggestions have been extensively 
helpful at all the stages of my study. 
Special thanks also go to Prof. Hiroyuki Sagawa and 
Mr. Takahito Maruyama, 
whose comments and collaborations 
made enormous contribution to this work. 

I thank 
Dr. Akira Ono, 
Dr. Shuichiro Ebata, 
Mr. Takeshi Yamamoto, 
Mr. Takaho Fujii, 
Mr. Yusuke Tanimura, 
for the meaningful discussions 
in various phases of this study. 
I thank also to 
Prof. Shoichi Sasaki, 
Prof. Masaaki Kimura and 
Prof. Kiyoshi Kato 
for their helpful comments. 

I would like to thank all the people 
who organized the Global COE Program 
``Weaving Science Web beyond Particle-Matter Hierarchy'' 
at Tohoku University, 
for a grant that made it possible to complete this study. 
This work was also supported in part by Grants for Excellent Graduate
Schools, MEXT, Japan. 

Finally, I would like to show my greatest appreciation to 
my family for their continuous support and encouragements. 
Their help throughout my time in the undergraduate 
and the graduate schools has been indispensable for 
the achievement of this work. 
\include{end}

\appendix
\pagestyle{fancy}
  \fancyhead{} 
  \fancyhead[LE,RO]{\leftmark}
  \renewcommand{\chaptermark}[1]{\markboth{Appendix~\thechapter\; \; #1}{}}
  \cfoot{\thepage}
\documentclass[a4paper,12pt]{report}
\include{begin}

\chapter{Numerov Method} \label{Ap_Numerov}
This is the numerical method to solve an ordinary differential equation 
in which only the zeroth and the second order terms are included, such as 
\beq
 \left[ \frac{d^2}{dx^2} + f(x) \right] U(x) = 0, \label{eq:Num0} 
\eeq 
where $f(x)$ is an arbitrary source function. 
The solution, $U(x)$, is sampled at equidistant points $x_n,(n=0 \sim N)$ 
where the distance between two sampling points is defined as $a$. 
With this method, starting from the solution values at 
two consecutive sampling points, 
namely $U_0 \equiv U(x_0)$ and $U_1 \equiv U(x_1)$, 
we can calculate the remaining solution values as 
\beq
 U_{n+2} = \frac{ (2-5a^2f_{n+1}/6)U_{n+1} - (1+a^2f_{n})U_{n} }
                { 1+a^2f_{n+2}/12 } + \mathcal{O}(a^6), \label{eq:Num1} 
\eeq
where we neglect $\mathcal{O}(a^6)$. 
The derivation of Eq.(\ref{eq:Num1}) is based on the discrete Taylor expansion 
for $U(x)$ until the fifth order. 
Considering the two sampling points, $x_{n-1}=x_n-a$ and $x_{n+1}=x_n+a$, 
Taylor expansions are given as 
\beqa
 U_{n+1} &\equiv& U(x_n+a) \nonumber \\
 &=& U_n+aU'_n+\frac{a^2}{2!}U''_n
        +\frac{a^3}{3!}U^{(3)}_n+\frac{a^4}{4!}U^{(4)}_n
        +\frac{a^5}{5!}U^{(5)}_n+\mathcal{O}(a^6), \\
 U_{n-1} &\equiv& U(x_n-a) \nonumber \\
 &=& U_n-aU'_n+\frac{a^2}{2!}U''_n
        -\frac{a^3}{3!}U^{(3)}_n+\frac{a^4}{4!}U^{(4)}_n
        -\frac{a^5}{5!}U^{(5)}_n+\mathcal{O}(a^6), 
\eeqa
where $U^{(m)}_n \equiv \left. d^mU(x)/dx^m \right|_{x=x_n}$. 
The sum of these two equations gives 
\beq
  U_{n-1} + U_{n+1} = 2U_n + a^2U''_n + \frac{a^4}{12}U^{(4)}_n + \mathcal{O}(a^6). 
\eeq
Solving this equation for $a^2U''_n$ leads to 
\beq
 -a^2U''_n = 2U_n - U_{n-1} - U_{n+1} + \frac{a^4}{12}U^{(4)}_n + \mathcal{O}(a^6). 
 \label{eq:Num4} 
\eeq
In this equation, we can replace $U''_n$ to $-f_n U_n$ because of Eq.(\ref{eq:Num0}). 
Similarly, for the fourth term in the right hand side, we can use 
\beq
 U^{(4)}(x) = \frac{d^2}{dx^2}[-f(x)U(x)]. 
\eeq
The numerical definition of the second derivative is given as 
the second order difference quotient, that is 
\beq
 \frac{d^2}{dx^2}[-f(x)U(x)] \Rightarrow 
 -\frac{f_{n-1}U_{n-1} - 2f_nU_n +f_{n+1}U_{n+1}}{a^2}. 
\eeq
After these replacements, Eq.(\ref{eq:Num4}) is transformed as 
\beq
 a^2f_nU_n = 
 2U_n - U_{n-1} - U_{n+1} - 
 \frac{a^4}{12} \frac{f_{n-1}U_{n-1} - 2f_nU_n +f_{n+1}U_{n+1}}{a^2} + 
 \mathcal{O}(a^6). 
\eeq
Finally, we solve this equation for $U_{n+1}$ to get 
\beq
 U_{n+1} = \frac{ (2-5a^2f_n/6)U_n - (1+a^2f_{n-1})U_{n-1} }
                { 1+a^2f_{n+1}/12 } + \mathcal{O}(a^6), 
\eeq
which is equivalent to Eq.(\ref{eq:Num1}). 

\include{end}
\documentclass[a4paper,12pt]{report}
\include{begin}

\chapter{Formalism of Many-Body Coordinates} \label{Ap_3body}
In this Chapter, we introduce the general formalism for the transformation of 
coordinates of many-particle systems. 
For $N$ particles in the three dimensional space, one needs $N$ 
coordinates of space, $\{ \bi{x}_i \},\; i=1 \sim N$. 
In general, these degrees of freedom are separated into the center-of-mass 
coordinate, $\bir _G$, and $N-1$ relative coordinates, 
${\bir_k},\; k=1 \sim N-1$. 
The V-coordinate used in this thesis is just one kind of definitions 
of ${\bir_k},\; k=1,2$, and its definition is detail in the following. 
The derivations of the three-body Hamiltonian in the V-coordinates 
is also introduced. 
For more general formulations and applications, see {\it e.g.} the 
textbook \cite{98Suzuki}. 

\section{Coordinates for Many-Body Problems}
We start our discussions from the original coordinates $\bi{x}_i$ and 
their conjugate momenta $\bi{\pi}_i$. 
These satisfy 
\beq
 [(\bi{x}_i)_{\mu}, (\bi{\pi}_j)_{\nu}] 
 = i\hbar \delta_{ij} \cdot \delta_{\mu \nu} \label{eq:ap101}
 \; \Longrightarrow \; \bi{\pi}_i 
 = -i\hbar \frac{\partial }{\partial \bi{x}_i}, 
\eeq
where $\mu$ and $\nu = x,y,z$. 
Here we define the column vector $\vec{X}$ and $\vec{\Pi}$, whose $i$-th 
component is $\bi{x}_i$ and $\bi{\pi}_i$, respectively. 
\beq
  \vec{X} \equiv 
  \left[ \begin{array}{c} 
  \bi{x}_1 \\ \vdots \\ \bi{x}_N \end{array} \right], \qquad 
  \vec{\Pi} \equiv 
  \left[ \begin{array}{c} 
  \bi{\pi}_1 \\ \vdots \\ \bi{x}_N \end{array} \right]. 
\eeq
Using the transform-matrix $U$, one can define the new set of 
coordinates, $\{ \bir_i \}$, as follows: 
\beq
  \vec{R} \equiv \left[ \begin{array}{c} 
  \bir_1 \\ \vdots \\ \bir_N  \end{array} \right] 
  = U \vec{X} \quad \Longleftrightarrow \quad 
    \vec{X} = U^{-1} \vec{R}, 
\eeq
or equivalently, 
\beq
  \bir_i = \sum_{j=1}^{N} U_{ij} \bi{x}_j \quad \Longleftrightarrow \quad 
  \bi{x}_i = \sum_{j=1}^{N} (U^{-1})_{ij} \bir_j. 
\eeq
The conjugate momenta are also written by applying the chain-rule: 
\beq
  \bi{\pi}_i = -i\hbar \frac{\partial}{\partial \bi{x}_i} 
  = -i\hbar \sum_{j} \frac{\partial \bir_j}{\partial \bi{x}_i} 
    \frac{\partial }{\partial \bir_j} 
  = \sum_{j} U_{ji} \bip_j. 
\eeq
Thus, indicating the transverse matrix of $U$ as ${}^{t}U$, 
the conjugate momenta of $\vec{R}$ can be defined as 
\beq
  \vec{P} \equiv \left[ \begin{array}{c} 
  \bip_1 \\ \vdots \\ \bip_N \end{array} \right] 
  = ({}^{t}U)^{-1} \vec{\Pi} \quad \Longleftrightarrow \quad 
  \vec{\Pi} = {}^{t}U \vec{P}, 
\eeq
or equivalently, 
\beq
  \bip_i = \sum_{j=1}^{N} ({}^{t}U)^{-1}_{ij} \bi{\pi}_j \quad \Longleftrightarrow \quad 
  \bi{\pi}_i = \sum_{j=1}^{N} {}^{t}U_{ij} \bip_j. 
\eeq
Note that the conjugate relation is still satisfied as 
\beq
 [(\bir_i)_{\mu}, (\bip_j)_{\nu}] = i\hbar \delta_{ij} \cdot \delta_{\mu \nu}. 
\eeq
The transform-matrix, $U$, can be chosen arbitrarily and there is no 
mathematical discrimination between different $U$s. 
However, in practice, there are two major coordinates used to solve 
the many-body problems. 
In this thesis, we choose so called ``core-center coordinates'' 
defined by the transform-matrix, 
\beq
 U \equiv \left( \begin{array}{ccccc cc} 
          1 & 0 & 0 & \cdots & 0 & 0 & -1 \\
          0 & 1 & 0 & \cdots & 0 & 0 & -1 \\
          \vdots &&&&&& \\
          0 & 0 & 0 & \cdots & 0      & 1 & -1 \\
          \frac{m_1}{M} & \frac{m_2}{M} & \frac{m_3}{M} & \cdots & \frac{m_{N-2}}{M} & \frac{m_{N-1}}{M} & \frac{m_N}{M} 
          \end{array} \right), 
\eeq
where its inverse matrix is given by 
\beq
 U^{-1} \equiv \left( \begin{array}{ccccc c} 
               1-\frac{m_1}{M} &  -\frac{m_2}{M} &  -\frac{m_3}{M} & \cdots &  -\frac{m_{N-1}}{M} & 1 \\ &&&&& \\
                -\frac{m_1}{M} & 1-\frac{m_2}{M} &  -\frac{m_3}{M} & \cdots &  -\frac{m_{N-1}}{M} & 1 \\ &&&&& \\
                -\frac{m_1}{M} &  -\frac{m_2}{M} & 1-\frac{m_3}{M} & \cdots &  -\frac{m_{N-1}}{M} & 1 \\
               \vdots &&&&& \\
                -\frac{m_1}{M} &  -\frac{m_2}{M} &  -\frac{m_3}{M} & \cdots & 1-\frac{m_{N-1}}{M} & 1 \\ &&&&& \\
                -\frac{m_1}{M} &  -\frac{m_2}{M} &  -\frac{m_3}{M} & \cdots &  -\frac{m_{N-1}}{M} & 1 \\
               \end{array} \right), 
\eeq
with $M\equiv \sum_{i=1}^{N} m_i$. 
In these coordinates, the vector $\bir_N$ indicates the center-of-mass motion, 
whereas $\bir_k$ with $k=1 \sim (N-1)$ indicates the relative motion between the central 
core and the $k$-th particle. 
We schematically indicate these coordinates in the case of three-body systems 
in Figure \ref{fig:A1_1}. 
\begin{figure}[t] \begin{center}
\fbox{ \includegraphics[width = 0.9\hsize]{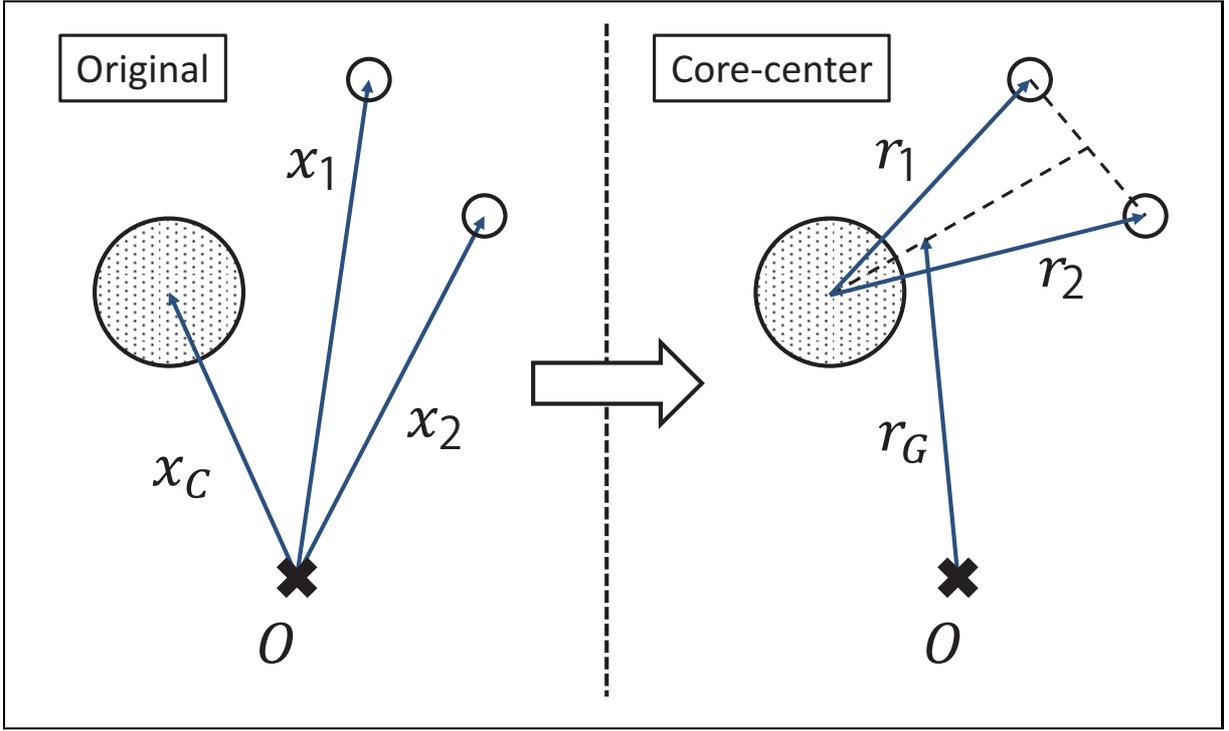}}
\caption{(left panel) The original coordinates for the three-body system. 
(right panel) The core-center coordinates. }
\label{fig:A1_1}
\end{center} \end{figure}

We briefly introduce another famous coordinates, 
namely ``Jacobi coordinates''. 
Those are defined by 
\beq
U_{J} \equiv \left( \begin{array}{ccccc cc} 
         1 & -1 & 0 & 0                                                        & \cdots & 0 & 0 \\ &&&&&& \\
         \frac{m_1}{m_{12}}  & \frac{m_2}{m_{12}}  & -1 & 0                    & \cdots & 0 & 0 \\ &&&&&& \\
         \frac{m_1}{m_{123}} & \frac{m_2}{m_{123}} & \frac{m_{23}}{m_{123}} & -1 & \cdots & 0 & 0 \\
         \vdots &&&&&& \\
         \frac{m_1}{m_{12\cdots (N-1)}} & \frac{m_2}{m_{12\cdots (N-1)}} & \cdots & \cdots & \cdots & \frac{m_{N-1}}{m_{12\cdots (N-1)}} & -1 \\
         &&&&&& \\
         \frac{m_1}{M} & \frac{m_2}{M} & \cdots & \cdots & \cdots & \frac{m_{N-1}}{M} & \frac{m_N}{M} 
         \end{array} \right), 
\eeq
and its inverse matrix is given by 
\beq
U_{J}^{-1} \equiv \left( \begin{array}{ccccc c} 
              \frac{m_2}{m_{12}} &     \frac{m_3}{m_{123}} & \frac{m_4}{m_{1234}} & \cdots & \frac{m_N}{M} & 1 \\ &&&&& \\
             -\frac{m_1}{m_{12}} &     \frac{m_3}{m_{123}} & \frac{m_4}{m_{1234}} & \cdots & \frac{m_N}{M} & 1 \\ &&&&& \\
             0                   & -\frac{m_{12}}{m_{123}} & \frac{m_4}{m_{1234}} & \cdots & \frac{m_N}{M} & 1 \\ &&&&& \\
             0 & 0 & -\frac{m_{123}}{m_{1234}} & \cdots & \frac{m_N}{M} & 1 \\
             \vdots &&&&& \\
             0 & 0 & 0 & \cdots & -\frac{m_{12\cdots (N-1)}}{M} & 1 
             \end{array} \right), 
\eeq
with $m_{12\; \cdots N'} \equiv \sum_{i=1}^{N'} m_i$. 
This set of coordinates has been used as well as the core-center coordinates 
for nuclear few-body models \cite{98Suzuki}. 
In this thesis, we use these coordinates only for 
the two relative momenta in the three-body system, 
in order to calculate $\Braket{h_{\rm c-NN}}$ and 
$\Braket{h_{\rm N-N}}$ in Chapters \ref{Ch_Results1} and \ref{Ch_Results2}. 

\section{Hamiltonian of Three-Body System}
In this thesis, we study the quantum three-body systems consisting of 
the core-nucleus and the two valence nucleons. 
We approximately use the same mass for a proton and a neutron. 
Thus, the masses of the core and a valence nucleon are $m_C=A_C m$ and 
$m_1=m_2=m$, respectively, where $A_C$ is the mass-number of the core. 
The total Hamiltonian written in the original coordinates, $\{ \bi{x}_i \}$, 
is given as 
\beqa
H \nonumber 
 &=& \frac{\bi{\pi}_1^2}{2m} + \frac{\bi{\pi}_2^2}{2m} + \frac{\bi{\pi}_C^2}{2A_C m} \\
 & & + V_{C-N_1}(\bi{x}_1 - \bi{x}_C) + V_{C-N_2}(\bi{x}_2 - \bi{x}_C) 
     + v_{N_1-N_2}(\bi{x}_1 - \bi{x}_2), 
\eeqa
where we assigned the third coordinate, ${\bi{x}_3}$, to the core-nucleus. 

To get the Hamiltonian in the core-center, or sometimes called 
``V-coordinates'' for three-body systems, 
we consider the transform-matrix, $U$, defined as 
\beq
U = \left( \begin{array}{ccc} 
    1 & 0 & -1 \\
    0 & 1 & -1 \\
    \frac{m}{M} & \frac{m}{M} & \frac{A_C m}{M} 
    \end{array} \right) 
\quad \Longleftrightarrow \quad 
U^{-1} = \left( \begin{array}{ccc} 
         1-\frac{m}{M} & -\frac{m}{M} & 1 \\
         -\frac{m}{M} & 1-\frac{m}{M} & 1 \\
         -\frac{m}{M}  & -\frac{m}{M} & 1 
         \end{array} \right), 
\eeq
where $M = m_1+m_2+m_C = (A_C+2)m$. 
The transformed coordinates are written as 
\beq
  \left[ \begin{array}{c}
  \bir_1 \\ \bir_2 \\ \bir_3 
  \end{array} \right] 
= U 
  \left[ \begin{array}{c}
  \bi{x}_1 \\ \bi{x}_2 \\ \bi{x}_C 
  \end{array} \right] 
= \left[ \begin{array}{c}
  \bi{x}_1 - \bi{x}_C \\
  \bi{x}_2 - \bi{x}_C \\
  \frac{m}{M}\bi{x}_1 + \frac{m}{M}\bi{x}_2 + \frac{A_C m}{M}\bi{x}_C 
  \end{array} \right], 
\eeq
where $\bir_3$ corresponds to the center-of-mass, $\bir_G$. 
Notice also that $\bi{x}_1-\bi{x}_2 = \bir_1-\bir_2$ and thus 
we can simply replace 
$v_{N_1-N_2}(\bi{x}_1-\bi{x}_2) \longrightarrow v_{N_1-N_2}(\bir_1-\bir_2) $. 
On the other hand, the conjugate momenta are transformed as 
\beq
\left[ \begin{array}{c}
\bi{\pi}_1 \\
\bi{\pi}_2 \\
\bi{\pi}_C 
\end{array} \right] 
= {}^{t}U 
\left[ \begin{array}{c}
\bip_1 \\
\bip_2 \\
\bip_G 
\end{array} \right] 
= 
\left[ \begin{array}{c}
\bip_1 + \frac{m}{M} \bip_G \\
\bip_2 + \frac{m}{M} \bip_G \\
- \bip_1 - \bip_2 + \frac{A_C m}{M} \bip_G 
\end{array} \right]. \label{equ_A1_pi1} 
\eeq
From Eq.(\ref{equ_A1_pi1}), the kinetic terms can be 
re-written as 
\beqa
&& \frac{\bi{\pi}_1^2}{2m} + \frac{\bi{\pi}_2^2}{2m} + \frac{\bi{\pi}_C^2}{2A_C m} \\
&& \qquad = \nonumber 
\frac{1}{2} \left( \frac{1}{m} + \frac{1}{A_C m} \right) \bip_1^2 + 
\frac{1}{2} \left( \frac{1}{m} + \frac{1}{A_C m} \right) \bip_2^2 + 
\frac{1}{A_C m} \bip_1 \cdot \bip_2 + \frac{1}{2M} \bip_G^2 \\
&& \qquad = 
\frac{\bip_1^2}{2\mu} + 
\frac{\bip_2^2}{2\mu} + 
\frac{\bip_1 \cdot \bip_2}{A_C m} + \frac{\bip_G^2}{2M} 
\eeqa
where $\mu = m (A_C+1)/A_C$. 
As the final result, the total Hamiltonian takes the form below. 
\beqa
H &=& \nonumber 
\frac{\bip_1^2}{2\mu} + \frac{\bip_2^2}{2\mu} + 
\frac{\bip_1 \cdot \bip_2}{A_C m} + \frac{\bip_G^2}{2M} \\
& & + V_{C-N_1}(\bir_1) + V_{C-N_2}(\bir_2) + v_{N_1 N_2}(\bir_1-\bir_2). 
\label{equ_A1_H2} 
\eeqa
Notice that in Eq.(\ref{equ_A1_H2}), the center-of-mass motion is separated 
from the three-body relative motion. 
Assuming that $\bip_G = \bi{0}$, 
the three-body Hamiltonian, $H_{\rm 3b}$, in Chapter \ref{Ch_3body} 
is correctly derived. 

\include{end}
\documentclass[a4paper,12pt]{report}
\include{begin}

\chapter{Scattering Problem with Contact Potential} \label{Ap_Scat_Contact}
In this Chapter, we discuss the nucleon-nucleon scattering 
problem with the phenomenological contact potential, 
$V(\bir)=V_0 \delta(\bir)$. 
The contact potential can provide physical meanings only within the 
truncated space defined by the energy cutoff, $E_{\rm cut}$. 
With an arbitrary $E_{\rm cut}$ value, one can determine $V_0$ to reproduce 
the scattering character, namely the scattering length or the phase shift 
at the lower energy limit. 
The \Schr equation of this scattering problem is written as 
\beq
\left[ -\frac{\hbar^2}{2\mu} \nabla^2_{\bir} + V_0\delta(\bir) \right] \phi (\bir) 
= E \phi (\bir), 
\eeq
with the incident energy $E$. 
Here $\mu=m/2$ is the relative mass for the 
two-nucleon system where $m$ is the one-nucleon mass.

We define the relative momentum, $k \equiv \frac{\sqrt{2\mu E}}{\hbar}$, and 
the converted potential, $v(\bir) \equiv \frac{2\mu}{\hbar^2} V_0 \delta(\bir) = v_0 \delta(\bir)$. 
Using these symbols, we can modify the \Schr equation as below. 
\beq
 \left[ \nabla^2_{\bir} + k^2 \right] \phi (\bir,k) = v(\bir) \phi (\bir,k). 
\eeq
The outgoing solution of this equation is formally represented with a Green's 
function \cite{03Fetter}: 
\begin{eqnarray}
  G^{(+)}(\bir,\bir',k) &\equiv & \label{equ_A2_G1} 
  \lim_{\eta \rightarrow 0} \int \frac{d^3\bip}{(2\pi)^3} 
  \frac{e^{i \bip \cdot (\bir-\bir')}}{p^2-k^2-i\eta} \\
  &=& \label{equ_A2_G2} \frac{1}{4\pi} \frac{e^{ik |\bir - \bir'|}}{|\bir - \bir'|}, 
\end{eqnarray}
which satisfies 
\beqa
 \left[ \nabla^2_{\bir} + k^2 \right] G^{(+)} (\bir,\bir',k) 
   &=& -\int \frac{d^3\bip}{(2\pi)^3} e^{i \bip \cdot (\bir-\bir')} \\
   &=& -\delta({\bf r}-{\bf r}'). 
\eeqa
The scattered wave function within the outgoing boundary condition, 
$\phi^{(+)}$, is formulated as 
\beq
 \label{equ_A2_phi1} \phi^{(+)}(\bir,k) = \phi^{(+)}_0(\bir,k) 
 - \int d^3\bir' G^{(+)} (\bir,\bir',k) v(\bir') \phi^{(+)}(\bir',k), 
\eeq
where $\phi^{(+)}_0(\bir,k)$ indicates the outgoing plane-wave. 
Substituting $v(\bir') = v_0 \delta(\bir')$, we can solve the 
$\phi^{(+)}(\bir',k)$ as 
\beqa
  \phi^{(+)}(\bir,k) 
  &=& \phi^{(+)}_0(\bir,k) - \int d^3\bir' G^{(+)} (\bir,\bir',k) v_0 \delta(\bir') 
      \phi^{(+)}(\bir',k), \\
  &=& \phi^{(+)}_0(\bir,k) - G^{(+)} (\bir,\bi{0},k) 
      v_0 \phi^{(+)}(\bi{0},k), \\
  &=& \phi^{(+)}_0(\bir,k) - \frac{v_0}{4\pi} 
      \phi^{(+)}(\bi{0},k) \frac{e^{ikr}}{r}. 
\eeqa
Here we used Eq.(\ref{equ_A2_G2}) to get the last formula. 
Now we can derive the well-known formula for the scattered wave, that is 
\beq
 \phi^{(+)}(\bir,k) = \phi^{(+)}_0(\bir,k) + f(k) \frac{e^{ikr}}{r}, 
\eeq
by defining the ``form factor'', $f(k)$, as 
\beq
 f(k) = - \frac{v_0}{4\pi} \phi^{(+)}(\bi{0},k). \label{eq:A2_f1} 
\eeq
It means that the scattered wave was disrupted only at $\bir=\bi{0}$, 
consistently to the infinitesimal range of the contact interaction. 

On the other hand, substituting Eq.(\ref{equ_A2_G1}) into 
Eq.(\ref{equ_A2_phi1}), we can derive an alternative formula for the scattered wave: 
\beq
  \phi^{(+)}(\bir,k) = \phi^{(+)}_0(\bir,k) - v_0 \phi^{(+)}(\bi{0},k) 
  \lim_{\eta \rightarrow 0} \int \frac{d^3\bip}{(2\pi)^3} 
  \frac{e^{i \bip \cdot \bir}}{p^2-k^2-i\eta}. 
\eeq
At $\bir = \bi{0}$, assuming the energy cutoff, 
$E_{C} \longleftrightarrow k_C$, we can apparently solve this equation. 
That is 
\beqa
 \phi^{(+)}(\bi{0},k) &=& \phi^{(+)}_0(\bi{0},k) - v_0 \phi^{(+)}(\bi{0},k) 
 \frac{1}{2\pi^2} \int_0^{k_C} dp \frac{p^2}{p^2-k^2} \\
 &=& 1 - \frac{v_0}{2\pi^2} \phi^{(+)}(\bi{0},k) 
 \left[ k_C+\frac{k}{2} \ln \left| \frac{k_C-k}{k_C+k} \right| \right]. 
\eeqa
Thus we get 
\beq
 \phi^{(+)}(\bi{0},k) = \left( 1 + \frac{v_0}{2\pi^2} 
 \left[ k_C+\frac{k}{2} \ln \left| \frac{k_C-k}{k_C+k} \right| \right] 
 \right)^{-1} \label{equ_A2_phi2}, 
\eeq
which is the complementary equation to Eq.(\ref{eq:A2_f1}). 

\section{Low Energy Limit}
In the following, we consider the s-wave at the 
low energy limit ($k \rightarrow 0$). 
As well known, the form factor can be written as 
\beq
 f_s(k) = \frac{1}{k} e^{i\delta_s} \sin \delta_s 
\eeq
where $\delta_s $ is the phase shift. 
It leads to an asymptotic formula below: 
\beqa
 |f_s(k)|^2 &=& \frac{\sin^2 \delta_s}{k^2} = \frac{1}{k^2 + k^2\cot^2 \delta_s} \\
 \Longrightarrow k\cot \delta_s &=& \left( \frac{1}{|f_s(k)|^2 }-k^2 \right)^{1/2} 
 \simeq \frac{1}{|f_s(k)|} \left( 1 - \frac{k^2|f_s(k)|^2}{2} \right). \label{equ_A2_kcotd1} 
\eeqa
From Eqs.(\ref{eq:A2_f1}), (\ref{equ_A2_phi2}) and (\ref{equ_A2_kcotd1}), 
the phase shift of the s-wave can be approximated as 
\beq
  k\cot \delta_s \simeq -\frac{4\pi}{v_0} 
  \left( 1 + \frac{v_0}{2\pi^2} \left[ k_C+\frac{k}{2} \ln \left| \frac{k_C-k}{k_C+k} \right| \right] 
  \right). \label{equ_A2_kcotd3} 
\eeq
Note that the logarithmic term is expanded as a polynomial of $k$: 
\beq
 \ln \left| \frac{k-k_C}{k+k_C} \right| \simeq 
 -2\frac{k}{k_C} + \mathcal{O}\left( \left( \frac{k}{k_C} \right)^3 \right), 
\eeq
where there are no terms on the order of $k^0$. 
On the other hand, 
there is an well-known empirical formula for $k\cot \delta_s$ at $k \rightarrow 0$, 
such as 
\beq
 k\cot \delta_s \simeq - \frac{1}{a_{\rm nn}} + \frac{r_{\rm nn}}{2} k^2. \label{equ_A2_kcotd2} 
\eeq
Here $a_{\rm nn}$ is the nucleon-nucleon scattering length whose empirical 
value is $-18.5$ fm, whereas the $r_{\rm nn}$ is the effective range. 
Comparing the leading terms in Eqs. (\ref{equ_A2_kcotd3}) and 
(\ref{equ_A2_kcotd2}), the strength $V_0$ can be defined within 
the energy-cutoff $k_C$ to reproduce the scattering length. 
That is 
\beqa
 -\frac{4\pi}{v_0} \left(1 + \frac{v_0 k_C}{2\pi^2} \right) 
 &=& -\frac{1}{a_{\rm nn}} \nonumber \\
 \Longrightarrow v_0 
 &=& 4\pi \left( \frac{1}{a_{\rm nn}} - \frac{2}{\pi} k_C \right)^{-1} 
     = 4\pi \left( \frac{\pi a_{\rm nn}}{\pi -2 a_{\rm nn} k_C} \right). 
\eeqa
Remembering $v_0/2 = \mu V_0/\hbar^2$, consequently we get the 
fitting formula for the contact potential: 
\beqa
  V_0 &=& \label{equ_A2_V03} \frac{\hbar^2}{2\mu} 
          \left( \frac{4\pi^2 a_{\rm nn}}{\pi -2 a_{\rm nn} k_C} \right). 
\eeqa
It should be noted that Eq. (\ref{equ_A2_V03}) is valid at $k \rightarrow 0$ limit. 
However, in practical cases, the parameter defined with 
$a_{\rm nn} = -18.5$ fm may be too strong especially for the valence nucleons. 
Thus the lower value, {\it e.g.} $a_{\rm nn}=15.0$ fm, is also used to 
prepare the appropriate paring attraction as often as the original value. 
In this thesis, we confirmed that our conclusions do not change even if 
we employ $a_{\rm nn} = 15.0$ fm instead of $a_{\rm nn} = 18.5$ fm. 

\include{end}
\documentclass[a4paper,12pt]{report}
\include{begin}

\chapter{Two-Body Scattering with Spherical Potential} \label{Ap_Scat_2body}
Our goal in this Appendix is to derive the fitting formula for the phase-shift 
of two-body scattering problems. 
For simplicity, we assume that the potential between 
two particles is spherical. 
For quantum resonances in two-body systems, one can usually 
solve the asymptotic waves analytically. 
The phase shift and its derivative can be computed by using these asymptotic 
waves, where it indicates the pole(s) of the S-matrix for the resonance. 
Even if one is interested in the scattering problem with three 
or more particles, 
it is often necessary to solve the partial two-body systems in order to, 
{\it e.g.} prepare the fine two-body interactions. 

\section{Solutions in Asymptotic Region}
Assuming the relative wave function as 
$\phi_{ljm}(\bir,\bis) = R_{lj}(r) \mathcal{Y}_{ljm}(\ubir,\bis)$, 
the radial equation of this problem reads 
\beq
  \left[ -\frac{\hbar^2}{2\mu}\left\{ \frac{d^2}{dr^2} - \frac{l(l+1)}{r^2} \right\} + V_{lj}(r) - E \right] U_{lj} (r,E) = 0, 
\eeq
where we defined $U_{lj}(r,E) \equiv rR_{lj}(r)$ from the radial wave function. 
The relative energy, $E$, for the scattering problem satisfies 
\beq
 E > \lim_{r \rightarrow \infty} V_{lj}(r) \equiv 0. 
\eeq
The equivalent but more convenient radial equation takes the form given by 
\beq
  \left[ \frac{d^2}{d\rho^2} - \frac{l(l+1)}{\rho^2} - \frac{V_{lj}(r)}{E} + 1 \right] 
  U_{lj} (\rho) = 0, \label{eq:apC01}
\eeq
where $\rho \equiv kr$ defined with the relative momentum, $k(E) \equiv \sqrt{2E\mu}/\hbar$. 
In numerical calculations, this type of equations can be solved with, 
{\it e.g.} Numerov method explained in Chapter \ref{Ch_3body}. 

To calculate the phase-shift and also other important quantities, 
asymptotic solutions of Eq.(\ref{eq:apC01}) are often necessary. 
In the following, we note these solutions for two major potentials 
frequently used in nuclear physics. 

\subsection{Short-Range Potential}
Short-range potentials, including nuclear interactions, are characterized as 
\beq
 \lim_{r \rightarrow \infty} V_{lj}(r) < \mathcal{O} (r^{-2}). 
\eeq
The asymptotic condition can be satisfied at $\rho \gg 1$. 
A general solution in this region can be written as 
\beq
 \frac{U_{lj}(\rho)}{\rho} = C_1 j_l(\rho) + C_2 n_l(\rho), 
\eeq
with spherical Bessel and Neumann functions, such as 
\beqa
 j_l (kr) &\longrightarrow & \frac{1}{kr}  \sin \left(kr-l\frac{\pi}{2} \right), \\
 n_l (kr) &\longrightarrow & \frac{-1}{kr} \cos \left(kr-l\frac{\pi}{2} \right). 
\eeqa
Or equivalently, the out-going and in-coming waves can be given as 
\beqa
 h^{(+)}_l (kr) &\equiv & j_l(kr) + i n_l(kr)
    \longrightarrow  \frac{1}{ikr}   e^{ i\left( kr-l\frac{\pi}{2} \right)}, \\
 h^{(-)}_l (kr) &\equiv & j_l(kr) - i n_l(kr)
    \longrightarrow  \frac{-1}{ikr}  e^{-i\left( kr-l\frac{\pi}{2} \right)}. 
\eeqa
Using the coefficients $A_{lj}$ and $B_{lj}$, a general solution takes the form of 
\beqa
 \frac{U_{lj}(kr)}{kr} &=& A_{lj}(E)h^{(+)}_l (kr) + B_{lj}(E)h^{(-)}_l (kr) \nonumber \\
 &=& B_{lj}(E) [S_{lj}(E)h^{(+)}_l (kr) + h^{(-)}_l (kr) ], 
\eeqa
with the S-matrix, $S_{lj}(E) \equiv A_{lj}(E)/B_{lj}(E)$. 
Note that $\abs{S_{lj}(E)}^2 =1$ from the conservation law of the flux. 
Introducing the phase-shift, $\delta_{lj}(E)$ as $S_{lj}(E) \equiv e^{2i\delta_{lj}(E)}$, 
we can get the well-known asymptotic form of $U_{lj}$. 
\beqa
  \frac{U_{lj}(kr)}{kr} &\longrightarrow& \frac{B_{lj}(E)}{ikr} \nonumber 
  \left[ S_{lj}(E) e^{ i\left( kr-l\frac{\pi}{2} \right)} - e^{-i\left( kr-l\frac{\pi}{2} \right)} \right] \\
  && = \frac{B_{lj}(E)e^{i\delta_{lj}(E)}}{ikr} \nonumber 
  \left[ e^{ i\left( kr-l\frac{\pi}{2}+\delta_{lj}(E) \right)} - e^{-i\left( kr-l\frac{\pi}{2}+\delta_{lj}(E) \right)} \right] \\
  && \propto \frac{1}{kr} \sin \left[ kr-l\frac{\pi}{2}+\delta_{lj}(E) \right]. \label{eq:apC05}
\eeqa
Note that $\delta_{lj}(E) \in \mathbb{R}$ since $\abs{S_{lj}(E)}^2 =1$.

\subsection{Coulomb Potential}
It is formulated as 
\beq
 V_{lj}(r) = V(r) = \alpha \hbar c \frac{Z_1 Z_2}{r}, \phantom{00} 
 \alpha \equiv \frac{e^2}{4\pi \epsilon_0 \cdot \hbar c}. 
\eeq
Defining Sommerfeld parameter, $\eta\equiv Z_1 Z_2 \alpha \mu c/\hbar k$, Eq.(\ref{eq:apC01}) 
can be written as 
\beq
  \left[ \frac{d^2}{d\rho^2} - \frac{l(l+1)}{\rho^2} - \frac{2\eta}{\rho} + 1 \right] U_{l} (\rho,\eta) = 0. 
\eeq
With this Coulomb potential, the asymptotic condition can be satisfied at $\rho \gg 2\eta$. 
A general solution takes the form as 
\beq
 \frac{U_{l}(\rho,\eta)}{\rho} = C_1\frac{F_l(\rho,\eta)}{\rho} + C_2\frac{G_l(\rho,\eta)}{\rho}, 
\eeq
where $F_l$ and $G_l$ are the Coulomb functions \cite{72Abramo}. 
Precise derivations of these functions are found in, {\it e.g.} textbook \cite{07Sasakawa}. 
Their asymptotic forms read 
\beqa
 \frac{1}{kr}F_l(kr,\eta) &\longrightarrow & \frac{1}{kr} \sin \left(kr-l\frac{\pi}{2}-\eta\ln 2kr+a_l(\eta) \right), \\
 \frac{1}{kr}G_l(kr,\eta) &\longrightarrow & \frac{1}{kr} \cos \left(kr-l\frac{\pi}{2}-\eta\ln 2kr+a_l(\eta) \right), 
\eeqa
with $a_l(\eta)=\arg \Gamma(l+1+i\eta)$, which is independent of $kr$. 
There is also an iterative formula for $a_l(\eta)$ as 
\beq
 a_{l+1}(\eta) = a_{l}(\eta) + \tan^{-1} \frac{\eta}{l+1}. 
\eeq
Eliminating these unimportant phases, the outgoing and incoming waves 
can be formulated as \cite{07Sasakawa}, 
\beqa
 u^{(+)}_l (kr,\eta) &\equiv & e^{-ia_l(\eta)} \left[G_l(kr,\eta) + i F_l(kr,\eta)\right]
    \longrightarrow e^{ i\left( kr-l\frac{\pi}{2}-\eta\ln 2kr \right)}, \\
 u^{(-)}_l (kr,\eta) &\equiv & e^{ ia_l(\eta)} \left[G_l(kr,\eta) - i F_l(kr,\eta)\right]
    \longrightarrow e^{-i\left( kr-l\frac{\pi}{2}-\eta\ln 2kr \right)}. 
\eeqa
By using these functions, a general solution can be replaced to 
\beqa
 U_{lj}(\rho,\eta) &=& A_{lj}(E,\eta) u^{(+)}_l (kr,\eta) + B_{lj}(E,\eta) u^{(-)}_l (kr,\eta) \\
 &\propto& \left[ S_{lj}(E,\eta) u^{(+)}_l (kr,\eta) + u^{(-)}_l (kr,\eta) \right], 
\eeqa
where we need an additional variable, $\eta$, in two coefficients. 
The S-matrix, $S_{lj}(E,\eta)$, and the phase-shift, $\delta_{lj}(E,\eta)$, can be defined 
similarly in the case with short-range potentials. 
The asymptotic solution is also given as 
\beq
 U_{lj}(\rho,\eta) \longrightarrow \propto 
 \sin \left[ \rho-l\frac{\pi}{2}-\eta \ln 2\rho + \delta_{lj}(E,\eta) \right]. \label{eq:apC06}
\eeq
In the following, however, we will not use Eqs.(\ref{eq:apC05}) and (\ref{eq:apC06}), 
although those are useful for analytic discussions. 

\section{Fitting Formula for Phase-Shift}
We explain how to compute the S-matrix within the numerical framework. 
First, we consider the position $r=R_b$ at which two particles 
can be separated sufficiently from each other. 
The radial mesh, $dr$, should be enough small compared with $R_b$. 
At this point, we assess the quantity $q$ defined as 
\beq
 q(X) \equiv \frac{U_{lj}(X)}{U_{lj}(X+d)} \label{eq:apC11}
\eeq
with $X\equiv k\cdot R_b$ and $d \equiv k\cdot dr$. 
Remember that the perturbed wave, $U_{lj}(X)$, is computed numerically. 
On the other hand, in the case with Coulomb potential for instance, 
$q(X)$ is also evaluated as 
\beq
 q(X) = \label{eq:apC12}
 \frac{S_{lj}(E,\eta)u^{(+)}_l(X,\eta)+u^{(-)}_l(X,\eta)}{S_{lj}(E,\eta)u^{(+)}_l(X+d,\eta)+u^{(-)}_l(X+d,\eta)}, 
\eeq
where $u_l^{(+)}$ and $u_l^{(-)}$ can be computed independently of $U_{lj}$. 
By solving Eq.(\ref{eq:apC11}) and Eq.(\ref{eq:apC12}) simultaneously for $S_{lj}(E,\eta)$, 
we can get 
\beq
 S_{lj}(E,\eta) = 
 \frac{ U_{lj}(X+d)u_l^{(-)}(X,\eta) - U_{lj}(X)u_l^{(-)}(X+d,\eta) }
      { U_{lj}(X)u_l^{(+)}(X+d,\eta) - U_{lj}(X+d)u_l^{(+)}(X,\eta) }, 
\eeq
and $2i \delta_{lj}(E,\eta) = \ln S_{lj}(E,\eta)$. 
This is the numerical formula for the S-matrix and the phase-shift. 
Notice that the similar formula can be derived in the case with short-range potentials. 

Practically, it is well known that the phase-shift can be fitted by the 
Breit-Wigner distribution. 
That is 
\beq
 \delta_{lj}(E) = \tan^{-1} \left[ \frac{\Gamma_0/2}{E_0-E} \right] + C_{lj}(E), 
\eeq
or equivalently, 
\beq
 \frac{d\delta_{lj}(E)}{dE} = \frac{\Gamma_0/2}{\Gamma_0^2/4 + (E_0-E)^2} + \frac{dC_{lj}(E)}{dE}, \label{eq:apcps}
\eeq
where $C_{lj}(E)$ is a smooth back-ground. 
The central value, $E_0$, and width, $\Gamma_0$, correspond to the complex pole of the S-matrix, 
locating at $E=E_0-i\Gamma_0/2$. 
Accordingly, we have got the fitting formula, which is equivalent to Eq.(\ref{eq:sigde}) in 
Chapter \ref{Ch_Results1}. 

\include{end}
\documentclass[a4paper,12pt]{report}
\include{begin}

\chapter{General Formalism of Time-Dependent Method} \label{Ap_TDM}
We briefly introduce the basic formalism of the time-dependent framework for 
quantum meta-stable phenomena in this Chapter. 
We mainly assume the (multi-)particle emissions. 
However, almost all formulas in the following can be generally applied to 
various kinds of quantum meta-stable phenomena, which has been described 
within several structure or reaction models. 

\section{Continuum Expansion}
First we assume the eigen-states of the total Hamiltonian, 
$H$, which is responsible for the time-evolution\footnote{For simplicity, 
the total Hamiltonian is assumed to be static, 
and is not dependent on the wave function self-consistently. 
The similar time-dependent theory with non-static Hamiltonian 
can be considered. 
However, it is over complicated and beyond the coverage of this thesis.}. 
Considering the degeneration, they can be formulated as 
\beqa
 && H \ket{E, i(E)} = E \ket{E, i(E)}, \\
 && \Braket{E',j(E')|E,i(E)} = \delta(E'-E) \delta_{ji}. 
\eeqa
Here the eigen-energy $E$ is real so that we consider the pure 
Hermite space, 
in contrast to other theoretical methods which employ the space with 
complex eigen-energies, such as Berggren space. 
The $i(E)$ identifies one of the degenerating states with the same energy $E$. 
However, as a basic rule in the following, we omit these labels for simplicity. 
If it is necessary to recount the degeneration, 
we remember these labels only for some important formulas. 

Adopting these eigen-states as bases, 
an arbitrary meta-stable state, $\ket{\psi_0}$, can be expanded as 
\beq
  \ket{\psi_0} = \int dE \mu (E) \ket{E}, 
\eeq
where $\left\{ \mu(E) \right\}$ are the expanding coefficients. 
The normalization is represented as 
\beq
  1 = \Braket{\psi_0|\psi_0} = \int dE \abs{\mu (E)}^2. 
\eeq
Physical properties of $\ket{\psi_0}$ are not clear at this moment. 
Those are characterized by the expanding coefficients. 
In the following, we discuss about physics described with $\ket{\psi_0}$.

\section{Time Evolution}
The quantum meta-stable phenomena, 
including particle(s)-decays and emissions, 
can be treated as the time-developments of meta-stable systems. 
Assuming $\ket{\psi_0}$ as the initial state, 
we consider the time-evolution via $H$. 
\beqa
 \ket{\psi (t)} 
 &=& e^{-itH/\hbar} \ket{\psi_0 } \\
 &=& \int dE \mu (E) e^{-itE/\hbar} \ket{E}. \label{eq:psi_t}
\eeqa
The expectational value of $H$, indicated as $E_0$, 
obviously conserves during the time-evolution. 
\beq
  E_0 \equiv \Braket{\psi_0|H|\psi_0} = \Braket{\psi(t)|H|\psi(t)} = \int dE E \abs{\mu (E)}^2. 
\eeq
This conservation coincide with that the energy-spectrum, 
defined by $\left\{ \abs{\mu(E)}^2 \right\}$, is invariant during 
the time-evolution. 
For a particle(s)-decay or emission, $E_0$ corresponds to the 
Q-value carried out by the emitted particle(s). 

The survival coefficient, $\beta(t)$, is defined as the overlap between 
the initial and the present states. 
\beqa
  \beta(t) &\equiv & \Braket{\psi_0 | \psi(t)} \\
  &=& \int dE' \mu (E') \int dE \mu (E) \Braket{E' | e^{-itE/\hbar} | E} \nonumber \\
  &=& \int dE \abs{\mu (E)}^2 e^{-itE/\hbar}. \label{eq:Krylov}
\eeqa
Note that $\beta(0) = 1$. 
In Eq.(\ref{eq:Krylov}), 
the survival coefficient can be given by the Fourier transformation of 
the invariant energy-spectrum. 
This is nothing but the ``Krylov-Fock theorem'' \cite{47Kry,89Kuku}. 
As one of the important observable properties, 
the survival probability can be given by 
\beq
  P_{\rm surv}(t) = \abs{\beta(t)}^2, 
\eeq
which leads to the decay-rule in this meta-stable process. 
In the next section, we discuss the correspondence between 
the actual decay-rule and the invariant energy-spectrum.

\section{Exponential Decay-Rule}
The exponential decay-rule has been popular especially in the radioactive processes. 
That is 
\beq
  P(t) = e^{-t/\tau} P(0), 
\eeq
where $P(t)$ means the probability of a radioactive nucleus to survive 
with its characteristic lifetime, $\tau$. 
As the first step to discuss the decay-rule, 
we proof that this exponential decay-rule is equivalent to 
the ideal Breit-Wigner (BW-) distribution in the energy-spectrum. 
The squared expanding coefficients, $\left\{ \abs{\mu (E)}^2 \right\}$, 
are assumed to have the BW-distribution, 
or equivalently, the form of Cauchy-Lorentz function whose center and 
full width at the half maximum (FWHM) are $E_0$ and $\Gamma_0$, respectively. 
That is 
\beq
  \abs{\mu (E)}^2 = \frac{1}{\pi} \frac{(\Gamma_0 /2)}{(E-E_0)^2 + (\Gamma_0 /2)^2} \label{eq:BW1}
\eeq
with $-\infty \leq E \leq \infty$. 
Or equivalently, 
\beq
  \ket{\psi_0} = \int_{-\infty}^{\infty} dE \sqrt{\frac{\Gamma_0}{2\pi}} 
  \frac{e^{ia(E)} }{(E_0 - i\Gamma_0/2) - E} \ket{E}, \label{eq:BW2}
\eeq
where $\left\{ e^{ia(E)} \right\}$ with $a(E) \in \mathbb{R}$ are 
arbitrary phase-factors. 
If we consider the degeneration, Eq.(\ref{eq:BW1}) is modified as 
\beq
  \abs{\mu (E)}^2 = \sum_{i(E)} \abs{\mu (E,i(E))}^2 
  = \frac{1}{\pi} \frac{(\Gamma_0 /2)}{(E-E_0)^2 + (\Gamma_0 /2)^2}. 
\eeq
The normalization is obviously given by 
\beq
 1 = \Braket{\psi (t) | \psi (t)} 
   = \int_{-\infty}^{+\infty} dE \frac{1}{\pi} \frac{(\Gamma_0 /2)}{(E-E_0)^2 + (\Gamma_0 /2)^2}. 
\eeq
For the ideal BW-distribution, however, 
how to define the expectational value of $H$ is not obvious. 
We should be careful for the range of the integration 
which is critical for the 1st moment of BW-distributions. 
At this moment, we assume the isotropic infinite range with 
the central value of $E_0$. 
\beq
  \int_I dE \equiv \lim_{R \rightarrow \infty} \int_{E_0-R}^{E_0+R} dE. 
\eeq
Thus, the 1st moment of the energy is identical to the Cauchy's 
principal value, namely the center of the distribution. 
\beqa
 \Braket{\psi_0 | H | \psi_0} 
 &=& \Braket{\psi (t) | H | \psi (t)} \\
 &=& \int_I dE' \mu (E') \int_I dE \mu (E) 
     \Braket{E' | H | E} \nonumber \\
 &=& \int_I dE' \mu (E') \int_I dE \mu (E) 
     \delta(E'-E) E \nonumber \\
 &=& \int_I dE \abs{\mu (E)}^2 E = E_0, 
\eeqa
In the following, we omit the subscript $I$. 
Substituting Eq.(\ref{eq:BW1}) into Eq.(\ref{eq:Krylov}), 
the survival coefficient can be derived by picking up the residue at 
the pole of $E = E_0 -i\Gamma_0 /2$, namely 
\beqa
 \beta(t) &=& \frac{1}{\pi} \int dE \frac{(\Gamma_0 /2)}{(E-E_0)^2 + (\Gamma_0 /2)^2} e^{-itE/\hbar} = \cdots \nonumber \\
 &=& e^{-it(E_0 -i\Gamma_0/2)/\hbar}. \label{eq:xpdr}
\eeqa
Then the survival probability yields the well-known exponential 
decay-rule, such that 
\beq
 P_{\rm surv}(t) = \abs{\beta(t)}^2 = e^{-t/\tau}, 
\eeq
where the $\tau = \hbar/\Gamma_0$ is the lifetime of this 
meta-stable state \cite{89Bohm, 09Konishi}.

\section{Practical Problems}
In practice, however, the situation is not so simple. 
First of all, there is the lower limit for the expansion on the energy space, 
consistently to the threshold of the emission. 
Fixing it as $E=0$, we should modify Eq.(\ref{eq:BW2}) as 
\beq
 \int_{-\infty}^{\infty} dE \longrightarrow 
 \int_{0}^{\infty} dE. 
\eeq
Second, the actual energy spectra are not limited 
to have the perfect BW-distributions. 
This discordance leads to the deviation from 
the exponential decay-rule \cite{12Pons}. 
Especially, if the decay width is comparably broad to 
the Q-value: $E_0 \approx \Gamma_0$, 
assuming the BW-distribution may diverge from reality. 

For the numerical calculations, 
we intuitively have to concern two additional affairs. 
The first is the discretization of the continuum space, and 
the second is the energy cutoff, $E_{\rm cut}$. 
Thus, Eq.(\ref{eq:psi_t}) should be modified as 
\beq
 \ket{\psi(t)} = \sum_{N} F_N (0) e^{-itE_N/\hbar} \ket{E_N}, 
\eeq
where $E_N \leq E_{\rm cut}$. 

Finally, we mention the effect of the initial configuration (IC). 
One cannot discuss the meta-stable process without concerning how the initial state 
should be defined. 
The initial state, especially of the particle(s)-emission, 
is usually characterized as, for instance, 
the state where the emitted particles are confined in the narrow region, and/or 
the state which obeys the outgoing boundary condition. 
However, even with these constraints, 
there may be different ICs which follow almost the same decay-rule. 
Possibly, obtained results after the time-evolution may significantly depend on 
the selection of the IC, even though the decay-rule itself hardly changes. 
In this thesis, we employed the phenomenological procedure 
with confining potentials to fix it. 
The more realistic way to fix the IC is, of course, considerable. 
Discussing this effect is, however, beyond the scope of this thesis.

\include{end}

\end{document}